# Condensed-matter analogs of the Sauter–Schwinger effect

Von der
**Fakultät für Physik**
der
**Universität Duisburg-Essen**
genehmigte

## Dissertation

zur Erlangung des akademischen Grades
Dr. rer. nat.

von
**Dipl.-Phys. Malte Friedrich Linder**
aus Essen

## Danksagung

Mein besonderer Dank gilt Herrn Prof. Dr. Ralf Schützhold, der mich während meines Promotionsstudiums und beim Abfassen dieser Dissertation betreut hat und damit diese Promotion ermöglicht hat. Durch seine beständige und freundliche Bereitschaft zur Diskussion über die hier behandelten Themen konnte er mir bei Problemen stets zu neuer Inspiration verhelfen und hat daher maßgeblich zum Gelingen dieser Arbeit beigetragen.

Herrn Dr. Nikodem Szpak danke ich für viele erhellende Diskussionen im Zusammenhang mit der Pulsformabhängigkeit im dynamisch assistierten Sauter-Schwinger-Effekt.

Ebenso danke ich Herrn Prof. Dr. Axel Lorke für seine wertvollen Beiträge insbesondere zu den experimentellen Aspekten der in dieser Arbeit behandelten Analogie in Halbleitern.

Den weiteren Mitgliedern der „Sauter-Schwinger-Crew“, Herrn Johannes Oertel und Herrn Christian Schneider, sowie meinem ehemaligen Büronachbarn, Herrn Dr. Nicolai ten Brinke, danke ich für den inspirierenden und sehr hilfreichen Austausch über den in dieser Dissertation behandelten Themenkomplex und alle anderen Aspekte des Promotionsstudiums. Auch möchte ich allen anderen jetzigen und ehemaligen Mitgliedern der Arbeitsgruppe an dieser Stelle für die stets angenehme und von Hilfsbereitschaft geprägte Arbeitsatmosphäre und die gute Zusammenarbeit danken.

Mein Dank gilt außerdem meiner Familie für die Unterstützung während meines Studiums und der Promotionszeit.

# List of publications

Most of the results developed in this thesis were published in two peer-reviewed journal articles:

[1] M. F. Linder, C. Schneider, J. Sicking, N. Szpak, and R. Schützhold, "Pulse shape dependence in the dynamically assisted Sauter-Schwinger effect," Phys. Rev. D **92**, 085009 (2015), arXiv:1505.05685.

[2] M. F. Linder, A. Lorke, and R. Schützhold, "Analog Sauter-Schwinger effect in semiconductors for spacetime-dependent fields," Phys. Rev. B **97**, 035203 (2018), arXiv:1503.07108.

The following article, which is unrelated to the topics covered in this thesis, was also prepared during my doctoral studies:

[3] M. F. Linder, R. Schützhold, and W. G. Unruh, "Derivation of Hawking radiation in dispersive dielectric media," Phys. Rev. D **93**, 104010 (2016), arXiv:1511.03900.

# Abstract

The Sauter–Schwinger effect predicts the creation of electron–positron pairs from the vacuum due to a quasiconstant electric field $E_{\text{strong}}$. The pair-creation yield can be exponentially enhanced without destroying the tunneling-like nature of this mechanism by adding a weaker temporal Sauter pulse $E_{\text{weak}}/\cosh^2(\omega t)$ with $\omega$ above a certain threshold $\omega_{\text{crit}}$. In this original form of the so-called dynamically assisted Sauter–Schwinger effect, $\omega_{\text{crit}}$ is independent of $E_{\text{weak}} \ll E_{\text{strong}}$. Via the semiclassical solution (contour integral) of the Riccati equation in 1+1 spacetime dimensions, we find that a Gaussian-shaped pulse $E_{\text{weak}} \exp[-(\omega t)^2]$ assists tunneling in a similar way but with $\omega_{\text{crit}}$ depending on $E_{\text{weak}}$. This remarkable sensitivity to the pulse shape arises due to the different pole structures of the vector potentials for complex times. We also study dynamical assistance by an oscillation $E_{\text{weak}} \cos(\omega t)$ as a model for counterpropagating laser beams and find another dependence $\omega_{\text{crit}}(E_{\text{weak}})$.

The largeness of the Schwinger limit $E_{\text{crit}}^{\text{QED}} \approx 10^{18}\,\text{V/m}$ has rendered the observation of this nonperturbative pair-creation mechanism impossible so far. In order to facilitate a better understanding of this effect and its dynamical assistance via experiments, we propose an analog of the many-body Dirac Hamiltonian in direct-bandgap semiconductors. The nonrelativistic Bloch-electron Hamiltonian is restricted to the valence and conduction bands in reciprocal space, which correspond to the two relativistic energy continua. Similar models have been considered before—but mainly for constant external fields. Here, we present a detailed derivation of the analogy between the long-wavelength parts of both Hamiltonians for spacetime-dependent electric fields $E(t, x)$ in 1+1 dimensions. Based on this analogy, we propose experimental simulations of the above-mentioned pair-creation mechanisms in gallium arsenide (GaAs), for example Landau–Zener tunneling assisted by a carbon dioxide laser. The electron mass and the vacuum speed of light take on much smaller, effective values in the semiconductor analog, which drastically reduces the equivalent of the Schwinger limit ($E_{\text{crit}}^{\text{GaAs}} = 565\,\text{MV/m}$), thus simplifying such experiments. As an outlook, we calculate the exact two-band Bloch-electron Hamiltonian in 2+1 dimensions for perpendicular, constant electric and magnetic fields and show that the corresponding local dispersion relation for long wavelengths approximately coincides with the relativistic form.

# Zusammenfassung

Der Sauter-Schwinger-Effekt bezeichnet die Erzeugung von Elektron-Positron-Paaren aus dem Vakuum durch quasikonstante elektrische Felder $E_{\text{strong}}$. Die Paarausbeute kann unter Beibehaltung des zugrundeliegenden Tunnelmechanismus exponentiell gesteigert werden, indem ein schwächerer zeitlicher Sauter-Puls $E_{\text{weak}}/\cosh^2(\omega t)$ mit einem $\omega$ oberhalb eines bestimmten Schwellwerts $\omega_{\text{crit}}$ hinzugefügt wird. In dieser ursprünglichen Form des sogenannten dynamisch assistierten Sauter-Schwinger-Effekts hängt $\omega_{\text{crit}}$ nicht von $E_{\text{weak}} \ll E_{\text{strong}}$ ab. Wir zeigen mithilfe der semiklassischen Lösung (Konturintegral) der Riccati-Gleichung in 1+1 Raumzeitdimensionen, dass ein gaußscher Puls $E_{\text{weak}} \exp[-(\omega t)^2]$ Tunneln in ähnlicher Weise verstärken kann, wobei $\omega_{\text{crit}}$ hier von $E_{\text{weak}}$ abhängt. Der Grund für diese auffällige Pulsformabhängigkeit liegt in den unterschiedlichen Polstrukturen der Vektorpotentiale im Komplexen. Wir behandeln außerdem eine assistierende Oszillation $E_{\text{weak}} \cos(\omega t)$ als Model gegenläufiger Laserstrahlen und erhalten eine andere Abhängigkeit $\omega_{\text{crit}}(E_{\text{weak}})$.

Aufgrund des hohen Wertes des Schwinger-Limits $E_{\text{crit}}^{\text{QED}} \approx 10^{18}\,\text{V/m}$ ist es bisher nicht gelungen, diese nichtperturbative Paarerzeugung zu beobachten. Um durch Experimente ein besseres Verständnis dieses Effekts und seiner dynamischen Verstärkung zu ermöglichen, stellen wir ein Analogon des Dirac'schen Vielteilchen-Hamiltonians in direkten Halbleitern vor. Der nichtrelativistische Hamiltonian der Bloch-Elektronen wird im reziproken Raum auf Valenz- und Leitungsband beschränkt, welche den zwei relativistischen Energiekontinua entsprechen. Ähnliche Modelle wurden bereits im Falle konstanter externer Felder betrachtet. Wir leiten die Analogie zwischen den beiden Hamiltonians im Langwellenbereich hier für raumzeitabhängige elektrische Felder $E(t, x)$ in 1+1 Dimensionen her. Auf dieser Grundlage schlagen wir experimentelle Simulationen der oben genannten Paarerzeugungsmechanismen in Galliumarsenid (GaAs) vor, zum Beispiel durch einen Kohlenstoffdioxidlaser assistiertes Landau-Zener-Tunneln. Die Elektronenmasse und die Vakuumlichtgeschwindigkeit nehmen im Halbleiteranalogon kleinere, effektive Werte an. Dadurch wird das Äquivalent des Schwinger-Limits drastisch reduziert ($E_{\text{crit}}^{\text{GaAs}} = 565\,\text{MV/m}$) und somit Experimente deutlich vereinfacht. Als Ausblick berechnen wir den exakten Zweiband-Hamiltonian für Bloch-Elektronen in 2+1 Dimensionen im Falle zueinander senkrechter, konstanter elektrischer und magnetischer Felder. Wir zeigen, dass die zugehörige lokale Dispersionsrelation für große Wellenlängen in guter Näherung mit der relativistischen Form übereinstimmt.

# Notations and conventions

### Abbreviations

| | |
|---|---|
| BZ | (first) Brillouin zone |
| Ch(s). | chapter(s) |
| Eq(s). | equation(s) |
| Fig(s). | figure(s) |
| GaAs | gallium arsenide |
| JWKB | Jeffreys–Wentzel–Kramers–Brillouin (approximation) |
| p. | page |
| pp. | pages |
| QED | quantum electrodynamics |
| Ref(s). | reference(s) |
| Sec(s). | section(s) |

### Units

We use SI units throughout, except in Secs. 2.4.3–2.4.4 and in Part II, in which we employ natural units with $c = \hbar = 1$.

### Mathematical notation

| | |
|---|---|
| $\mathbb{1}$ | identity matrix or operator |
| $\boldsymbol{v}$ | vector in $\mathbb{R}^3$ |
| $\boldsymbol{r}$ | position vector $\boldsymbol{r} = x\boldsymbol{e}_x + y\boldsymbol{e}_y + z\boldsymbol{e}_z$ |
| $\vec{v}$ | More general vector in $\mathbb{R}^n$.<br>In Sec. 2.1: vector in Euclidean spacetime ($\mathbb{R}^4$).<br>In Ch. 9: vector in two-dimensional space ($\mathbb{R}^2$). |
| $\mathcal{C}$ | contour of integration in the complex plane |
| $z^*$ | complex conjugate of $z \in \mathbb{C}$ |
| $\arg z \in (-\pi, \pi]$ | argument (principal value) of a complex number $z$ |
| $\sqrt{z}$ | principal value $\sqrt{\lvert z\rvert}\mathrm{e}^{\mathrm{i}\arg(z)/2}$ of the complex square root |
| $M^\mathsf{T}$ | transpose of the matrix $M$ |
| $\hat{A}^\dagger$ | Hermitian conjugate of $\hat{A}$ |
| $\dot{f}(t) = \mathrm{d}f(t)/\mathrm{d}t$ | time derivative |
| $\{\hat{A}, \hat{B}\}$ | $= \hat{A}\hat{B} + \hat{B}\hat{A}$ (anticommutator of operators/matrices) |
| $I_0(x)$, $I_1(x)$ | modified Bessel functions of the first kind |

### Temporal Fourier transform

$$\tilde{f}(\omega) = \mathcal{F}_t[f(t)](\omega) = \frac{1}{\sqrt{2\pi}} \int_{-\infty}^{\infty} f(t)\mathrm{e}^{+\mathrm{i}\omega t}\,\mathrm{d}t$$

$$f(t) = \mathcal{F}_\omega^{-1}[\tilde{f}(\omega)](t) = \frac{1}{\sqrt{2\pi}} \int_{-\infty}^{\infty} \tilde{f}(\omega) \mathrm{e}^{-\mathrm{i}\omega t} \, \mathrm{d}\omega$$

**Spatial Fourier transform (one dimensional)**

$$\tilde{f}(k) = \mathcal{F}_x[f(x)](k) = \frac{1}{\sqrt{2\pi}} \int_{-\infty}^{\infty} f(x) \mathrm{e}^{-\mathrm{i}kx} \, \mathrm{d}x$$

$$f(x) = \mathcal{F}_k^{-1}[\tilde{f}(k)](x) = \frac{1}{\sqrt{2\pi}} \int_{-\infty}^{\infty} \tilde{f}(k) \mathrm{e}^{+\mathrm{i}kx} \, \mathrm{d}k$$

**Electric and magnetic fields and the corresponding potentials**

$$\boldsymbol{E}(t,\boldsymbol{r}) = \boldsymbol{\nabla}\Phi(t,\boldsymbol{r}) + \frac{\mathrm{d}\boldsymbol{A}(t,\boldsymbol{r})}{\mathrm{d}t} \quad \text{(electric field)}$$

$$\boldsymbol{B}(t,\boldsymbol{r}) = -\boldsymbol{\nabla} \times \boldsymbol{A}(t,\boldsymbol{r}) \quad \text{(magnetic field)}$$

**Fundamental physical quantities**

| | |
|---|---|
| $\varepsilon_0$ | vacuum permittivity |
| $\mu_0$ | vacuum permeability |
| $c = 1/\sqrt{\varepsilon_0\mu_0}$ | speed of light in vacuum |
| $h$ | Planck constant |
| $\hbar = h/(2\pi)$ | reduced Planck constant |
| $k_B$ | Boltzmann constant |
| $m$ | electron/positron rest mass |
| $q$ | absolute value of the electric charge of an electron or positron |
| $\lambda_C = h/(mc)$ | Compton wavelength of an electron or positron |
| $\bar{\lambda}_C = \hbar/(mc)$ | reduced Compton wavelength |
| $E_{\mathrm{crit}}^{\mathrm{QED}}$ | Schwinger limit [$E_{\mathrm{crit}}^{\mathrm{QED}} = m^2c^3/(\hbar q)$] |
| $B_{\mathrm{crit}}^{\mathrm{QED}}$ | critical magnetic field strength [$B_{\mathrm{crit}}^{\mathrm{QED}} = E_{\mathrm{crit}}^{\mathrm{QED}}/c$] |

**Four-vector notation**

| | |
|---|---|
| $\eta_{\mu\nu}$ | Minkowski metric [$\eta_{\mu\nu} = \mathrm{diag}(-1,1,1,1)$] |
| $x^\mu = (ct,\boldsymbol{r})$ | contravariant four-vector components |
| $x_\mu = (-ct,\boldsymbol{r})$ | covariant components |
| $A^\mu = (\Phi/c,\boldsymbol{A})$ | contravariant components of the four-potential ($\Phi$ and $\boldsymbol{A}$ generally depend on $t$ and $\boldsymbol{r}$) |
| $A_\mu = (-\Phi/c,\boldsymbol{A})$ | covariant components |
| $\partial_\mu = (\partial_t/c,\boldsymbol{\nabla})$ | four-gradient |

**Minimal coupling for electrons (with charge $-q$)**

| | |
|---|---|
| $\hat{\boldsymbol{p}} = -\mathrm{i}\hbar\boldsymbol{\nabla}$ | $\to \hat{\boldsymbol{p}} + q\boldsymbol{A}(t,\boldsymbol{r})$ (momentum operator) |
| $\mathrm{i}\hbar\partial_t$ | $\to \mathrm{i}\hbar\partial_t + q\Phi(t,\boldsymbol{r})$ (energy operator) |
| $\mathrm{i}\hbar\partial_\mu$ | $\to \mathrm{i}\hbar\partial_\mu - qA_\mu(t,\boldsymbol{r})$ (four-vector notation) |

**Other physical quantities**

| | |
|---|---|
| $c_\star$ | effective speed of light in the semiconductor analog |
| $\chi = E/E_{\text{crit}}^{\text{QED}}$ | dimensionless substitute for a constant electric field $E$ |
| $\mathcal{E}$ | energy |
| $\mathcal{E}_g$ | bandgap |
| $\varepsilon = E_{\text{weak}}/E_{\text{strong}}$ | dimensionless substitute for the amplitude of a weak time-dependent electric field which is added to a strong quasistatic field ("background field") |
| $f_n(\boldsymbol{K},\boldsymbol{r})$ | wave function of the Bloch state in the $n$th energy band with the quasi-wave vector $\boldsymbol{K}$ |
| $\varphi_k(t)$ | phase function appearing in the integrand's exponential part $\mathrm{e}^{2\mathrm{i}\varphi_k(t)}$ in the integral representation of $R_k^{\text{out}}$ |
| $\gamma_\omega$ | temporal Keldysh parameter $\gamma_\omega = mc\omega/(qE_{\text{max}})$ associated with an inverse timescale $\omega$ |
| $\gamma_k$ | spatial Keldysh parameter $\gamma_k = mc^2k/(qE_{\text{max}})$ associated with an inverse length scale $k$ |
| $\gamma_c$ | combined Keldysh parameter $\gamma_c = mc\omega/(qE_{\text{strong}})$ of a time-dependent ($\omega$) weak electric field plus a quasiconstant strong field ("background field") |
| $\hat{H}_D$ | Many-body Hamiltonian of the quantized Dirac field coupled to an external electromagnetic field |
| $\hat{H}_S^{\text{full}}$ | Many-body Hamiltonian of the quantized Schrödinger field (describing spinless Bloch electrons) coupled to an external electromagnetic field |
| $\hat{H}_S$ | $\hat{H}_S^{\text{full}}$ restricted to the valence band and the conduction band (two-band Hamiltonian) |
| $\boldsymbol{K}$ | quasi-wave vector of a Bloch electron in a periodic lattice (modulo $\hbar$: crystal momentum or quasimomentum) |
| $\kappa = \hbar k/(mc)$ | dimensionless substitute for the canonical wave vector $k$ (one spatial dimension) |
| $\varkappa$ | complex off-diagonal element of the momentum matrix (divided by $m$) in the Bloch-wave basis for a two-band semiconductor |
| $\varkappa_0 > 0$ | positive, real value (without loss of generality) of $\varkappa$ at the center of the Brillouin zone $\boldsymbol{K} = 0$ |
| $\ell$ | lattice constant |
| $m_\star$ | effective rest mass of an electron in the semiconductor analog |
| $m_{\star,e}$ | effective electron mass at the conduction-band minimum (band curvature) |
| $m_{\star,h}$ | effective light-hole mass at the valence-band maximum (band curvature) |
| $\mathcal{N}_{\mathrm{e}^+\mathrm{e}^-}$ | number of created electron–positron pairs per unit volume (density) |
| $\dot{\mathcal{N}}_{\mathrm{e}^+\mathrm{e}^-}$ | pair-creation rate per unit volume |
| $\dot{\mathcal{N}}_{\mathrm{e}^-\text{–hole}}$ | electron–hole pair-creation rate per unit volume in a semiconductor |
| $\Omega_k(t)$ | positive instantaneous relativistic energy of an electron with a canonical wave vector $k$ (one spatial dimension) coupled to the vector potential $A(t)$ |
| $P_{\mathrm{e}^+\mathrm{e}^-}$ | (total) electron–positron pair-creation probability |
| $P_k^{\mathrm{e}^+\mathrm{e}^-}$ | pair-creation probability for the mode given by the conserved canonical wave vector $k$ (one dimensional) |
| $\mathcal{P}_{\mathrm{e}^+\mathrm{e}^-}$ | pair-creation probability per unit four-volume |
| $\varpi = \hbar\omega/(2mc^2)$ | dimensionless substitute for the angular frequency $\omega$ |
| $R_k^{\mathrm{out}}$ | outgoing value ($t \to \infty$) of the solution $R_k(t)$ of the Riccati equation in 1+1 spacetime dimensions for the mode given by the canonical wave vector $k$ |
| $\mathcal{T} = \mathrm{i}t$ | imaginary-time coordinate in Euclidean spacetime |
| $\tau$ | dimensionless substitute for the time $t$ (definition depends on context) |
| $t^\star$, $\tau^\star$ | "temporal turning points" in the complex plane [singularities, most of them zeros of $\Omega_k(t)$] |
| $u_n(\boldsymbol{K}, \boldsymbol{r})$ | lattice-periodic Bloch factor for the state in the $n$th energy band with the quasi-wave vector $\boldsymbol{K}$ |
| $V$ | potential energy (in Part III: lattice-periodic crystal potential) |
| $v^{\mathrm{gr}}$ | group velocity |
| $\mathcal{V}$ | volume |
| $\mathcal{V}_{\mathrm{cell}}$ | volume of a unit cell in a crystal lattice |
| $x_\star$ | classical turning point (one spatial dimension) |
| $\Xi_k(t)$ | nonexponential part of the integrand in the integral representation of $R_k^{\mathrm{out}}$ |

**Bra–ket notation for Bloch states (Part III)**

$$\langle n, \boldsymbol{K} | \hat{A} | n', \boldsymbol{K'} \rangle = \iiint\limits_{\mathbb{R}^3} f_n^*(\boldsymbol{K}, \boldsymbol{r}) \hat{A} f_{n'}(\boldsymbol{K'}, \boldsymbol{r}) \, \mathrm{d}^3 r$$

$$\langle n, K | \hat{A} | n', K' \rangle_{\text{cell}} = \frac{2\pi}{\ell} \int\limits_0^\ell u_n^*(K, x) \hat{A} u_{n'}(K', x) \, \mathrm{d}x$$

(unit-cell scalar product in 1+1 spacetime dimensions;
$\ell$: lattice constant)

$$\langle n, \vec{K} | \hat{A} | n', \vec{K}' \rangle_{\text{cell}} = \frac{4\pi^2}{\mathcal{V}_{\text{cell}}} \iint\limits_{\text{cell}} u_n^*(\vec{K}, \vec{r}) \hat{A} u_{n'}(\vec{K}', \vec{r}) \, \mathrm{d}^2 r$$

(unit-cell scalar product in 2+1 spacetime dimensions;
$\mathcal{V}_{\text{cell}}$: volume of a unit cell)

# Contents

# Part I.

# Introduction and basics

# 1. Historical development

The early 20th century was a fruitful time for our understanding of physics. During this period, the fundamentals of the two pillars of modern physics were developed: the general theory of relativity and quantum theory. Broadly speaking, Einstein's general relativity describes the large-scale structure of the universe (e.g., the formation and movement of stars, planets, galaxies, etc.) while the laws of quantum theory dominate the physics of the smallest constituents of matter like elementary particles, atoms, molecules, etc. If gravitation is weak compared to other forces in a certain system under consideration (which is assumed throughout this thesis), the phenomenon of spacetime curvature in general relativity can be neglected. The resulting "smaller" theory of special relativity still predicts significant deviations from Newton's laws of physics when parts of a system move with a velocity of the order of $c$, the speed of light in vacuum, such as Lorentz contraction and time dilation.

**Origins of quantum physics**

The new concepts introduced by quantum physics into our understanding of nature are (arguably) even more counterintuitive than relativistic effects. One of these concepts is wave–particle duality, the idea that all quantum objects exhibit both wavelike and particlelike behaviors or properties. As a consequence, a single quantum object (like an electron) is described by a time-dependent **wave function** $\psi(t, \boldsymbol{r})$ to account for its wavelike propagation through space. Erwin Schrödinger published a famous equation in 1926 which governs the evolution of such a wave function [4]:

**Schrödinger equation**

$$\mathrm{i}\hbar\partial_t\psi(t, \boldsymbol{r}) = \left[-\frac{\hbar^2\boldsymbol{\nabla}^2}{2m} + V(\boldsymbol{r})\right]\psi(t, \boldsymbol{r}), \tag{1.1}$$

where $\hbar$ denotes the reduced Planck constant, $m$ is the mass of the quantum object (always electrons in the following), and $V(\boldsymbol{r})$ is an (optional) external potential. The particle character of the quantum object manifests itself, for example, in the **conservation of the norm**

$$\langle\psi|\psi\rangle = \iiint\limits_{\mathbb{R}^3} |\psi(t, \boldsymbol{r})|^2\,\mathrm{d}^3r, \tag{1.2}$$

which, as a consequence, allows for the statistical (Copenhagen) interpretation according to which $|\psi(t,\boldsymbol{r})|^2$ is the probability density to find the object at the location $\boldsymbol{r}$ at the time $t$.

The wave equation published by Schrödinger has two drawbacks: First, it is **not Lorentz invariant**, that is, not compatible with the laws of special relativity. This is because the equation was inspired by the energy formula $\mathcal{E} = \mathcal{E}_{\text{kin}} + \mathcal{E}_{\text{pot}}$ for a *nonrelativistic* classical particle ($\mathcal{E}_{\text{kin}} = mv^2/2$). The second drawback of Schrödinger's original equation is that it **ignores the spin** of the quantum object. This property is another quantum-theoretical phenomenon which has no counterpart in classical physics; however, its existence had already been postulated for electrons in 1925 by Uhlenbeck and Goudsmit [5, 6] based on experimental results (Stern–Gerlach experiment, anomalous Zeeman effect, etc.) before Schrödinger published his equation. This inspired Wolfgang Pauli to derive a refined version of the Schrödinger equation for spin-1/2 particles (such as electrons), which he published in 1927 [7]. Although the Pauli equation could successfully explain the observations in the Stern–Gerlach experiment, for example, it is a bit unsatisfactory from a theoretical point of view that the spin had to be incorporated explicitly into the equation, purely based on phenomenological reasons.

### Relativistic quantum physics

**Dirac equation**

Paul Dirac showed in 1928 that this whole situation becomes much more elegant when quantum theory is formulated in accordance with special relativity [8, 9]. Due to the formal similarity between time and space coordinates in special relativity, he chose a linear combination of first-order time and space derivatives plus a mass term as an ansatz for his **relativistic wave equation** for a free quantum object. This wave equation can be written in the covariant form

$$\left(\mathrm{i}\hbar\gamma^\mu\partial_\mu - mc\right)\underline{\psi}(t,\boldsymbol{r}) = 0, \tag{1.3}$$

where $\partial_\mu = (\partial_t/c, \boldsymbol{\nabla})$ is the four-gradient, and we sum over $\mu \in \{0,1,2,3\}$ (Einstein notation). In order to find the right "coefficients" $\gamma^\mu$, Dirac required the squared equation to be compatible with the **relativistic energy–momentum relation**

$$\mathcal{E}^2 = m^2c^4 + c^2\boldsymbol{p}^2, \tag{1.4}$$

with the electron rest mass $m$ and its relativistic momentum $\boldsymbol{p}$. This assumption led to the conclusion that the "coefficients" must be quadratic matrices which obey the **Clifford algebra**

$$\{\gamma^\mu, \gamma^\nu\} = \gamma^\mu \cdot \gamma^\nu + \gamma^\nu \cdot \gamma^\mu = -2\eta^{\mu\nu}\mathbb{1}. \tag{1.5}$$

In this equation, we have defined the anticommutator $\{A, B\} = A \cdot B + B \cdot A$ and the Minkowski metric $\eta^{\mu\nu} = \mathrm{diag}(-1, 1, 1, 1)$. As a consequence of the fact that the $\gamma^\mu$ are matrices, the wave function $\underline{\psi}$ (Dirac spinor) has multiple components. In the wave equation (1.3), these matrices automatically account for the spin of the quantum object. The minimal possible matrix size ($4 \times 4$) yields a wave equation for spin-1/2 objects (such as electrons), which is now known as the **Dirac equation**, the relativistic version of the Pauli equation. A well-known representation of the corresponding **gamma matrices**, written with $2 \times 2$ submatrices, is

$$\gamma^0 = \begin{pmatrix} \mathbb{1} & 0 \\ 0 & -\mathbb{1} \end{pmatrix}, \qquad \gamma^n = \begin{pmatrix} 0 & \sigma_n \\ -\sigma_n & 0 \end{pmatrix} \tag{1.6}$$

with the Pauli matrices

$$\sigma_1 = \sigma_x = \begin{pmatrix} 0 & 1 \\ 1 & 0 \end{pmatrix}, \quad \sigma_2 = \sigma_y = \begin{pmatrix} 0 & -\mathrm{i} \\ \mathrm{i} & 0 \end{pmatrix}, \quad \text{and} \quad \sigma_3 = \sigma_z = \begin{pmatrix} 1 & 0 \\ 0 & -1 \end{pmatrix}. \tag{1.7}$$

The fact that the spin character naturally comes into play when constructing a quantum-theoretical wave equation which is consistent with special relativity can be interpreted as a hint for the reasonability of this ansatz from a theoretical standpoint.

**Two energy continua**

However, many physicists were confused by some puzzling features of the Dirac equation. Dirac himself found that the dispersion relation of his wave equation consists of both the positive and the negative solutions of the relativistic energy–momentum relation:

$$\mathcal{E}_\pm(\boldsymbol{p}) = \pm\sqrt{m^2c^4 + c^2\boldsymbol{p}^2} \tag{1.8}$$

(for a free electron) [10]. According to that formula, electron energy has no lower bound, so each electron could be a source of arbitrary amounts of energy—a feature which is manifestly not observed in reality. Dirac proposed a solution to this problem by assuming that the state we know and perceive as vacuum is actually the state in which precisely all negative-energy states are occupied with electrons. An additional positive-energy electron on top of this **Dirac sea** thus cannot get into the negative energy continuum due to the Pauli exclusion principle. Although the Dirac sea is not an entirely satisfactory solution to the problem of the negative-energy states (since it also implies infinite charge and mass densities in vacuum, for example), it is a useful picture to understand physical phenomena like electron–positron pair creation from the vacuum. We will thus refer to this picture frequently throughout this thesis.

**Vacuum state**

### Positrons and pair creation

**Positrons: holes in the Dirac sea**

The Dirac-sea picture immediately suggested the possibility to **excite an electron from the Dirac sea** into the positive energy continuum by providing the required amount of energy via a high-energy photon, for example [10]. The remaining "**hole**" in the Dirac sea effectively behaves like an electron but with the opposite sign of the electric charge. This can be understood as follows: If the Lorentz force exerted by an electromagnetic field pushes a Dirac-sea electron into the hole (which is actually a free electron state in this picture), the hole effectively moves to the former position of this electron, that is, exactly into the opposite direction. The hole thus "feels" the inverted Lorentz force of the electron, which corresponds to the Lorentz force on a particle with the inverted electric charge. Furthermore, Dirac-sea holes have the remarkable feature that they can be filled up with electrons from the positive energy continuum ("real" electrons). This **recombination** process restores the vacuum state (completely filled Dirac sea) and goes hand in hand with the emission of the energy the electron loses during the transition from the positive to the negative energy continuum. All in all, the holes comply with the common conception of antiparticles of electrons (positrons): they share the same properties except for the inverted electric charge; pairs of electrons and positrons can be created from the vacuum if the required energy is provided (via, e.g., a high-energy photon[1]), and electrons and positrons can annihilate each other under the emission of radiation energy. Dirac thus predicted the existence of the positron (although he erroneously identified the holes with protons at first, which can be seen in the title of his article [10]). Only two years later, Carl Anderson was the first one to report on the detection of positrons in the laboratory [13, 14], which marked one of the great successes of Dirac's theory.

**Variable number of real particles**

One important consequence of these findings was that a consistent relativistic theory of electrons (and positrons) cannot be formulated for a fixed number of *real* particles because the electric charges of these particles give rise to electromagnetic fields, which could excite pairs from the vacuum. Ironically, this contradicts Dirac's initial motivation to work on his wave equation: he aimed to derive a relativistic wave equation with a conserved current which could be interpreted as a probability current (i.e., with a non-negative Noether charge, like $|\psi|^2$ in Schrödinger's nonrelativistic theory) in order to describe a single, real particle. (The relativistic, second-order Klein–Gordon equation [15, 16], which had already been known before, does not meet this condition and furthermore describes spin-0 particles.) The **conserved probability current** of the Dirac equation reads

$$\begin{aligned} 0 &= c\partial_\mu \left[\underline{\psi}^\dagger(t,\boldsymbol{r})\gamma^0\gamma^\mu\underline{\psi}(t,\boldsymbol{r})\right] \\ &= \partial_t \Big[\underbrace{\underline{\psi}^\dagger(t,\boldsymbol{r})\underline{\psi}(t,\boldsymbol{r})}_{\text{probability density}}\Big] + \boldsymbol{\nabla}\cdot\Big[\underbrace{c\underline{\psi}^\dagger(t,\boldsymbol{r})\gamma^0\boldsymbol{\gamma}\underline{\psi}(t,\boldsymbol{r})}_{\text{probability current density}}\Big] \end{aligned} \tag{1.9}$$

with $\boldsymbol{\gamma} = \gamma^1\boldsymbol{e}_x + \gamma^2\boldsymbol{e}_y + \gamma^3\boldsymbol{e}_z$. However, interpreting $\underline{\psi}^\dagger(t,\boldsymbol{r})\underline{\psi}(t,\boldsymbol{r})$ as the local probability density to find a single, real electron is wrong in the presence of electric fields due to the possibility of pair creation. This led to some confusion shortly afterwards Dirac had published his equation: When Oskar Klein applied the Dirac equation to calculate the transmission and reflection coefficients for (positive-energy) electrons at a potential step [17], he encountered some unexpected results like a nonvanishing transmission coefficient for potential steps much larger than the kinetic energy of the incoming electron. Fritz Sauter tried to "resolve" this so-called **Klein paradox** a few years later by replacing the delta-peak-shaped electric field (potential step) with a constant, finite electric field [18] or a smooth and finite electric peak $E_{\max}/\cosh^2(kx)$ [19], which is also known as (spatial) Sauter pulse. However, Sauter's results confirmed the puzzling effects found by Klein. It was not until the existence of negative-energy states, positrons, and pair creation by the electric field in Dirac theory were taken into account that these results could be understood. See, e.g., Ref. [20] for a recent discussion of the Klein paradox.

[1] A single on-shell photon is in fact not sufficient to excite an electron–positron pair from the vacuum since both the total momentum and the total energy must be conserved by this process. One possible solution is pair creation by a high-energy photon within the Coulomb field of an atomic nucleus (Bethe–Heitler pair creation [11]). Another way, the simplest pair-creation mechanism relying on photons only, is the Breit–Wheeler process [12], which requires two (or more) photons.

### Quantum field theory

The modern framework to treat problems involving a variable number of particles is quantum field theory; see, for example, Ref. [21] for a detailed introduction to this topic. Quantum field theory unifies the classical concepts of particles and fields by describing every elementary physical entity by a wave equation (wave character) but interprets the former wave function $\underline{\psi}$ as a **field operator**, which creates ($\hat{\underline{\Psi}}^\dagger$) or annihilates ($\hat{\underline{\Psi}}$) particles (particle character). Like the classical wave function, the field operators satisfy a wave equation (e.g., the Dirac equation) and furthermore obey **canonical equal-time commutation or anticommutation relations**, depending on whether the quantum field is bosonic or fermionic.

**Quantized Dirac field**

In the case of electrons (fermions), we assume the canonical anticommuta-

tion relations

$$\{\hat{\Psi}_i(t,\boldsymbol{r}_1),\hat{\Psi}_j(t,\boldsymbol{r}_2)\} = 0,$$
$$\left\{\hat{\Psi}_i^\dagger(t,\boldsymbol{r}_1),\hat{\Psi}_j^\dagger(t,\boldsymbol{r}_2)\right\} = 0, \quad \text{and}$$
$$\left\{\hat{\Psi}_i(t,\boldsymbol{r}_1),\hat{\Psi}_j^\dagger(t,\boldsymbol{r}_2)\right\} = \delta^{(3)}(\boldsymbol{r}_1-\boldsymbol{r}_2)\,\delta_{ij}, \tag{1.10}$$

where $\hat{\Psi}_i$ with $i \in \{1,2,3,4\}$ is the $i$th component of the **Dirac field operator** $\underline{\hat{\Psi}}$. Such a field operator can be expanded in the classical solutions of the Dirac equation, also in the presence of an external electromagnetic four-potential $A_\mu = (-\Phi/c, \boldsymbol{A})$, which couples to the four-gradient $\partial_\mu = (\partial_t/c, \boldsymbol{\nabla})$ via **minimal coupling**

$$\mathrm{i}\hbar\partial_\mu \to \mathrm{i}\hbar\partial_\mu - qA_\mu(t,\boldsymbol{r}) \qquad \Rightarrow \qquad \begin{aligned} \mathrm{i}\hbar\partial_t &\to \mathrm{i}\hbar\partial_t + q\Phi(t,\boldsymbol{r}), \\ -\mathrm{i}\hbar\boldsymbol{\nabla} &\to -\mathrm{i}\hbar\boldsymbol{\nabla} + q\boldsymbol{A}(t,\boldsymbol{r}) \end{aligned} \tag{1.11}$$

with $q$ denoting the absolute value of the electron charge. By multiplying the explicitly covariant form of the Dirac equation (1.3) with $\gamma^0 = \mathrm{diag}(1,1,-1,-1)$, inserting the covariant derivatives (1.11), and rearranging, we arrive at the Schrödinger form of the classical Dirac equation in an external electromagnetic field:

**Dirac equation in Schrödinger form**

$$\mathrm{i}\hbar\partial_t\underline{\psi}(t,\boldsymbol{r}) = \Big\{c\boldsymbol{\alpha}\cdot[-\mathrm{i}\hbar\boldsymbol{\nabla} + q\boldsymbol{A}(t,\boldsymbol{r})] + \gamma^0 mc^2 - q\Phi(t,\boldsymbol{r})\Big\}\underline{\psi}(t,\boldsymbol{r}), \tag{1.12}$$

where the vector $\boldsymbol{\alpha}$ of $4\times 4$ matrices is defined by

$$\boldsymbol{\alpha} = \begin{pmatrix}\gamma^0\cdot\gamma^1\\ \gamma^0\cdot\gamma^2\\ \gamma^0\cdot\gamma^3\end{pmatrix} \qquad \Rightarrow \qquad \alpha_n = \begin{pmatrix}0 & \sigma_n\\ \sigma_n & 0\end{pmatrix} \text{ for } n \in \{1,2,3\} \tag{1.13}$$

with the Pauli matrices (1.7). Say we have a **complete set of classical solutions** $\underline{\psi}_{\vec{s}}(t,\boldsymbol{r})$ of the Dirac equation (1.12), $\vec{s}$ denoting a vector of suitable quantum numbers. By means of the (time-independent) scalar product

$$\langle\underline{\psi}_{\vec{s}_1}|\underline{\psi}_{\vec{s}_2}\rangle = \iiint_{\mathbb{R}^3} \underline{\psi}_{\vec{s}_1}^\dagger(t,\boldsymbol{r})\underline{\psi}_{\vec{s}_2}(t,\boldsymbol{r})\,\mathrm{d}^3r \tag{1.14}$$

between two solutions of the Dirac equation, we assume that our basis of classical solutions is orthonormalized according to

$$\langle\underline{\psi}_{\vec{s}_1}|\underline{\psi}_{\vec{s}_2}\rangle = \delta(\vec{s}_1-\vec{s}_2), \tag{1.15}$$

where $\delta(\vec{s}_1 - \vec{s}_2)$ is meant to denote a product of delta distributions for the continuous quantum numbers and Kronecker deltas for discrete numbers (e.g., the spin state).The **Dirac field operator** can now be expanded in these classical solutions:

**Creation and annihilation operators**

$$\underline{\hat{\Psi}}(t, \boldsymbol{r}) = \sum\!\!\!\!\!\!\int_{\vec{s}} \underline{\psi}_{\vec{s}}(t, \boldsymbol{r})\, \hat{a}_{\vec{s}}. \tag{1.16}$$

Projecting the field operator onto a classical mode thus isolates the corresponding operator

$$\hat{a}_{\vec{s}} = \left\langle \underline{\psi}_{\vec{s}} \middle| \underline{\hat{\Psi}} \right\rangle \qquad \Rightarrow \qquad \hat{a}^\dagger_{\vec{s}} = \left\langle \underline{\hat{\Psi}}^\dagger \middle| \underline{\psi}_{\vec{s}} \right\rangle \tag{1.17}$$

due to the orthonormality relation (1.15). Using the canonical anticommutation relations (1.10) of the field operator, we find the anticommutation relations of the $\hat{a}_{\vec{s}}$ operators:

$$\left\{\hat{a}_{\vec{s}_1}, \hat{a}_{\vec{s}_2}\right\} = \left\{\hat{a}^\dagger_{\vec{s}_1}, \hat{a}^\dagger_{\vec{s}_2}\right\} = 0 \text{ and } \left\{\hat{a}_{\vec{s}_1}, \hat{a}^\dagger_{\vec{s}_2}\right\} = \left\langle \underline{\psi}_{\vec{s}_1} \middle| \underline{\psi}_{\vec{s}_2} \right\rangle = \delta(\vec{s}_1 - \vec{s}_2). \tag{1.18}$$

These are the canonical anticommutation relations of **fermionic creation and annihilation operators**. The operator $\hat{a}^\dagger_{\vec{s}}$ creates an electron in the mode $\underline{\psi}_{\vec{s}}(t, \boldsymbol{r})$ while $\hat{a}_{\vec{s}}$ removes an electron from this mode.

**Advantages of quantum field theory**

This whole approach turned out to be very useful to deal with the problems of the older quantum-theoretical constructs ("first quantization") above:

- It allows for a **variable number of particles**. The creation and annihilation operators (as well as the field operators) act on the Hilbert-space vector representing the state of the quantized Dirac field, which can hold an arbitrary number of excitations (particles). Pair production is thus not problematic within this framework.

- The interpretation of $\underline{\psi}^\dagger \underline{\psi}$ as a probability density to find a particular considered electron (which is not a valid interpretation in the case of the classical Dirac equation anyway as indicated by the Klein paradox, for example) is not required in quantum field theory, which is a many-body framework. Ironically, the original motivation of Dirac to search for his wave equation was his dissatisfaction with the lack of a *positive* Noether charge density in the case of the Klein–Gordon equation. But within quantum field theory, the Klein–Gordon equation [15, 16]

$$\left[-\hbar^2 \partial^\mu \partial_\mu + m^2 c^2\right] \phi(t, \boldsymbol{r}) = 0 \tag{1.19}$$

is a perfectly valid, relativistic wave equation for a scalar wave function $\phi(t, \boldsymbol{r})$, which describes spin-0 particles.

- The introduction of the **Dirac sea can be avoided** by exchanging the roles of creation and annihilation operators for electrons in the negative energy continuum; that is, the sum/integral in the expansion (1.16) of the Dirac field operator is split into two parts corresponding to the upper and lower energy continuum, respectively, and $\hat{a}_{\vec{s}}$ is replaced by $\hat{b}^{\dagger}_{\vec{s}}$ for the negative-energy/-frequency states. This way, **positrons become a second type of particle** within quantized Dirac theory (corresponding to the *absence* of negative-energy electrons in the Dirac-sea picture). There is no need for a Dirac sea in the resulting theory of electrons and positrons anymore because both particle sorts have a positive energy.
- Introducing antiparticles in order to solve the problem of negative energies (which occurs in *every* relativistic theory, not just in the case of spin-1/2 fermions) **does also work in the bosonic case**, in which a Dirac-sea-like concept would not help due to the lack of Pauli's exclusion principle.

A well-known quantum field theory is **quantum electrodynamics** (QED), which incorporates the quantized electromagnetic field and a quantum field of charged particles coupled to the electromagnetic field. In our case, the charged field is the quantized Dirac field of electrons and positrons (**spinor QED**). Another possible option is the spin-0 (i.e., bosonic) Klein–Gordon field (**scalar QED**).

### QED effects in strong electromagnetic fields

One outstanding feature of quantum theory is the phenomenon of vacuum fluctuations. In a simple, intuitive picture, the vacuum (i.e., the ground state, in which no real particles are present) is filled with so-called **virtual pairs** of electrons and positrons, which are constantly created and annihilated again. Since the Dirac field is coupled to the electromagnetic field, real excitations of the quantized electromagnetic field (photons) can interact with these virtual electron–positron pairs. In simple words, pair creation then corresponds to a virtual pair gaining sufficient energy during an interaction with the electromagnetic field to become a real pair. However, even if no real electron–positron pairs are ultimately created[2], photons may interact with virtual pairs which then interact with other photons and so on. What we observe in this case is thus an effective **photon–photon interaction** in the vacuum. This is one prominent example for a pure QED effect, which does not occur in the classical theory of electromagnetism and has been observed recently in the ATLAS experiment at the CERN [24].

**Light-by-light scattering**

[2]This is a reasonable assumption if only photons with energies much smaller than the rest energy $2mc^2 \approx 1\,\mathrm{MeV}$ of an electron–positron pair are present [22, 23].

The effective interaction between photons and the Dirac vacuum (see, e.g., the reviews [25, 26]) can be described via additional, **nonlinear terms in the Maxwell equations** which correspond to terms of higher order than $\boldsymbol{E}^2 - c^2\boldsymbol{B}^2$ (and including also the other Lorentz invariant $\boldsymbol{E}\cdot\boldsymbol{B}$) appearing in an effective Lagrangian density of the electromagnetic field. In their famous paper [22] from 1936, Werner Heisenberg and his student Hans Euler derived this effective Lagrangian $\mathcal{L}_{\mathrm{eff}}^{\mathrm{HE}}$ in the limit of low photon energies (quasiconstant $\boldsymbol{E}$ and $\boldsymbol{B}$ fields) and for only one virtual electron–positron pair mediating the effective interaction (in simple words). This **one-loop effective Lagrangian density for spinor QED** they found reads

**Heisenberg–Euler effective Lagrangian**

$$\mathcal{L}_{\mathrm{eff}}^{\mathrm{HE}}(\boldsymbol{E},\boldsymbol{B}) = \frac{\varepsilon_0}{2}\mathfrak{F} + \frac{q^2}{hc}\int_0^\infty \frac{\mathrm{e}^{-\xi}}{\xi^3}\left[\xi^2\mathfrak{G}\frac{\operatorname{Re}\cos\left(\xi\sqrt{\mathfrak{F}+2\mathrm{i}\mathfrak{G}}/E_{\mathrm{crit}}^{\mathrm{QED}}\right)}{\operatorname{Im}\cos\left(\xi\sqrt{\mathfrak{F}+2\mathrm{i}\mathfrak{G}}/E_{\mathrm{crit}}^{\mathrm{QED}}\right)} + \left(E_{\mathrm{crit}}^{\mathrm{QED}}\right)^2 - \frac{\xi^2}{3}\mathfrak{F}\right]\mathrm{d}\xi \quad (1.20)$$

with the vacuum permittivity $\varepsilon_0$, the Planck constant $h = 2\pi\hbar$, the **critical electric field strength**

$$E_{\mathrm{crit}}^{\mathrm{QED}} = \frac{m^2c^3}{\hbar q} \approx 1.3\times 10^{18}\,\frac{\mathrm{V}}{\mathrm{m}} \quad (1.21)$$

(also known as **Schwinger limit** today and often denoted by $E_S$ in the literature), and the two Lorentz invariants

$$\mathfrak{F} = \boldsymbol{E}^2 - c^2\boldsymbol{B}^2 \qquad \text{and} \qquad \mathfrak{G} = c\boldsymbol{E}\cdot\boldsymbol{B}. \quad (1.22)$$

The invariant $\mathfrak{F}$ is given by the contraction of the electromagnetic field tensor

$$F_{\mu\nu} = \partial_\mu A_\nu - \partial_\nu A_\mu \quad \Rightarrow \quad (F_{\mu\nu}) = \begin{pmatrix} 0 & E_x/c & E_y/c & E_z/c \\ -E_x/c & 0 & -B_z & B_y \\ -E_y/c & B_z & 0 & -B_x \\ -E_z/c & -B_y & B_x & 0 \end{pmatrix} \quad (1.23)$$

[with $A_\mu = (-\Phi/c, \boldsymbol{A})$ denoting the covariant four-potential] via

$$\mathfrak{F} = -\frac{c^2}{2}F^{\mu\nu}F_{\mu\nu}, \quad (1.24)$$

and

$$\mathfrak{G} = -\frac{c^2}{4}G^{\mu\nu}F_{\mu\nu} \quad (1.25)$$

is related to the dual field tensor

$$G^{\mu\nu} = \frac{1}{2}\varepsilon^{\mu\nu\varrho\varsigma}F_{\varrho\varsigma} \quad \Rightarrow \quad (G^{\mu\nu}) = \begin{pmatrix} 0 & -B_x & -B_y & -B_z \\ B_x & 0 & E_z/c & -E_y/c \\ B_y & -E_z/c & 0 & E_x/c \\ B_z & E_y/c & -E_x/c & 0 \end{pmatrix} \tag{1.26}$$

with the Levi-Civita symbol $\varepsilon^{\mu\nu\varrho\varsigma}$.

**Nonlinear QED effects in slow fields require high field intensities**

Shortly afterwards, Weisskopf calculated the counterpart of $\mathcal{L}_{\text{eff}}^{\text{HE}}$ in scalar QED [23]. Both papers [22, 23] show that the nonlinear QED vacuum effects act like an $\boldsymbol{E}$- and $\boldsymbol{B}$-field-dependent permittivity/permeability of the vacuum, at least for the low photon frequencies under consideration. Furthermore, the nonlinearities are negligibly small for fields far below the critical limit (1.21). The huge value of $E_{\text{crit}}^{\text{QED}}$ shows that nonlinear vacuum effects due to virtual electron–positron pairs do not play any significant role in our everyday life. They only become important at extreme field strengths of the order of $E_{\text{crit}}^{\text{QED}}$. Such intense fields are composed of a huge number ($\gg 1$) of low-energy photons, and thus these electromagnetic fields may be treated as **classical fields** in this realm of QED (like in [22, 23], for example).

# 2. Nonperturbative pair creation in QED: the Sauter–Schwinger effect

The critical field strength $E_{\text{crit}}^{\text{QED}}$ in Eq. (1.21) had already been identified by Sauter in his works [18, 19] on the Klein paradox as the required slope (approximately) for an electric potential variation in space to give rise to what Sauter interpreted as an incoming electron traversing the reflecting barrier to arrive at the region of "negative momentum" or "negative kinetic energy". Later, when holes in the Dirac sea were identified with positrons, it was realized [22] that this effect was actually a pair-creation process induced by a constant [18] or Sauter-pulse-shaped [19] electric field.

While Sauter studied the classical Dirac wave equation, Schwinger [27] reconsidered nonlinear QED phenomena two decades later from the point of view of quantum field theory. He also addressed the constant-field case (i.e., constant $\boldsymbol{E}$ and $\boldsymbol{B}$ fields) and thus rederived the Heisenberg–Euler effective Lagrangian $\mathcal{L}_{\text{eff}}^{\text{HE}}$ in Eq. (1.20). As had already been suggested by Heisenberg and Euler [22], he realized that this effective Lagrangian has a nonvanishing imaginary part in the case of a **constant $\boldsymbol{E}$ field** (plus the $\boldsymbol{B}$ field under certain restrictions, but let us concentrate on the pure electric case first for simplicity). This imaginary part gives rise to electron–positron pair creation because $\mathcal{L}_{\text{eff}}^{\text{HE}}$ is related to the **vacuum persistence amplitude** $\langle 0_{\text{out}}|0_{\text{in}}\rangle$ of the Dirac field (which is coupled to the external electromagnetic field) via

**Im $\mathcal{L}_{\text{eff}}^{\text{HE}} \neq 0$ leads to pair creation**

$$\langle 0_{\text{out}}|0_{\text{in}}\rangle = \mathrm{e}^{\mathrm{i}\int \mathrm{d}t \iiint \mathrm{d}^3r\, \mathcal{L}_{\text{eff}}^{\text{HE}}/\hbar} = \mathrm{e}^{\mathrm{i}\mathcal{V}_4 \mathcal{L}_{\text{eff}}^{\text{HE}}/\hbar} \tag{2.1}$$

according to Schwinger [27], where $|0_{\text{in/out}}\rangle$ denotes the initial/outgoing vacuum state of the Dirac field[1], and $\mathcal{V}_4$ is some finite four-volume within which we consider the **effective action** $\int \mathrm{d}t \iiint \mathrm{d}^3r\, \mathcal{L}_{\text{eff}}^{\text{HE}}$ here in the constant-field case in order to avoid infinities. Since only the vacuum state of the Dirac field is empty of real particles, vacuum decay (i.e., $|\langle 0_{\text{out}}|0_{\text{in}}\rangle| < 1$) is associated with the creation of electron–positron pairs. The **probability for pair creation**

[1] Initial and outgoing Dirac vacua can in fact only be defined properly if the external field vanishes asymptotically for $t \to \pm\infty$. This condition is *not* met in the constant-field case considered here, but we may instead focus on the decay of the Dirac vacuum within a unit four-volume instead (i.e., vacuum decay per unit time and unit volume).

therefore reads

$$P_{\mathrm{e}^+\mathrm{e}^-} = 1 - |\langle 0_{\mathrm{out}}|0_{\mathrm{in}}\rangle|^2 = 1 - \mathrm{e}^{-2\mathcal{V}_4 \operatorname{Im}\mathcal{L}_{\mathrm{eff}}^{\mathrm{HE}}/\hbar} \approx 2\frac{\mathcal{V}_4}{\hbar}\operatorname{Im}\mathcal{L}_{\mathrm{eff}}^{\mathrm{HE}}, \tag{2.2}$$

where the approximation is justified by the fact that $|\mathcal{V}_4 \operatorname{Im}\mathcal{L}_{\mathrm{eff}}^{\mathrm{HE}}/\hbar|$ is a very small number for typical, realistic quasiconstant electric fields, which always have a finite spatial and temporal extent measured by $\mathcal{V}_4$. The pair-creation probability per unit time and volume is thus given by the imaginary part of the Heisenberg–Euler Lagrangian density (1.20) (with $\mathfrak{F} = \boldsymbol{E}^2 = E^2$ and $\mathfrak{G} = 0$) as stated above:

$$\mathcal{P}_{\mathrm{e}^+\mathrm{e}^-} = \frac{P_{\mathrm{e}^+\mathrm{e}^-}}{\mathcal{V}_4} \approx \frac{2}{\hbar}\operatorname{Im}\mathcal{L}_{\mathrm{eff}}^{\mathrm{HE}} = \frac{q^2E^2}{4\pi^3\hbar^2 c}\sum_{n=1}^{\infty}\frac{\mathrm{e}^{-n\pi E_{\mathrm{crit}}^{\mathrm{QED}}/E}}{n^2}. \tag{2.3}$$

This expression for $\operatorname{Im}\mathcal{L}_{\mathrm{eff}}^{\mathrm{HE}}$ was calculated by Schwinger [27][2]. One way to derive this formula (or at least the exponents) is via the worldline-instanton method, which will be introduced in Sec. 2.1 below.

**Interpretation of the pair-creation probability**

The series in Eq. (2.3) reflects the infinite number of possible ways for the considered unit four-volume of vacuum to decay: via the creation of $n = 1$ pair, or two pairs, etc.; see Ref. [28] for a detailed discussion on the physical meaning of the expression (2.3). However, the contributions from the terms with $n > 1$ are negligible even for relatively strong fields: for example, the $n = 2$ term is suppressed by the factor $\exp(-10\pi)/4 < 10^{-14}$ with respect to the $n = 1$ term for $E = E_{\mathrm{crit}}^{\mathrm{QED}}/10$. We may thus **approximate the pair-creation probability** per unit volume by

$$\mathcal{P}_{\mathrm{e}^+\mathrm{e}^-} \approx \frac{q^2E^2}{4\pi^3\hbar^2 c}\mathrm{e}^{-\pi E_{\mathrm{crit}}^{\mathrm{QED}}/E}. \tag{2.4}$$

Note that the first term ($n = 1$) in the series (2.3) does furthermore coincide with the **expected number of pairs** $\dot{\mathcal{N}}_{\mathrm{e}^+\mathrm{e}^-}$ created per unit time and volume in a constant electric field, so

$$\dot{\mathcal{N}}_{\mathrm{e}^+\mathrm{e}^-} = \frac{q^2E^2}{4\pi^3\hbar^2 c}\mathrm{e}^{-\pi E_{\mathrm{crit}}^{\mathrm{QED}}/E}. \tag{2.5}$$

This fact was found by Nikishov [29] and is explained in much more detail in Ref. [28].

**Nonperturbative pair creation**

In conclusion, equation (2.3) describes the creation of electron–positron pairs by a static and homogeneous electric field. This process is called the

[2] There is a typo in Schwinger's paper [27] in the second line of Eq. (6.41): $\alpha^2$ should be $\alpha$.

**Sauter–Schwinger effect** (other common names in the literature are *Schwinger effect* or *Schwinger mechanism*); see Ref. [30] for a recent review. One important characteristic of this effect is the exponential function in $\mathcal{P}_{\mathrm{e^+e^-}}$ with $E_{\mathrm{crit}}^{\mathrm{QED}}/E = m^2c^3/(\hbar qE)$ in its argument: this term cannot be reproduced by a Taylor series in $E$ or $q$ (the coupling constant between the electromagnetic field and the Dirac field) around the unperturbed case $qE = 0$, in which the Dirac vacuum remains stable. Hence, the Sauter–Schwinger effect cannot be described by any perturbation series including a finite number of powers of the small quantity $E/E_{\mathrm{crit}}^{\mathrm{QED}}$ only and is thus a **nonperturbative** QED effect. This fact can also be understood nicely as follows [30]: in frequency space, a static electric field only has a time-independent component $\propto \exp(-\mathrm{i}t \times 0) = 1$, so the field can be thought of as being composed of zero-energy photons only ($\mathcal{E} = \hbar\omega$). A finite number of these photons cannot provide the energy $2mc^2$ required to create an electron–positron pair, and thus we have to take an infinite number of photons into account—which means that pair creation in a static field is a nonperturbative phenomenon.

#### Required electric field strengths

As a consequence of the nonperturbative term $\exp(-\pi E_{\mathrm{crit}}^{\mathrm{QED}}/E)$ in $\dot{\mathcal{N}}_{\mathrm{e^+e^-}}$, the pair-creation rate decreases rapidly to negligible values for field strengths $E$ far below the Schwinger limit. Let us do a simple estimate to illustrate this point.

A realistic electric field always has a finite spatial (and temporal) extent, so we can estimate its total pair-creation rate by integrating the density (2.5) over the spatial field volume $\mathcal{V}$, which simply yields the product $\dot{\mathcal{N}}_{\mathrm{e^+e^-}}\mathcal{V}$. Say the pair-creation rate must be of the order of $1\,\mathrm{s}^{-1}$ in order to be measurable. We rewrite Eq. (2.5) as

$$\begin{aligned}\dot{\mathcal{N}}_{\mathrm{e^+e^-}} &= 4\pi\frac{c}{\lambda_C^4}\left(\frac{E}{E_{\mathrm{crit}}^{\mathrm{QED}}}\right)^2 \mathrm{e}^{-\pi E_{\mathrm{crit}}^{\mathrm{QED}}/E} \\ &\approx \frac{1}{\lambda_C^3} \times 10^{20} \times 4\pi \left(\frac{E}{E_{\mathrm{crit}}^{\mathrm{QED}}}\right)^2 \mathrm{e}^{-\pi E_{\mathrm{crit}}^{\mathrm{QED}}/E}\,\frac{1}{\mathrm{s}}\end{aligned} \tag{2.6}$$

and now see the dependence of $\dot{\mathcal{N}}_{\mathrm{e^+e^-}}$ on $E/E_{\mathrm{crit}}^{\mathrm{QED}}$ more clearly. The quantity

$$\lambda_C = \frac{h}{mc} \approx 2.4 \times 10^{-12}\,\mathrm{m} \tag{2.7}$$

is the Compton wavelength of electrons and positrons. Equation (2.6) tells us that $\mathcal{V}$, measured in units of $\lambda_C^3$ ("Compton volumes"), must compensate the smallness of $4\pi \times 10^{20} \times (E/E_{\mathrm{crit}}^{\mathrm{QED}})^2 \times \exp(-\pi E_{\mathrm{crit}}^{\mathrm{QED}}/E)$ for the field to produce a measurable pair-creation rate; that is, the weaker the field, the larger its spatial extent must be. This factor is plotted in Fig. 2.1. We see that one Compton volume $\lambda_C^3$ produces approximately one pair per second at $E \approx 0.075 E_{\mathrm{crit}}^{\mathrm{QED}}$. However, at $E = 0.01 E_{\mathrm{crit}}^{\mathrm{QED}} \approx 10^{16}\,\mathrm{V/m}$, which is still an enormous field strength, we already need $10^{120}$ Compton volumes to attain the same total pair-creation rate. This is the volume of a cube with an edge length of $10^{40}\lambda_C \approx 10^{28}\,\mathrm{m}$—which is larger than the diameter of the observable universe! An experimental realization of the (pure) Sauter–Schwinger effect thus requires very strong electric fields between one and two orders of magnitude below $E_{\mathrm{crit}}^{\mathrm{QED}}$.

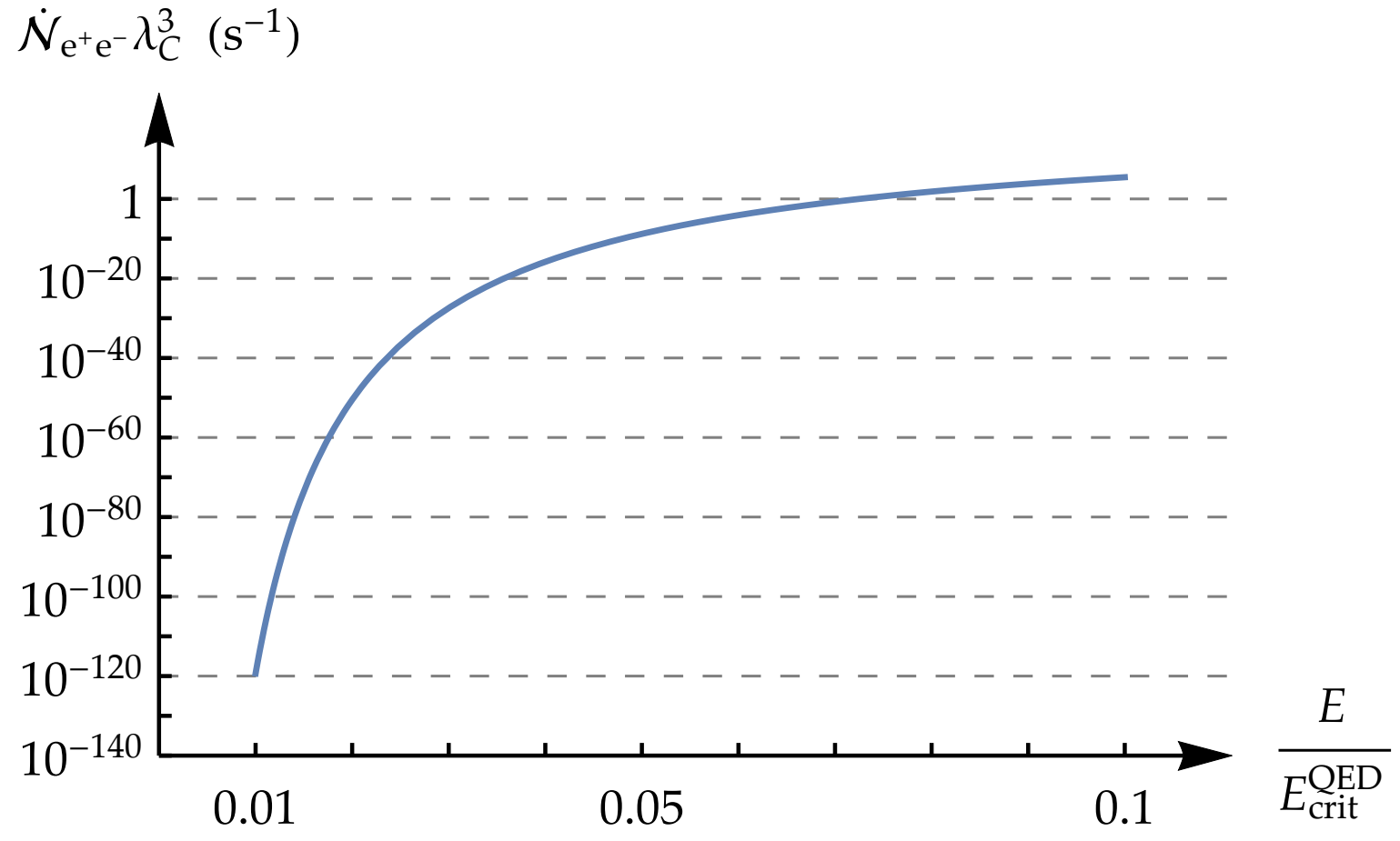

**Figure 2.1.:** Pair-creation rate (pairs per second) of one Compton volume $\lambda_C^3$ in a constant electric field as a function of the field strength $E$ over $E_{\mathrm{crit}}^{\mathrm{QED}}$; see Eq. (2.6).

## 2.1. Worldline-instanton technique for scalar QED

One way to calculate the vacuum persistence amplitude $\langle 0_{\mathrm{out}}|0_{\mathrm{in}}\rangle$ of the Dirac field [and thus the pair-creation probability according to Eq. (2.2)] under the influence of an external electromagnetic four-potential $A_\mu = (-\Phi/c, \boldsymbol{A})$ is the worldline-instanton technique [31, 32, 33, 34, 35, 36, 37, 38]. Since we will refer to results obtained via this method at some points in this thesis, we will give a brief introduction to this method in this section. See, e.g., [39, 40] for more-detailed derivations. Furthermore, this method is suitable to derive Schwinger's result (2.3) for a constant electric field as a simple example (see Sec. 2.1.1).

Note that the nonexponential prefactors in the pair-creation probability (2.3) will change if we **neglect the spin** of electrons/positrons by applying scalar QED instead of spinor QED, but the exponents are not sensitive to this approximation [34]. Since the exponent in the leading-order term in $\mathcal{P}_{e^+e^-}$ is our main interest (it dominates the behavior of $\mathcal{P}_{e^+e^-}$), we will present the worldline-instanton method for **scalar QED** here for simplicity.

**Focus on exponents in the effective action**

## Derivation

The derivation starts with the action functional of the scalar (Klein–Gordon) field $\phi(t, \boldsymbol{r})$, which is coupled to the external four-potential via the covariant derivatives (1.11) [we omit to write the dependence of $\phi$ and $A^\mu$ on the spacetime coordinates $x^\mu = (ct, \boldsymbol{r})$ explicitly here for brevity]:

$$\begin{aligned} \mathcal{A}_{\mathrm{KG}}[\phi, A_\mu] &= \int_{-\infty}^{\infty} \iiint_{\mathbb{R}^3} \frac{c}{\hbar} \left[ -|(\mathrm{i}\hbar\partial_\mu - qA_\mu)\phi|^2 - m^2c^2|\phi|^2 \right] \mathrm{d}^3r\,\mathrm{d}t \\ &= \iiiint_{\mathbb{R}^4} \phi^* \left[ \hbar c \underbrace{\left(\partial^\mu + \mathrm{i}\frac{q}{\hbar}A^\mu\right)\left(\partial_\mu + \mathrm{i}\frac{q}{\hbar}A_\mu\right)}_{\Box_{A_\mu}} - \frac{m^2c^3}{\hbar} \right] \phi \, \frac{\mathrm{d}^4x}{c}, \end{aligned} \tag{2.8}$$

where we have shifted a covariant derivative from $\phi^*$ to $\phi$ in the second line via integration by parts. Note that $\Box_{A_\mu}$ coincides with the usual d'Alembert operator in the case of a vanishing external potential:

$$\Box_{A_\mu=0} = -\frac{\partial_t^2}{c^2} + \boldsymbol{\nabla}^2. \tag{2.9}$$

By means of the action (2.8), the **vacuum persistence amplitude** of the Klein–Gordon field can be expressed as a path integral over the two independent fields $\phi$ and $\phi^*$ with the boundary condition that the initial and the final state is the vacuum:

**Path-integral expression for $\langle 0_{\mathrm{out}} | 0_{\mathrm{in}} \rangle$**

$$\begin{aligned} \langle 0_{\mathrm{out}} | 0_{\mathrm{in}} \rangle &= \iint_{\mathit{in}\,\mathrm{vac}}^{\mathit{out}\,\mathrm{vac}} \mathrm{e}^{\mathrm{i}\mathcal{A}_{\mathrm{KG}}[\phi, A_\mu]/\hbar} \, \mathcal{D}\phi\,\mathcal{D}\phi^* \\ &= \iint \mathrm{e}^{\mathrm{i}\iiiint \phi^* [\Box_{A_\mu} - m^2c^2/\hbar^2]\phi\,\mathrm{d}^4x} \, \mathcal{D}\phi\,\mathcal{D}\phi^* \\ &= \iint \mathrm{e}^{\mathrm{i}\iiiint \phi^* [\Box_{A_\mu} - \bar{\lambda}_C^{-2}]\phi\,\mathrm{d}^4x} \, \mathcal{D}\phi\,\mathcal{D}\phi^*, \end{aligned} \tag{2.10}$$

where we have defined the reduced Compton wavelength

$$\bar{\lambda}_C = \frac{\hbar}{mc} = \frac{\lambda_C}{2\pi} \tag{2.11}$$

of electrons/positrons.

**Imaginary time**

We now perform a **Wick rotation**; that is, we transform to the imaginary time coordinate[3]

$$\mathcal{T} = \mathrm{i}t \qquad \Rightarrow \qquad \mathrm{d}\mathcal{T} = \mathrm{i}\,\mathrm{d}t \qquad \Rightarrow \qquad \partial_t = \mathrm{i}\partial_{\mathcal{T}}, \tag{2.12}$$

and we combine the resulting spacetime coordinates $\mathcal{T}$ and $\boldsymbol{r}$ to form a four-dimensional vector

$$\vec{X} = \begin{pmatrix} c\mathcal{T} \\ \boldsymbol{r} \end{pmatrix} = \begin{pmatrix} c\mathcal{T} \\ x \\ y \\ z \end{pmatrix} = X_i \vec{e}_i \qquad \text{with} \qquad i \in \{1,2,3,4\}. \tag{2.13}$$

This is a useful notation in this context because the squared Euclidean norm of $\vec{X}$ coincides with the relativistic invariant $x^\mu x_\mu$:

$$\vec{X}^2 = c^2\mathcal{T}^2 + \boldsymbol{r}^2 = -c^2t^2 + \boldsymbol{r}^2 = x^\mu x_\mu. \tag{2.14}$$

For this reason, the $X_i$'s are also known as the **coordinates of Euclidean spacetime**. Expressing the d'Alembertian in Eq. (2.8) in terms of these coordinates yields

$$\begin{aligned} \Box_{A_\mu} &= -\left(\frac{\partial_t}{c} + \mathrm{i}\frac{q}{\hbar}A_0\right)\left(\frac{\partial_t}{c} + \mathrm{i}\frac{q}{\hbar}A_0\right) + \left(\boldsymbol{\nabla} + \mathrm{i}\frac{q}{\hbar}\boldsymbol{A}\right)\cdot\left(\boldsymbol{\nabla} + \mathrm{i}\frac{q}{\hbar}\boldsymbol{A}\right) \\ &= -\left(\frac{\mathrm{i}\partial_{\mathcal{T}}}{c} + \mathrm{i}\frac{q}{\hbar}A_0\right)^2 + \left(\boldsymbol{\nabla} + \mathrm{i}\frac{q}{\hbar}\boldsymbol{A}\right)^2 \\ &= \left[\frac{\partial_{\mathcal{T}}}{c} + \frac{q}{\hbar}\underbrace{A_0(t = -\mathrm{i}\mathcal{T}, \boldsymbol{r})}_{-\Phi/c}\right]^2 + \left(\boldsymbol{\nabla} + \mathrm{i}\frac{q}{\hbar}\boldsymbol{A}\right)^2 \\ &= \left[\frac{\partial_{\mathcal{T}}}{c} + \mathrm{i}\frac{q}{\hbar}\mathrm{i}\frac{\Phi(t = -\mathrm{i}\mathcal{T}, \boldsymbol{r})}{c}\right]^2 + \left(\boldsymbol{\nabla} + \mathrm{i}\frac{q}{\hbar}\boldsymbol{A}\right)^2 \\ &= \left[\vec{\nabla}_{\vec{X}} + \mathrm{i}\frac{q}{\hbar}\vec{A}(\vec{X})\right]^2 \\ &= \Delta_{\vec{A}} \end{aligned} \tag{2.15}$$

with

$$\vec{\nabla}_{\vec{X}} = \vec{e}_i \partial_{X_i} = \begin{pmatrix} \partial_{\mathcal{T}}/c \\ \partial_x \\ \partial_y \\ \partial_z \end{pmatrix} \tag{2.16}$$

[3]Note that we omit carefully taking care of the rotation of the integration contour for the resulting $\mathcal{T}$ integral here for brevity. See, e.g., Ref. [36] for more details on that point.

and the **Euclidean four-potential**

$$\vec{A}(\vec{X}) = \begin{pmatrix} \mathrm{i}\Phi(-\mathrm{i}\mathcal{T}, \boldsymbol{r})/c \\ \boldsymbol{A}(-\mathrm{i}\mathcal{T}, \boldsymbol{r}) \end{pmatrix}, \tag{2.17}$$

so $\Box_A$ coincides with the **Laplace operator in Euclidean spacetime**, with $\vec{\nabla}_{\vec{X}}$ being minimally coupled to the four-dimensional Euclidean vector potential $\vec{A}(\vec{X})$, which is related to the "ordinary" four-potential according to Eq. (2.17).

**Euclidean-spacetime expression for $\langle 0_{\text{out}}|0_{\text{in}}\rangle$**

The vacuum persistence amplitude in Eq. (2.10) thus becomes ($\mathrm{d}^4x = c\,\mathrm{d}t\,\mathrm{d}^3x = -\mathrm{i}\,\mathrm{d}^4X$)

$$\langle 0_{\text{out}}|0_{\text{in}}\rangle = \iint \mathrm{e}^{-\iiiint \phi^*[-\Delta_{\vec{A}}+\bar{\lambda}_C^{-2}]\phi\,\mathrm{d}^4X}\,\mathcal{D}\phi\,\mathcal{D}\phi^* \tag{2.18}$$

after the Wick rotation. Since $-\Delta_{\vec{A}} + \bar{\lambda}_C^{-2}$ is a **positive operator** in Euclidean spacetime, the path integrals are analogous to a finite-dimensional Gaussian double integral of the form

$$\iint \mathrm{e}^{-\vec{v}^\dagger M \vec{v}}\,\mathrm{d}^N v\,\mathrm{d}^N v^* \propto \frac{1}{\det M}, \tag{2.19}$$

where $M$ is a Hermitian, positive-definite $N \times N$ matrix. In analogy to this formula, the vacuum persistence amplitude in Eq. (2.18) is proportional to the reciprocal of the **functional determinant** of the positive operator:

$$\langle 0_{\text{out}}|0_{\text{in}}\rangle \propto \frac{1}{\det(-\Delta_{\vec{A}} + \bar{\lambda}_C^{-2})}. \tag{2.20}$$

In order to cancel the constant of proportionality, we consider the ratio of $\langle 0_{\text{out}}|0_{\text{in}}\rangle$ to the vacuum persistence amplitude for $A_\mu = 0$ (in which case we do not expect any pair creation) in the following:

$$\frac{\langle 0_{\text{out}}|0_{\text{in}}\rangle}{\langle 0_{\text{out}}|0_{\text{in}}\rangle_{A_\mu=0}} = \frac{\det(-\Delta_{\vec{A}=0} + \bar{\lambda}_C^{-2})}{\det(-\Delta_{\vec{A}} + \bar{\lambda}_C^{-2})}. \tag{2.21}$$

Note that this is equivalent to adding a constant (which is not of any physical importance) to the resulting effective action below.

**Calculations of the functional determinants**

Both functional determinants are the products of the (positive) eigenvalues $\lambda_i$ of the respective operator, so the logarithm of the determinant in the denominator reads

$$\ln\det(-\Delta_{\vec{A}} + \bar{\lambda}_C^{-2}) = \sum\!\!\!\!\!\!\int_i \ln\lambda_i = \operatorname{tr}\ln(-\Delta_{\vec{A}} + \bar{\lambda}_C^{-2}), \tag{2.22}$$

and we thus get

$$\frac{\langle 0_{\text{out}}|0_{\text{in}}\rangle}{\langle 0_{\text{out}}|0_{\text{in}}\rangle_{A_\mu=0}} = \frac{\mathrm{e}^{\operatorname{tr}\ln(-\Delta_{\vec{A}=0}+\bar{\lambda}_C^{-2})}}{\mathrm{e}^{\operatorname{tr}\ln(-\Delta_{\vec{A}}+\bar{\lambda}_C^{-2})}}$$
$$= \exp\left[-\operatorname{tr}\ln\left(\frac{-\Delta_{\vec{A}}+\bar{\lambda}_C^{-2}}{-\Delta_{\vec{A}=0}+\bar{\lambda}_C^{-2}}\right)\right]. \tag{2.23}$$

The relation between the vacuum persistence amplitude and the effective action/Lagrangian density of the external electromagnetic field has been introduced in Eq. (2.1) for spinor QED in Minkowskian spacetime; cf. [27]. Since $\mathrm{d}t \to -\mathrm{i}\,\mathrm{d}\mathcal{T}$ in Euclidean spacetime, the Euclidean form $\Gamma_{\text{eff}}^{\text{Eucl}}$ of the effective action is given by

$$\frac{\langle 0_{\text{out}}|0_{\text{in}}\rangle}{\langle 0_{\text{out}}|0_{\text{in}}\rangle_{A_\mu=0}} = \mathrm{e}^{\mathrm{i}\Gamma_{\text{eff}}[A_\mu]/\hbar} = \mathrm{e}^{\mathrm{i}\iiiint \mathcal{L}_{\text{eff}}[A_\mu]\,\mathrm{d}^3r\,\mathrm{d}t/\hbar} = \mathrm{e}^{\Gamma_{\text{eff}}^{\text{Eucl}}[\vec{A}]/\hbar}. \tag{2.24}$$

Comparing this equation to the result (2.23) above yields an expression for the **Euclidean effective action**:

$$\Gamma_{\text{eff}}^{\text{Eucl}}[\vec{A}] = -\hbar\operatorname{tr}\ln\left(\frac{-\Delta_{\vec{A}}+\bar{\lambda}_C^{-2}}{-\Delta_{\vec{A}=0}+\bar{\lambda}_C^{-2}}\right). \tag{2.25}$$

The real part of $\Gamma_{\text{eff}}^{\text{Eucl}}$ is thus related to the **pair-creation probability** since it determines the modulus of the vacuum persistence amplitude.

**Calculation of the effective action**

In order to calculate this effective action, we make use of the integral representation

$$-\ln\left(\frac{a}{b}\right) = \int_0^\infty \frac{\mathrm{e}^{-a\xi}-\mathrm{e}^{-b\xi}}{\xi}\,\mathrm{d}\xi \qquad \text{for} \qquad a,b>0, \tag{2.26}$$

so we get

$$\Gamma_{\text{eff}}^{\text{Eucl}}[\vec{A}] \to \hbar\operatorname{tr}\int_0^\infty \frac{\mathrm{e}^{-\xi(-\bar{\lambda}_C^2\Delta_{\vec{A}}+1)}}{\xi}\,\mathrm{d}\xi = \hbar\int_0^\infty \frac{\mathrm{e}^{-\xi}}{\xi}\operatorname{tr}\left(\mathrm{e}^{\xi\bar{\lambda}_C^2\Delta_{\vec{A}}}\right)\mathrm{d}\xi, \tag{2.27}$$

where we have omitted the $\Delta_{\vec{A}=0}$ term since it does not contribute to pair creation. As the next step, we have to evaluate the trace. Note that

$$\hat{H}_4 = -\frac{\hbar^2\Delta_{\vec{A}}}{2m} = \frac{\left[-\mathrm{i}\hbar\vec{\nabla}_{\vec{X}}+q\vec{A}(\vec{X})\right]^2}{2m} \tag{2.28}$$

[cf. Eq. (2.15)] can be interpreted as the nonrelativistic **Hamilton operator** of an electron coupled to the space-dependent vector potential $\vec{A}(\vec{X})$ in *four* space dimensions (i.e., the imaginary-time coordinate $X_1 = c\mathcal{T}$ is also interpreted as a spatial coordinate). We choose the position eigenstates $|\vec{X}\rangle$ in four-dimensional Euclidean space, which satisfy

$$\langle \vec{X}_1 | \vec{X}_2 \rangle = \delta^{(4)}(\vec{X}_1 - \vec{X}_2), \tag{2.29}$$

as a basis in order to calculate the trace in Eq. (2.27). Since this trace is the integral over all diagonal elements of

$$e^{\xi \bar{\lambda}_C^2 \Delta_{\vec{A}}} = \exp\left[-\frac{i}{\hbar}\hat{H}_4\left(-2i\frac{\bar{\lambda}_C}{c}\xi\right)\right] \tag{2.30}$$

with respect to this basis, we get

$$\Gamma_{\text{eff}}^{\text{Eucl}}[\vec{A}] = \hbar \int_0^\infty \frac{e^{-\xi}}{\xi} \iiiint\limits_{\mathbb{R}^4} \underbrace{\left\langle \vec{X} \middle| \exp\left[-\frac{i}{\hbar}\hat{H}_4\left(-2i\frac{\bar{\lambda}_C}{c}\xi\right)\right] \middle| \vec{X} \right\rangle}_{\text{"propagator"}} d^4X \, d\xi. \tag{2.31}$$

**Nonrelativistic propagator in three-dimensional space**

Let us turn to a standard 3+1-dimensional scenario for a moment: We consider the *nonrelativistic* motion of an electron in an external vector potential $\boldsymbol{A}(\boldsymbol{r})$. Say the electron is located at $\boldsymbol{r}'$ at the time $t = 0$. The probability amplitude to find this electron at $\boldsymbol{r}'$ again after the time $\Delta t$ has elapsed is given by the matrix element $\langle \boldsymbol{r}' | \hat{U}(\Delta t) | \boldsymbol{r}' \rangle$, where

$$\hat{U}(\Delta t) = \exp\left(-\frac{i}{\hbar}\hat{H}_3 \Delta t\right) \qquad \text{with} \qquad \hat{H}_3 = \frac{[-i\hbar\boldsymbol{\nabla} + q\boldsymbol{A}(\boldsymbol{r})]^2}{2m} \tag{2.32}$$

is the time-evolution operator, and $\hat{H}_3$ is the Hamilton operator. This matrix element (**propagator**) can be expressed as an integral over all classical paths $\boldsymbol{r}(t)$ which satisfy the boundary conditions $\boldsymbol{r}(0) = \boldsymbol{r}(\Delta t) = \boldsymbol{r}'$. Each path is weighted by its classical action via

$$\langle \boldsymbol{r}' | \hat{U}(\Delta t) | \boldsymbol{r}' \rangle = \int_{\boldsymbol{r}(0)=\boldsymbol{r}'}^{\boldsymbol{r}(\Delta t)=\boldsymbol{r}'} \exp\left[\frac{i}{\hbar}\int_0^{\Delta t} L(\dot{\boldsymbol{r}}(t), \boldsymbol{r}(t))\, dt\right] \mathcal{D}[\boldsymbol{r}(t)] \tag{2.33}$$

with the Lagrange function

$$L(\dot{\boldsymbol{r}}, \boldsymbol{r}) = \frac{m}{2}\dot{\boldsymbol{r}}^2 - q\dot{\boldsymbol{r}} \cdot \boldsymbol{A}(\boldsymbol{r}). \tag{2.34}$$

Substituting the **dimensionless time variable** $u = t/\Delta t$ yields

$$\langle \boldsymbol{r}' | \hat{U}(\Delta t) | \boldsymbol{r}' \rangle = \int\limits_{\boldsymbol{r}(0)=\boldsymbol{r}'}^{\boldsymbol{r}(1)=\boldsymbol{r}'} \exp\left[ \int\limits_0^1 \frac{\mathrm{i}m}{2\hbar\Delta t} \left( \frac{\mathrm{d}\boldsymbol{r}(u)}{\mathrm{d}u} \right)^2 - \frac{\mathrm{i}q}{\hbar} \frac{\mathrm{d}\boldsymbol{r}(u)}{\mathrm{d}u} \cdot \boldsymbol{A}(\boldsymbol{r}(u))\, \mathrm{d}u \right] \mathcal{D}[\boldsymbol{r}(u)]. \tag{2.35}$$

**Calculation of the Euclidean effective action**

Since $\hat{H}_4$ in Eq. (2.28) is a formal generalization of $\hat{H}_3$ in Eq. (2.32) to four spatial dimensions (with $\boldsymbol{r} \to \vec{X}$, $\boldsymbol{\nabla} \to \vec{\nabla}_{\vec{X}}$, and $\boldsymbol{A} \to \vec{A}$), we can directly infer a path-integral expression for the "propagator" in Eq. (2.31) by analogy with Eq. (2.35) (setting $\Delta t \to -2\mathrm{i}\bar{\lambda}_C \xi / c$):

$$\Gamma_{\text{eff}}^{\text{Eucl}}[\vec{A}] = \hbar \int\limits_0^\infty \mathrm{d}\xi \, \frac{\mathrm{e}^{-\xi}}{\xi} \iiiint\limits_{\mathbb{R}^4} \mathrm{d}^4 X' \int\limits_{\vec{X}(0)=\vec{X}'}^{\vec{X}(1)=\vec{X}'} \mathcal{D}[\vec{X}(u)] \times \exp\left[ -\int\limits_0^1 \frac{1}{4\bar{\lambda}_C^2 \xi} \left( \frac{\mathrm{d}\vec{X}(u)}{\mathrm{d}u} \right)^2 + \frac{\mathrm{i}q}{\hbar} \frac{\mathrm{d}\vec{X}(u)}{\mathrm{d}u} \cdot \vec{A}(\vec{X}(u))\, \mathrm{d}u \right]. \tag{2.36}$$

This expression can be simplified since

$$\iiiint\limits_{\mathbb{R}^4} \mathrm{d}^4 X' \int\limits_{\vec{X}(0)=\vec{X}'}^{\vec{X}(1)=\vec{X}'} \mathcal{D}[\vec{X}(u)] \ldots = \int\limits_{\vec{X}(0)=\vec{X}(1)} \mathcal{D}[\vec{X}(u)] \ldots; \tag{2.37}$$

that is, we integrate over *all* closed paths in Euclidean spacetime, the paths being parameterized by the dimensionless variable $u$, and obtain the final **exact expression**

$$\Gamma_{\text{eff}}^{\text{Eucl}}[\vec{A}] = \hbar \int\limits_0^\infty \mathrm{d}\xi \, \frac{\mathrm{e}^{-\xi}}{\xi} \int\limits_{\vec{X}(0)=\vec{X}(1)} \mathcal{D}[\vec{X}(u)] \times \exp\left[ -\int\limits_0^1 \frac{1}{4\bar{\lambda}_C^2 \xi} \left( \frac{\mathrm{d}\vec{X}(u)}{\mathrm{d}u} \right)^2 + \frac{\mathrm{i}q}{\hbar} \frac{\mathrm{d}\vec{X}(u)}{\mathrm{d}u} \cdot \vec{A}(\vec{X}(u))\, \mathrm{d}u \right]. \tag{2.38}$$

**Semiclassical approximation of the Euclidean effective action**

We now apply two (semiclassical) approximations in order to calculate $\Gamma_{\text{eff}}^{\text{Eucl}}[\vec{A}]$. First, note that the dominating contribution to the path integral (2.38) comes from the closed path $\vec{X}_{\mathrm{dom}}(u)$ with the minimal (real part of the) classical action ($u$ integral) due to the exponential suppression of the contributions from the other paths. Since the action is necessarily stationary at an extremum, $\vec{X}_{\mathrm{dom}}(u)$ must satisfy the **Euler–Lagrange equations**, the $u$ integrand in Eq. (2.38) being the Lagrange function, and $u$ corresponds to the time in classical mechanics. The resulting classical equations of motion in Euclidean spacetime are

$$\frac{\mathrm{d}^2 X_i(u)}{\mathrm{d}u^2} = \frac{2\mathrm{i}q\bar{\lambda}_C\xi}{mc} F_{ij}^{\mathrm{Eucl}}(\vec{X}(u))\frac{\mathrm{d}X_j(u)}{\mathrm{d}u} \tag{2.39}$$

with $i, j \in \{1,2,3,4\}$, where we have defined the **Euclidean electromagnetic field tensor**

$$F_{ij}^{\mathrm{Eucl}}(\vec{X}) = \partial_{X_i} A_j(\vec{X}) - \partial_{X_j} A_i(\vec{X})$$

$$\Rightarrow \quad \big(F_{ij}^{\mathrm{Eucl}}\big)(\vec{X}) = \left.\begin{pmatrix} 0 & -\mathrm{i}E_x/c & -\mathrm{i}E_y/c & -\mathrm{i}E_z/c \\ \mathrm{i}E_x/c & 0 & -B_z & B_y \\ \mathrm{i}E_y/c & B_z & 0 & -B_x \\ \mathrm{i}E_z/c & -B_y & B_x & 0 \end{pmatrix}\right|_{\substack{t=-\mathrm{i}X_1/c \\ \boldsymbol{r}=(X_2,X_3,X_4)^\intercal}} \tag{2.40}$$

in analogy to Eq. (1.23). Every closed [i.e., $\vec{X}(0) = \vec{X}(1)$] solution of the Eqs. (2.39) is called a **worldline instanton**, and we consequently call the Eqs. (2.39) the instanton equations. Note that the (dimensionless) "speed"

$$v_{\mathrm{inst}}[\vec{X}(u)] = \frac{1}{\bar{\lambda}_C}\left|\frac{\mathrm{d}\vec{X}(u)}{\mathrm{d}u}\right| \tag{2.41}$$

of each instanton is constant since

$$\frac{\mathrm{d}}{\mathrm{d}u}\left(\frac{\mathrm{d}\vec{X}(u)}{\mathrm{d}u}\right)^2 = 2\frac{\mathrm{d}\vec{X}(u)}{\mathrm{d}u}\cdot\frac{\mathrm{d}^2\vec{X}(u)}{\mathrm{d}u^2} \propto \frac{\mathrm{d}X_i(u)}{\mathrm{d}u}F_{ij}^{\mathrm{Eucl}}(\vec{X}(u))\frac{\mathrm{d}X_j(u)}{\mathrm{d}u} = 0 \tag{2.42}$$

due to the antisymmetry of $F_{ij}^{\mathrm{Eucl}}$. By means of the **saddle-point approximation**, the instanton $\vec{X}_{\mathrm{dom}}(u)$ with the minimal action

**First saddle-point approximation**

$$\int_0^1 \frac{1}{4\bar{\lambda}_C^2\xi}\left(\frac{\mathrm{d}\vec{X}_{\mathrm{dom}}(u)}{\mathrm{d}u}\right)^2 + \frac{\mathrm{i}q}{\hbar}\frac{\mathrm{d}\vec{X}_{\mathrm{dom}}(u)}{\mathrm{d}u}\cdot\vec{A}(\vec{X}_{\mathrm{dom}}(u))\,\mathrm{d}u$$

$$= \frac{v_{\mathrm{inst}}^2[\vec{X}_{\mathrm{dom}}(u)]}{4\xi} + \frac{\mathrm{i}q}{\hbar}\int_0^1 \frac{\mathrm{d}\vec{X}_{\mathrm{dom}}(u)}{\mathrm{d}u}\cdot\vec{A}(\vec{X}_{\mathrm{dom}}(u))\,\mathrm{d}u \tag{2.43}$$

dominates the path integral in Eq. (2.38), so we get

$$\Gamma_{\text{eff}}^{\text{Eucl}}[\vec{A}] \propto \hbar \int_0^\infty \frac{1}{\xi} \exp\left\{ -v_{\text{inst}}^2[\vec{X}_{\text{dom}}] \left( \frac{\xi}{v_{\text{inst}}^2[\vec{X}_{\text{dom}}]} + \frac{1}{4\xi} \right) \right\} \mathrm{d}\xi$$
$$\times \exp\left[ -\frac{\mathrm{i}q}{\hbar} \int_0^1 \frac{\mathrm{d}\vec{X}_{\text{dom}}(u)}{\mathrm{d}u} \cdot \vec{A}(\vec{X}_{\text{dom}}(u))\, \mathrm{d}u \right]. \tag{2.44}$$

We did not write the (nonexponential) prefactor from the saddle-point approximation explicitly here since we will focus on the exponent governing the size of $\Gamma_{\text{eff}}^{\text{Eucl}}$ in the following for simplicity[4].

**Second saddle-point approximation**

If we furthermore assume that $v_{\text{inst}}^2[\vec{X}_{\text{dom}}]$ is a large number—that is,

$$v_{\text{inst}}^2[\vec{X}_{\text{dom}}(u)] \gg 1 \tag{2.45}$$

(one has to check the validity of this inequation later when considering instantons for concrete field profiles in order to ensure that the results are consistent with the assumptions made here)—we may solve the remaining $\xi$ integral in Eq. (2.44) using the **saddle-point approximation** as well, the relevant saddle point being $\xi = v_{\text{inst}}[\vec{X}_{\text{dom}}]/2$. Thus, we get our end result

**Result**

$$\Gamma_{\text{eff}}^{\text{Eucl}}[\vec{A}] \propto \hbar\, \mathrm{e}^{-\mathcal{A}_{\text{inst}}[\vec{X}_{\text{dom}}(u)]} \tag{2.46}$$

with the (dimensionless) **instanton action**

$$\mathcal{A}_{\text{inst}}[\vec{X}(u)] = v_{\text{inst}}[\vec{X}(u)] + \frac{\mathrm{i}q}{\hbar} \int_0^1 \frac{\mathrm{d}\vec{X}(u)}{\mathrm{d}u} \cdot \vec{A}(\vec{X}(u))\, \mathrm{d}u, \tag{2.47}$$

and the **instanton equations** (2.39) become

$$\frac{\mathrm{d}^2 X_i(u)}{\mathrm{d}u^2} = \mathrm{i}\frac{q\bar{\lambda}_C v_{\text{inst}}[\vec{X}(u)]}{mc} F_{ij}^{\text{Eucl}}(\vec{X}(u)) \frac{\mathrm{d}X_j(u)}{\mathrm{d}u}. \tag{2.48}$$

The Euclidean effective action determines the vacuum persistence amplitude of the Klein–Gordon field according to Eq. (2.24).

[4]The prefactor has been considered in, e.g., Refs. [36, 41] within the worldline-instanton formalism. Note that we cannot reproduce the exact prefactors in Schwinger's result (2.3) here anyway since we consider the worldline-instanton technique for scalar QED for simplicity.

### 2.1.1. Example: constant electric field (Sauter–Schwinger effect)

Let us apply the worldline-instanton method to a constant electric field $\boldsymbol{E} = E\boldsymbol{e}_x$ with $E > 0$ in the following (cf. [34]). This calculation reproduces the exponents of all terms in Schwinger's result (2.3) for the imaginary part of the effective action in spinor QED.

We choose the **temporal gauge** $\boldsymbol{A} = Et\boldsymbol{e}_x$ ($\Phi = 0$) here, so the Euclidean-spacetime vector potential (2.17) reads ($X_1 = c\mathcal{T} = \mathrm{i}ct$)

$$\vec{A}(X_1) = (0, -\mathrm{i}EX_1/c, 0, 0)^\mathsf{T} \tag{2.49}$$

in this example, which leads to the instanton equations [cf. Eq. (2.48)]

$$\begin{aligned}\frac{\mathrm{d}^2}{\mathrm{d}u^2}\begin{pmatrix}X_1\\X_2\\X_3\\X_4\end{pmatrix} &= \mathrm{i}\frac{q\bar{\lambda}_C v_\mathrm{inst}[\vec{X}(u)]}{mc}\begin{pmatrix}0 & -\mathrm{i}E/c & 0 & 0\\ \mathrm{i}E/c & 0 & 0 & 0\\ 0 & 0 & 0 & 0\\ 0 & 0 & 0 & 0\end{pmatrix}\frac{\mathrm{d}}{\mathrm{d}u}\begin{pmatrix}X_1\\X_2\\X_3\\X_4\end{pmatrix}\\ &= D\frac{\mathrm{d}}{\mathrm{d}u}\begin{pmatrix}X_2\\0\\0\\-X_1\end{pmatrix} \qquad \text{with} \qquad D = \frac{E}{E_\mathrm{crit}^\mathrm{QED}}v_\mathrm{inst}[\vec{X}(u)]\end{aligned} \tag{2.50}$$

in matrix form. Since a worldline instanton must be a closed path [i.e., $\vec{X}(0) = \vec{X}(1)$], we set $\mathrm{d}X_3/\mathrm{d}u = \mathrm{d}X_4/\mathrm{d}u = 0$ and choose $X_3 = X_4 = 0$ for simplicity (i.e., the resulting instanton trajectories lie in the $\mathcal{T}$–$x$ plane). The two remaining coupled differential equations can be combined into one harmonic-oscillator equation and are thus solved by a circular path parameterized by

$$X_1(u) = R\sin(Du) \qquad \text{and} \qquad X_2(u) = R\cos(Du) \tag{2.51}$$

with a radius $R > 0$. Note that we have centered this **instanton trajectory** at the origin—this arbitrary choice does not have any influence on the corresponding instanton action. The radius is fixed by the equation

$$\begin{aligned} v_\mathrm{inst}^2[\vec{X}(u)] &\overset{\text{(def.)}}{=} \frac{(\mathrm{d}\vec{X}/\mathrm{d}u)^2}{\bar{\lambda}_C^2} = \frac{R^2D^2}{\bar{\lambda}_C^2} = \frac{R^2}{\bar{\lambda}_C^2}\left(\frac{E}{E_\mathrm{crit}^\mathrm{QED}}\right)^2 v_\mathrm{inst}^2[\vec{X}(u)]\\ \Rightarrow \qquad R &= \frac{E_\mathrm{crit}^\mathrm{QED}}{E}\bar{\lambda}_C = \frac{mc^2}{qE}.\end{aligned} \tag{2.52}$$

In order to satisfy $\vec{X}(0) = \vec{X}(1)$, we demand

$$D \overset{!}{=} 2\pi n \qquad \text{with} \qquad n \in \mathbb{N}, \tag{2.53}$$

which fixes the instanton "velocity"

$$v_{\text{inst}}[\vec{X}(u)] = 2\pi n \frac{E_{\text{crit}}^{\text{QED}}}{E}. \tag{2.54}$$

Each of these possible instanton solutions has the same circular trajectory, but $n$ determines the number of (full) turns around the origin.

**Validity of the semiclassical approximation**

Remember that we have assumed in Eq. (2.45) that $v_{\text{inst}}[\vec{X}(u)]$ is a large number, so the electric field must be far below the Schwinger limit (semiclassical range) in principle—however, the worldline-instanton method yields the correct exponents governing the vacuum persistence amplitude in this simple example for any value of $E$.

Insertion of the instanton trajectory into Eq. (2.47) yields the resulting **instanton actions**:

$$\begin{aligned}\mathcal{A}_{\text{inst}}[\vec{X}(u)] &= 2\pi n \frac{E_{\text{crit}}^{\text{QED}}}{E} + \frac{qE}{\hbar c} \int_0^1 \frac{\mathrm{d}X_2(u)}{\mathrm{d}u} X_1(u)\,\mathrm{d}u \\ &= 2\pi n \frac{E_{\text{crit}}^{\text{QED}}}{E} - \frac{qE}{\hbar c} \underbrace{\frac{m^2c^4}{q^2E^2}}_{R^2} \underbrace{2\pi n}_{D} \underbrace{\int_0^1 \sin^2(2\pi n u)\,\mathrm{d}u}_{1/2} \\ &= \pi n \frac{E_{\text{crit}}^{\text{QED}}}{E}. \end{aligned} \tag{2.55}$$

We see that $n = 1$ is the dominating instanton solution (minimal instanton action) and corresponds to the leading-order exponent $-\pi E_{\text{crit}}^{\text{QED}}/E$ in the imaginary part of the Heisenberg–Euler effective action (2.3) calculated by Schwinger. Furthermore, even the subleading exponents are reproduced correctly by the higher-order instanton solutions ($n > 1$) in this example.

## 2.2. Tunneling picture of nonperturbative pair creation

**Single-electron picture**

In this section, we introduce the well-known tunneling picture of nonperturbative pair creation from the Dirac vacuum via a given electromagnetic field. Since we ignore the **interaction** between the created electrons and positrons as well as their **backreaction** on the electromagnetic field throughout this thesis, we may apply the classical Dirac equation (1.12) for a single electron ("first quantization") with an external electromagnetic field. Within this formalism, the creation of a pair manifests itself in the transition of a Dirac-sea electron to the upper energy continuum, leaving behind a hole in the Dirac sea (positron) [30].

**Constant $E$ field**

The special case of a static, homogeneous electric field leads to the **Sauter–Schwinger effect**. Such a field $\boldsymbol{E} = E\boldsymbol{e}_x$ can be described by the scalar potential $\Phi(x) = Ex$ via $\boldsymbol{E} = \boldsymbol{\nabla}\Phi$ (with zero vector potential). In a simple, semiclassical picture, $\Phi(x)$ gives rise to a position-dependent potential-energy contribution $q\Phi(x)$ to the two energy continua $\mathcal{E}_\pm = \pm\sqrt{m^2c^4 + c^2\boldsymbol{p}^2}$ of the free Dirac equation. The edges of both continua are given by the electron states with zero momentum ($\boldsymbol{p} = 0$). Due to the constant $E$ field, the energy levels of these edge states are tilted in space and read $\mathcal{E}_\pm^{p=0} = \pm mc^2 - qEx$ (see Fig. 2.2). The other states with $\boldsymbol{p} \neq 0$ lie above/below these two borders. Now, how can an electron from the lower continuum (Dirac sea) get into the upper continuum? Electron energy is conserved due to the time-independent gauge, so electrons can only move along paths of constant energy (constant height in Fig. 2.2). The **quantum-mechanical tunneling effect** allows a Dirac-sea electron on the edge of the lower continuum (at some spatial point $x_\star^-$) to tunnel through the classically forbidden region against the electric force until it eventually arrives at the lower edge of the upper energy continuum at some other position $x_\star^+$; see Fig. 2.2. The endpoints $x_\star^\pm$ of the tunneling region are usually referred to as the **(classical) turning points**, because this is where the classical motion of a point charge stops and changes its direction. The length of the tunneling region is given by

**Turning points and tunneling length**

$$\Delta x_\star = |x_\star^+ - x_\star^-| = \frac{2mc^2}{qE} = \frac{\lambda_C}{\pi}\frac{E_{\text{crit}}^{\text{QED}}}{E} \tag{2.56}$$

(cf. Fig. 2.2). We have estimated above that a measurable pair-creation rate is expected in the range $E/E_{\text{crit}}^{\text{QED}} = 0.01$–$0.1$, which thus roughly corresponds to tunneling lengths $\Delta x_\star$ of 3–30 Compton wavelengths.

In order to calculate the tunneling probability, one has to find the energy eigenfunctions of the Dirac equation within the linear potential $qEx$. This is precisely what Sauter studied in his work [18]: the wave function of a Dirac-sea electron decays exponentially along the way through the forbidden region, which leads to the nonperturbative, exponential factor $\exp(-\pi E_{\text{crit}}^{\text{QED}}/E)$ in Eq. (2.4) suppressing the pair-creation rate/tunneling probability [30].

Hence, the Sauter–Schwinger effect can be understood as the quantum tunneling phenomenon depicted in Fig. 2.2 according to the single-electron picture. In order to derive also the (nonexponential) prefactor in $\mathcal{P}_{\mathrm{e^+e^-}}$ within this formalism, one has to integrate over all possible electron states which contribute to the total tunneling current (all **transverse momenta $\boldsymbol{p}_\perp \perp \boldsymbol{E} \parallel \boldsymbol{e}_x$**). A nonzero transverse momentum $\boldsymbol{p}_\perp$ (which is conserved since $\Phi$ only depends

**Effective mass for states with transverse momenta**

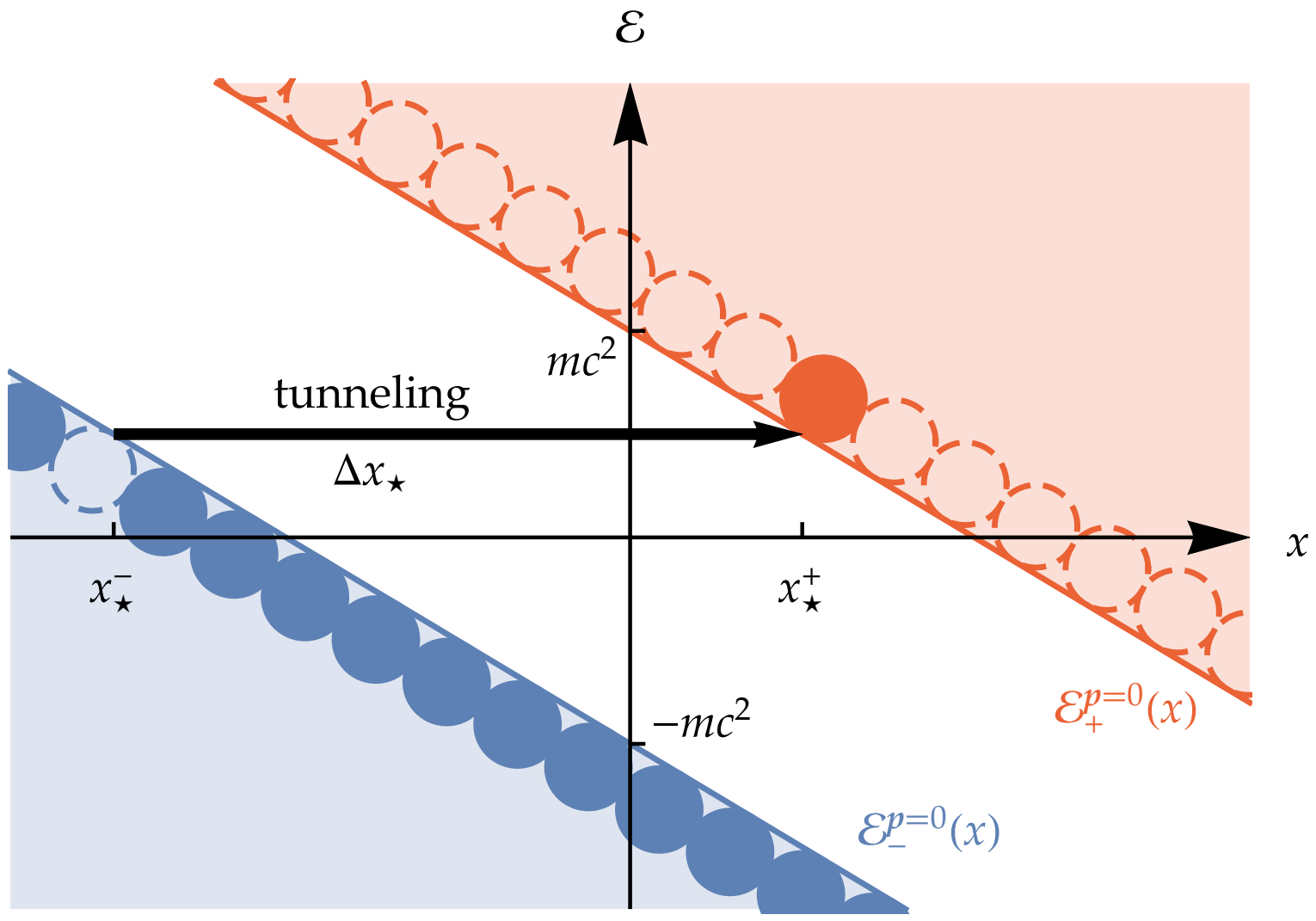

**Figure 2.2.:** Relativistic energy continua (blue and red areas) in space, tilted by a constant electric field $\boldsymbol{E} = E\boldsymbol{e}_x$. The edges $\mathcal{E}_\pm^{p=0} = \pm mc^2 - qEx$ are separated by the mass gap $2mc^2$. The filled (empty) circles represent occupied (free) electron states on these edges. Dirac-sea electrons may tunnel through the classically forbidden gap along a line of constant energy (e.g., the bold, horizontal arrow). The borders of the tunneling region (for a given $\mathcal{E}$) are the so-called (classical) turning points $x_\star^\pm$.

on $x$) lowers the tunneling rate because such an electron cannot move directly from $x_\star^-$ to $x_\star^+$ parallel to the $x$ axis—it also moves into the direction of $\boldsymbol{p}_\perp$ during tunneling, which increases the tunneling length and thus decreases the corresponding probability (exponentially). This whole effect can be easily incorporated into the single-electron formulas by substituting the electron rest mass $m$ with an **effective mass**

$$m_{p_\perp} = \sqrt{m^2 + \frac{\boldsymbol{p}_\perp^2}{c^2}} \geq m, \tag{2.57}$$

which depends on the transverse momentum of the considered electron state [18, 29, 42, 43, 44]. We may thus concentrate on electron states with $\boldsymbol{p}_\perp = 0$ in the following for simplicity.

## 2.3. Space-dependent electric fields

Since the spatial extent of any realistic electric field is finite, it is appropriate to study the effect of space-dependencies of the electric field strength on tunneling pair creation. We restrict ourselves to the 1+1-dimensional case here; that is, $\boldsymbol{E} = E(x)\boldsymbol{e}_x$, so we focus on the effect of a variable field strength along a straight electric field line. This static field can still be derived from a suitable scalar potential via $\boldsymbol{E} = \boldsymbol{\nabla}\Phi(x)$, which gives rise to the same tunneling interpretation of pair creation as above in Sec. 2.2, but with a more complicated shape of the $\mathcal{E}_\pm^{p=0}(x)$ curves than the straight lines in Fig. 2.2.

**Constant-field approximation**

First, note that if the electric field strength changes very slowly in space, we may approximate $E(x)$ at each point in space by a locally constant field and simply reuse the Sauter–Schwinger result (2.5) with $E \to E(x)$ to calculate a local density of the pair-creation rate at $x$ (which can then be integrated over the field volume to find the total pair-creation rate). This **constant-field approximation** should provide good results if the function $E(x)$ varies negligibly over the tunneling length (2.56) associated with $E(x)$; see, e.g., Ref. [45].

### 2.3.1. Spatial Sauter pulse

If the constant-field approximation is not justified, the full Dirac equation (1.12) with the potential $q\Phi(x)$ has to be considered in principle. One well-known example, in which case the Dirac equation is exactly solvable in terms of hypergeometric functions, is the electric field profile

$$E(x) = \frac{E_{\max}}{\cosh^2(kx)} \tag{2.58}$$

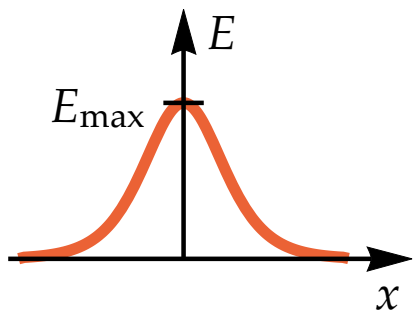

that was studied by Sauter [19]. This profile is therefore also known as (spatial) **Sauter pulse** in the literature. The peak value $E_{\max}$ of the electric field is located at $x = 0$, and the pulse width is described by the quantity $k > 0$ ($k = 0$ is the limiting case of a homogeneous field). The corresponding scalar potential

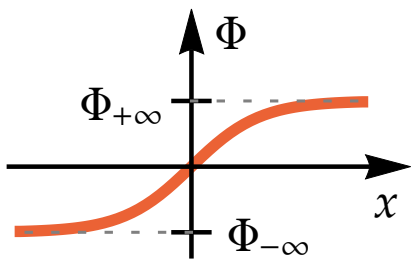

$$\Phi(x) = \frac{E_{\max}}{k} \tanh(kx) \tag{2.59}$$

approaches the values $\Phi_{\pm\infty} = \pm E_{\max}/k$ far away from the maximum ($x \to \pm\infty$), so this function is like a smooth potential step of height

$$\Delta\Phi = \Phi_{+\infty} - \Phi_{-\infty} = \frac{2E_{\max}}{k}. \tag{2.60}$$

The plots of the space-dependent upper and lower edges of the two relativistic energy continua in the case of an external Sauter-pulse field,

$$\mathcal{E}_{\pm}^{p=0} = \pm mc^2 - \frac{qE_{\max}}{k} \tanh(kx) \tag{2.61}$$

(semiclassical picture), in Fig. 2.3 show that it makes sense to distinguish between three different regimes in the magnitude of $\Delta\Phi$ [2] (see also Ref. [35] for a related discussion).

**Three regimes**

If the energy step $q\Delta\Phi$ is much larger than the mass gap $2mc^2$ [see Fig. 2.3(a)], the minimal possible tunneling length, which is located around the pulse center, falls well within the range $|kx| \ll 1$ where the Sauter pulse is approximately constant. We consider the minimal tunneling length because tunneling occurs there with the highest probability. Hence, this tunneling length $\Delta x_\star = |x_\star^+ - x_\star^-|$ approximately coincides with the constant-field result (2.56) with $E \to E_{\max}$, and so this is the case in which the **constant-field approximation** works well (in the vicinity of the pulse maximum at least, which, however, generates the dominant contribution to the total pair-creation rate).

If $q\Delta\Phi$ is close to $2mc^2$ but still greater than the mass gap [see Fig. 2.3(b)], the turning points $x_\star^\pm$ will be significantly greater than in the quasiconstant case. We also see in Fig. 2.3(b) that the bent parts of the energy curves $\mathcal{E}_\pm^{p=0}$ are now part of the tunneling region, so the constant-field approximation is no longer applicable here, and thus the **full Dirac equation** must be taken into account in this regime. In the limit $q\Delta\Phi \searrow 2mc^2$, the turning points diverge to infinity: $x_\star^\pm \to \pm\infty$.

This is precisely the transition to the third regime $q\Delta\Phi < 2mc^2$, within which **tunneling (pair creation) is impossible** because there is no constant-energy line connecting the upper Dirac-sea edge $\mathcal{E}_-^{p=0}$ with the lower edge $\mathcal{E}_+^{p=0}$ of the positive energy continuum anymore [see Fig. 2.3(c)]. This result is manifestly incompatible with the constant-field approximation, according to which a nonvanishing electric field always gives rise to a nonvanishing pair-creation rate [46].

All these conclusions, which have been drawn from the simple, semiclassical picture, are confirmed by more exact studies: Nikishov's work [29], for example, is based on the exact solutions of the Dirac equation with an external, spatial Sauter pulse. Other analytic studies like [33, 34, 35, 36, 47, 45] apply the worldline-instanton technique.

**Effect of space-dependent electric fields on pair creation**

The above explanation of tunneling pair creation induced by a spatial Sauter pulse serves as an example for the effect of space-dependencies of the electric field: if the field varies too rapidly in space, tunneling pair creation may no longer be understood as a local process (approximation of the local field strength as constant) [46], but the full Dirac equation must be considered, which can result in a much smaller tunneling rate than the constant-field approximation would predict (even zero if the field does not provide sufficient electrostatic energy $q\Delta\Phi$ to create a pair). This seems to be a general phenomenon in the context of the Sauter–Schwinger effect [34, 36, 48].

## 2.4. Time-dependent electric fields

Let us now turn to the effect of (pure) time dependencies of the electric field strength on tunneling pair creation; that is, we now assume that there is a homogeneous, time-dependent external field $\boldsymbol{E} = \boldsymbol{E}(t)$ with a fixed direction, say $\boldsymbol{e}_x$, as before. It is often advantageous to use the **temporal gauge** when treating this kind of fields, so the field is given by

$$\boldsymbol{E}(t) = \frac{\mathrm{d}\boldsymbol{A}(t)}{\mathrm{d}t} = \dot{A}(t)\boldsymbol{e}_x \tag{2.62}$$

with a time-dependent vector potential $\boldsymbol{A}$, while the scalar potential $\Phi$ is zero. For simplicity, we ignore all electron states with nonvanishing transverse momenta $\boldsymbol{p}_\perp$ again (quasi-1+1-dimensional case) since these can be covered easily by substituting the effective electron mass (2.57).

**No simple tunneling picture**

One important difference to the space-dependent case above is that the simple tunneling picture used before does *not* work here since energy is not a conserved quantity in a time-dependent scenario, and we do not have a space-dependent potential energy here. However, we still get the same result (2.4) for a constant $E$ field of course, which corresponds to $A(t) = Et$ in temporal

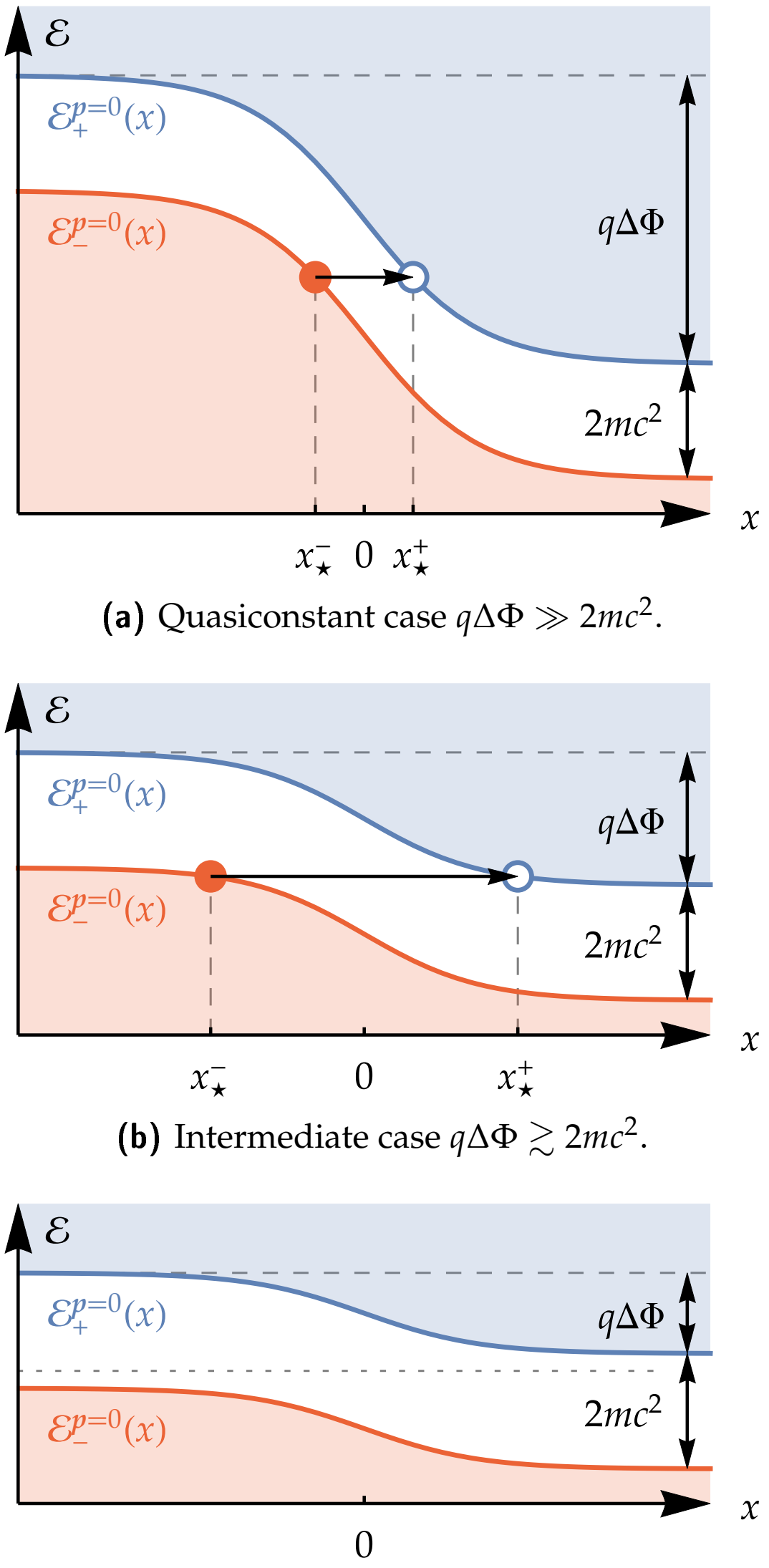

**(a)** Quasiconstant case $q\Delta\Phi \gg 2mc^2$.

**(b)** Intermediate case $q\Delta\Phi \gtrsim 2mc^2$.

**(c)** No-tunneling case $q\Delta\Phi \leq 2mc^2$.

**Figure 2.3.:** Space-dependent edges (2.61) of the relativistic energy continua plotted for three different values of $\Delta\Phi$ [(a)–(c)]. The turning points $x_\star^\pm$ corresponding to the minimal possible tunneling lengths (horizontal arrows) are marked in (a) and (b), respectively. The dotted line in (c) indicates that tunneling between the continua is impossible.

gauge [42, 49]—but the appearance of the nonperturbative, exponential factor $\exp(-\pi E_{\text{crit}}^{\text{QED}}/E)$ in $\mathcal{P}_{e^+e^-}$ is (arguably) not as easy to understand as in the tunneling picture.

### 2.4.1. Nonperturbative and perturbative regimes

**Constant-field approximation for time-dependent fields?**

As in the space-dependent case, one might ask under which circumstances the constant-field approximation (2.5) is appropriate for a time-dependent electric field[5]. This approximation works well in a space-dependent field which hardly varies over the local tunneling length (2.56) [45]. The problem is that this condition cannot be simply transferred to the time-dependent case because this would require us to specify a tunneling time—a concept which is not well defined.

**Pair-formation time**

So how can we find a **critical time/frequency scale for tunneling-type pair creation** in a constant $E$ field described in temporal gauge? A first guess could be the QED timescale set by the electron/positron mass: $\hbar/(mc^2)$; however, this corresponds to the photon energy $mc^2$, so electric fields with frequencies of this order will predominantly produce pairs via multiphoton processes—which is not the nonperturbative effect we are interested in here. The simplest timescale which can be associated with a constant electric field strength is obtained by multiplying $E$ with other physical constants appearing in QED such that we get a quantity with the dimension of time:

$$t_{\text{form}} = \frac{mc}{qE} = \frac{E_{\text{crit}}^{\text{QED}}}{E}\frac{\hbar}{mc^2}. \tag{2.63}$$

This timescale decreases with increasing $E$ and diverges in the limit $E \to 0$ just like the tunneling length $\Delta x_\star \propto 1/E$. It therefore seems to be a good candidate for a timescale associable with the **formation of a pair via tunneling**.

**Keldysh parameter**

This was indeed confirmed in the works [43, 50, 51] (see also [34] for an alternative approach using the worldline-instanton technique), in which the authors studied pair creation by a **sinusoidal electric field** $\boldsymbol{E}(t) = E_{\max}\cos(\omega t)\boldsymbol{e}_x$, that is, a field with a definite frequency $\omega$. Inspired by Keldysh's famous paper [52] on the ionization of atoms in an electromagnetic wave, it was found that the dimensionless so-called **Keldysh (adiabaticity) parameter**

$$\gamma_\omega = \frac{mc\omega}{qE_{\max}} = \frac{E_{\text{crit}}^{\text{QED}}}{E_{\max}}\frac{\hbar\omega}{mc^2} \tag{2.64}$$

[5] In this case, the field strength $E(t)$ inserted in $\dot{\mathcal{N}}_{e^+e^-}$ yields an instantaneous pair-creation rate, which can then be integrated over time to calculate the number of pairs created by the time-dependent field (per unit volume).

indicates whether the temporal variation of $E(t)$ affects tunneling pair creation in QED or not. Interestingly, $\gamma_\omega$ coincides with the ratio of the field frequency $\omega$ to the frequency scale $2\pi/t_{\mathrm{form}}$ corresponding to the formation time (2.63), a frequency scale associated with tunneling in the field $E_{\mathrm{max}}$; cf. [52, 43, 50, 51]. There are thus two well-distinguishable asymptotic regimes: If the field remains approximately constant during the "tunneling timescale" (in which case $\gamma_\omega \ll 1$), the time dependence of $E(t)$ will not interfere with tunneling, so the pair-creation process will still be tunneling-like and can be described well by the constant-field approximation (instantaneous pair-creation rate) [43, 50, 51].

**Tunneling regime $\gamma_\omega \ll 1$**

**Multiphoton regime $\gamma_\omega \gg 1$**

On the other hand, a large Keldysh parameter $\gamma_\omega \gg 1$ implies that the field varies much more rapidly than "tunneling can happen", so the process will be manifestly non-quasistatic. The resulting pair-creation probability (per unit four-volume) can be approximated by the leading-order term with respect to the small quantity $1/\gamma_\omega$ in this case, which leads to

$$\mathcal{P}_{\mathrm{e^+e^-}} \propto E_{\mathrm{max}}^2 \left( \frac{qE_{\mathrm{max}}}{mc\omega} \right)^{4mc^2/(\hbar\omega)} \qquad \text{for} \qquad \gamma_\omega \gg 1 \tag{2.65}$$

according to [43, 50, 51] and also [34]. Note that half the exponent, $N \approx 2mc^2/(\hbar\omega)$, equals the number of photons required to create an electron–positron pair. We can interpret the expression (2.65) as a **perturbation series** of $N$th order in $[qA_{\mathrm{max}}/(mc)]^2$, with $A_{\mathrm{max}} = E_{\mathrm{max}}/\omega$ denoting the amplitude of $A(t)$. Such a perturbation series arises when the external potential $A(t)$ is treated as a small perturbation to the free Dirac equation, and $N$ is the minimal order required to describe pair creation in the sinusoidal field [43, 50, 51, 34]. This leads to the conclusion that pairs are created via the absorption of a finite number of photons (**multiphoton pair creation**) in this $\gamma_\omega$ range—which is not the nonperturbative, tunneling-like Sauter–Schwinger effect with $\mathcal{P}_{\mathrm{e^+e^-}} \propto \exp(-\pi E_{\mathrm{crit}}^{\mathrm{QED}}/E)$ we are interested in.

Since every realistic electric field involves a time dependence of $\boldsymbol{E}$, which can be associated with a finite value of $\gamma_\omega$, every real pair-creation process will be a mixture of tunneling and multiphoton pair creation, especially in the intermediate regime $\gamma_\omega \approx 1$, which is referred to as "nonperturbative multiphoton regime" in [53]. However, the nonperturbative dependence of the (pure) Sauter–Schwinger effect on $qE$ is preserved in slowly varying fields ($\gamma_\omega \ll 1$). This is the regime we are primarily interested in throughout this thesis because tunneling pair creation from the Dirac vacuum has not been observed directly yet [54]—in contrast to pair creation in the perturbative regime (multiphoton absorption), which has been experimentally verified in a famous experiment [55] at the Stanford Linear Accelerator Center (SLAC).

**Temporal Sauter pulse**

Note that the concept of an adiabaticity parameter, which measures the "quasi-staticness" or "nonperturbativeness" of a time-dependent $E$ field, is not only applicable in the case of a sinusoidal (i.e., a single-frequency) field: another well-known example is the time-dependent version of the spatial Sauter pulse (2.58), a **temporal Sauter pulse**

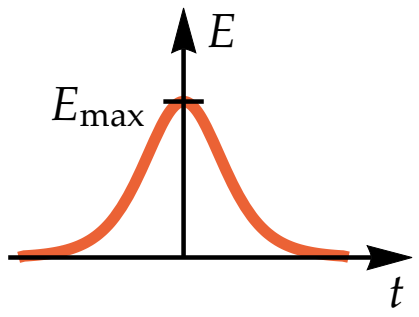

$$E(t) = \frac{E_{\max}}{\cosh^2(\omega t)}. \tag{2.66}$$

In this context, $\omega > 0$ specifies an inverse timescale which is related to the pulse width/duration. Figure 2.4 shows that this parameter $\omega$ also determines the order of the highest frequencies in the pulse. The low-frequency limit $\omega \to 0$ consequently leads to a static field (pure tunneling pair creation). Deviations from this result for finite values of $\omega$ were studied in Refs. [44, 34, 56]. It turned out that the **same Keldysh parameter** (2.64) as in the sinusoidal case above [but with $E_{\max}$ and $\omega$ from Eq. (2.66)] is appropriate to describe the adiabaticity of the temporal Sauter pulse. Hence, there is a nonperturbative regime $\gamma_\omega \ll 1$ and a perturbative regime $\gamma_\omega \gg 1$ (as well as a mixed intermediate regime) also for this pulsed field profile; see also [57]. In the nonperturbative regime, tunneling in a quasistatic electric field is the dominant pair-creation mechanism.

### 2.4.2. Scattering picture

**Time-dependent energies**

The problem we have to solve when considering the (classical) Dirac equation with a time-dependent vector potential in order to calculate pair-creation probabilities is as follows (see [42, 43, 44, 50, 58]): The vector potential becomes part of the relativistic electron energy levels via minimal coupling $\hat{\boldsymbol{p}} \to \hat{\boldsymbol{p}} + q\boldsymbol{A}(t)$, so the resulting energies are time dependent. We use the separation ansatz $\underline{\psi}(t, \boldsymbol{r}) = \underline{\phi}(t) \exp(\mathrm{i}\boldsymbol{k} \cdot \boldsymbol{r})$ for the (multicomponent) wave function in the Dirac equation, which corresponds to states with a defined canonical momentum ($\hat{\boldsymbol{p}} \to \boldsymbol{p} = \hbar\boldsymbol{k}$), and we can thus write the instantaneous electron energies of these states as

$$\mathcal{E}_\pm(t) = \pm\sqrt{m^2c^4 + c^2[\hbar\boldsymbol{k} + q\boldsymbol{A}(t)]^2}. \tag{2.67}$$

***In* and *out* frequencies**

An **unambiguous particle interpretation** is only possible for a vanishing external field (constant energies) [30]. Let us therefore assume that the field vanishes asymptotically for $t \to \pm\infty$, which corresponds to $A(t)$ and thus $\mathcal{E}_\pm(t)$ becoming asymptotically constant:

$$A(t \to \pm\infty) = A_{\pm\infty} = \text{const.} \tag{2.68}$$

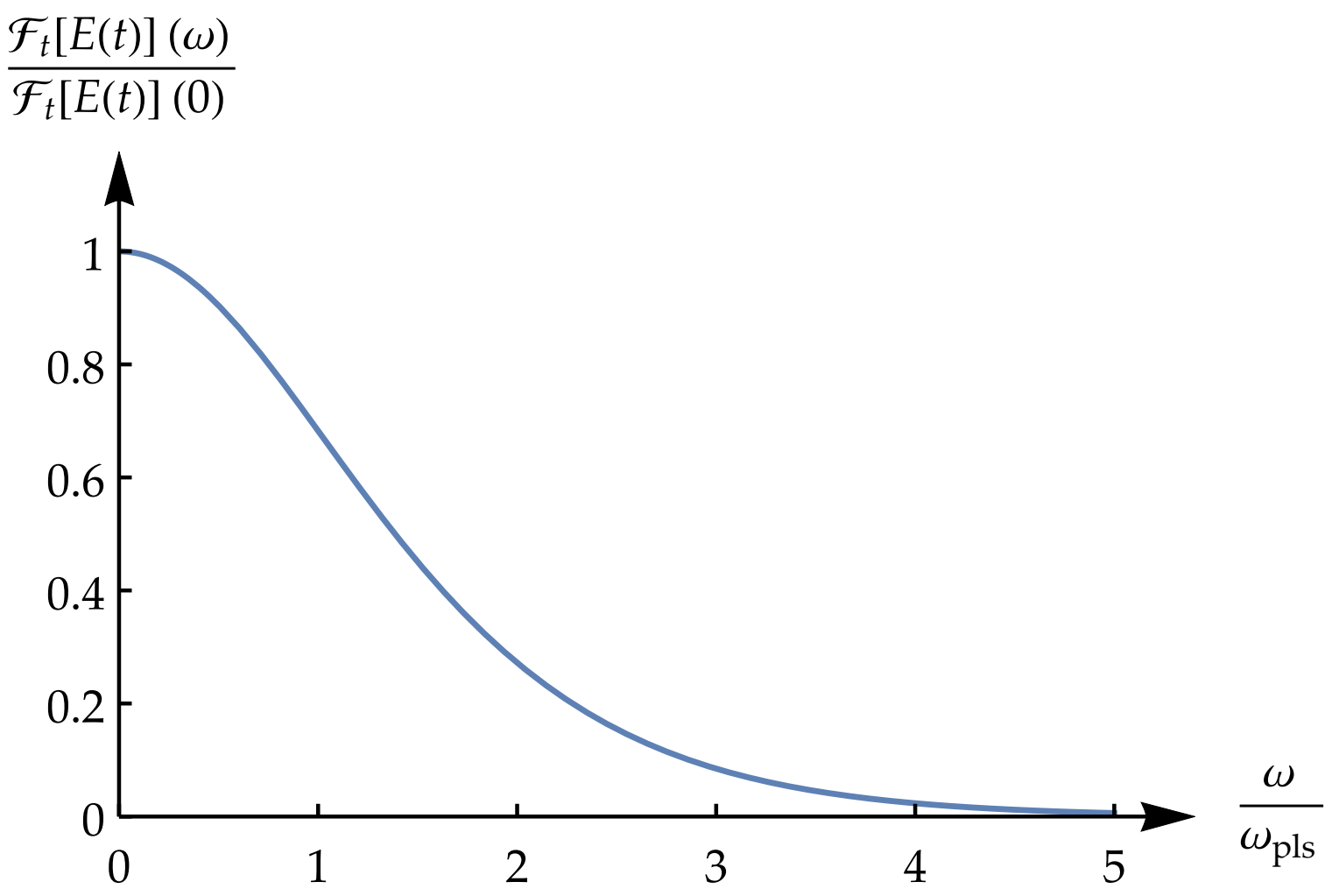

**Figure 2.4.:** Fourier transform $\sqrt{\pi/2}E_{\max}/\omega_{\text{pls}}^2 \times \omega/\sinh[\pi\omega/(2\omega_{\text{pls}})]$ of the temporal Sauter pulse $E(t) = E_{\max}/\cosh^2(\omega_{\text{pls}}t)$ plotted over $\omega$. We see that the highest frequencies which contribute significantly to the pulse are of the order of the pulse's frequency parameter $\omega_{\text{pls}}$.

There are thus two allowed frequencies

$$\omega_{\pm}^{\text{in}} = \frac{\mathcal{E}_{\pm}(t \to -\infty)}{\hbar} = \pm\frac{1}{\hbar}\sqrt{m^2c^4 + c^2(\hbar\boldsymbol{k} + qA_{-\infty}\boldsymbol{e}_x)^2} \gtrless 0 \tag{2.69}$$

for an electron at early times, and we can uniquely identify the lower, **negative frequency** $\omega_{-}^{\text{in}}$ with a **Dirac-sea state**. Since our initial condition is the Dirac vacuum (all electrons are in the Dirac sea), the time-dependent part of the wave function of each electron (which are indexed by $\boldsymbol{k}$) must satisfy $\underline{\phi}(t) \propto \exp(-\mathrm{i}\omega_{-}^{\text{in}}t)$ for $t \to -\infty$.

At late times, we can then decompose the wave function into the two possible outgoing frequencies

$$\omega_{\pm}^{\text{out}} = \frac{\mathcal{E}_{\pm}(t \to +\infty)}{\hbar} = \pm\frac{1}{\hbar}\sqrt{m^2c^4 + c^2(\hbar\boldsymbol{k} + qA_{+\infty}\boldsymbol{e}_x)^2} \gtrless 0. \tag{2.70}$$

The ratio of the $\omega_{+}^{\text{out}}$ component to the amplitude of the initial wave function $\propto \exp(-\mathrm{i}\omega_{-}^{\text{in}}t)$ gives the probability (via squaring the absolute value of this ratio) of finding the electron in the upper energy continuum after the electric field has been "active", so this corresponds to the **pair-creation probability** of the considered electron state. In conclusion, the time dependence of $A(t)$ has the potential to scatter a certain amount of the incoming negative-frequency wave into the outgoing positive-energy state (pair creation).

**Harmonic-oscillator picture for $|E(t)| \ll E_{\text{crit}}^{\text{QED}}$**

It is possible to derive a quite intuitive picture for this scattering process: if the electric field strength $E(t) = \dot{A}(t)$ is always well below $E_{\text{crit}}^{\text{QED}}$ (subcritical), which we assume throughout this thesis, the equations for the upper and lower components[6] of $\underline{\phi}(t)$ decouple (approximately), and both equations can be combined, yielding a harmonic-oscillator equation for one of the components[7] (say the upper, $\phi_+$),

$$\frac{\mathrm{d}^2\phi_+(t)}{\mathrm{d}t^2} + \omega_{\boldsymbol{k}}^2(t)\phi_+(t) = 0, \tag{2.71}$$

with the time-dependent eigenfrequency [43, 44, 50, 58]

$$\omega_{\boldsymbol{k}}(t) = \frac{1}{\hbar}\sqrt{m^2c^4 + c^2[\hbar\boldsymbol{k} + q\boldsymbol{A}(t)]^2}. \tag{2.72}$$

**Scattering picture**

Such a problem can be formulated equivalently as a nonrelativistic, **one-dimensional Schrödinger scattering problem** in space by interpreting $t$ as a spatial coordinate [44, 58, 59]. Comparing Eq. (2.71) to the time-independent Schrödinger equation

$$-\frac{\hbar^2}{2m}\frac{\mathrm{d}^2\psi(x)}{\mathrm{d}x^2} + V(x)\psi(x) = \mathcal{E}\psi(x), \tag{2.73}$$

we see that the harmonic-oscillator equation (2.71) describes scattering for the state with the "energy $\mathcal{E} = 0$" by the "space-dependent potential[8]" $-\hbar^2\omega_{\boldsymbol{k}}^2(t)/(2m)$; see Fig. 2.5 for an example. Our initial condition $\phi_+(t \to -\infty) \propto \exp(-\mathrm{i}\omega_-^{\text{in}}t)$ corresponds to a wave escaping to minus infinity with the "wave vector" $-\omega_-^{\text{in}}$ at the left end of the $t$ axis. On the right ($t \to +\infty$), there is an "incoming" wave $\mathfrak{T}\exp(-\mathrm{i}\omega_-^{\text{out}}t)$ from plus infinity, which is partly reflected at the inhomogeneous "potential" $-\hbar^2\omega_{\boldsymbol{k}}^2(t)/(2m)$ and thus gives rise to the third wave $\mathfrak{R}\exp(-\mathrm{i}\omega_+^{\text{out}}t)$ with the **reflection coefficient** $\mathfrak{R}$. Note that these waves oscillate for all times since our "potential" is always negative while we consider the state "$\mathcal{E} = 0$", so our waves can never meet a classically forbidden region (no tunneling). Let us assume that we adjust the coefficient $\mathfrak{T}$ in a way that the initial wave $\phi_+(t \to -\infty) \propto \exp(-\mathrm{i}\omega_-^{\text{in}}t)$ on the left-hand side is normalized. Then, the initial condition (electron starts in the Dirac sea) is satisfied and $|\mathfrak{R}|^2$ is the **pair-creation probability**. This whole process is depicted in Fig. 2.5.

[6] We assume scalar upper and lower components of the wave function here (no spin), which corresponds to the Dirac equation in 1+1 [Eq. (2.74)] or 2+1 [cf. Ch. 9] spacetime dimensions.

[7] In scalar QED, we have a scalar wave function and the Klein–Gordon equation is equivalent to the harmonic-oscillator equation (2.71) in the first place.

[8] Note that the physical dimension of this "potential" is not that of an energy since $t$ and $x$ do not have the same dimension.

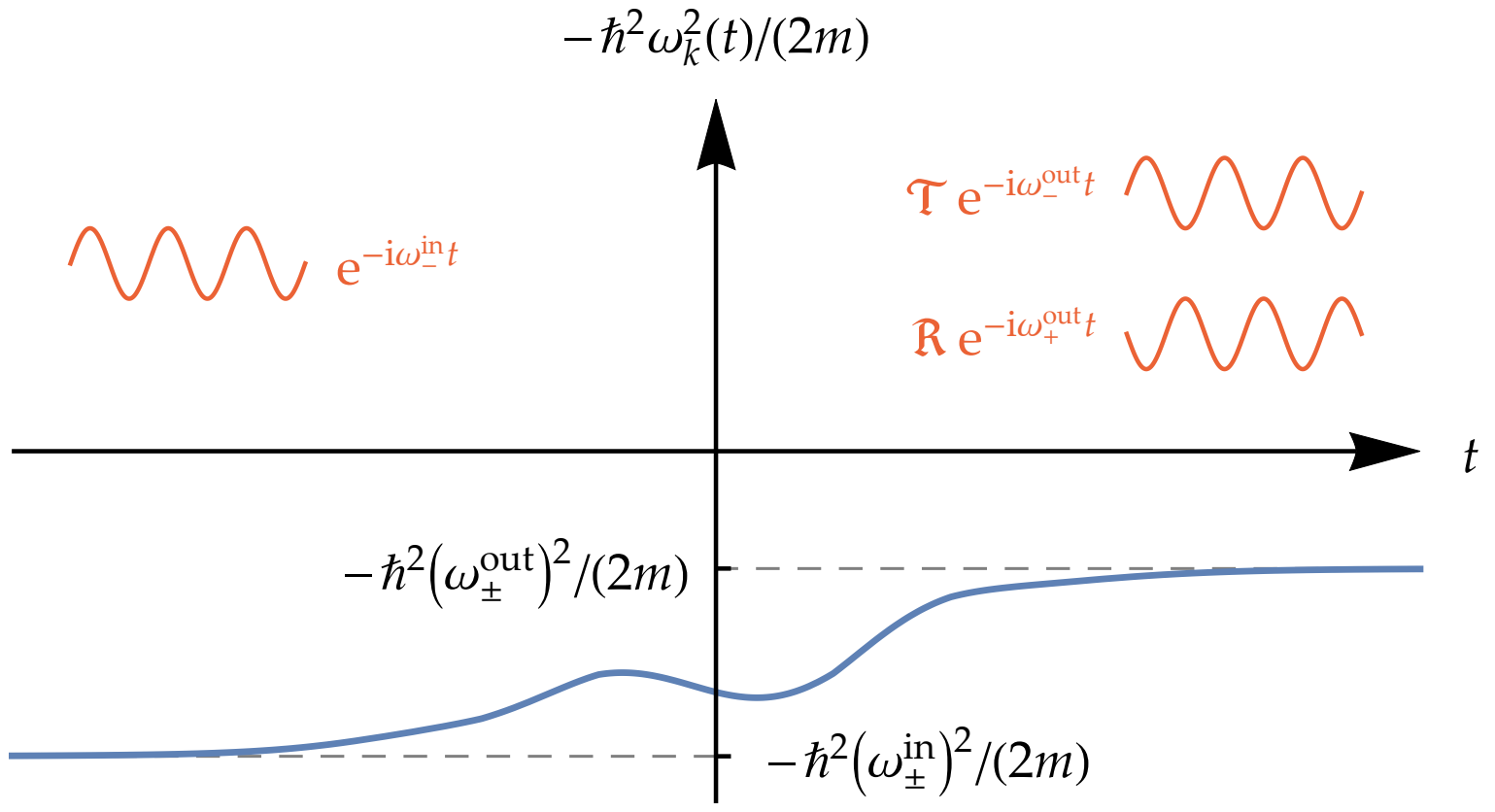

**Figure 2.5.:** Scattering picture of pair creation in a homogeneous, time-dependent electric field of the form $\boldsymbol{E}(t) = \dot{A}(t)\boldsymbol{e}_x$. The normalized wave $\propto \exp(-\mathrm{i}\omega_-^{\mathrm{in}}t)$ at early times corresponds to a Dirac-sea electron. Due to the electric field (time-dependent $\omega_k$), the final state is a mixture of the two possible frequencies $\omega_\pm^{\mathrm{out}}$, with coefficients $\mathfrak{T}$ and $\mathfrak{R}$. The probability $|\mathfrak{R}|^2$ of finding the electron in the positive energy continuum in the final state is the pair-creation probability.

### 2.4.3. Riccati equation in 1+1 spacetime dimensions

The problem of **finding the reflection coefficient** $\mathfrak{R}$ (see Fig. 2.5) by solving the second-order harmonic-oscillator equation (2.71) can be reformulated in a way to obtain $\mathfrak{R}$ more directly: it is possible to derive a first-order and nonlinear Riccati equation which is equivalent to the Dirac equation. The outgoing value ($t \to +\infty$) of the solution of this Riccati equation yields the reflection coefficient and thus the **pair-creation probability**. The Riccati-equation formalism was developed and used in [44, 58, 59, 60, 61, 62, 63, 1], for example, and we will present it here for a 1+1-dimensional spacetime with an external electric field $\boldsymbol{E}(t) = \dot{A}(t)\boldsymbol{e}_x$ (there are no magnetic fields in 1+1 dimensions).

**Natural units**

In order to keep the following equations concise, we use natural units in accordance with $c = \hbar = 1$ in this subsection.

#### Derivation

**Dirac equation in 1+1 dimensions**

The derivation starts with the Dirac equation (1.3) in 1+1 spacetime dimen-sions (i.e., with a single spatial coordinate $x$). The vector potential couples to the electron-momentum operator via **minimal coupling** (1.11), so the resulting classical **Dirac equation** (single-electron picture) reads

$$\mathrm{i}\gamma^0\partial_t\underline{\psi}(t,x) = \{\gamma^1[\hat{p}_x + qA(t)] + m\}\underline{\psi}(t,x), \tag{2.74}$$

where $\underline{\psi}$ denotes the Dirac spinor

$$\underline{\psi}(t,x) = \begin{pmatrix}\psi_+(t,x)\\ \psi_-(t,x)\end{pmatrix}, \tag{2.75}$$

which has two scalar components in 1+1 dimensions (no spin). For the two gamma matrices in this case, we choose the representation

$$\gamma^0 = \sigma_z = \begin{pmatrix}1 & 0\\ 0 & -1\end{pmatrix}, \quad \gamma^1 = \mathrm{i}\sigma_y = \begin{pmatrix}0 & 1\\ -1 & 0\end{pmatrix} \tag{2.76}$$

in terms of the Pauli matrices (1.7). Multiplying Eq. (2.74) by $\gamma^0$ yields the Dirac equation in the Schrödinger form:

$$\mathrm{i}\partial_t\underline{\psi}(t,x) = \begin{pmatrix}m & -\mathrm{i}\partial_x + qA(t)\\ -\mathrm{i}\partial_x + qA(t) & -m\end{pmatrix}\underline{\psi}(t,x). \tag{2.77}$$

**Transformation to momentum space**

We now express this Dirac equation in momentum space by inserting the inverse **Fourier transform** of $\underline{\tilde{\psi}}(t,k)$, the Dirac spinor in $k$ space:

$$\underline{\psi}(t,x) = \frac{1}{\sqrt{2\pi}}\int_{-\infty}^{\infty}\underline{\tilde{\psi}}(t,k)\,\mathrm{e}^{\mathrm{i}kx}\,\mathrm{d}k. \tag{2.78}$$

The derivatives $\partial_x$ in the Dirac equation (2.77) only act on the factor $\exp(\mathrm{i}kx)$ appearing in this inverse Fourier transform, so inserting Eq. (2.78) into Eq. (2.77) gives

$$\int_{-\infty}^{\infty}\left[\mathbb{1}\mathrm{i}\partial_t - \begin{pmatrix}m & k + qA(t)\\ k + qA(t) & -m\end{pmatrix}\right]\underline{\tilde{\psi}}(t,k)\,\mathrm{e}^{\mathrm{i}kx}\,\mathrm{d}k = 0. \tag{2.79}$$

Since this equation must be satisfied for all $x$ and the functions $\exp(\mathrm{i}kx)$ are linearly independent for different values of $k$, the integrand must vanish for each $k$, so each Fourier component satisfies an independent **Dirac equation in momentum space**:

$$\mathrm{i}\partial_t\underline{\tilde{\psi}}(t,k) = \underbrace{\begin{pmatrix}m & k + qA(t)\\ k + qA(t) & -m\end{pmatrix}}_{H_k(t)}\underline{\tilde{\psi}}(t,k). \tag{2.80}$$

The fact that the $k$ modes decouple implies that the canonical momentum $k$ is a **conserved quantity** here (as expected in the case of a purely time-dependent equation of motion). We may thus study the pair-creation probability for each $k$ (specifying an independent electron state) separately, so $k$ becomes a parameter in our following consideration of a single mode.

**Diagonalization of the Hamiltonian**

The single-electron Hamiltonian $H_k(t)$ in the momentum-space Dirac equation (2.80) above is a simple, $k$- and $t$-dependent $2 \times 2$ matrix, which can easily be diagonalized in order to obtain the **instantaneous energy eigenvalues** $\mathcal{E}_k(t)$ of the considered $k$ mode. Solving the characteristic-polynomial equation $\det[H_k(t) - \mathcal{E}_k(t)\mathbb{1}] = 0$ yields two different energies

$$\mathcal{E}_k^{\pm}(t) = \pm \underbrace{\sqrt{m^2 + [k + qA(t)]^2}}_{\Omega_k(t)} \gtreqless \pm m, \tag{2.81}$$

where the $\pm$ solution corresponds to a state in the upper/lower relativistic continuum [cf. Eq. (2.67) in the previous subsection]. The associated normalized **Dirac eigenspinors** read

$$\underline{u}_k^{+}(t) = \frac{1}{\sqrt{1 + d_k^2(t)}} \begin{pmatrix} 1 \\ d_k(t) \end{pmatrix} \quad \text{and} \quad \underline{u}_k^{-}(t) = \frac{1}{\sqrt{1 + d_k^2(t)}} \begin{pmatrix} -d_k(t) \\ 1 \end{pmatrix} \tag{2.82}$$

with the abbreviation

$$d_k(t) = \frac{k + qA(t)}{m + \Omega_k(t)}, \tag{2.83}$$

and they satisfy the (instantaneous) eigenvalue equation

$$H_k(t)\underline{u}_k^{\pm}(t) = \mathcal{E}_k^{\pm}(t)\underline{u}_k^{\pm}(t) = \pm\Omega_k(t)\underline{u}_k^{\pm}(t). \tag{2.84}$$

**Projection onto the eigenspinors**

We now want to project the Fourier component $\underline{\tilde{\psi}}(t,k)$ appearing in the Dirac equation (2.80) onto the eigenspinors (2.82). To this end, we expand $\underline{\tilde{\psi}}(t,k)$ in terms of $\underline{u}_k^{\pm}(t)$. During times of constant $A(t)$ (no electric field), the eigenspinors merely oscillate in time according to the Dirac equation (2.80), with the frequencies $\pm\Omega_k(t)$. Since $\Omega_k(t) \geq m$, these oscillations are much more rapid than the frequencies appearing in $A(t)$ for electric fields incorporating only photon energies (frequencies) much smaller than the mass gap ($|\omega| \ll 2m$). In our expansion, we will separate the **rapid oscillations** due to $\Omega_k(t)$ from the slower time development caused by the slowly varying $A(t)$.

**Phase function $\varphi_k(t)$**

To this end, we define the **phase function**

$$\varphi_k(t) = \int_{t_0}^{t} \Omega_k(t')\,\mathrm{d}t' \tag{2.85}$$

(with an arbitrary $t_0$) and write the corresponding phase factors explicitly in the **expansion**:

$$\underline{\tilde{\psi}}(t,k) = \alpha_k(t)\mathrm{e}^{-\mathrm{i}\varphi_k(t)}\underline{u}_k^+(t) + \beta_k(t)\mathrm{e}^{+\mathrm{i}\varphi_k(t)}\underline{u}_k^-(t). \tag{2.86}$$

This way, only the slowly varying $A(t)$ will cause the expansion coefficients $\alpha_k(t)$ and $\beta_k(t)$ to change in time[9].

**Evolution of $\alpha_k(t)$ and $\beta_k(t)$**

When inserting this expansion into the Dirac equation (2.80) and using $\mathrm{i}\partial_t \exp[\mp\mathrm{i}\varphi_k(t)] = \pm\Omega_k(t)\exp[\mp\mathrm{i}\varphi_k(t)]$ and the eigenvalue equation (2.84), the remaining terms are

$$\left[\dot{\alpha}_k(t)\underline{u}_k^+(t) + \alpha_k(t)\underline{\dot{u}}_k^+(t)\right]\mathrm{e}^{-\mathrm{i}\varphi_k(t)} + \left[\dot{\beta}_k(t)\underline{u}_k^-(t) + \beta_k(t)\underline{\dot{u}}_k^-(t)\right]\mathrm{e}^{+\mathrm{i}\varphi_k(t)} = 0. \tag{2.87}$$

Since the two eigenspinors are orthogonal, we can obtain two independent equations by projecting this equation onto $\underline{u}_k^\pm(t)$, respectively. The fact that our $\underline{u}_k^\pm(t)$ are normalized implies that

$$\underline{u}_k^\pm(t)\cdot\underline{u}_k^\pm(t) = 1 \quad \overset{\mathrm{d}/\mathrm{d}t}{\Rightarrow} \quad \underline{u}_k^\pm(t)\cdot\underline{\dot{u}}_k^\pm(t) = 0, \tag{2.88}$$

and from the orthogonality of $\underline{u}_k^\pm(t)$ follows that

$$\underline{u}_k^+(t)\cdot\underline{u}_k^-(t) = 0 \quad \overset{\mathrm{d}/\mathrm{d}t}{\Rightarrow} \quad \underline{\dot{u}}_k^+(t)\cdot\underline{u}_k^-(t) = -\underline{u}_k^+(t)\cdot\underline{\dot{u}}_k^-(t) = \Xi_k(t). \tag{2.89}$$

**Calculation of $\Xi_k(t)$**

We calculate $\Xi_k(t)$ by using this orthogonality and by inserting the eigenspinors (2.82) and the abbreviations therein:

$$\begin{aligned}
\Xi_k(t) = \underline{\dot{u}}_k^+(t)\cdot\underline{u}_k^-(t) &= \frac{\dot{d}_k(t)}{1+d_k^2(t)} = \frac{\mathrm{d}}{\mathrm{d}t}\arctan[d_k(t)] \\
&= \frac{\mathrm{d}}{\mathrm{d}t}\arctan\left[\frac{\frac{k+qA(t)}{m}}{1+\sqrt{1+\left(\frac{k+qA(t)}{m}\right)^2}}\right] \\
&= \frac{\mathrm{d}}{\mathrm{d}t}\frac{1}{2}\arctan\left[\frac{k+qA(t)}{m}\right] = \frac{\frac{q\dot{A}(t)}{2m}}{1+\left(\frac{k+qA(t)}{m}\right)^2} = \frac{mqE(t)}{2\Omega_k^2(t)}.
\end{aligned} \tag{2.90}$$

Taking the dot product of Eq. (2.87) with $\underline{u}_k^\pm(t)$, respectively, using the relations (2.88)–(2.89), then yields the **evolution equations**

$$\dot{\alpha}_k(t) = \Xi_k(t)\mathrm{e}^{2\mathrm{i}\varphi_k(t)}\beta_k(t) \qquad \text{and} \qquad \dot{\beta}_k(t) = -\Xi_k(t)\mathrm{e}^{-2\mathrm{i}\varphi_k(t)}\alpha_k(t) \tag{2.91}$$

[9] Note, however, that this is merely one possible way to define the coefficients and does not involve any approximations/assumptions yet.

of the expansion coefficients (also known as **Bogoliubov coefficients** in the literature). These two equations are completely equivalent to the Dirac equation.

**Conserved norm**

Using the evolution equations (2.91), it is easy to show that

$$|\alpha_k(t)|^2 + |\beta_k(t)|^2 = \text{const.} \tag{2.92}$$

is a **conserved quantity** since

$$\begin{aligned}
&\frac{\mathrm{d}}{\mathrm{d}t}\left[|\alpha_k(t)|^2 + |\beta_k(t)|^2\right] \\
&= \dot{\alpha}_k^*(t)\alpha_k(t) + \alpha_k^*(t)\dot{\alpha}_k(t) + \dot{\beta}_k^*(t)\beta_k(t) + \beta_k^*(t)\dot{\beta}_k(t) \\
&= \Xi_k(t)\left[\mathrm{e}^{-2\mathrm{i}\varphi_k(t)}\beta_k^*(t)\alpha_k(t) + \alpha_k^*(t)\mathrm{e}^{2\mathrm{i}\varphi_k(t)}\beta_k(t)\right. \\
&\qquad \left. - \mathrm{e}^{2\mathrm{i}\varphi_k(t)}\alpha_k^*(t)\beta_k(t) - \beta_k^*(t)\mathrm{e}^{-2\mathrm{i}\varphi_k(t)}\alpha_k(t)\right] \\
&= 0.
\end{aligned} \tag{2.93}$$

**Initial condition: Dirac vacuum**

Assuming that the external electric field vanishes in the initial state [i.e., $A(t \to -\infty) = \text{const.}$], the negative energy eigenvalue $-\Omega_k(t)$ can be clearly identified with the Dirac-sea state of the considered electron at early times, so our **initial condition** is

$$\alpha_k(t \to -\infty) = \alpha_k^{\text{in}} \overset{!}{=} 0 \tag{2.94}$$

because only then the initial state is proportional to $\underline{u}_k^-(t)$, the eigenspinor corresponding to the negative eigenvalue. This initial condition is valid for each mode, so every electron is initially located in the Dirac sea (Dirac vacuum). Furthermore, we **normalize the initial state** by setting

$$\beta_k(t \to -\infty) = \beta_k^{\text{in}} \overset{!}{=} 1, \tag{2.95}$$

so that

$$|\alpha_k(t)|^2 + |\beta_k(t)|^2 = |\alpha_k^{\text{in}}|^2 + |\beta_k^{\text{in}}|^2 = 1 \quad \forall t \tag{2.96}$$

due to the conservation law (2.92).

***Out* state**

At late times $t \to \infty$, when the external field vanishes (asymptotically) again, there is a simple relation between the outgoing value $\alpha_k^{\text{out}} = \alpha_k(t \to \infty)$ of the coefficient $\alpha_k$ and the probability $P_k^{\mathrm{e}^+\mathrm{e}^-}$ of finding the considered electron in the positive-energy state (**pair-creation probability** for this mode), due to the conserved norm (2.96):

$$P_k^{\mathrm{e}^+\mathrm{e}^-} = |\alpha_k^{\text{out}}|^2 = \frac{|\alpha_k^{\text{out}}|^2}{|\alpha_k^{\text{out}}|^2 + |\beta_k^{\text{out}}|^2} = \frac{\left|\frac{\alpha_k^{\text{out}}}{\beta_k^{\text{out}}}\right|^2}{\left|\frac{\alpha_k^{\text{out}}}{\beta_k^{\text{out}}}\right|^2 + 1} = \frac{|R_k^{\text{out}}|^2}{|R_k^{\text{out}}|^2 + 1}, \tag{2.97}$$

where we have defined the outgoing value $R_k^{\text{out}} = R_k(t \to \infty)$ of the ratio

$$R_k(t) = \frac{\alpha_k(t)}{\beta_k(t)}. \tag{2.98}$$

**Scattering picture**

Comparing our definitions here to the scattering picture for pair creation introduced in the previous subsection (see Fig. 2.5), we see that $\alpha_k^{\text{out}}$ corresponds to the **reflection coefficient** $\mathfrak{R}$ (whose absolute value squared gives the pair-creation probability), while $\beta_k^{\text{out}}$ coincides with the transmission coefficient $\mathfrak{T}$. Despite the symbol $R$, the outgoing value $R_k^{\text{out}}$ is thus *different* from the reflection coefficient in general. However, for small outgoing values $|R_k^{\text{out}}| \ll 1$, we may approximate the pair-creation probability (2.97):

$$P_k^{\mathrm{e^+e^-}} = |R_k^{\text{out}}|^2 + \mathcal{O}(|R_k^{\text{out}}|^4) \approx |R_k^{\text{out}}|^2, \tag{2.99}$$

so $R_k^{\text{out}}$ approximately equals the reflection coefficient $\mathfrak{R}$ in this case.

**Riccati equation**

According to Eq. (2.97), it is sufficient to integrate the ratio $R_k(t)$ from the initial to the final state in order to obtain the pair-creation probability, instead of integrating both coefficients $\alpha_k(t)$ and $\beta_k(t)$. The differential equation which describes the evolution of $R_k(t)$ can be derived by differentiating Eq. (2.98) with respect to $t$ and then inserting the evolution equations (2.91) of $\alpha_k(t)$ and $\beta_k(t)$. This way, we get

$$\begin{aligned} \dot{R}_k(t) &= \frac{\dot{\alpha}_k(t)\beta_k(t) - \alpha_k(t)\dot{\beta}_k(t)}{\beta_k^2(t)} \\ &= \frac{\Xi_k(t)\mathrm{e}^{2\mathrm{i}\varphi_k(t)}\beta_k^2(t) + \Xi_k(t)\mathrm{e}^{-2\mathrm{i}\varphi_k(t)}\alpha_k^2(t)}{\beta_k^2(t)} \end{aligned} \tag{2.100}$$

and thus arrive at the **Riccati equation**

$$\dot{R}_k(t) = \Xi_k(t)\left[\mathrm{e}^{2\mathrm{i}\varphi_k(t)} + \mathrm{e}^{-2\mathrm{i}\varphi_k(t)}R_k^2(t)\right], \tag{2.101}$$

which is a nonlinear, first-order differential equation for $R_k(t)$. The **initial conditions** (2.94) and (2.95) translate to

$$R_k(t \to -\infty) = 0. \tag{2.102}$$

The corresponding solution of the Riccati equation gives us the **pair-creation probability** for the considered electron state ($k$) in the external field $E(t) = \dot{A}(t)$. To this end, we have to evaluate $R_k(t \to +\infty) = R_k^{\text{out}}$ and insert this value into Eq. (2.97).

This is the exact Riccati-equation formalism, which allows us to treat pair-creation problems in time-dependent electric fields (here: in 1+1 spacetime dimensions), and this approach is completely equivalent to obtaining the pair-creation probability by solving the Dirac equation.

### 2.4.4. Linearized Riccati equation (semiclassical approximation) and its solution

The fact that the exact Riccati equation (2.101) is nonlinear makes it difficult to find analytic solutions, even for relatively simple vector potentials $A(t)$. So let us look for an approximation which is appropriate for the field profiles we are interested in. We continue to use **natural units** with $c = \hbar = 1$ in this subsection.

**Assumption:** $|R_k(t)|^2 \ll 1 \,\forall t$

In subcritical electric fields $|E(t)| \ll E_{\text{crit}}^{\text{QED}}$ we expect **small pair-creation probabilities** $P_k^{\text{e}^+\text{e}^-} \ll 1$, which correspond to small values $|R_k^{\text{out}}|^2 \ll 1$ according to Eq. (2.99). Unfortunately, small outgoing values $|R_k^{\text{out}}|^2 \ll 1$ do not necessarily require $|R_k(t)|^2$ to be small for all times—however, we assume that many typical field profiles with $|R_k^{\text{out}}|^2 \ll 1$ do also fulfill $|R_k(t)|^2 \ll 1$ for all $t$.

#### Numerical example

Let us consider a **temporal Sauter pulse**

$$E(t) = \frac{E_{\max}}{\cosh^2(\omega t)} \qquad \Leftarrow \qquad A(t) = \frac{E_{\max}}{\omega}\tanh(\omega t) \tag{2.103}$$

as an example to test this assumption. We introduce the **dimensionless quantities**

$$\chi = \frac{E_{\max}}{E_{\text{crit}}^{\text{QED}}}, \qquad \varpi = \frac{\omega}{2m}, \qquad \text{and} \qquad \tau = \omega t \tag{2.104}$$

for brevity in calculations, where $\varpi$ (the ratio of the frequency scale to the mass gap) can also be expressed via the temporal Keldysh parameter (2.64), which measures the "nonperturbativeness" of the pair-creation process in the time-dependent electric field:

$$\gamma_\omega = \frac{m\omega}{qE_{\max}} = \frac{2\varpi}{\chi} \qquad \Leftrightarrow \qquad \varpi = \frac{\chi\gamma_\omega}{2}. \tag{2.105}$$

We only consider the mode $k = 0$ in this example for simplicity. The corresponding Riccati equation (2.101) (in terms of the dimensionless time variable $\tau = \omega t$) reads

$$\frac{\mathrm{d}R_0(\tau)}{\mathrm{d}\tau} = \frac{\mathrm{e}^{2\mathrm{i}\varphi_0(\tau)} + \mathrm{e}^{-2\mathrm{i}\varphi_0(\tau)}R_0^2(\tau)}{2\gamma_\omega\cosh^2(\tau)\left[1 + \left(\frac{\tanh\tau}{\gamma_\omega}\right)^2\right]} \tag{2.106}$$

with the rapid-phase function

$$\varphi_0(\tau) = \frac{1}{\chi\gamma_\omega}\int\limits_0^\tau \sqrt{1 + \left(\frac{\tanh\tau'}{\gamma_\omega}\right)^2}\,\mathrm{d}\tau' \tag{2.107}$$

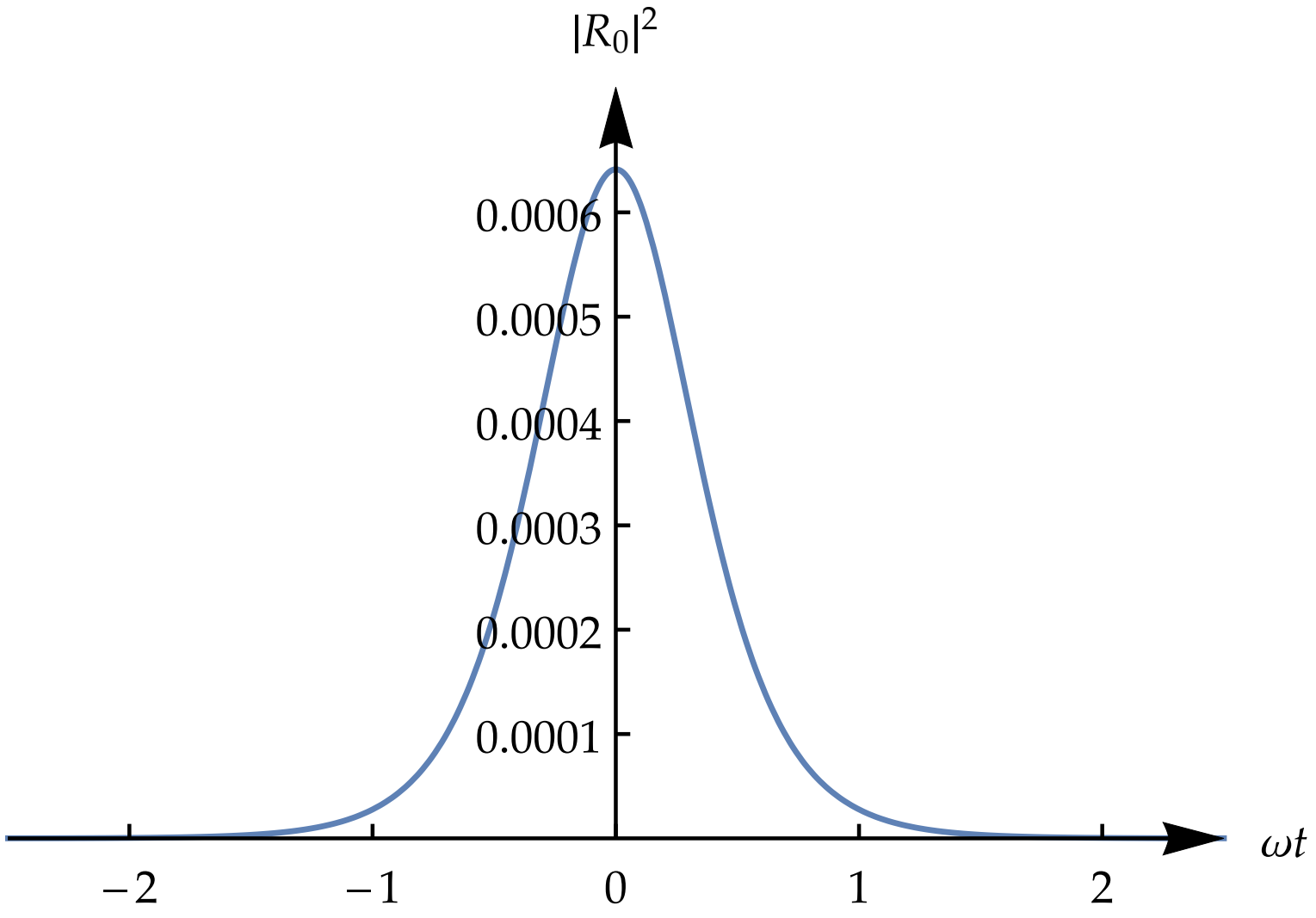

**Figure 2.6.**: Numerical solution (squared modulus) of the nonlinear Riccati equation (2.106) for the temporal-Sauter-pulse profile $E(t) = E_{\max}/\cosh^2(\omega t)$ and the mode $k = 0$. The parameter values are $\chi = E_{\max}/E_{\text{crit}}^{\text{QED}} = 0.1$ (strong but subcritical) and $\varpi = \omega/(2m) = 0.05$, which corresponds to the Keldysh parameter $\gamma_\omega = 1$ (i.e., between the tunneling regime and the multiphoton regime). Note that the initial condition $R_0 = 0$ has been imposed at the finite value $-\tau_{\max} = -10$ in this numerical example, and the outgoing value $|R_0^{\text{out}}|^2 \approx 5 \times 10^{-12}$ was evaluated at $\tau_{\max}$.

[where we have set $t_0 = 0$ in Eq. (2.85)] for this field profile. The **numerical solution** of this Riccati equation (2.106) is plotted in Fig. 2.6 for the parameter values $\chi = E_{\max}/E_{\text{crit}}^{\text{QED}} = 0.1$ and $\gamma_\omega = 1$ [corresponding to $\varpi = \omega/(2m) = 0.05$]. These parameter values are rather high [yet they still satisfy $|E(t)| \ll E_{\text{crit}}^{\text{QED}}$ and $\omega \ll 2m$ fairly well], so we expect larger values of $R_0(\tau)$ in this example than for weaker field configurations. We see in Fig. 2.6 that $|R_0(\tau)|^2$ is indeed much larger than the outgoing value $|R_0^{\text{out}}|^2 \approx 5 \times 10^{-12}$ (pair-creation probability) during intermediate times $|\tau| \lesssim 1$, but the assumption $|R_0(\tau)|^2 \ll 1$ is still justified there (and thus for all $\tau$).

#### Linearized Riccati equation

**Approximation: neglect the nonlinearity**

Inspired by the numerical example, we assume that $|R_k(t)|^2 \ll 1\, \forall t$ is a valid statement for many realistic field configurations within our boundaries (subcritical fields, frequencies below the electron mass). In this case, the nonlinear term $\propto R_k^2(t)$ in the Riccati equation (2.101) is always tiny in comparison to $\exp[2\mathrm{i}\varphi_k(t)]$, which has an absolute value of 1. Neglecting the nonlinear term should thus be a good approximation. Note that this approximation is known to leave the exponential dependence of $R_k^{\text{out}}$ on parameters (such as $E_{\text{max}}$ appearing in, e.g., $|R_k^{\text{out}}|^2 \propto \mathrm{e}^{-\pi E_{\text{crit}}^{\text{QED}}/E_{\text{max}}}$ for a quasistatic field) invariant, but the **nonexponential prefactor changes** due to this approximation [64, 65, 66, 1]. However, we are mainly interested in the **exponents** governing the pair-creation probability in this thesis, so this approximation is appropriate for our purpose.

After neglecting the nonlinearity, we arrive at the **linearized form of the Riccati equation** [1]

$$\dot{R}_k(t) \approx \Xi_k(t)\mathrm{e}^{2\mathrm{i}\varphi_k(t)}. \tag{2.108}$$

**Solution of the linearized equation**

This differential equation can be integrated directly [with the initial condition (2.102)] in order to calculate $R_k^{\text{out}}$:

$$\int_{-\infty}^{\infty} \dot{R}_k(t)\,\mathrm{d}t = \underbrace{R_k(t\to\infty)}_{R_k^{\text{out}}} - \underbrace{R_k(t\to-\infty)}_{=0}, \tag{2.109}$$

and thus we obtain an **integral representation** for $R_k^{\text{out}}$ according to the above approximation, which is essentially a form of the semiclassical Jeffreys–Wentzel–Kramers–Brillouin (**JWKB**) approximation within the Riccati-equation formalism:

$$\begin{aligned} R_k^{\text{out}} &\approx \int_{-\infty}^{\infty} \Xi_k(t)\mathrm{e}^{2\mathrm{i}\varphi_k(t)}\,\mathrm{d}t \\ &= \int_{-\infty}^{\infty} \frac{mq\dot{A}(t)}{2\Omega_k^2(t)} \exp\left[2\mathrm{i}\int_{t_0}^{t} \Omega_k(t')\,\mathrm{d}t'\right] \mathrm{d}t. \end{aligned} \tag{2.110}$$

In the second line of this expression, we have inserted the functions $\Xi_k(t)$ and $\varphi_k(t)$ from Eqs. (2.90) and (2.85). The quantity

$$\Omega_k(t) = \sqrt{m^2 + [k + qA(t)]^2}, \tag{2.111}$$

which has been introduced in Eq. (2.81), denotes the (positive) instantaneous eigenfrequency/relativistic energy of the considered electron.

In conclusion, given the vector potential $A(t)$ of an electric field, finding the pair-creation probability (for a mode $k$) by means of the semiclassical approximation comes down to the calculation of the integral (2.110).

#### 2.4.4.1. Contour-integration method (complex times)

**Deformation of the integration contour**

When calculating $R_k^{\text{out}}$ according to Eq. (2.110), the integration from the initial state ($t \to -\infty$) to the outgoing state ($t \to +\infty$) along the $t$ axis may be shifted into the complex $t$ plane by means of **Cauchy's integral theorem**; that is, while holding the endpoints fixed, we may arbitrarily deform the **integration contour** connecting these points as long as we do not cross any **singularities** or **branch cuts** in the integrand during this (continuous) deformation process. The underlying motivation to shift the contour of integration is to make the (outer) integrand $\Xi_k(t) \exp[2\mathrm{i}\varphi_k(t)]$ exponentially small, which becomes possible through the phase function $\varphi_k(t)$ acquiring a nonzero imaginary part for complex $t$. However, the singularities of the (analytic continuation of the) integrand in the complex plane prevent us to shift the contour arbitrarily far away from the real $t$ axis—and thus these singularities ultimately determine the value of the integral (2.110).

It makes sense to distinguish between **two groups of singularities** of the outer integrand $\Xi_k(t) \exp[2\mathrm{i}\varphi_k(t)]$:

**Singularities: in $A(t)$ or ...**

- On the one hand, the analytic continuation of the **vector potential** $A(t)$ could have singularities in the complex plane, which then automatically become singularities of the integrand. In this thesis, singularities of this type only occur in Sec. 2.4.6 on tunneling in a constant $E$ field assisted by a temporal Sauter pulse (the original form of the dynamically assisted Sauter–Schwinger effect [67]).

**... at $t_k^\star$ given by $\Omega_k(t_k^\star) = 0$**

- On the other hand, the nonexponential prefactor $\Xi_k(t)$ in the integrand clearly has **poles at the zeros of $\Omega_k(t)$**, which we denote by $t_k^\star$. These zeros are thus given by the solutions of the equation

$$\Omega_k(t_k^\star) = 0 \qquad \Leftrightarrow \qquad A(t_k^\star) = \frac{-k \pm \mathrm{i}m}{q}, \tag{2.112}$$

  so these singularities are totally independent of the singularities (if any) of the complex continuation of $A(t)$.

**Branch cuts originating from $t_k^\star$**

Note that the singularities at $t_k^\star$ are not simply poles of the integrand (2.110) since $\Omega_k(t)$ itself has a square-root-type branch point at each $t_k^\star$, which gives rise to a **branch cut** (or even multiple branch cuts in the case of higher-order zeros of the function underneath the square root in $\Omega_k$) originating from the respective $t_k^\star$. As a consequence, the phase function $\varphi_k(t)$ will have branch cuts along the very same paths in the complex $t$ plane, as will the whole integrand. In conclusion, the integrand (2.110) diverges at the zeros $t_k^\star$ of $\Omega_k(t)$, and branch cuts originate from these singularities. When deforming the integration contour, we thus have to circumvent the singularities $t_k^\star$ as well as the associated branch cuts.

**Paths of the branch cuts**

The paths the branch cuts take in the complex plane ultimately depend on the **definition of the square root** in $\Omega_k(t)$. Throughout this thesis, we consistently define the square root of a complex number $z$ by its **principal value**

$$\sqrt{z} = \sqrt{|z|}\mathrm{e}^{\mathrm{i}\arg(z)/2} \qquad \text{with} \qquad \arg z \in (-\pi, \pi]. \tag{2.113}$$

According to this definition, $\operatorname{Im}\sqrt{z}$ is discontinuous on the **negative real axis** (branch cut), and $\operatorname{Re}\sqrt{z} \geq 0\,\forall z$, where the case of equality ($\operatorname{Re}\sqrt{z} = 0$) is only true on the branch cut (apart from the branch point $z = 0$). Since $\Omega_k(t)$ *is* a square root, we may infer that

$$\Omega_k(t) = \underbrace{\operatorname{Re}\Omega_k(t)}_{>0} + \mathrm{i}\operatorname{Im}\Omega_k(t) \tag{2.114}$$

as long as we stay away from the branch points ($t_k^\star$) and cuts of $\Omega_k(t)$.

**Making the integrand exponentially small**

Now let us consider how the exponential factor $\exp[2\mathrm{i}\int\Omega_k(t')\,\mathrm{d}t']$ in the integrand (2.110) changes when we go **upwards in the complex plane** by an infinitesimal step, that is, $\mathrm{d}t' = \mathrm{i}\,\mathrm{d}\xi$ with $\mathrm{d}\xi > 0$: during this step, $2\mathrm{i}\Omega_k(t')\,\mathrm{d}t'$ is added to the argument of the exponential function, and since

$$\begin{aligned}\operatorname{Re}\Big[2\mathrm{i}\Omega_k(t')\,\mathrm{d}t'\Big] &= \operatorname{Re}\Big\{-2\Big[\underbrace{\operatorname{Re}\Omega_k(t')}_{>0} + \mathrm{i}\operatorname{Im}\Omega_k(t')\Big]\mathrm{d}\xi\Big\} \\ &= -2\operatorname{Re}[\Omega_k(t')]\,\mathrm{d}\xi < 0,\end{aligned} \tag{2.115}$$

going upwards in the complex plane suppresses the integrand (2.110) exponentially.

**Choosing the integration contour**

Our general approach when calculating $R_k^{\mathrm{out}}$ is thus as follows: We shift the integration contour from the real $t$ axis into the **upper complex half-plane** as far as possible (up to infinity) because the integrand (2.110) decays exponentially there and thus does not yield any finite contribution to the integral. But we have to **integrate around all singularities** $t_k^\star$ and the corresponding **branch cuts in the upper half-plane**, which generates nonvanishing contributions to $R_k^{\mathrm{out}}$ (we do not need to consider the integrand in the lower half-plane in the following). If a particular vector potential $A(t)$ introduces **additional singularities** in the upper half-plane as well, we also have to circumvent these nonholomorphic regions, of course.

### 2.4.5. Singularity structure in the complex plane

While understanding nonperturbative pair creation in the case of slowly varying time-dependent electric fields via the "over-the-top" scattering picture (see Fig. 2.5 on page 58) is not as intuitive as in the case of space-dependent fields, to which the tunneling picture applies, the semiclassical contour-integration method introduced in the previous subsection makes it much easier to understand the appearance of the nonperturbative factor $\exp(-\pi E_{\text{crit}}^{\text{QED}}/E)$ in the pair-creation probability in the case of a constant $E$ field (Sauter–Schwinger effect).

According to this method, every field profile $\boldsymbol{E}(t) = \dot{A}(t)\boldsymbol{e}_x$ has a characteristic **singularity structure** in the upper complex half-plane (the lower half-plane is not interesting for us), consisting of the zeros $t_k^\star$ of $\Omega_k(t)$ [see Eq. (2.112)] plus the singularities of $A(t)$ itself at complex times. When calculating $R_k^{\text{out}}$ via the integral (2.110), these singularities limit the way we can shift the integration contour upwards into the complex plane where the integrand is exponentially suppressed due to the term $\exp[2\mathrm{i}\varphi_k(t)]$—and thus the **minimal possible exponential suppression** of the integrand at a singularity is given by the (absolute value) of $\exp[2\mathrm{i}\varphi_k(t)]$ at this singularity.

**Example 1: constant $E$ field**

Let us consider a subcritical constant electric field $|E| \ll E_{\text{crit}}^{\text{QED}}$ (such that we may apply semiclassical methods) as an example [1]. The vector potential $A(t) = Et$ itself does not have any singularities in this case, and there is just one zero of $\Omega_k(t)$ in the upper complex half-plane for each mode $k$:

$$t_k^\star = \frac{-\hbar k + \mathrm{i}mc}{qE} \quad \Rightarrow \quad \Omega_k(t_k^\star) = \frac{1}{\hbar}\sqrt{m^2c^4 + c^2[\hbar k + qA(t_k^\star)]^2} = 0. \tag{2.116}$$

The exponential suppression of the integrand (2.110) at this singularity is given by

$$\left|\mathrm{e}^{2\mathrm{i}\varphi_k(t_k^\star)}\right| = \exp\left[-2\,\mathrm{Im}\int_0^{t_k^\star} \frac{1}{\hbar}\sqrt{m^2c^4 + c^2(\hbar k + qEt')^2}\,\mathrm{d}t'\right] = -\frac{\pi}{2}\frac{E_{\text{crit}}^{\text{QED}}}{E}, \tag{2.117}$$

so we may conclude

$$P_k^{\mathrm{e}^+\mathrm{e}^-} \approx |R_k^{\text{out}}|^2 \propto \mathrm{e}^{-\pi E_{\text{crit}}^{\text{QED}}/E}, \tag{2.118}$$

which means that the exponent appearing in the pair-creation probability of each mode coincides with the leading-order exponent in Schwinger's result (2.4). This example illustrates the origin of the **nonperturbative factor** $\exp(-\pi E_{\rm crit}^{\rm QED}/E)$ in a time-dependent setting within the Riccati-equation formalism. We will calculate the contour-integral representation of $R_k^{\rm out}$ for a constant $E$ field in much more detail in Ch. 3.

**Analogy to turning points**

Note that the complex singularity positions can be considered as the analog of the spatial turning points $x_\star$ in the tunneling picture (introduced in Sec. 2.2) [62]. The analogy becomes especially clear within the worldline-instanton formalism: For a constant $E$ field, the instanton trajectory in Euclidean spacetime is a circle around the origin with radius $mc^2/(qE)$ [see Eqs. (2.51)–(2.52) in Sec. 2.1.1], so it intersects the imaginary-time axis at $\pm\mathcal{T}_0 = \pm \mathrm{i}t_0 = mc/(qE)$—exactly at ($\pm$) the imaginary part of the singularity $t_k^\star$ above in Eq. (2.116).

**Example 2: temporal Sauter pulse**

Another field profile which can serve as an example for this analogy is the temporal Sauter pulse $A(t) = E_{\rm max}\tanh(\omega t)/\omega$ with the corresponding Keldysh parameter $\gamma_\omega = mc\omega/(qE_{\rm max})$. The resulting instanton trajectory [34] is parameterized by

$$x(u) = \frac{mc^2}{qE}\frac{\operatorname{arsinh}[\gamma_\omega\cos(2\pi u)]}{\gamma_\omega\sqrt{1+\gamma_\omega^2}} \quad \text{and}$$
$$\mathcal{T}(u) = \frac{mc}{qE}\frac{1}{\gamma_\omega}\arcsin\left[\frac{\gamma_\omega}{\sqrt{1+\gamma_\omega^2}}\sin(2\pi u)\right] \tag{2.119}$$

with $u \in [0,1]$. It is ellipse-like shaped, centered at the origin of the $\mathcal{T}$–$x$ plane, and intersects the $\mathcal{T}$ axis at

$$\pm\mathcal{T}_0 = \pm\frac{mc}{qE_{\rm max}}\frac{1}{\gamma_\omega}\arcsin\left(\frac{\gamma_\omega}{\sqrt{1+\gamma_\omega^2}}\right) = \pm\frac{\arctan\gamma_\omega}{\omega}. \tag{2.120}$$

When treating the same field profile within the Riccati-equation formalism (we consider only the mode $k=0$ here for simplicity), the zero of $\Omega_0(t)$ which lies closest to the real axis in the upper half-plane is given by

$$t_0^\star = \mathrm{i}\frac{\arctan\gamma_\omega}{\omega}. \tag{2.121}$$

**Singularities of $A(t)$**

So, again, the extent of the instanton trajectory with respect to the $\mathcal{T}$ axis coincides with the imaginary $t$ region confined by ($\pm$) the singularity (2.121). Note that the vector potential $A(t)$ itself has singularities at $t = \pm\mathrm{i}\pi(1/2+n)/\omega$ with $n \in \mathbb{N}_0$ in the case of a temporal Sauter pulse; however, all poles of the tanh function in the upper half-plane are always located above $t_0^\star$ since $\arctan\gamma_\omega < \pi/2$, so the position of $t_0^\star$ determines the exponential suppression of $R_0^{\rm out}$ when solving the contour integral (2.110).

**Singularities determine exponents**

In conclusion, we have briefly discussed the singularity structures of two simple field profiles. The singularities determine the **exponents** in $R_k^{\text{out}}$ and are thus highly relevant for us when studying pair creation in time-dependent electric fields. More sophisticated field profiles generally lead to more complicated singularity structures. It may also happen that several singularities lie similarly close to the real time axis, in which case all of those singularities must be considered since no singularity clearly dominates—this can lead to **interference** patterns in the momentum ($\hbar k$) spectrum of the pair-creation probability [68, 69, 60, 70, 61, 38, 62, 63].

**"Movement" of the singularities**

Furthermore, the way singularities depend on the parameters of the field profile (like $E_{\text{max}}$ and $\omega$) can tell us something about the associated **pair-creation mechanism**—consider the temporal Sauter pulse above again: For small Keldysh parameters $\gamma_\omega \ll 1$ (quasistatic regime), we can approximate the dominating singularity (2.121) by

$$t_0^\star \overset{\gamma_\omega \ll 1}{\approx} \mathrm{i}\frac{\gamma_\omega}{\omega} = \mathrm{i}\frac{mc}{qE_{\text{max}}}, \tag{2.122}$$

which equals the imaginary part of the singularity (2.116) in the constant-field case (with $E \leftrightarrow E_{\text{max}}$), so tunneling pair creation dominates in this regime. For large $\gamma_\omega \gg 1$ (multiphoton regime), on the other hand, we may use the identity $\arctan(1/x) = \pi/2 - \arctan x$ for $x > 0$ to approximate the singularity (2.121). This yields

$$t_0^\star \overset{\gamma_\omega \gg 1}{\approx} \frac{\mathrm{i}}{\omega}\left(\frac{\pi}{2} - \frac{1}{\gamma_\omega}\right) = \mathrm{i}\left(\frac{\pi}{2\omega} - \frac{qE_{\text{max}}}{mc\omega^2}\right) \sim qE_{\text{max}} \tag{2.123}$$

and leads to pair-creation probabilities depending perturbatively on $q$ and $E_{\text{max}}$ [49, 44, 58, 34]. In conclusion, the way in which (crucial) singularities "move" in the complex plane when parameters are varied is related to the pair-creation mechanism.

### 2.4.6. Dynamically assisted Sauter–Schwinger effect

The problem with observing nonperturbative pair creation out of the Dirac vacuum, which has not succeeded yet [54], is the huge critical field strength (1.21): $E_{\text{crit}}^{\text{QED}} \approx 10^{18}\,\text{V/m}$. Time dependencies of the field can significantly reduce the exponential suppression of pair creation [49, 44, 50, 58, 34, 36]—but for too large adiabaticity parameters ($\gamma_\omega \approx 1$ or higher), the underlying process can no longer be identified with the Sauter–Schwinger effect, like in the SLAC experiment [55].

In order to render also the verification of the nonperturbative aspect of pair creation in the laboratory possible (despite the limited available field strengths), the **dynamically assisted Sauter–Schwinger effect** was proposed [67]. The specific field profile considered in Ref. [67] consists of a strong and slow electric Sauter pulse combined with a weak and fast one:

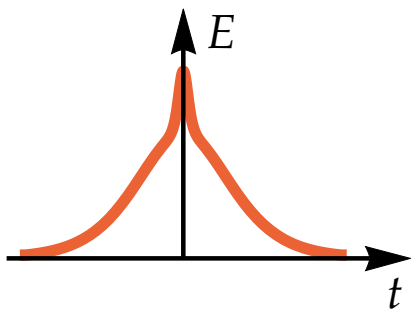

$$E(t) = \frac{E_{\text{strong}}}{\cosh^2(\omega_{\text{slow}} t)} + \frac{E_{\text{weak}}}{\cosh^2(\omega_{\text{fast}} t)}. \tag{2.124}$$

The Keldysh parameter of the first pulse should be small ($\gamma_{\omega_{\text{slow}}} \ll 1$); that is, this pulse is supposed to induce tunneling pair creation. Its peak field strength $E_{\text{strong}}$ must be subcritical ($E_{\text{strong}} \ll E_{\text{crit}}^{\text{QED}}$) to allow for semiclassical techniques (contour-integral representation of $R_k^{\text{out}}$, worldline-instanton method, etc.), but in principle $E_{\text{strong}}$ should be as strong as practically possible, which, unfortunately, does not lead to a measurable number of created pairs with present-day technology. The second pulse is intended to assist the tunneling process by **adding a component stimulating multiphoton processes**. The time dependence of this pulse should therefore be much faster ($\omega_{\text{fast}} \gg \omega_{\text{slow}}$) but still sufficiently far below $2mc^2/\hbar$ in order to suppress pair creation via this pulse alone. Furthermore, the pulse amplitude $E_{\text{weak}}$ is assumed to be very small ($E_{\text{weak}} \ll E_{\text{strong}}$), which is also realistic from an experimental point of view, but it must still be large enough to allow for the classical-field picture. All in all, the second pulse pulse alone cannot create a significant number of pairs as well, and its Keldysh parameter will be large ($\gamma_{\omega_{\text{fast}}} \gg 1$), which corresponds to the multiphoton regime.

**Effect of the combined pulses**

The outstanding finding in Ref. [67] is that the combination of the two pulses (2.124) will produce a much larger number of pairs than just the sum of the pairs created by the individual pulses (which would be a tiny number). In the context of this dynamically assisted Sauter–Schwinger effect, the so-called **combined Keldysh (adiabaticity) parameter**

$$\gamma_c = \frac{mc\omega_{\text{fast}}}{qE_{\text{strong}}} = \frac{E_{\text{crit}}^{\text{QED}}}{E_{\text{strong}}} \frac{\hbar\omega_{\text{fast}}}{mc^2} \tag{2.125}$$

determines whether pair creation will be enhanced due to this effect or not. For small values $\gamma_c < \gamma_c^{\text{crit}}$ with the **critical threshold**

$$\gamma_c^{\text{crit}} = \frac{\pi}{2}, \tag{2.126}$$

the effect of the fast pulse is negligible, so we essentially just get Sauter–Schwinger pair creation due to the first pulse then, which leads to a density of created pairs $\mathcal{N}_{e^+e^-} \propto \exp(-\pi E_{\text{crit}}^{\text{QED}}/E_{\text{strong}})$; cf. Eq. (2.4). But for $\gamma_c$ above the critical threshold value (2.126), the fast pulse actively enhances pair creation exponentially by (effectively) **modifying the pure Sauter–Schwinger exponent** $-\pi E_{\text{crit}}^{\text{QED}}/E_{\text{strong}}$ [67]:

$$\mathcal{N}_{e^+e^-} \propto \exp\left\{ -\frac{\pi E_{\text{crit}}^{\text{QED}}}{E_{\text{strong}}} \underbrace{\frac{2}{\pi}\left[\arcsin\left(\frac{\gamma_c^{\text{crit}}}{\gamma_c}\right) + \frac{\gamma_c^{\text{crit}}}{\gamma_c}\sqrt{1-\left(\frac{\gamma_c^{\text{crit}}}{\gamma_c}\right)^2}\right]}_{\text{factor (see marginal plot)}} \right\} \quad (2.127)$$

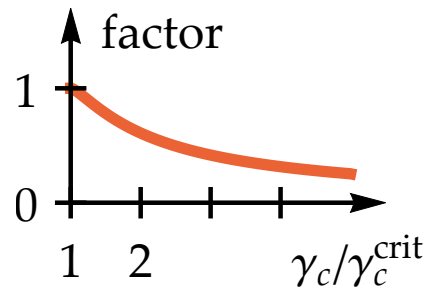

for $\gamma_c \geq \gamma_c^{\text{crit}}$. Note that our primary interest in the context of assisted tunneling pair creation is always the change of the exponent in $\mathcal{N}_{e^+e^-}$ since it describes the main effect (exponential enhancement of $\mathcal{N}_{e^+e^-}$). This simplification has recently been proven to be justified [41] by showing that the (non-exponential) prefactor in $\mathcal{N}_{e^+e^-}$ does not interfere negatively with the main features of various forms of dynamical assistance (via a weak, assisting Sauter pulse as above, as well as for an assisting oscillation or Gauss pulses etc., which will be studied in Part II).

The enhancement of the pure Sauter–Schwinger exponent due to the additional factor in Eq. (2.127) can have a huge effect on the pair-creation yield. For example, the factor is $1/2$ for $\gamma_c \approx 3.9$, which corresponds to an effective doubling of $E_{\text{strong}}$. For $E_{\text{strong}} = E_{\text{crit}}^{\text{QED}}/40$, for example, dynamical assistance thus has the potential to make the difference between $\exp(-\pi E_{\text{crit}}^{\text{QED}}/E_{\text{strong}}) \approx 10^{-55}$ and $\exp[-\pi E_{\text{crit}}^{\text{QED}}/(2E_{\text{strong}})] \approx 10^{-27}$.

A very simple picture illustrating the exponential enhancement of tunneling pair creation due to an additional multiphoton component (the fast pulse) is that its photons "lift" the Dirac-sea electrons up into the mass gap, which decreases the remaining tunneling length (see Fig. 2.7) [67]. However, this picture is incomplete because it cannot explain, for example, why the exponential amplification of $\mathcal{N}_{e^+e^-}$ only works for fast pulses with $\omega_{\text{fast}}$ above a certain critical value corresponding to $\gamma_c = \gamma_c^{\text{crit}}$ (for a fixed $E_{\text{strong}}$).

#### 2.4.6.1. Singularity structure

A better approach to understand the dynamically assisted Sauter–Schwinger effect is to consider the singularity structure of the field profile (2.124) in the complex $t$ plane; see [67, 62, 1]. Since the slow pulse is approximately constant during the time the fast pulse is "active" (because $\omega_{\text{slow}} \ll \omega_{\text{fast}}$), we may treat the slow pulse as a **constant background field** $E_{\text{strong}}$ when trying to understand the impact of the fast pulse on tunneling, so we study the **field profile** given by

$$A(t) = E_{\text{strong}}t + \frac{E_{\text{weak}}}{\omega_{\text{fast}}}\tanh(\omega_{\text{fast}}t) \quad (2.128)$$

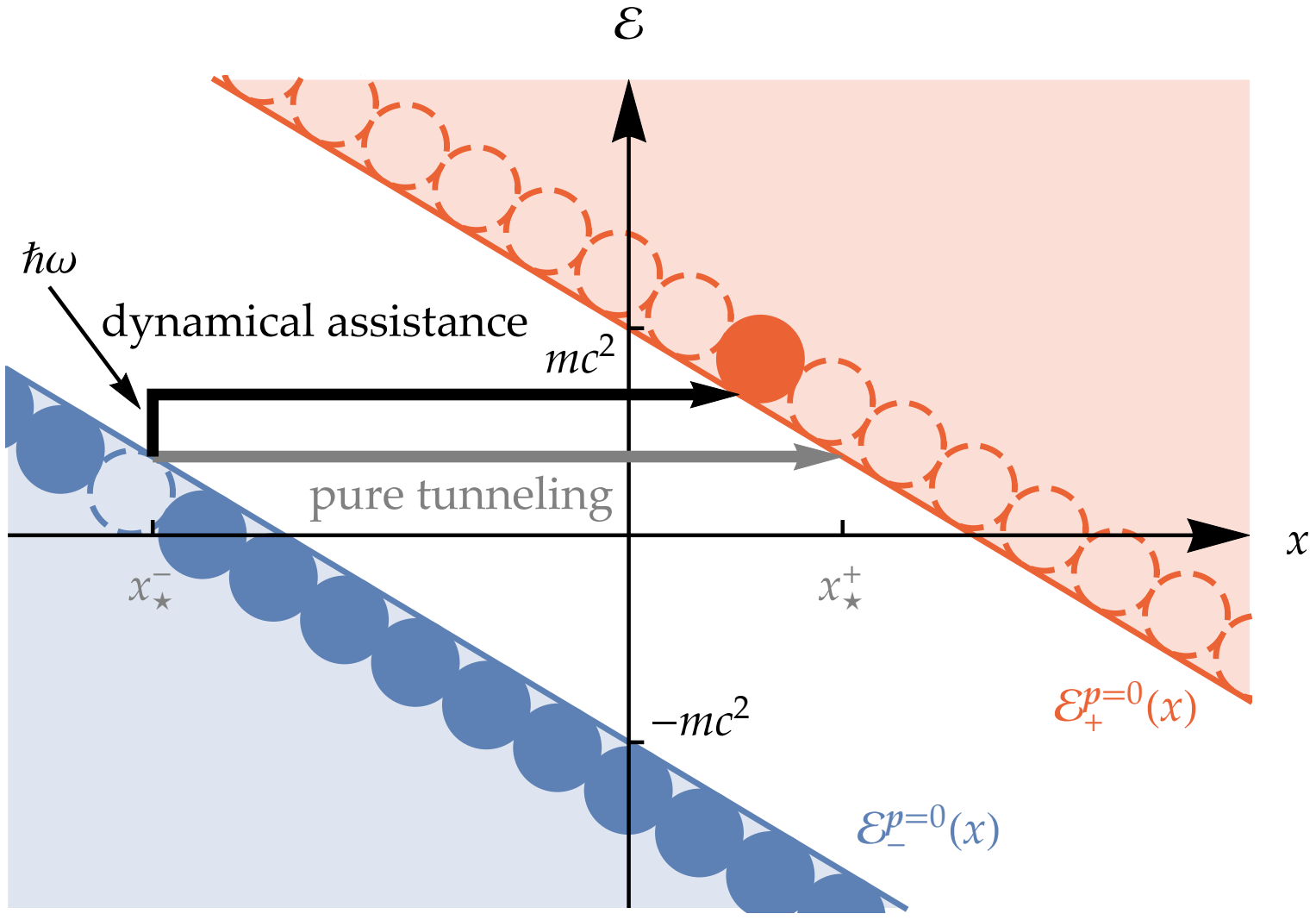

**Figure 2.7.:** Simplified picture of the dynamically assisted Sauter–Schwinger effect. The slow temporal Sauter pulse in Eq. (2.124) tilts the relativistic energy continua, which allows for tunneling pair creation along the gray arrow (ordinary Sauter–Schwinger effect). We can think of a photon with frequency $\omega$ provided by the additional fast pulse as lifting a Dirac-sea electron up by $\hbar\omega$ into the mass gap (vertical part of the black arrow), which reduces the remaining tunneling length (horizontal part). A shorter tunneling length leads to an exponential increase of the tunneling probability.

in the following.

The singularities $t_k^\star$ of the integrand in Eq. (2.110), the contour-integral representation of $R_k^{\text{out}}$, fulfill $\Omega_k(t_k^\star) = 0$ with

$$\Omega_k(t) = \frac{1}{\hbar}\sqrt{m^2c^4 + c^2\left[\hbar k + qE_{\text{strong}}t + \frac{qE_{\text{weak}}}{\omega_{\text{fast}}}\tanh(\omega_{\text{fast}}t)\right]^2}. \tag{2.129}$$

After introducing the **dimensionless quantities**

$$\varepsilon = \frac{E_{\text{weak}}}{E_{\text{strong}}}, \qquad \kappa = \frac{\hbar k}{mc}, \qquad \text{and} \qquad \tau = \omega_{\text{fast}}t, \tag{2.130}$$

the equation $\Omega_k(t_k^\star) = 0$ [see Eq. (2.112)] can be cast into the compact form

**Singularity equation**

$$\tau^\star + \varepsilon \tanh \tau^\star = (-\kappa \pm \mathrm{i})\gamma_c \tag{2.131}$$

(we omit to add a mode index to a solution $\tau^\star$ of this equation for brevity). Unfortunately, this is a transcendental equation which cannot be solved for $\tau^\star$ in closed form, but we can utilize our assumption $\varepsilon \ll 1$ to solve this equation approximately. Let us just consider the dominating mode $\kappa = k = 0$ here for simplicity[10]. Due to the symmetry $\Omega_{-k}(-t) = \Omega_k(t)$, we look for the crucial singularities on the imaginary $\tau$ axis, so we make the ansatz $\tau^\star = \mathrm{i}v$ in Eq. (2.131), which leads us to the real equation

$$v + \varepsilon \tan v = \gamma_c, \tag{2.132}$$

where we skipped the minus case in Eq. (2.131) since we are only interested in singularities in the upper half-plane ($v > 0$). Due to the periodicity of the tangent function, this equation has infinitely many solutions—however, the singularity with the smallest imaginary part determines the **exponential suppression** of the leading-order term in $R_0^{\text{out}}$, so we focus on that **"main" singularity**.

**Subcritical regime: $\gamma_c < \gamma_c^{\text{crit}}$**

For very small $\varepsilon$, the term $\varepsilon \tan v$ in Eq. (2.132) can be neglected, provided we stay away from the pole of the tangent at $v = \pi/2$ (see Fig. 2.8). Hence, if $\gamma_c < \pi/2 = \gamma_c^{\text{crit}}$, the crucial solution of the singularity equation is $v = \gamma_c$ (in the limit $\varepsilon \to 0$), so

$$\tau^\star_{\text{main}} \overset{\gamma_c<\pi/2}{\approx} \mathrm{i}\gamma_c = \frac{\mathrm{i}mc\omega_{\text{fast}}}{qE_{\text{strong}}} \qquad \Rightarrow \qquad t^\star_{0,\text{main}} \overset{\gamma_c<\pi/2}{\approx} \frac{\mathrm{i}mc}{qE_{\text{strong}}}. \tag{2.133}$$

[10] The fact that $A(t)$ in Eq. (2.128) is an odd function leads to a $k \leftrightarrow -k$ symmetry, so $k = 0$ is a preferred canonical momentum for this reason. It has been shown in [62] that this momentum makes the largest contribution to the total pair-creation yield for the field profile (2.128).

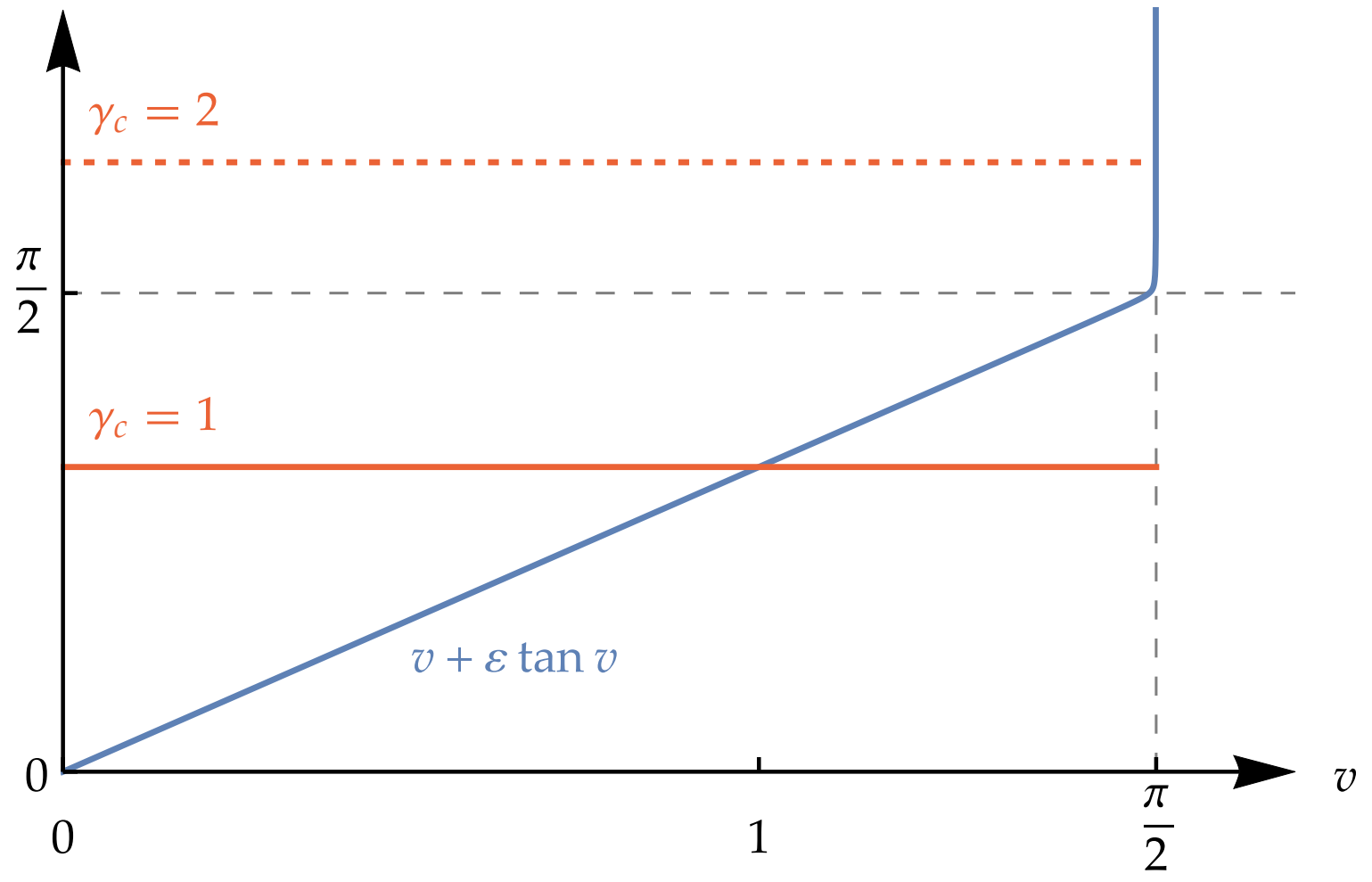

**Figure 2.8.:** Graphical solution of the singularity equation (2.132) for $\varepsilon = 10^{-4}$. We see that the left-hand side of the equation, $v + \varepsilon \tan v$ (blue plot line), is approximated well by the straight line $v$ for $v < \pi/2$, so the intersection point with a $\gamma_c < \pi/2$ (e.g., the horizontal red line) is located at $v \approx \gamma_c$ (ordinary Sauter–Schwinger effect). At $v = \pi/2$, however, the tangent has a pole which makes the blue function diverge abruptly. Thus, the blue graph intersects a $\gamma_c > \pi/2$ (such as the red dotted line) always at $v \approx \pi/2$, which leads to an effective reduction of the tunneling exponent (dynamical assistance).

This result coincides with the (single) singularity (2.116) from the constant-field case ($E_{\text{strong}}$ only), which gives rise to the **ordinary exponential Schwinger factor**: $R_0^{\text{out}} \propto \exp(-\pi E_{\text{crit}}^{\text{QED}}/E_{\text{strong}})$. In conclusion, the influence of the fast Sauter pulse on pair creation can be neglected for subcritical values of the combined Keldysh parameter ($\gamma_c < \gamma_c^{\text{crit}}$), and thus we get the pure tunneling result due to the strong, (quasi-)static field $E_{\text{strong}}$ in this range (ordinary Sauter–Schwinger effect).

**Supercritical regime: $\gamma_c \geq \gamma_c^{\text{crit}}$**

Above the threshold ($\gamma_c \geq \gamma_c^{\text{crit}}$), however, $\operatorname{Im} \tau_{\text{main}}^\star$ stops increasing with $\gamma_c$ because the pole of the tangent function in Eq. (2.132) acts like a "wall" in the limit $\varepsilon \to 0$ (see Fig. 2.8), so

$$\tau_{\text{main}}^\star \overset{\gamma_c \geq \pi/2}{\approx} \frac{\mathrm{i}\pi}{2} \qquad \Rightarrow \qquad t_{0,\text{main}}^\star \overset{\gamma_c \geq \pi/2}{\approx} \frac{\mathrm{i}\pi}{2\omega_{\text{fast}}}. \tag{2.134}$$

That is, the fast pulse manifestly influences the position of the main singularity in this $\gamma_c$ range (**dynamical assistance**). The leading-order exponential suppression of $R_0^{\text{out}}$ is given by $\exp[-2 \operatorname{Im} \varphi_0(t_{0,\text{main}}^\star)]$. We may neglect the contribution from the fast pulse to $\Omega_0(t)$ [i.e., the tanh function in Eq. (2.129)] when evaluating the phase function at $t_{0,\text{main}}^\star$ since this approximation hardly changes the result. We get

$$\begin{aligned}
& -2 \operatorname{Im} \varphi_0(t_{0,\text{main}}^\star) \\
&= -2 \operatorname{Im} \int_0^{t_{0,\text{main}}^\star} \Omega_0(t')\, \mathrm{d}t' \\
&\approx -2 \operatorname{Im} \int_0^{\mathrm{i}\pi/(2\omega_{\text{fast}})} \frac{1}{\hbar} \sqrt{m^2c^4 + c^2(qE_{\text{strong}}t')^2}\, \mathrm{d}t' \\
&= -2 \frac{E_{\text{crit}}^{\text{QED}}}{E_{\text{strong}}} \operatorname{Re} \int_0^{\pi/(2\gamma_c)} \sqrt{1-\xi^2}\, \mathrm{d}\xi \\
&= -\frac{E_{\text{crit}}^{\text{QED}}}{E_{\text{strong}}} \left[ \arcsin\left(\frac{\gamma_c^{\text{crit}}}{\gamma_c}\right) + \frac{\gamma_c^{\text{crit}}}{\gamma_c} \sqrt{1 - \left(\frac{\gamma_c^{\text{crit}}}{\gamma_c}\right)^2} \right]
\end{aligned} \tag{2.135}$$

for $\gamma_c \geq \gamma_c^{\text{crit}} = \pi/2$. From $R_0^{\text{out}} \sim \exp[-2 \operatorname{Im} \varphi_0(t_{0,\text{main}}^\star)]$ and the fact that $k = 0$ is the dominating mode, we conclude that $\mathcal{N}_{\mathrm{e}^+\mathrm{e}^-} \sim \exp[-4 \operatorname{Im} \varphi_0(t_{0,\text{main}}^\star)]$, which coincides with the **reduced Schwinger exponent** (2.127) from [67].

## Additional remarks

**Perturbativity with respect to the fast field**

Note that the resulting number (density) of created pairs (2.127) is still non-perturbative in the slow field $E_{\text{strong}}$; that is, the tunneling character of pair creation due to the slow field is preserved, but this process is assisted by the additional fast field. It has recently been shown that dynamical assistance by a fast Sauter pulse can be described well using first-order perturbation theory in the small quantity $E_{\text{weak}}/E_{\text{strong}}$ around the exact tunneling solution for the slow background field only [71]. However, for other types of time-dependent assisting fields (such as a Gauss pulse or a harmonic oscillation, which will be considered in Part II), first-order perturbation theory is not sufficient in general according to that reference.

**Interference effects**

For other modes ($k \neq 0$), the singularity associated with tunneling in the constant background field $E_{\text{strong}}$ is not located on the imaginary $t$ axis [see Eq. (2.116)], but rather has a nonvanishing real part $\propto k$. Close to the critical threshold ($\gamma_c \approx \pi/2$), this singularity is approximately at the same "height" ($\mathrm{i}\pi/2$ in the $\tau$ representation) as the first pole of the hyperbolic tangent $\varepsilon \tanh \tau$ in the upper half-plane, which is always located on the imaginary axis. Hence, there are two competing singularities in this situation, which leads to an interference pattern in the momentum spectrum in this $\gamma_c$ range [70, 62].

#### Goals in this thesis

The goal in Part II of this thesis is to study dynamically assisted tunneling for other fast-field profiles than a temporal Sauter pulse (see also, e.g., [72, 73, 74, 75, 76, 77, 78]). We will consider a **temporal Gauss pulse** $\propto \exp[-(\omega_{\text{fast}}t)^2]$ and see that, although it "looks" very similar to a Sauter pulse, it behaves differently in the context of dynamical assistance because the corresponding vector potential has no singularities in the complex plane—in contrast to the Sauter-pulse potential $\propto \tanh(\omega_{\text{fast}}t)$.

We will also study an **oscillating electric field** $\propto \cos(\omega_{\text{fast}}t)$, which should be good model for dynamical assistance via continuous laser beams. This profile is especially interesting when transferred to the analog of QED pair creation in semiconductors (electron–hole pair creation), which is the topic of Part III.

## 2.5. Spacetime-dependent electric fields

In the previous two sections, the external field depended either on the space coordinate (we consider a 1+1-dimensional spacetime here for simplicity) or on the time coordinate. Unfortunately, the simple physical pictures known from these cases—the tunneling picture (see Fig. 2.3 on page 52) for space-dependent fields and the scattering picture (see Fig. 2.5 on page 58) for time-dependent fields—are not applicable if the field is genuinely spacetime dependent[11]. Hence, studying nonperturbative pair creation in spacetime-dependent electric (or electromagnetic) fields is considerably more difficult, and thus most of the existing studies [86, 87, 88, 89, 90, 91, 92, 93] on this topic apply **numerical methods**. In the recent paper [90], the dynamically assisted Sauter–Schwinger effect (with an assisting temporal Sauter pulse) is derived analytically for an inhomogeneous, static background field (spatial Sauter pulse) via the **worldline-instanton technique**; see below for some results of this paper.

**Goals in this thesis**

Since spacetime-dependent fields are a more realistic model when it comes to possible experimental realizations of nonperturbative QED pair creation via laser beams/pulses, for example, it is desirable to learn more about the relation between pair creation and the shape of the stimulating field. However, the real QED effect is still a great challenge to observe experimentally [54], so condensed-matter analogs of QED in external fields could be a valuable "playground" to study nonperturbative pair creation in the laboratory since those analogs typically exhibit much less extreme scales. Such an analogy for QED in semiconductors which also works for spacetime-dependent field profiles will be derived in Part III of this thesis, and various results for the dynamically assisted Sauter–Schwinger effect in QED will be transferred to this analog.

### 2.5.1. Dynamically assisted Sauter–Schwinger effect with a spatial Sauter pulse as background field

Let us consider the analytic results in Ref. [90] a bit closer because we aim to adapt these results from QED to the semiconductor analog later (in Sec. 8.3).

In [90] as well as in the original paper [67] on the dynamically assisted Sauter–Schwinger effect, the **worldline-instanton method** (see Sec. 2.1) is applied to calculate pair-creation probabilities for specific electric-field profiles. Both references [67, 90] employ a temporal Sauter pulse $E_{\text{weak}}/\cosh^2(\omega_{\text{fast}} t)$ to assist tunneling, but the (approximately static) background field is homogeneous in [67], while [90] assumes a localized electric field with the shape of

[11] Note that if the external field depends only on one composed coordinate, such as on one of the light-cone coordinates $\propto ct \pm x$ as in [79, 80, 81, 82, 83, 84, 85] (see also [48]), for example, the pair-creation problem can still be formulated as an ordinary differential equation. As a consequence, useful techniques (such as the JWKB approximation) and intuitive pictures known from the space- or time-dependent cases remain applicable in a properly adjusted way.

a Sauter pulse (we denote $\omega_{\text{fast}}$ by $\omega$ here for brevity):

$$E(t,x) = \frac{E_{\text{strong}}}{\cosh^2(kx)} + \frac{E_{\text{weak}}}{\cosh^2(\omega t)}, \tag{2.136}$$

where $k > 0$ is an inverse wavelength scale describing the width of the spatial Sauter pulse (strong background field), and $\omega > 0$ is the frequency scale of the weak, assisting Sauter pulse.

**Keldysh parameters: $\gamma_\omega$, $\gamma_k$, $\gamma_c$**

Note that there are three different Keldysh parameters related to that field profile: The assisting temporal Sauter pulse is associated with a Keldysh parameter $\gamma_\omega = mc\omega/(qE_{\text{weak}})$ as defined in Eq. (2.64). The idea is that this pulse provides a weak-intensity ($E_{\text{weak}} \ll E_{\text{strong}}$) multiphoton contribution to the total pair-creation mechanism, so this Keldysh parameter should be large ($\gamma_\omega \gg 1$). The spatial Sauter pulse, in contrast, has a vanishing *temporal* Keldysh parameter because this field is static (pure tunneling regime).

**Spatial Keldysh parameter $\gamma_k$**

However, we can define a dimensionless **spatial Keldysh parameter** for this background field in analogy to the temporal Keldysh parameter in Eq. (2.64):

$$\gamma_k = \frac{mc^2 k}{qE_{\text{strong}}}. \tag{2.137}$$

The third interesting dimensionless quantity is the combined Keldysh parameter $\gamma_c = mc\omega/(qE_{\text{strong}})$ known from Eq. (2.125), which compares the frequency scale of the assisting temporal pulse to the maximum field strength of the static (or, in practice, very slow) and strong background field.

**Three regimes of $\gamma_k$**

Let us discuss the meaning of $\gamma_k$. We assume that the maximum field strength of the spatial Sauter pulse is fixed and subcritical ($E_{\text{strong}} \ll E_{\text{crit}}^{\text{QED}}$), so that we may apply semiclassical techniques (such as the worldline-instanton method). The limit $k \to 0$ and thus $\gamma_k \to 0$ corresponds to an infinite pulse width; that is, the background field becomes constant, so the temporal pulse will assist tunneling for $\omega > \omega_{\text{crit}}$ with $\omega_{\text{crit}}$ corresponding to $\gamma_c^{\text{crit}} = \pi/2$ [see Eq. (2.126)] according to the "ordinary" dynamically assisted Sauter–Schwinger effect [67]. For $k > 0$, the spatial pulse length scale is finite and we can distinguish between three regimes, which have already been discussed in Sec. 2.3.1 on the spatial Sauter pulse (see also Fig. 2.3 on page 52): If the pulse is broad ($\gamma_k \ll 1$) as in Fig. 2.3(a), there are many possible tunneling transitions near the pulse center which may be treated with the constant-field approximation. As $\gamma_k$ increases (i.e., the pulse becomes narrower), the number of possible transitions gets lower, so the associated tunneling current decreases. For $\gamma_k \lesssim 1$, the remaining tunneling transitions are significantly influenced by the exponential tails of the tanh potential corresponding to the spatial pulse [see Fig. 2.3(b)], so the tunneling lengths are larger than they would be in a constant field $E_{\text{strong}}$. The minimal tunneling length approaches infinity in the limit $\gamma_k \nearrow 1$, so this is precisely the no-tunneling limit, above which ($\gamma_k > 1$) tunneling due to the spatial background pulse is not possible [see Fig. 2.3(c)]. Since **assisted tunneling** is only a meaningful concept if the background field allows for tunneling without the temporal Sauter pulse, we focus on the range $\gamma_k < 1$ in the context of the dynamically assisted Sauter–Schwinger effect in inhomogeneous background fields.

**Dynamical assistance for $\gamma_k = 0$**

In order to understand how a localized background field affects dynamical assistance within the worldline-instanton formalism, we start with the homogeneous-field limit [67] (here: $k = \gamma_k = 0$). The instanton trajectory of the constant background field $E_{\text{strong}}$ alone is a circle with radius $mc^2/(qE_{\text{strong}})$ in the $c\mathcal{T}$–$x$ plane of Euclidean spacetime, centered at the origin [34] [see Eqs. (2.51)–(2.52)], so it intersects the $x$ axis at the spatial turning points $\pm x_\star(\gamma_k = 0)$ with

$$x_\star(\gamma_k = 0) = \frac{mc^2}{qE_{\text{strong}}} \tag{2.138}$$

[cf. Eq. (2.56)] and the $\mathcal{T}$ axis at $\pm\mathcal{T}_0(\gamma_k = 0)$ with

$$\mathcal{T}_0(\gamma_k = 0) = \frac{mc}{qE_{\text{strong}}} \tag{2.139}$$

[this coincides with the imaginary part of the constant-field singularity $t_k^\star$ (2.116) which appears within the Riccati-equation formalism]. The vector potential $E_{\text{weak}} \tanh(\omega t)/\omega$ of the additional temporal Sauter pulse becomes singular in Euclidean spacetime at $\pm\mathcal{T}_{\text{sing}}$ with

$$\mathcal{T}_{\text{sing}} = \frac{\pi}{2\omega} \tag{2.140}$$

because $\tanh(\omega t) = -\mathrm{i}\tan(\omega\mathcal{T})$ has poles there. Provided that its amplitude is small ($E_{\text{weak}} \ll E_{\text{strong}}$), the temporal pulse only has a negligible influence on the instanton trajectory of the background field (here: the circle) as long as the trajectory stays away from the singularities $\pm\mathcal{T}_{\text{sing}}$ (just as in Sec. 2.4.6). We may thus think of the assisting pulse as establishing "walls" which are parallel to the $x$ axis and located at $\pm\mathcal{T}_{\text{sing}}$ in Euclidean spacetime (see Fig. 2.9 and cf. Fig. 2.8 on page 76). Only when the **instanton trajectory touches these walls**, it will be "reflected" and thus be influenced by the temporal Sauter pulse—which is the effect of **dynamical assistance** in the worldline-instanton picture [90]. Figure 2.9 allows us to easily identify the onset frequency scale

$\omega_{\text{crit}}$ for dynamical assistance (for the present case $\gamma_k = 0$):

$$\frac{\pi}{2\omega_{\text{crit}}(\gamma_k=0)} \stackrel{!}{=} \mathcal{T}_0(\gamma_k=0) = \frac{mc}{qE_{\text{strong}}}$$
$$\Rightarrow \qquad \gamma_c^{\text{crit}}(\gamma_k=0) = \frac{mc\omega_{\text{crit}}(\gamma_k=0)}{qE_{\text{strong}}} = \frac{\pi}{2}, \tag{2.141}$$

which is precisely the result of [67].

**$\gamma_k > 0$ case: $\omega_{\text{crit}}(\gamma_k)$ decreases**

If we now increase $k$ (and thus $\gamma_k$), the instanton trajectory in Fig. 2.9 becomes that of a spatial Sauter pulse (provided it does not touch the "walls"), which is parameterized by [34]

$$x(u) = \underbrace{\frac{mc^2}{qE_{\text{strong}}}}_{x_\star(\gamma_k=0)} \frac{1}{\gamma_k} \operatorname{arsinh}\left[\frac{\gamma_k}{\sqrt{1-\gamma_k^2}}\sin(2\pi u)\right] \quad \text{and}$$
$$\mathcal{T}(u) = \underbrace{\frac{mc}{qE_{\text{strong}}}}_{\mathcal{T}_0(\gamma_k=0)} \frac{1}{\gamma_k\sqrt{1-\gamma_k^2}} \arcsin[\gamma_k\cos(2\pi u)] \tag{2.142}$$

with $u \in [0,1]$. This instanton loop is ellipse-like shaped because the intersection points with the axes, $\mathcal{T}_0$ and $x_\star$, depend differently on $\gamma_k$:

$$\mathcal{T}_0(\gamma_k) = \mathcal{T}_0(\gamma_k=0)\frac{\arcsin\gamma_k}{\gamma_k\sqrt{1-\gamma_k^2}}, \tag{2.143}$$

$$x_\star(\gamma_k) = x_\star(\gamma_k=0)\frac{1}{\gamma_k}\operatorname{arsinh}\left(\frac{\gamma_k}{\sqrt{1-\gamma_k^2}}\right). \tag{2.144}$$

Both values are strictly increasing with respect to $\gamma_k$ and diverge in the no-tunneling limit $\gamma_k \nearrow 1$. This **growth of the instanton trajectory** immediately makes clear (in view of Fig. 2.9 again) that $\omega_{\text{crit}}(\gamma_k)$ decreases as we increase $\gamma_k$ because

$$\frac{\pi}{2\omega_{\text{crit}}(\gamma_k)} \stackrel{!}{=} \mathcal{T}_0(\gamma_k) \quad \Leftrightarrow \quad \gamma_c^{\text{crit}}(\gamma_k) = \frac{mc\omega_{\text{crit}}(\gamma_k)}{qE_{\text{strong}}} = \frac{\pi}{2}\frac{\gamma_k\sqrt{1-\gamma_k^2}}{\arcsin\gamma_k}; \tag{2.145}$$

cf. Ref. [90]. For a background field close to the tunneling threshold ($\gamma_k \nearrow 1$ limit), we thus have

$$\gamma_c^{\text{crit}}(\gamma_k) \propto \omega_{\text{crit}}(\gamma_k) \sim \sqrt{1-\gamma_k^2}, \tag{2.146}$$

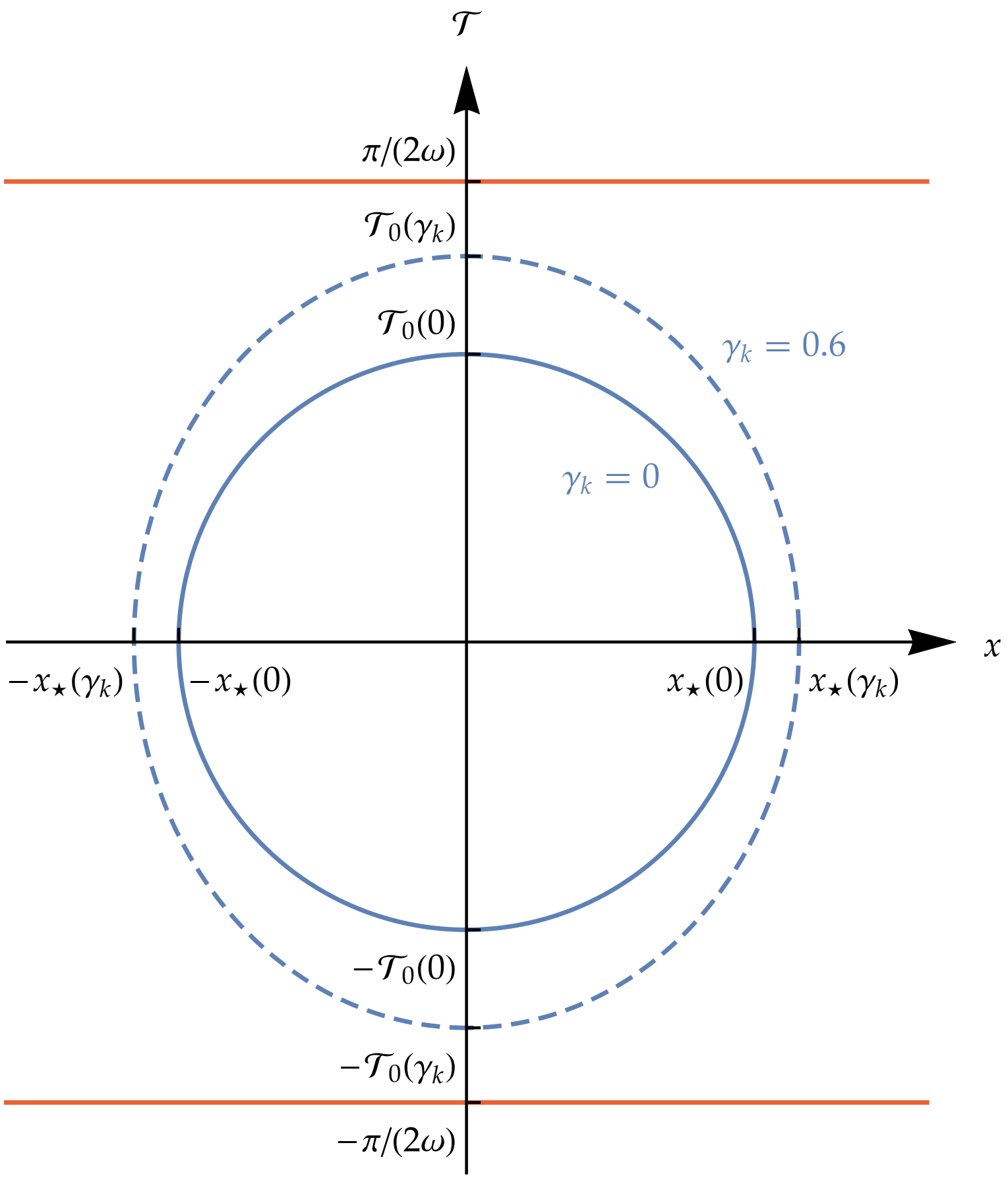

**Figure 2.9.:** Instanton trajectory (blue loops) of a static electric field $E_{\text{strong}}/\cosh^2(kx)$ in Euclidean spacetime [see Eq. (2.142) for the parameterization]. The red horizontal lines represent the "walls" [singularities at $\pm\mathcal{T}_{\text{sing}} = \pm\pi/(2\omega)$] introduced by an additional temporal Sauter pulse $E_{\text{weak}}/\cosh^2(\omega t)$ with a weak amplitude $E_{\text{weak}} \ll E_{\text{strong}}$. For a constant background field ($\gamma_k = 0$), the instanton trajectory is a circle (solid loop) around the origin with radius $c\mathcal{T}_0 = x_\star = mc^2/(qE_{\text{strong}})$. For $\gamma_k > 0$, the instanton trajectory expands (dashed loop) and thus $\mathcal{T}_0$ and $x_\star$ increase according to Eqs. (2.143) and (2.144).

so $\omega_{\text{crit}}(\gamma_k)$ approaches zero in the no-tunneling limit.

In conclusion, the physical effect of an inhomogeneous and localized background field (instead of a constant field as in the purely time-dependent case [67]) on the dynamically assisted Sauter–Schwinger effect is that the critical frequency scale $\omega_{\text{crit}}$ for dynamical assistance via the temporal Sauter pulse decreases and can even become arbitrarily small for static fields close to the tunneling threshold $q\Delta\Phi = 2mc^2 \Leftrightarrow \gamma_k = 1$. However, keep in mind that the tunneling current induced by the static field also decreases considerably when the spatial pulse is narrowed (by increasing $k$) [34], so even the assisted tunneling current may become too tiny to observe when approaching the tunneling threshold too close.

Just like $\omega_{\text{crit}}(\gamma_k)$ approaches zero according to a characteristic scaling law (2.146) in the limit $\gamma_k \nearrow 1$, the pure tunneling current (without dynamical assistance) also exhibits a characteristic functional form in this limit. This form depends on the way the static field approaches zero asymptotically for $x \to \pm\infty$—see Refs. [94, 95] for more information on this kind of universal phenomena in the no-tunneling limit.

**Goals in this thesis**

One of the goals in this thesis is to transfer this QED scenario to the semiconductor analog and to show that—even though the spatial Sauter pulse in the field profile (2.136) considered here is a rather specific background-field profile—the scaling law (2.146) for $\omega_{\text{crit}}$ in the no-tunneling limit should also be valid for such an analog (see Sec. 8.3.1).

Furthermore, we will estimate how dynamical assistance via a temporal harmonic oscillation $E_{\text{weak}} \cos(\omega t)$—which is considered in Ch. 5 with a constant background field—is influenced by choosing a spatial Sauter pulse as background field instead (see Sec. 8.3.2). A harmonic oscillation is a better model for continuous, counterpropagating laser beams, for example, than a temporal Sauter pulse, and thus this profile could be useful to describe tunneling in semiconductors assisted by such beams.

## 2.6. Magnetic fields

Going from 1+1 to 2+1 or 3+1 spacetime dimensions makes it possible to add also magnetic components to the external field, and thus the question arises of how magnetic fields $\boldsymbol{B}$ influence the process of pair creation via tunneling induced by electric fields $\boldsymbol{E}$.

**Constant $\boldsymbol{E}$ and $\boldsymbol{B}$ fields**

Heisenberg and Euler [22] as well as Schwinger [27] studied constant external $\boldsymbol{E}$ and $\boldsymbol{B}$ fields (low-energy limit, i.e., fields varying very slowly in space and time). In this simple case, the (homogeneous) effective Lagrangian density $\mathcal{L}_{\text{eff}}^{\text{HE}}$ (1.20) can be expressed as a function of two numbers, the Lorentz invariants $\mathfrak{F} = \boldsymbol{E}^2 - c^2\boldsymbol{B}^2$ and $\mathfrak{G} = c\boldsymbol{E} \cdot \boldsymbol{B}$, which were introduced in Eq. (1.22). An interesting special case is $\boldsymbol{E} \cdot \boldsymbol{B} = 0$, that is, perpendicular ("crossed") electric and magnetic fields. Note that if the fields are crossed in one coordinate system, they will be crossed in any inertial frame since $\boldsymbol{E} \cdot \boldsymbol{B}$ is invariant under Lorentz transformations. If the other invariant, $\boldsymbol{E}^2 - c^2\boldsymbol{B}^2$, is positive, we can always find a Lorentz-boosted frame with respect to which there is only an electric field $\boldsymbol{E}'$ and no magnetic field ("electric-type" crossed fields) [27]. The strength of this purely electric field is reduced [96] since

**Crossed fields ($\boldsymbol{E} \cdot \boldsymbol{B} = 0$)**

$$|\boldsymbol{E}'| = \sqrt{\boldsymbol{E}^2 - c^2\boldsymbol{B}^2} = |\boldsymbol{E}| \underbrace{\sqrt{1 - \left(\frac{c|\boldsymbol{B}|}{|\boldsymbol{E}|}\right)^2}}_{<1 \text{ for } \boldsymbol{B} \neq 0}. \tag{2.147}$$

This reduced field will produce a certain number of pairs via the Sauter–Schwinger effect, which can also be observed in the original frame since the number of particles is Lorentz invariant. In conclusion, a not too strong ($|\boldsymbol{B}| < |\boldsymbol{E}|/c$) **perpendicular magnetic field lowers the effective pair-creating field strength** $\sqrt{|\boldsymbol{E}|^2 - c^2|\boldsymbol{B}|^2}$, and tunneling vanishes completely in the limit $|\boldsymbol{B}| \nearrow |\boldsymbol{E}|/c$ [25].

**Perpendicular $\boldsymbol{B}$ diminishes tunneling**

In a simple, semiclassical picture, this effect can be understood as follows: Imagine a Dirac-sea electron tunneling through the forbidden region, against the force exerted by the $\boldsymbol{E}$ field (as depicted in Fig. 2.2 on page 48). An additional, perpendicular $\boldsymbol{B}$ field will prevent the particle from moving straight through the barrier but instead force it to move along a circular path (around the vector $\boldsymbol{B}$), thus **increasing the tunneling length**—which corresponds to an effective decrease of the pair-creating field strength. The stronger the $\boldsymbol{B}$ field, the smaller the radius of the circular motion, so tunneling becomes impossible when the circular paths fit completely into the forbidden region, which happens at the threshold $|\boldsymbol{B}| = |\boldsymbol{E}|/c$.

A convenient way to explain the suppression of tunneling in a perpendicular $\boldsymbol{B}$ field more precisely (still in the semiclassical picture) is to consider the **local energy levels** of electrons in a crossed-fields setup. Say our constant, crossed fields are $\boldsymbol{E} = E\boldsymbol{e}_x$ and $\boldsymbol{B} = B\boldsymbol{e}_z$. We can choose the corresponding potentials in a way such that they do only depend on $x$:

**Dispersion relation in crossed fields**

$$\Phi(x) = Ex \qquad \text{and} \qquad \boldsymbol{A}(x) = -Bx\boldsymbol{e}_y. \tag{2.148}$$

The unperturbed relativistic energy levels are $\mathcal{E}_\pm = \pm\sqrt{m^2c^4 + c^2\boldsymbol{p}^2}$ (observables are treated like numbers/vectors in this semiclassical picture). Coupling the scalar potential to this relation via $\mathcal{E}_\pm \to \mathcal{E}_\pm + q\Phi(x)$ tilts the energy levels in space, which leads to the tunneling picture of the Sauter–Schwinger effect (see Fig. 2.2). The vector potential couples to the momentum, $\boldsymbol{p} \to \boldsymbol{p} + q\boldsymbol{A}(x)$, and thus gives rise to an additional space dependence of the energy levels. We set $\boldsymbol{p} = 0$ and thereby obtain the local edges of the two energy continua:

$$\mathcal{E}_\pm^{\boldsymbol{p}=0}(x) = -qEx \pm \sqrt{m^2c^4 + c^2q^2B^2x^2}. \tag{2.149}$$

These levels are plotted in Fig. 2.10 for different magnetic field strengths. Asymptotically ($|cqBx| \gg mc^2$), these energy levels have the functional form

$$\mathcal{E}_\pm^{\boldsymbol{p}=0}(x) \sim -qEx \pm cqB|x| = -qE|x| \left( \operatorname{sgn} x \mp \frac{cB}{E} \right), \tag{2.150}$$

so they are linear for large $|x|$. However, we see that a strong magnetic field ($cB/E > 1$) changes the asymptotic direction (decreasing versus increasing with respect to $x$) of $\mathcal{E}_\pm^{\boldsymbol{p}=0}$ on one side, respectively, and thus prevents tunneling completely as a consequence (cf. Fig. 2.10). Such strong magnetic fields correspond to $\boldsymbol{E}^2 - c^2\boldsymbol{B}^2 < 0$, the "magnetic-type" regime of crossed fields. For this type of fields, there is always an inertial observer that "sees" a purely magnetic field—which does not generate pairs since the corresponding effective Lagrangian density is real [27].

**$\boldsymbol{E} \cdot \boldsymbol{B} \neq 0$: effectively parallel fields**

The remaining class of constant fields is given by $\boldsymbol{E} \cdot \boldsymbol{B} \neq 0$. Such fields cannot be perpendicular in any Lorentz frame. Instead, it is always possible to find a frame according to which the transformed electric and magnetic field vectors are (anti-)parallel [25]. The effect of a parallel $B$ field on tunneling due to an $E$ field was studied in Refs. [42, 44, 35, 47, 45] (QED) and [97] (interband tunneling in semiconductors), for example. The $B$ field does not impede the tunneling motion against the electric-field force in this case, but the transverse motion becomes quantized (Landau levels) in this setting—whereas the transverse electron momentum $\boldsymbol{p}_\perp$ is a conserved quantity and can take any value in the absence of a parallel $B$ field ($\boldsymbol{p}_\perp$ has been introduced in Sec. 2.2). Since a parallel $B$ field **merely increases the prefactor** in the number of produced pairs per unit time and volume in QED a bit,

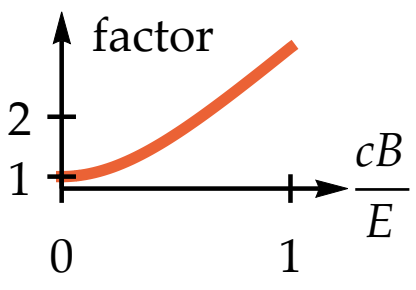

$$\dot{\mathcal{N}}_{e^+e^-} = \frac{q^2E^2}{4\pi^3\hbar^2 c} \underbrace{\frac{\pi cB}{E} \coth\left(\frac{\pi cB}{E}\right)}_{\text{additional factor due to the parallel } B \text{ field (see marginal plot)}} e^{-\pi E_{\text{crit}}^{\text{QED}}/E} \tag{2.151}$$

(see [42, 35]), but leaves the exponent unchanged, we will not consider this field profile further in this thesis.

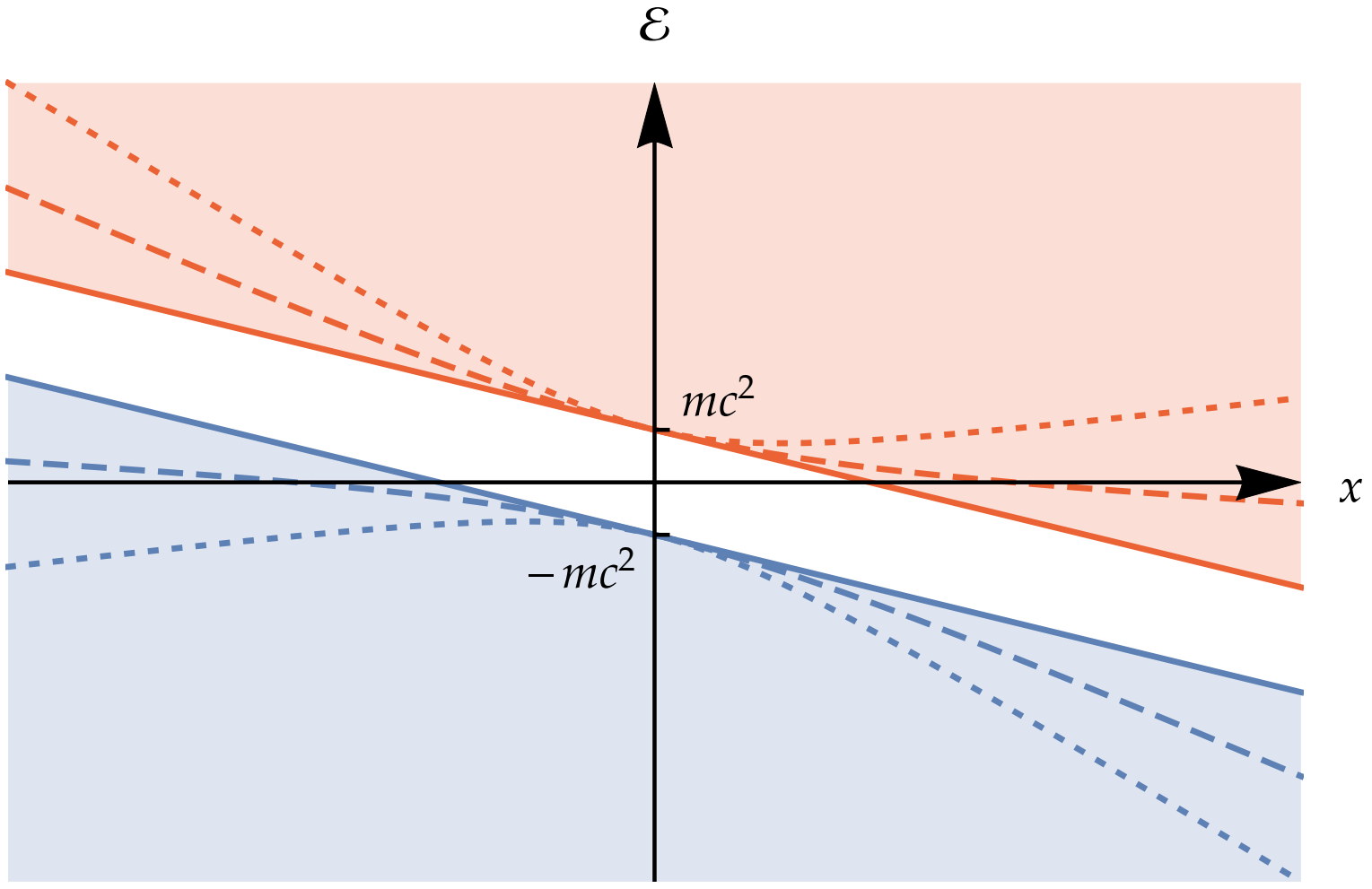

**Figure 2.10.:** Upper (red graphs) and lower (blue graphs) relativistic local electron energies (2.149) for $\boldsymbol{p} = 0$ (edges of the energy continua) in crossed fields $E\boldsymbol{e}_x$ and $B\boldsymbol{e}_z$, plotted for three different magnetic field strengths. The solid graphs represent the pure Sauter–Schwinger case ($B = 0$; coincides with Fig. 2.2). The dashed graphs correspond to $0 < B < E/c$, the regime in which tunneling is still possible but reduced by the $\boldsymbol{B}$ field; consider, e.g., the $x$ axis, which is also a line of constant energy: the red as well as the blue dashed graphs intersect the $x$ axis, so tunneling transitions between the two relativistic continua are possible, but the tunneling length is larger than in the case $B = 0$. The third case is $B > E/c$ (dotted graphs). We see in the plot that $\mathcal{E}_+^{p=0}(x)$ is always above the $x$ axis in this case, while the corresponding lower energy curve is always below this axis, so there are no allowed tunneling transitions.

**Nonconstant fields**

Beyond the constant-field approximation, there are some exact results for the effective Lagrangian density/action in nonconstant magnetic fields, such as spatial Sauter pulses $B\boldsymbol{e}_x / \cosh^2(kx)$ [98, 99] (which do not give rise to pair creation, however). Another important exact result was given by Schwinger [27]: a **plane electromagnetic wave** in vacuum cannot create any pairs, no matter how high its intensity or frequency is, so pair production by a single laser beam is not possible. This result is interesting because it cannot be explained without taking the magnetic component of the wave into account since a time-dependent and purely electric field will always create pairs [86]. Hence, the presence of magnetic fields and their spacetime dependence can have a significant influence on pair creation; see, e.g., [33, 35, 47, 45, 86, 100, 92] for studies on that topic.

**Counterpropagating laser beams**

To circumvent the problem that single laser beams cannot create pairs, an alternative setup consisting of two coherent, counterpropagating beams is often considered; see, e.g., [101] and also [102, 103], which deal with pair production via X-ray free-electron lasers. In the region where the two beams overlap, they form a standing wave (which can create pairs), and the magnetic components even cancel each other completely (under ideal conditions). For this reason, $\boldsymbol{B}$ fields are often ignored in the context of pair creation from the vacuum—an approximation which seems to be justified for low laser frequencies (tunneling regime) according to numerical studies like [86, 92].

### Goals in this thesis

The major part of this thesis is devoted to QED pair creation and its analogy in semiconductors for spacetime-dependent external *electric* fields in 1+1 spacetime dimensions, that is, without magnetic fields. However, we will also begin to generalize the analogy to 2+1 dimensions in Ch. 9, with special emphasis on pair creation in crossed, constant $\boldsymbol{E}$ and $\boldsymbol{B}$ fields—which is a simple yet interesting field profile illustrating how perpendicular magnetic fields can interfere with tunneling in QED (see [100]), and there are also existing results regarding interband tunneling in semiconductors for crossed fields [104, 96, 105, 97, 106].

# Part II.

# Dynamically assisted Sauter–Schwinger effect in time-dependent electric fields in 1+1 spacetime dimensions

In this part, we study electron–positron pair creation in time-dependent, subcritical electric fields $E(t) = \dot{A}(t) \ll E_{\text{crit}}^{\text{QED}}$ in 1+1 spacetime dimensions by means of the Riccati-equation formalism introduced in Secs. 2.4.3–2.4.4. Within the semiclassical (JWKB) approximation, $R_k^{\text{out}}$ is given by the integral (2.110). The singularities $t_k^\star$ of the integrand in the upper complex half-plane determine the exponents in $R_k^{\text{out}}$. Each of the following chapters is devoted to a specific field profile, respectively: we study

1. pure tunneling in a constant electric field,
2. assisted tunneling via a temporal Gauss pulse $E_{\text{weak}} \exp[-(\omega t)^2]$, and
3. assisted tunneling via an oscillation $E_{\text{weak}} \cos(\omega t)$.

Most of the results presented in this part have been published in the article [1], which is partly based on the thesis [107].

We choose units with $c = \hbar = 1$ throughout this part for brevity. **Natural units**

# 3. Pure tunneling in a constant electric field

In this chapter, we evaluate the integral representation (2.110) of $R_k^{\text{out}}$ for a **constant electric field**, so

$$A(t) = Et \qquad \Rightarrow \qquad E(t) = E > 0 \tag{3.1}$$

with $E \ll E_{\text{crit}}^{\text{QED}}$ (in order for the underlying semiclassical approximation to be valid). This profile tilts the Dirac sea as depicted in Fig. 2.2 on page 48 and thus gives rise to the ordinary **Sauter–Schwinger effect**.

We treat this simple case since the integral (2.110) is easier to calculate thoroughly (using the contour-integration method explained in Sec. 2.4.4.1) than for the field profiles considered later. In particular, we will calculate the contribution from the branch cut originating from the single crucial singularity in this case.

## 3.1. Singularities and integration contour

**Singularities**

- The vector potential (3.1) is an **entire function**, so there are no singularities originating from $A(t)$ itself.
- Equation (2.112) determining the zeros of

$$\Omega_k(t) = m\sqrt{1 + \left(\frac{k + qEt}{m}\right)^2} \tag{3.2}$$

  only has **one solution in the upper complex half-plane** here:

$$t_k^\star = \frac{-k + \mathrm{i}m}{qE}. \tag{3.3}$$

**Branch cut**

The corresponding **branch cut** originating from $t_k^\star$ runs upwards in the complex plane, parallel to the imaginary $t$ axis (see Fig. 3.1), according to the definition (2.113) of the square root in $\Omega_k(t)$. Hence, we choose the **integration contour** depicted in Fig. 3.1 here, which circumvents $t_k^\star$ and the branch cut

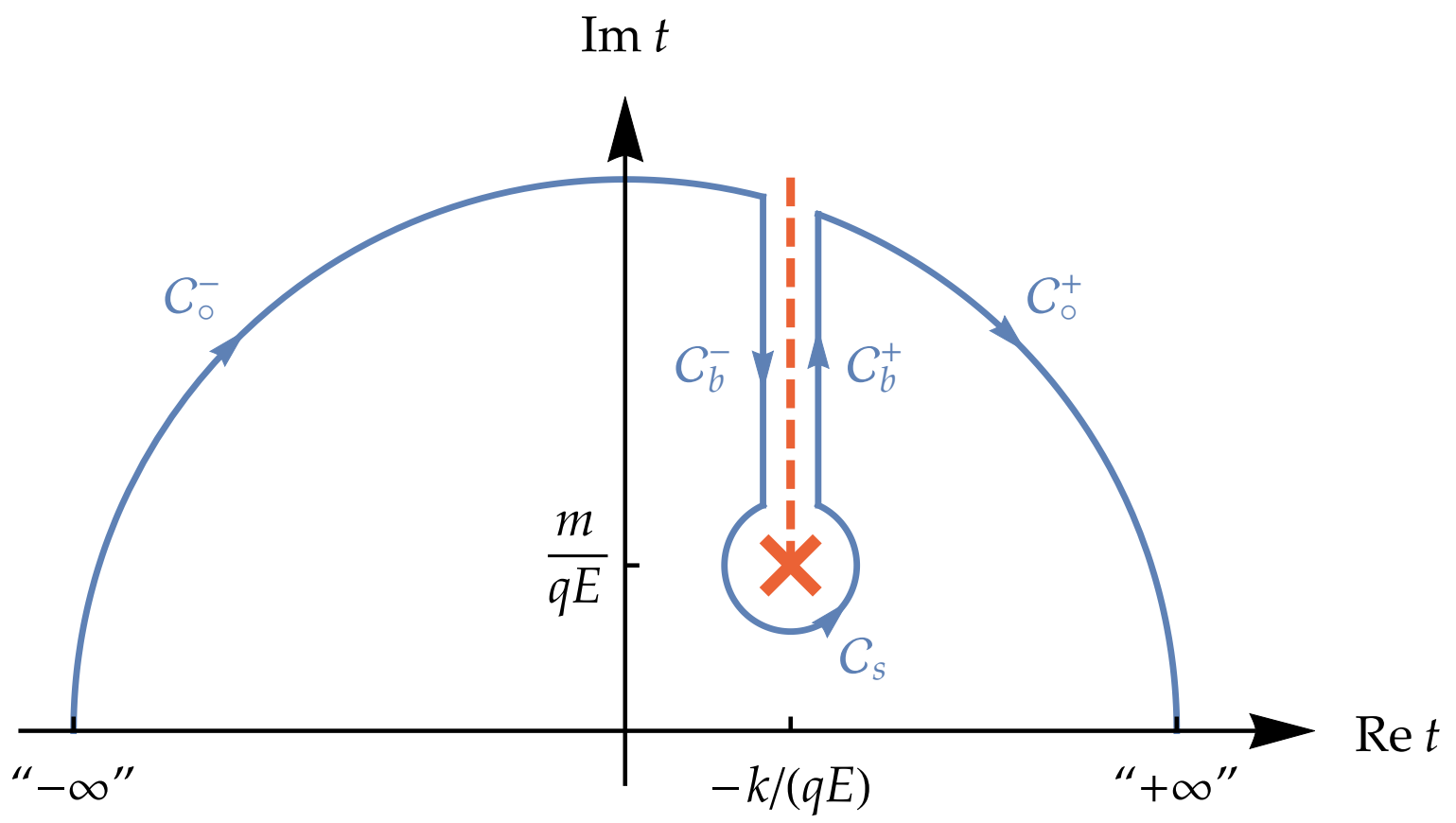

**Figure 3.1.:** Sketch of the integration contour for the calculation of $R_k^{\text{out}}$ according to Eq. (2.110) in the constant-field case $A(t) = Et$. The contour starts at $t \to -\infty$ (initial state), and we basically want to integrate to $t \to +\infty$ (final state) along an arc of infinite radius through the upper complex half-plane, where the integrand is exponentially suppressed. However, the first part of this arc ($\mathcal{C}_\circ^-$) hits the branch cut (dashed line) originating from the singularity $t_k^\star$ (red cross) at $\operatorname{Re} t = -k/(qE)$. We circumvent this "barrier" by integrating along the left side of the branch cut ($\mathcal{C}_b^-$) all the way down to the singularity, then around the singularity ($\mathcal{C}_s$), and finally upwards again ($\operatorname{Im} t \to \infty$), along the right side of the branch cut ($\mathcal{C}_b^+$). Then, the contour continues on the path of the large arc ($\mathcal{C}_\circ^+$) until it arrives at the endpoint.

(our strategy of how to choose the integration contour has been explained in Sec. 2.4.4.1).

## 3.2. Calculation of the contour integral

In order to actually perform the contour integration for a constant field, let us first write the integral (2.110) in a more explicit and convenient way.

We start with the phase function $\varphi_k(t) = \int_0^t \Omega_k(t')\, \mathrm{d}t'$ and substitute the **dimensionless "time" variable**

$$\tau = \frac{k + qEt}{m} \qquad \Rightarrow \qquad \mathrm{d}t = \frac{m}{qE}\, \mathrm{d}\tau, \tag{3.4}$$

which is, in this form, quite useful in the constant-field case. We get

$$\begin{aligned}
\varphi_k(t) &= \int_0^t m\sqrt{1+\left(\frac{k+qEt'}{m}\right)^2}\,\mathrm{d}t' \\
&= \frac{E_{\mathrm{crit}}^{\mathrm{QED}}}{E}\int_{k/m}^{(k+qEt)/m}\sqrt{1+\tau'^2}\,\mathrm{d}\tau' \\
&= \frac{E_{\mathrm{crit}}^{\mathrm{QED}}}{2E}\left[\phi\left(\frac{k+qEt}{m}\right)-\phi\left(\frac{k}{m}\right)\right],
\end{aligned} \tag{3.5}$$

where we have defined the auxiliary function

$$\phi(z) = z\sqrt{1+z^2} + \operatorname{arsinh} z \tag{3.6}$$

(cf., e.g., Refs. [67, 62, 1]). The prefactor function (2.90) reads

$$\Xi_k(t) = \frac{mqE}{2\left[m^2+(k+qEt)^2\right]} = \frac{m}{2}\frac{E}{E_{\mathrm{crit}}^{\mathrm{QED}}}\frac{1}{1+\left(\frac{k+qEt}{m}\right)^2}. \tag{3.7}$$

Inserting Eqs. (3.5) and (3.7) into the general **integral representation** (2.110) and substituting $\tau$ from Eq. (3.4) again then yields

**Integral representation of $R_k^{\mathrm{out}}$**

$$\begin{aligned}
R_k^{\mathrm{out}} &\approx \int_{-\infty}^{\infty}\frac{m}{2}\frac{E}{E_{\mathrm{crit}}^{\mathrm{QED}}}\frac{\exp\left\{\mathrm{i}\frac{E_{\mathrm{crit}}^{\mathrm{QED}}}{E}\left[\phi\left(\frac{k+qEt}{m}\right)-\phi\left(\frac{k}{m}\right)\right]\right\}}{1+\left(\frac{k+qEt}{m}\right)^2}\,\mathrm{d}t \\
&= \frac{1}{2}\mathrm{e}^{-\mathrm{i}\phi(\kappa)/\chi}\int_{-\infty}^{\infty}\frac{\mathrm{e}^{\mathrm{i}\phi(\tau)/\chi}}{1+\tau^2}\,\mathrm{d}\tau \\
&= \frac{1}{2}\mathrm{e}^{-\mathrm{i}\phi(\kappa)/\chi}\int_{-\infty}^{\infty}\frac{\mathrm{e}^{\mathrm{i}\left(\tau\sqrt{1+\tau^2}+\operatorname{arsinh}\tau\right)/\chi}}{1+\tau^2}\,\mathrm{d}\tau,
\end{aligned} \tag{3.8}$$

where we have introduced the **dimensionless parameter values**

$$\chi = \frac{E}{E_{\mathrm{crit}}^{\mathrm{QED}}} \qquad \text{and} \qquad \kappa = \frac{k}{m} \tag{3.9}$$

for clarity.

We see in Eq. (3.8) that the substitution of $t$ by $\tau$ allowed us to write the re-

**Integration contour**

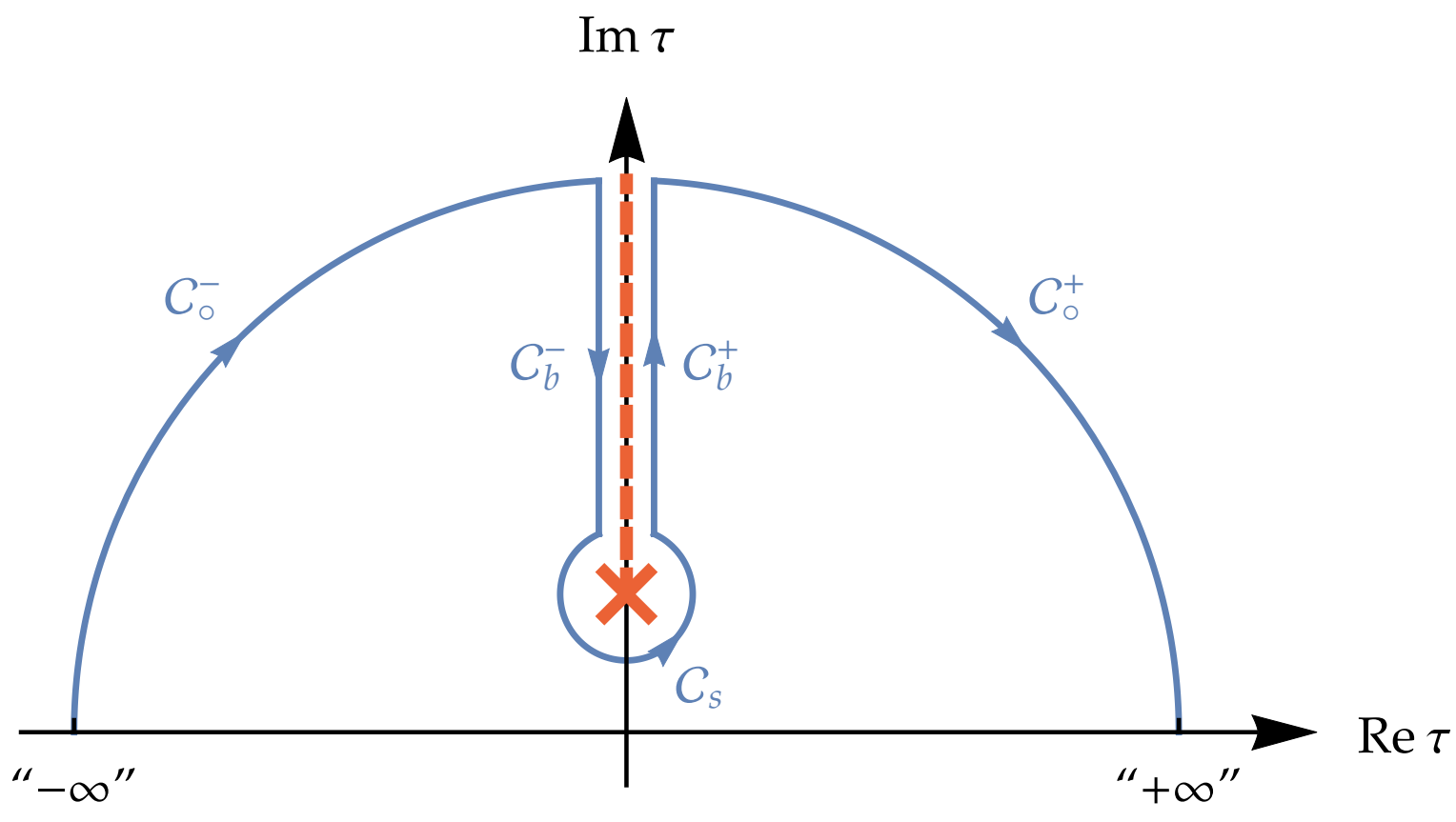

**Figure 3.2.:** Integration contour from Fig. 3.1 after the substitution of $t$ by the dimensionless variable $\tau$ defined in Eq. (3.4). In the $\tau$ representation, the singularity $t_k^\star$ is fixed at the position i (for all $k$ and $E$), and the branch cut runs along the imaginary axis.

maining integral in a way that is independent of the considered longitudinal conserved momentum (i.e., $k$ or $\kappa$). Furthermore, the singularity (3.3) becomes fixed at the position $\tau(t_k^\star) = \mathrm{i}$ due to the substitution, which makes the parameterization of the integration contour a little bit easier; see Fig. 3.2 for our integration contour in the complex $\tau$ plane.

In the following subsections, we will evaluate the integral (3.8) along the individual (named) sections of this contour.

### 3.2.1. Integration along the arcs

**Parameterizations** The arcs in the contour in Fig. 3.2 are parameterized by

$$\mathcal{C}_\circ^\pm : \quad \vartheta \mapsto \lim_{r\to\infty} r\mathrm{e}^{\mathrm{i}\vartheta} \quad \text{with} \quad \begin{cases} \vartheta : & \vartheta_1^+ = \frac{\pi}{2} \searrow 0 = \vartheta_2^+ , \\ \vartheta : & \vartheta_1^- = \pi \searrow \frac{\pi}{2} = \vartheta_2^- . \end{cases} \tag{3.10}$$

**Contributions to the contour integral** Calculating the contributions from these parts of the contour to the $\tau$ integral (3.8) yields

$$\int\limits_{\mathcal{C}_\circ^\pm} \frac{\mathrm{e}^{\mathrm{i}\phi(\tau)/\chi}}{1+\tau^2}\,\mathrm{d}\tau = \int\limits_{\vartheta_1^\pm}^{\vartheta_2^\pm} \lim_{r\to\infty} \frac{\mathrm{e}^{\mathrm{i}\left[r\mathrm{e}^{\mathrm{i}\vartheta}\sqrt{1+r^2\mathrm{e}^{2\mathrm{i}\vartheta}}+\operatorname{arsinh}(r\mathrm{e}^{\mathrm{i}\vartheta})\right]/\chi}}{1+r^2\mathrm{e}^{2\mathrm{i}\vartheta}}\,\mathrm{i}r\mathrm{e}^{\mathrm{i}\vartheta}\,\mathrm{d}\vartheta. \tag{3.11}$$

For large $r$, we may approximate $1 + r^2 e^{2i\vartheta} \approx r^2 e^{2i\vartheta}$. Furthermore, the absolute value of

$$\operatorname{arsinh}\tau = \ln\left(\tau + \sqrt{1+\tau^2}\right) \tag{3.12}$$

grows merely logarithmically with respect to $|\tau|$ , much slower than $\tau(1+\tau^2)^{1/2} \sim |\tau|^2$, so we may neglect arsinh(...) in the limit $r \to \infty$. This way, Eq. (3.11) becomes

$$\int_{\mathcal{C}_\circ^\pm} \frac{e^{i\phi(\tau)/\chi}}{1+\tau^2}\, d\tau = i \int_{\vartheta_1^\pm}^{\vartheta_2^\pm} \lim_{r\to\infty} \frac{e^{i\left(r^2 e^{i\vartheta}\sqrt{e^{2i\vartheta}}\right)/\chi}}{r e^{i\vartheta}}\, d\vartheta. \tag{3.13}$$

We see that $r$ in the denominator suppresses the integrand. However, we also have to check how the exponential function in the nominator behaves for large $r$.[1] The absolute value of this function reads

$$\left| e^{i\left(r^2 e^{i\vartheta}\sqrt{e^{2i\vartheta}}\right)/\chi} \right| = \exp\left[-\frac{r^2}{\chi}\operatorname{Im}\left(e^{i\vartheta}\sqrt{e^{2i\vartheta}}\right)\right]. \tag{3.14}$$

On the path $\mathcal{C}_\circ^\pm$, the principal value (2.113) of the square root is $\sqrt{\exp(2i\vartheta)} = \pm e^{i\vartheta}$, so

$$\left| e^{i\left(r^2 e^{i\vartheta}\sqrt{e^{2i\vartheta}}\right)/\chi} \right| = \exp\left[-\frac{r^2}{\chi}\operatorname{Im}\left(\pm e^{2i\vartheta}\right)\right] = \exp\left[\mp\frac{r^2}{\chi}\sin(2\vartheta)\right]. \tag{3.15}$$

Since $\sin(2\vartheta) \geq 0$ on $\mathcal{C}_\circ^+$ and $\leq 0$ on $\mathcal{C}_\circ^-$, the exponent is always $\leq 0$, and thus the absolute value of the exponential function in the nominator in the integrand (3.13) has an upper bound of 1. The integral therefore vanishes in the limit $r \to \infty$, so the **arcs do not contribute** to $R_k^{\text{out}}$: **Result**

$$\int_{\mathcal{C}_\circ^\pm} \frac{e^{i\phi(\tau)/\chi}}{1+\tau^2}\, d\tau = 0. \tag{3.16}$$

This result was expected because of the exponential suppression of the integrand (2.110) in the upper complex half-plane. Only the circumvention of branch cuts and singularities is expected to generate nonvanishing contributions to the contour integral.

### 3.2.2. Integration around the singularity ("residue")

As the next step, we perform the integration around the singularity at $\tau = i$. **Parameterization**

[1] In fact, we have already shown in Sec. 2.4.4.1 that the integrand will always be suppressed exponentially when going upwards in the complex plane (as long as we avoid singularities and branch cuts). But we do the explicit check here nonetheless because it is easy to carry out in the constant-field case.

We do this along a **circle of infinitesimal radius** (see Fig. 3.2), so our parameterization of $\mathcal{C}_s$ is

$$\mathcal{C}_s: \quad \vartheta \mapsto \mathrm{i} + \lim_{r\searrow 0} r\mathrm{e}^{\mathrm{i}\vartheta} \quad \text{with} \quad \vartheta: -\frac{3}{2}\pi \nearrow \frac{\pi}{2}. \tag{3.17}$$

**Contribution to the contour integral**

The corresponding contribution to the integral (3.8) reads

$$\begin{aligned}\int\limits_{\mathcal{C}_s} \frac{\mathrm{e}^{\mathrm{i}\phi(\tau)/\chi}}{1+\tau^2}\,\mathrm{d}\tau &= \int\limits_{-3\pi/2}^{\pi/2} \lim_{r\searrow 0} \frac{\mathrm{e}^{\mathrm{i}\left[(\mathrm{i}+r\mathrm{e}^{\mathrm{i}\vartheta})\sqrt{2\mathrm{i}r\mathrm{e}^{\mathrm{i}\vartheta}+r^2\mathrm{e}^{2\mathrm{i}\vartheta}}+\operatorname{arsinh}(\mathrm{i}+r\mathrm{e}^{\mathrm{i}\vartheta})\right]/\chi}}{2\mathrm{i}r\mathrm{e}^{\mathrm{i}\vartheta}+r^2\mathrm{e}^{2\mathrm{i}\vartheta}}\,\mathrm{i}r\mathrm{e}^{\mathrm{i}\vartheta}\,\mathrm{d}\vartheta \\ &= \mathrm{i}\int\limits_{-3\pi/2}^{\pi/2} \frac{\mathrm{e}^{\mathrm{i}\overbrace{\operatorname{arsinh}(\mathrm{i})}^{\mathrm{i}\pi/2}/\chi}}{2\mathrm{i}}\,\mathrm{d}\vartheta = \pi\mathrm{e}^{-\pi/(2\chi)}.\end{aligned} \tag{3.18}$$

This result is proportional to the square root of the characteristic, exponential **Schwinger factor** $\exp[-\pi/(2\chi)] = \sqrt{\exp(-\pi E_{\text{crit}}^{\text{QED}}/E)}$, so the approximate pair-creation probability $|R_k^{\text{out}}|^2$ will therefore be proportional to $\exp(-\pi E_{\text{crit}}^{\text{QED}}/E)$ (at least, according to this singularity contribution). This is exactly what we expect in a constant electric field.

**Relation to a residue**

Note that the only reason why we get a nonvanishing result in Eq. (3.18) is because we **integrate around a pole**, a zero of the denominator $1+\tau^2$—the argument $\propto \phi(\tau)$ [defined in Eq. (3.6)] of the exponential function in the nominator has a branch point at the singularity $\tau = \mathrm{i}$, but $\phi(\tau)$ is **continuous** at this point in the sense that $\phi(\mathrm{i}+r\mathrm{e}^{\mathrm{i}\vartheta})$ approaches the well-defined value $\mathrm{i}\pi/2$ in the limit $r \searrow 0$, independently of the direction $\vartheta$. Hence, we may effectively set the exponential function to its value at the position around which we integrate, $\exp[\mathrm{i}\phi(\mathrm{i})/\chi]$, and then write this constant factor before the integral:

$$\int\limits_{\mathcal{C}_s} \frac{\mathrm{e}^{\mathrm{i}\phi(\tau)/\chi}}{1+\tau^2}\,\mathrm{d}\tau = \mathrm{e}^{\mathrm{i}\phi(\mathrm{i})/\chi} \int\limits_{\mathcal{C}_s} \frac{\mathrm{d}\tau}{1+\tau^2}. \tag{3.19}$$

The remaining integrand is free of branch cuts; it merely has a pole at $\tau = \mathrm{i}$ (an isolated singularity), and it is thus manifestly no longer important to let the integration along $\mathcal{C}_s$ begin at $\vartheta = -3\pi/2$ and to let it end at $\pi/2$—the essential information is that we integrate once around the pole in a counterclockwise manner, which corresponds (up to a factor of $2\pi\mathrm{i}$) to the **residue of the remaining integrand** at the singularity $\tau = \mathrm{i}$, so

$$\int\limits_{\mathcal{C}_s} \frac{\mathrm{e}^{\mathrm{i}\phi(\tau)/\chi}}{1+\tau^2}\,\mathrm{d}\tau = \mathrm{e}^{\mathrm{i}\phi(\mathrm{i})/\chi} \int\limits_{\mathcal{C}_s} \frac{\mathrm{d}\tau}{1+\tau^2} = 2\pi\mathrm{i}\,\mathrm{e}^{\mathrm{i}\phi(\mathrm{i})/\chi}\operatorname{Res}_{\mathrm{i}} \frac{1}{1+\tau^2}. \tag{3.20}$$

#### 3.2.2.1. Exact contribution from the integration around a singularity for a general vector potential

The above argumentation leading to Eq. (3.20) can be generalized for a large class of vector potentials (including the field profiles treated in the following chapters), which even allows us to calculate the exact result for the integration around a singularity in the general case. Suppose that

**Assumptions**

- we consider a particular zero $t_k^\star$ of $\Omega_k(t)$ in the upper complex half-plane,
- and our vector potential $A(t)$ is holomorphic in a domain containing $t_k^\star$
- with $\dot{A}(t_k^\star) = E(t_k^\star) \neq 0$.

**One branch cut originating from $t_k^\star$**

Since $\Omega_k(t_k^\star) = 0$, we have $[k + qA(t_k^\star)]/m = \pm\mathrm{i}$ [i.e., plus *or* minus; cf. Eq. (2.112)]. The condition $\dot{A}(t_k^\star) \neq 0$ then leads to

$$\frac{\mathrm{d}}{\mathrm{d}t}\left[1 + \left(\frac{k+qA(t)}{m}\right)^2\right]\Bigg|_{t=t_k^\star} = 2\underbrace{\frac{k+qA(t_k^\star)}{m}}_{\pm\mathrm{i}}\frac{q\dot{A}(t_k^\star)}{m} \neq 0, \tag{3.21}$$

which means that the expression underneath the square root in $\Omega_k(t)$ [see Eq. (2.111)] has a first-order zero at $t_k^\star$, and thus there is only **one branch cut originating from $t_k^\star$** in the complex $t$ plane[2].

**Integration around $t_k^\star$**

The integration contour we choose according to Sec. 2.4.4.1 for the calculation of the integral (2.110) representing $R_k^{\mathrm{out}}$ must consequently go around $t_k^\star$ completely (e.g., along a full, infinitesimal circle again) in order to get from one side of this branch cut to the other side. Since $\Omega_k(t)$ is a square root, it has a branch point at $t_k^\star$; however, $\Omega_k(t)$ is nonetheless continuous at $t_k^\star$ in the above sense that $\Omega_k(t \to t_k^\star) \to 0$ independently of the direction, so the the phase function $\varphi_k(t) = \int_{t_0}^{t}\Omega_k(t')\,\mathrm{d}t'$ must also be continuous at $t_k^\star$ in this sense. Say $\mathcal{C}_s^{t_k^\star}$ is the infinitesimal circle around $t_k^\star$, with the parameterization

$$\mathcal{C}_s^{t_k^\star}: \quad \vartheta \mapsto t_k^\star + \lim_{r\searrow 0} r\mathrm{e}^{\mathrm{i}\vartheta} \quad \text{with} \quad \vartheta:\ \vartheta_0 \nearrow \vartheta_0 + 2\pi \tag{3.22}$$

and an arbitrary $\vartheta_0 \in \mathbb{R}$. Then, the phase function has a well-defined, constant value $\varphi_k(t_k^\star)$ on that path, and the corresponding contribution to the contour

[2] A higher-order zero of the radicand leads to multiple branch cuts in the principal value (2.113) of the square root which all originate from this particular zero. For example, $\sqrt{z^3}$ has three different branch cuts originating from $z = 0$.

integral thus reads

$$\int_{\mathcal{C}_s^{t_k^\star}} \Xi_k(t) e^{2i\varphi_k(t)}\, dt = e^{2i\varphi_k(t_k^\star)} \int_{\mathcal{C}_s^{t_k^\star}} \Xi_k(t)\, dt = 2\pi i\, e^{2i\varphi_k(t_k^\star)} \operatorname{Res}_{t_k^\star} \Xi_k(t), \tag{3.23}$$

which is the generalization of Eq. (3.20).

**Calculation of the residue**

The residue integral can be calculated exactly. We start by inserting the general expression for $\Xi_k(t)$ from Eq. (2.90), which has a pole at $t_k^\star$, and the parameterization (3.22):

$$\int_{\mathcal{C}_s^{t_k^\star}} \Xi_k(t) e^{2i\varphi_k(t)}\, dt = \frac{e^{2i\varphi_k(t_k^\star)}}{2m} \int_{\vartheta_0}^{\vartheta_0+2\pi} \lim_{r\searrow 0} \frac{q\dot{A}(t_k^\star + re^{i\vartheta})}{1 + \left[\frac{k+qA(t_k^\star+re^{i\vartheta})}{m}\right]^2}\, ire^{i\vartheta}\, d\vartheta. \tag{3.24}$$

Since $A(t)$ is per assumption analytic at $t_k^\star$, we may Taylor expand $A$ and $\dot{A}$ around $t_k^\star$, which allows us to perform the limit and to obtain the **exact and general result**

$$\begin{aligned}
&\int_{\mathcal{C}_s^{t_k^\star}} \Xi_k(t) e^{2i\varphi_k(t)}\, dt \\
&= \frac{i}{2m} e^{2i\varphi_k(t_k^\star)} \int_{\vartheta_0}^{\vartheta_0+2\pi} \lim_{r\searrow 0} \frac{q\overbrace{\dot{A}(t_k^\star)}^{\neq 0 \text{ per assumption}} + \mathcal{O}(r)}{1 + \Big[\underbrace{\frac{k+qA(t_k^\star)}{m}}_{\pm i \text{ since } \Omega_k(t_k^\star)=0} + \frac{q\dot{A}(t_k^\star)}{m} re^{i\vartheta} + \mathcal{O}(r^2)\Big]^2}\, re^{i\vartheta}\, d\vartheta \\
&= \frac{i}{2m} e^{2i\varphi_k(t_k^\star)} \int_{\vartheta_0}^{\vartheta_0+2\pi} \lim_{r\searrow 0} \frac{q\dot{A}(t_k^\star) + \mathcal{O}(r)}{\pm 2i\frac{q\dot{A}(t_k^\star)}{m} re^{i\vartheta} + \mathcal{O}(r^2)}\, re^{i\vartheta}\, d\vartheta \\
&= \pm \frac{1}{4} e^{2i\varphi_k(t_k^\star)} \int_{\vartheta_0}^{\vartheta_0+2\pi} \lim_{r\searrow 0} \frac{1 + \mathcal{O}(r)}{1 + \mathcal{O}(r)}\, d\vartheta \\
&= \pm \frac{\pi}{2} e^{2i\varphi_k(t_k^\star)},
\end{aligned} \tag{3.25}$$

where the sign is the same as in the equation

$$\frac{k + qA(t_k^\star)}{m} = \pm i \qquad \Leftrightarrow \qquad \Omega_k(t_k^\star) = 0, \tag{3.26}$$

[cf. Eq. (2.112)], which $t_k^\star$ satisfies per assumption.

**Comparison with the known constant-field result**

Let us check whether the general formula (3.25) yields the correct result for the only singularity $t_k^\star$ [see Eq. (3.3)] with $\operatorname{Im} t_k^\star > 0$ in the case of a constant electric field. This singularity satisfies the plus case of Eq. (3.26), so the general formula (3.25) yields [$\varphi_k(t)$ and $\phi(z)$ have been defined in Eqs. (3.5) and (3.6)]

$$\begin{aligned}\int\limits_{\mathcal{C}_s^{t_k^\star}} \Xi_k(t)\mathrm{e}^{2\mathrm{i}\varphi_k(t)}\,\mathrm{d}t &= \frac{\pi}{2}\mathrm{e}^{2\mathrm{i}\varphi_k(t_k^\star)} \\ &= \frac{\pi}{2}\exp\Biggl\{\mathrm{i}\underbrace{\frac{E_{\text{crit}}^{\text{QED}}}{E}}_{1/\chi}\Biggl[\phi\Bigl(\underbrace{\frac{k+qEt_k^\star}{m}}_{\mathrm{i}}\Bigr)-\phi\Bigl(\underbrace{\frac{k}{m}}_{\kappa}\Bigr)\Biggr]\Biggr\} \\ &= \frac{\pi}{2}\mathrm{e}^{-\mathrm{i}\phi(\kappa)/\chi}\mathrm{e}^{\mathrm{i}\overbrace{\operatorname{arsinh}(\mathrm{i})}^{\mathrm{i}\pi/2}/\chi} \\ &= \frac{\pi}{2}\mathrm{e}^{-\mathrm{i}\phi(\kappa)/\chi}\mathrm{e}^{-\pi/(2\chi)}. \end{aligned} \tag{3.27}$$

This **coincides with our result** in Eq. (3.18) when the prefactors from Eq. (3.8) are also taken into account:

$$\frac{1}{2}\mathrm{e}^{-\mathrm{i}\phi(\kappa)/\chi}\int\limits_{\mathcal{C}_s}\frac{\mathrm{e}^{\mathrm{i}\phi(\tau)/\chi}}{1+\tau^2}\,\mathrm{d}\tau = \frac{\pi}{2}\mathrm{e}^{-\mathrm{i}\phi(\kappa)/\chi}\mathrm{e}^{-\pi/(2\chi)}. \tag{3.28}$$

### 3.2.3. Branch-cut integral

**Parameterizations**

The last step in our evaluation of the constant-field integral (3.8) is the calculation of the contributions generated while integrating along both sides of the branch cut (see the contour in Fig. 3.2). The corresponding parameterizations are

$$\mathcal{C}_b^\pm: \quad \xi \mapsto (1+\xi)\mathrm{i} \pm \lim_{\epsilon\searrow 0}\epsilon \quad \text{with} \quad \begin{cases}\xi: & 0 \nearrow \infty, \\ \xi: & \infty \searrow 0.\end{cases} \tag{3.29}$$

**Contributions to the contour integral**

Remember that the branch cut in the integrand originates from the phase function $\varphi_k(t)$ only, which is related to the auxiliary function $\phi(\tau)$ via Eq. (3.5) in the constant-field case, so we may simply set $\epsilon = 0$ in the integrand's denom-

inator:

$$\int_{\mathcal{C}_b^\pm} \frac{\mathrm{e}^{\mathrm{i}\phi(\tau)/\chi}}{1+\tau^2}\,\mathrm{d}\tau = \pm \int_0^\infty \frac{\lim_{\epsilon\searrow 0} \mathrm{e}^{\mathrm{i}\phi[(1+\xi)\mathrm{i}\pm\epsilon]/\chi}}{1-(1+\xi)^2}\,\mathrm{i}\,\mathrm{d}\xi$$
$$= \mp\mathrm{i} \int_0^\infty \frac{\lim_{\epsilon\searrow 0} \mathrm{e}^{\mathrm{i}\phi[(1+\xi)\mathrm{i}\pm\epsilon]/\chi}}{\xi\,(2+\xi)}\,\mathrm{d}\xi. \tag{3.30}$$

**Limit $\epsilon \searrow 0$**

In order to perform the limit $\epsilon \searrow 0$, we express the term $\operatorname{arsinh}\tau$ in $\phi(\tau)$ [defined in Eq. (3.6)] via a complex logarithm [see Eq. (3.12)], which yields

$$\phi(\tau) = \tau\sqrt{1+\tau^2} + \ln\left(\tau + \sqrt{1+\tau^2}\right). \tag{3.31}$$

We see that $\phi(\tau)$ depends on the principal value of $\sqrt{1+\tau^2}$, which is sensitive to the side of the branch cut considered. On the paths of $\mathcal{C}_b^\pm$, this square root becomes

$$\lim_{\epsilon\searrow 0}\sqrt{1+[(1+\xi)\mathrm{i}\pm\epsilon]^2} = \lim_{\epsilon\searrow 0}\sqrt{1-(1+\xi)^2+\epsilon^2\pm 2(1+\xi)\epsilon\mathrm{i}}$$
$$= \lim_{\epsilon\searrow 0}\sqrt{\underbrace{-\xi\,(2+\xi)+\epsilon^2}_{<0 \text{ for } \xi>0 \text{ when } \epsilon\searrow 0} \pm \underbrace{2(1+\xi)\epsilon}_{>0}\,\mathrm{i}}$$
$$= \pm\mathrm{i}\sqrt{\xi\,(2+\xi)} \tag{3.32}$$

since the radicand approaches the negative real axis from above/below in the limit $\epsilon \searrow 0$. The argument of the logarithm in Eq. (3.31) along $\mathcal{C}_b^\pm$ is thus

$$(1+\xi)\mathrm{i} \pm \lim_{\epsilon\searrow 0}\epsilon \pm \mathrm{i}\sqrt{\xi\,(2+\xi)} = \Big[\underbrace{1+\xi\pm\sqrt{(1+\xi)^2-1}}_{>0}\Big]\mathrm{i} \pm \lim_{\epsilon\searrow 0}\epsilon. \tag{3.33}$$

Since the principal value of the complex logarithm [according to $\arg(\ldots)$ defined in Eq. (2.113)] is continuous on the positive imaginary axis, we may simply set $\epsilon = 0$ in the above equation. With that, we have taken care of all possible discontinuities of $\phi(\tau)$ at $\epsilon = 0$. By means of the last three equations,

we therefore get

$$\begin{aligned}\lim_{\epsilon\searrow 0}\phi[(1+\xi)\mathrm{i}\pm\epsilon] &= (1+\xi)\mathrm{i}\times(\pm\mathrm{i})\sqrt{\xi\,(2+\xi)}\\ &\quad+\ln\Big\{\mathrm{i}\Big[1+\xi\pm\sqrt{\xi\,(2+\xi)}\Big]\Big\}\\ &= \mp\,(1+\xi)\sqrt{\xi\,(2+\xi)}\\ &\quad+\ln\underbrace{\Big[1+\xi\pm\sqrt{\xi\,(2+\xi)}\Big]}_{>0}+\frac{\mathrm{i}\pi}{2}\end{aligned}\tag{3.34}$$

along $\mathcal{C}_b^{\pm}$.

Having performed the limit $\epsilon\searrow 0$, we insert the above equation into the integral contributions (3.30) and continue with the calculation:

**Simplification of the branch-cut contributions**

$$\begin{aligned}&\int\limits_{\mathcal{C}_b^{\pm}}\frac{\mathrm{e}^{\mathrm{i}\phi(\tau)/\chi}}{1+\tau^2}\,\mathrm{d}\tau\\ &=\mp\,\mathrm{i}\int\limits_0^{\infty}\frac{\mathrm{e}^{\mathrm{i}\left\{\mp(1+\xi)\sqrt{\xi\,(2+\xi)}+\ln\left[1+\xi\pm\sqrt{\xi\,(2+\xi)}\right]+\mathrm{i}\pi/2\right\}/\chi}}{\xi\,(2+\xi)}\,\mathrm{d}\xi\\ &=\mp\,\mathrm{i}\mathrm{e}^{-\pi/(2\chi)}\int\limits_0^{\infty}\frac{\mathrm{e}^{\mathrm{i}\left\{\mp(1+\xi)\sqrt{\xi\,(2+\xi)}+\ln\left[1+\xi\pm\sqrt{\xi\,(2+\xi)}\right]\right\}/\chi}}{\xi\,(2+\xi)}\,\mathrm{d}\xi.\end{aligned}\tag{3.35}$$

Note that both branch-cut contributions are proportional to the square root $\exp[-\pi/(2\chi)]$ of the **Schwinger factor** [just like the singularity contribution (3.18)], which arises from the imaginary part $\pi/2$ of the complex logarithm in the phase function. In order to simplify the remaining integrals a bit, we **substitute**

$$\mu=\sqrt{\xi\,(2+\xi)}=\sqrt{(1+\xi)^2-1}\in[0,\infty)\qquad\Leftrightarrow\qquad 1+\xi=\sqrt{1+\mu^2}$$
$$\Rightarrow\quad \mathrm{d}\xi=\frac{\mu}{\sqrt{1+\mu^2}}\,\mathrm{d}\mu,\tag{3.36}$$

so the integral contributions become

$$\int\limits_{\mathcal{C}_b^{\pm}}\frac{\mathrm{e}^{\mathrm{i}\phi(\tau)/\chi}}{1+\tau^2}\,\mathrm{d}\tau=\mp\mathrm{i}\mathrm{e}^{-\pi/(2\chi)}\int\limits_0^{\infty}\frac{\mathrm{e}^{\mathrm{i}\left[\mp\mu\sqrt{1+\mu^2}+\ln\left(\sqrt{1+\mu^2}\pm\mu\right)\right]/\chi}}{\mu\sqrt{1+\mu^2}}\,\mathrm{d}\mu.\tag{3.37}$$

Since

$$\ln\left(\sqrt{1+\mu^2}-\mu\right) = \ln\left(\frac{1+\mu^2-\mu^2}{\sqrt{1+\mu^2}+\mu}\right) = -\ln\left(\sqrt{1+\mu^2}+\mu\right), \tag{3.38}$$

we may write

$$\ln\left(\sqrt{1+\mu^2}\pm\mu\right) = \pm\ln\left(\sqrt{1+\mu^2}+\mu\right) \tag{3.39}$$

in Eq. (3.37) and thus get

$$\int_{\mathcal{C}_b^\pm} \frac{e^{i\phi(\tau)/\chi}}{1+\tau^2}\,d\tau = \mp i e^{-\pi/(2\chi)} \int_0^\infty \frac{e^{\mp i\left[\mu\sqrt{1+\mu^2}-\ln\left(\sqrt{1+\mu^2}+\mu\right)\right]/\chi}}{\mu\sqrt{1+\mu^2}}\,d\mu. \tag{3.40}$$

**Total contribution from the branch cut**

By adding both integrals together, we obtain the **exact expression** for the total contribution from the branch cut to the contour integral:

$$\begin{aligned}
&\int_{\mathcal{C}_b^+} \frac{e^{i\phi(\tau)/\chi}}{1+\tau^2}\,d\tau + \int_{\mathcal{C}_b^-} \frac{e^{i\phi(\tau)/\chi}}{1+\tau^2}\,d\tau \\
&= i e^{-\pi/(2\chi)} \int_0^\infty \frac{e^{i\left[\mu\sqrt{1+\mu^2}-\ln\left(\sqrt{1+\mu^2}+\mu\right)\right]/\chi} - \text{c.c.}}{\mu\sqrt{1+\mu^2}}\,d\mu \\
&= -2e^{-\pi/(2\chi)} \int_0^\infty \frac{\sin\left\{\frac{1}{\chi}\left[\mu\sqrt{1+\mu^2}-\ln\left(\sqrt{1+\mu^2}+\mu\right)\right]\right\}}{\mu\sqrt{1+\mu^2}}\,d\mu \\
&= -2e^{-\pi/(2\chi)} \underbrace{\int_0^\infty \frac{\sin\left[\frac{1}{\chi}\left(\mu\sqrt{1+\mu^2}-\operatorname{arsinh}\mu\right)\right]}{\mu\sqrt{1+\mu^2}}\,d\mu}_{I_b(\chi)}\,.
\end{aligned} \tag{3.41}$$

For a given $\chi = E/E_{\text{crit}}^{\text{QED}}$, the remaining, real integral $I_b(\chi)$ has to be evaluated, which can be done numerically. The integrand of $I_b(\chi)$ oscillates rapidly for large $\mu$ and is suppressed by $1/\mu^2$ asymptotically. Note that the integrand does *not* diverge at $\mu = 0$, but rather vanishes there because the nominator approaches zero like $\mu^3$ (modulo constant factors) for $\mu \searrow 0$ [since $\mu\sqrt{1+\mu^2} - \operatorname{arsinh}\mu = \mathcal{O}(\mu^3)$; see Eq. (3.42) below] while the denominator approaches zero like $\mu$. The parameter $\chi \ll 1$ appears as an inverse factor in the sine function and thus allows us to scale the overall rapidness of the oscillation. The integrand is plotted for two different values of $\chi$ in Fig. 3.3.

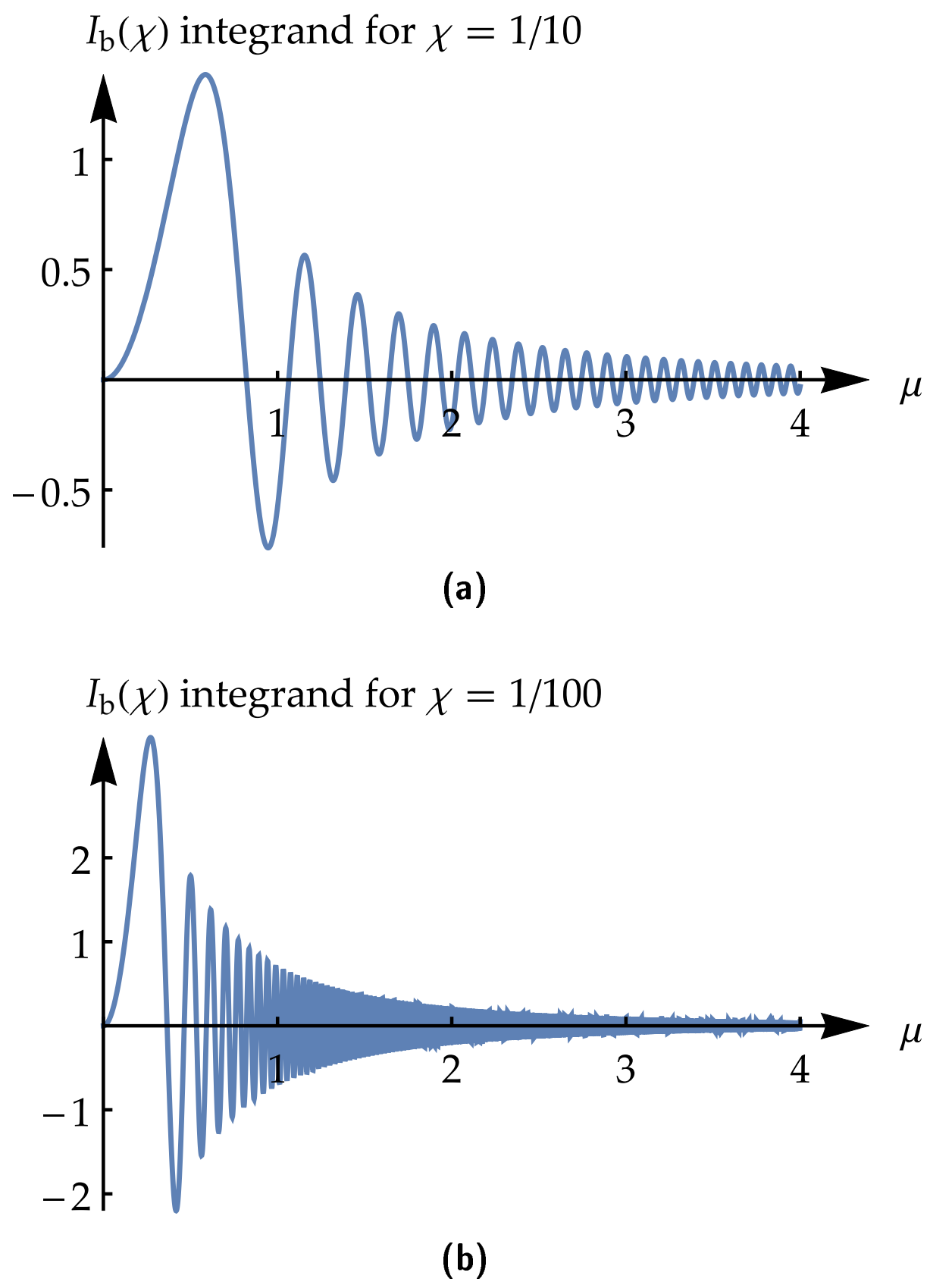

**Figure 3.3.:** Integrand of $I_b(\chi)$ defined in Eq. (3.41) plotted as a function of $\mu$ for the electric field strengths (a) $\chi = E/E_{\text{crit}}^{\text{QED}} = 1/10$ and (b) $\chi = 1/100$.

**Approximation of $I_\mathrm{b}(\chi)$ for small $\chi$**

In view of these plots, the major contribution to the integral $I_\mathrm{b}(\chi)$ seems to be generated over the **first few cycles of the integrand**, where the oscillation is still slow and the amplitude (comparatively) high. The smaller $\chi$, the smaller values of $\mu$ are required to describe this essential range. Hence, for sufficiently small $\chi$, the first cycles will be covered by very small values $\mu \ll 1$ already. This motivates the following approximation of $I_\mathrm{b}(\chi)$: We construct a simpler integrand of the same form (sine function over a suppressing denominator) by requiring this approximate integrand to coincide with the original integrand of $I_\mathrm{b}(\chi)$ for $\mu \ll 1$. Technically, we do this by Taylor expanding the argument of the sine function and the denominator in the original integrand with respect to $\mu$ (around zero) and neglecting everything except for the leading order, respectively. The resulting integral can be solved analytically:

$$\begin{aligned} I_\mathrm{b}(\chi) &= \int_0^\infty \frac{\sin\left[\frac{1}{\chi}\Big(\overbrace{\mu\sqrt{1+\mu^2} - \operatorname{arsinh}\mu}^{2\mu^3/3+\mathcal{O}(\mu^5)}\Big)\right]}{\underbrace{\mu\sqrt{1+\mu^2}}_{\mu+\mathcal{O}(\mu^3)}}\,\mathrm{d}\mu \\ &\overset{\chi\ll 1}{\approx} \int_0^\infty \frac{\sin\left(\frac{2\mu^3}{3\chi}\right)}{\mu}\,\mathrm{d}\mu = \frac{\pi}{6} \approx 0.524, \end{aligned} \tag{3.42}$$

which should be a good **approximation of $I_\mathrm{b}(\chi)$** for sufficiently small $\chi$. Interestingly, the approximated value is independent of $\chi$.

**Quality of the approximation**

In order to evaluate the accuracy of this approximation, we compare $\pi/6$ to **numerically calculated values** of $I_\mathrm{b}(\chi)$. For a relatively strong electric field with $\chi = E/E_\mathrm{crit}^\mathrm{QED} = 1/10$, $\pi/6$ deviates by about 6.3% from the numerically calculated value $I_\mathrm{b}(1/10) \approx 0.493$. This deviation can be explained by reconsidering the integrand in Fig. 3.3(a) again: the assumption that the first cycles of the integrand are completed over small $\mu \ll 1$ is manifestly not correct for $\chi = 1/10$. We see in Fig. 3.3(b) that $\chi = 1/100$ is in better accordance with this assumption, in which case the approximation deviates from the numerical result $I_\mathrm{b}(1/100) \approx 0.517$ by about 1.4% only.

## 3.3. Results and conclusion

**Resulting $R_k^\mathrm{out}$**

In this section, we have evaluated the integral representation (3.8) of $R_k^\mathrm{out}$ for a constant electric field along the contour depicted in Figs. 3.1 and 3.2. Putting our results in Eqs. (3.16), (3.18), and (3.41) together, we get

$$\begin{aligned} R_k^{\text{out}} &\approx \frac{1}{2}\mathrm{e}^{-\mathrm{i}\phi(\kappa)/\chi}\left[\int\limits_{\mathcal{C}_\circ^-}\frac{\mathrm{e}^{\mathrm{i}\phi(\tau)/\chi}}{1+\tau^2}\,\mathrm{d}\tau+\int\limits_{\mathcal{C}_b^-}\ldots\,\mathrm{d}\tau\right.\\ &\qquad\left.+\int\limits_{\mathcal{C}_s}\ldots\,\mathrm{d}\tau+\int\limits_{\mathcal{C}_b^+}\ldots\,\mathrm{d}\tau+\int\limits_{\mathcal{C}_\circ^+}\ldots\,\mathrm{d}\tau\right]\\ &=\frac{1}{2}\mathrm{e}^{-\mathrm{i}\phi(\kappa)/\chi}\Big[\pi-2\underbrace{I_\mathrm{b}(\chi)}_{\approx\pi/6}\Big]\,\mathrm{e}^{-\pi/(2\chi)}\\ &\approx\frac{\pi}{3}\mathrm{e}^{-\mathrm{i}\phi(\kappa)/\chi}\mathrm{e}^{-\pi/(2\chi)} \end{aligned} \tag{3.43}$$

where we have inserted $I_\mathrm{b}(\chi) \approx \pi/6$ from Eq. (3.42) in the last line since this approximation gives the correct order of magnitude of $I_\mathrm{b}(\chi)$, even for relatively strong fields (less than 7% deviation from the numerical result for $\chi = E/E_\text{crit}^\text{QED} = 1/10$, for example). According to Eq. (2.99), the corresponding pair-creation probability for the considered $k$ reads

$$P_k^{\mathrm{e}^+\mathrm{e}^-} \approx |R_k^\text{out}|^2 \approx \frac{\pi^2}{9}\mathrm{e}^{-\pi/\chi} = \frac{\pi^2}{9}\mathrm{e}^{-\pi E_\text{crit}^\text{QED}/E}. \tag{3.44}$$

This probability is independent of $k$, as expected in a constant electric field since a shift in $k$ simply corresponds to a shift in time in this case—see [42], for example, and the $\tau$ substitution in Eq. (3.4). Furthermore, our analysis has reproduced the **exact leading-order exponential Schwinger factor** $\exp(-\pi E_\text{crit}^\text{QED}/E)$, which is known from the total Schwinger pair-creation probability (2.3) per unit four-volume.

**Branch-cut contribution**

Note that the contribution from the branch cut to $R_k^\text{out}$ is of the same order as that from the circumvention of the singularity (interestingly, the branch-cut contribution even counteracts the singularity's contribution a bit, at least in this case). Since the branch-cut integrals are much more difficult to handle for the following field profiles, we take our result here as a hint that the error introduced into $R_k^\text{out}$ by **neglecting the branch cuts** is merely a factor of order $\mathcal{O}(1)$—especially the exponents in $R_k^\text{out}$ are not affected by this simplification.

**Prefactor in $P_k^{\mathrm{e}^+\mathrm{e}^-}$**

Even though we have calculated the full integral representing $R_k^\text{out}$ here, the prefactor in the resulting pair-creation probability (3.44) is nevertheless incorrect due to the semiclassical approximation (linearization of the Riccati equation), even in the limit $E \to 0$—according to [64, 65, 66], the prefactor should be 1 in the adiabatic limit (corresponding to $E \to 0$ here) instead of $\pi^2/9 \approx 1.097$. The same artifact ($\pi^2/9$ instead of 1 in the prefactor) does also occur when calculating the rate of electron–hole pair creation in insulating crystals exposed to constant electric fields (Landau–Zener tunneling) by means of the JWKB approximation [see Eq. (6.13) and the text below this equation in Sec. 6.2].

# 4. Dynamical assistance by a Gauss pulse

In this chapter, we study tunneling in a strong, constant ("background") field $E_{\text{strong}}$ assisted by a weak (i.e., $E_{\text{weak}} \ll E_{\text{strong}}$) **temporal Gauss pulse**

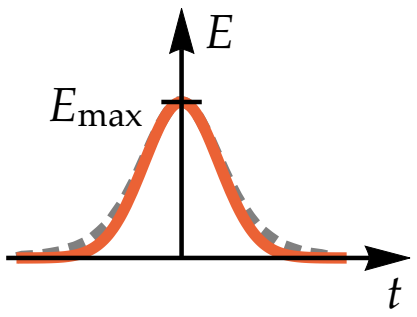

$$E(t) = E_{\text{weak}} \mathrm{e}^{-(\omega t)^2} \tag{4.1}$$

with an associated frequency scale of $\omega > 0$ (as in the case of a temporal Sauter pulse, $\omega$ determines the order of the highest frequencies which significantly contribute to the pulse; see the Fourier spectrum in Fig. 4.1). Compared to the temporal Sauter pulse $E_{\text{weak}} / \cosh^2(\omega t)$ (the dotted graph in the marginal plot), the Gauss pulse looks very similar; it just decays faster and is thus a little bit thinner (considering the full width at half maximum). Intuitively, one would probably not expect that this fact makes a significant physical difference—however, we will show in this chapter that there are remarkable differences between the Sauter pulse and the Gauss pulse in the context of the dynamically assisted Sauter–Schwinger effect; that is, tunneling pair creation seems to be very **sensitive to the pulse shape** of the assisting time-dependent field.

Our analysis in this chapter is analogous to that in Sec. 2.4.6.1 on the "original" dynamically assisted Sauter–Schwinger effect (with a temporal Sauter pulse assisting tunneling).

## 4.1. Main singularity

At first, let us focus on the **main singularity** $t^{\star}_{k,\text{main}}$ of the field profile considered here (constant field plus Gauss pulse), which can be described by the vector potential

$$A(t) = E_{\text{strong}} t + \frac{\sqrt{\pi}}{2} \frac{E_{\text{weak}}}{\omega} \operatorname{erf}(\omega t) \tag{4.2}$$

with the (Gauss) error function

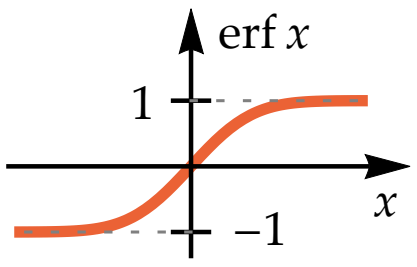

$$\operatorname{erf} x = \frac{2}{\sqrt{\pi}} \int_0^x \mathrm{e}^{-(x')^2} \, \mathrm{d}x'. \tag{4.3}$$

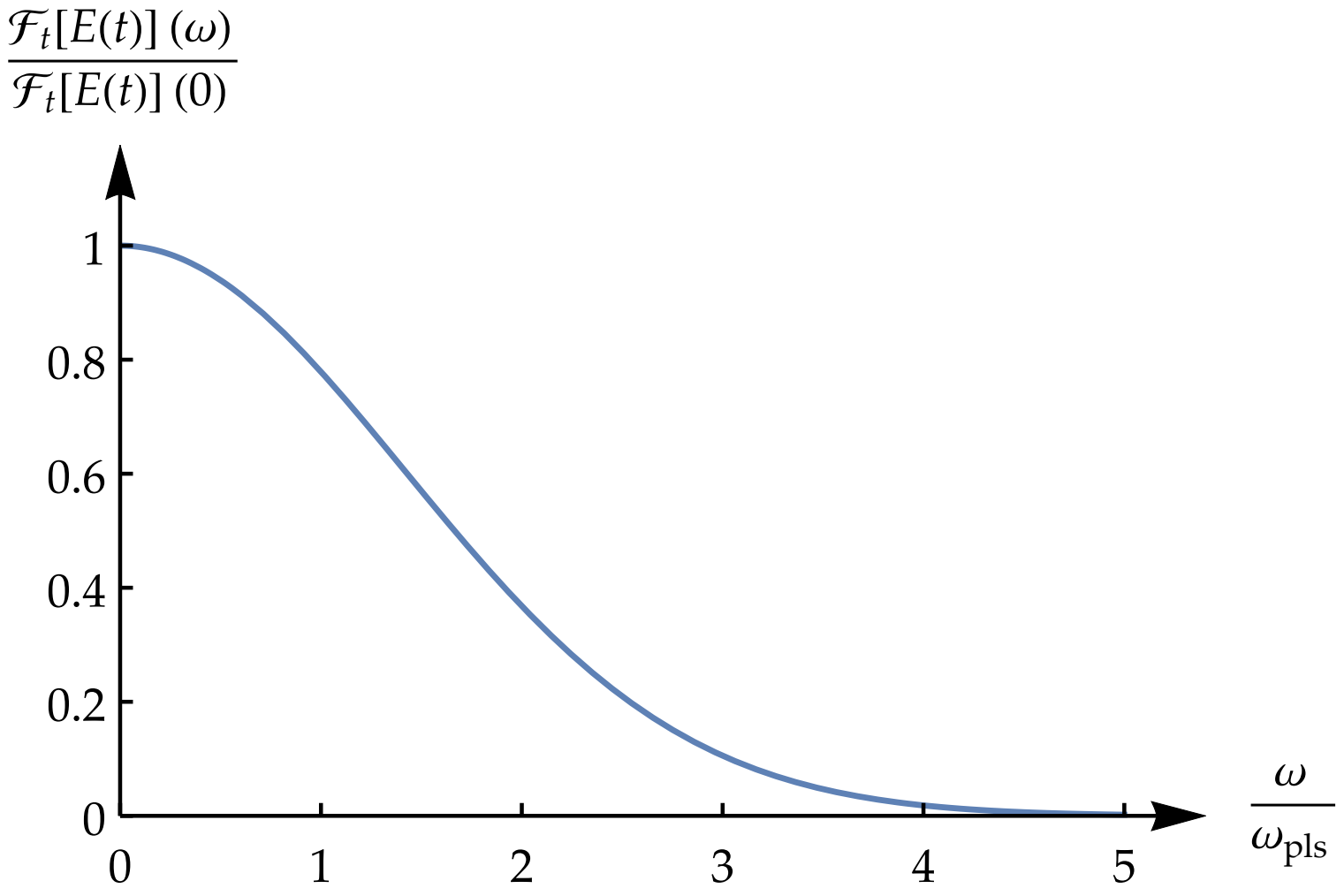

**Figure 4.1.:** Fourier transform $E_{\text{weak}} e^{-(\omega/\omega_{\text{pls}})^2/4}/(\sqrt{2}\omega_{\text{pls}})$ of the temporal Gauss pulse $E(t) = E_{\text{weak}} e^{-(\omega_{\text{pls}} t)^2}$ plotted over $\omega$. The highest significant frequencies are of the order of the pulse's frequency parameter $\omega_{\text{pls}}$, as in the case of a temporal Sauter pulse (cf. the spectrum in Fig. 2.4 on page 56).

The main singularity is the only singularity which remains for $E_{\text{weak}} = 0$, in which case it is given by Eq. (3.3) and corresponds to the pure Sauter–Schwinger effect. Increasing $E_{\text{weak}}$ affects the position of the main singularity in the complex $t$ plane and gives rise to **additional singularities** which do not exist in the constant-field case $E_{\text{weak}} = 0$ (these will be treated in the next section).

As always, the singularities $t_k^\star$ (for a given canonical momentum $k$) are determined by the **singularity equation** (2.112), which can be written as

$$\tau^\star + \frac{\sqrt{\pi}}{2} \varepsilon \operatorname{erf} \tau^\star = (-\kappa \pm \mathrm{i})\gamma_c \tag{4.4}$$

here, with the dimensionless quantities

$$\tau = \omega t, \qquad \varepsilon = \frac{E_{\text{weak}}}{E_{\text{strong}}} \ll 1, \qquad \kappa = \frac{k}{m}, \tag{4.5}$$

and the usual combined Keldysh parameter $\gamma_c = m\omega/(qE_{\text{strong}})$. As in Sec. 2.4.6.1, we concentrate on the **dominating momentum** $\kappa = k = 0$ in the following for simplicity. This canonical momentum is preferred due to the fact that $A(t)$ is an odd function (cf. Sec. 2.4.6.1), and it generates the leading-order exponent (the quantity we are interested in) in the pair-creation probability. As in the case of an assisting Sauter pulse, we suppose the main singularity to be located **on the imaginary axis** for symmetry reasons, which yields a real equation for the imaginary part by setting $\tau^\star_{\text{main}} = \mathrm{i}v$ in Eq. (4.4):

$$v + \frac{\sqrt{\pi}}{2}\varepsilon \operatorname{erfi} v = \gamma_c \tag{4.6}$$

with the imaginary error function

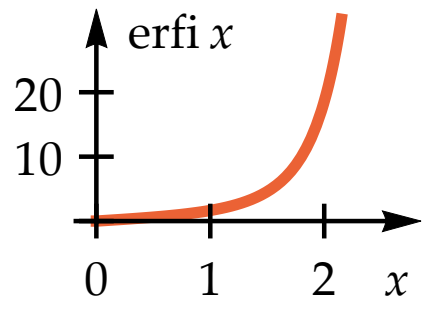

$$\operatorname{erfi} v = \frac{2}{\sqrt{\pi}} \int_0^v \mathrm{e}^{x^2}\, \mathrm{d}x = -\mathrm{i} \operatorname{erf}(\mathrm{i}v). \tag{4.7}$$

We omitted the minus case from Eq. (4.4) here because we are interested in singularities in the upper complex half-plane only.

**Onset of dynamical assistance**

The transcendental equation (4.6) can be solved graphically for given values of $\varepsilon$ and $\gamma_c$; see Fig. 4.2 and cf. Eq. (2.132) and Fig. 2.8 on page 76 from the Sauter-pulse case. The essential difference between both cases is that $\tan v$ in Eq. (2.132) (Sauter-pulse case) has a pole at the fixed position $v = \pi/2$ while $\operatorname{erfi} v$ in Eq. (4.6) is always finite. As a consequence, the point $v_{\text{crit}}$ where the (exponentially increasing) **Gauss-pulse contribution begins to dominate** the linear term $v$ on the left-hand side of Eq. (4.6), which happens (roughly) at

$$\frac{\varepsilon \operatorname{erfi} v_{\text{crit}}}{v_{\text{crit}}} = \mathcal{O}(1), \tag{4.8}$$

becomes arbitrarily large ($v_{\text{crit}} \to \infty$) in the limit $\varepsilon \to 0$—in contrast to the Sauter-pulse case, in which the pole at $v = \pi/2$ acts like a fixed "wall" in this limit[1]. Due to this fact, the critical threshold $\gamma_c^{\text{crit}}$ for dynamical assistance by the Gauss pulse scales with $\varepsilon$ for the following reason: The leading-order term in the asymptotic expansion of the imaginary error function reads [108]

$$\operatorname{erfi} x \overset{x\to\infty}{\sim} \frac{\mathrm{e}^{x^2}}{\sqrt{\pi}x}, \tag{4.9}$$

so in the limit $\varepsilon \to 0$ (i.e., when $v_{\text{crit}}$ grows large), we have

$$\mathcal{O}(1) = \frac{\varepsilon \operatorname{erfi} v_{\text{crit}}}{v_{\text{crit}}} \sim \varepsilon \frac{\mathrm{e}^{v_{\text{crit}}^2}}{v_{\text{crit}}^2} = \varepsilon \exp\Big(v_{\text{crit}}^2 - 2\underbrace{\ln v_{\text{crit}}}_{\text{subdominant}}\Big) \sim \varepsilon \mathrm{e}^{v_{\text{crit}}^2} \tag{4.10}$$

[1]Note that Eq. (4.8) is merely intended to provide an order-of-magnitude estimate—it is not meant to be a precise definition.

and thus

$$v_{\text{crit}}^2 \sim \underbrace{\ln \mathcal{O}(1)}_{\text{constant}} - \ln \varepsilon \qquad \Rightarrow \qquad v_{\text{crit}} \sim \sqrt{-\ln \varepsilon} = \sqrt{|\ln \varepsilon|} \tag{4.11}$$

since $\varepsilon \ll 1$. The value of the left-hand side of Eq. (4.6) at $v_{\text{crit}}$ (i.e., where the exponential Gauss-pulse contribution begins to dominate the linear, pure-tunneling term $v$) is a possible definition for $\gamma_c^{\text{crit}}$ (at least to find out how $\gamma_c^{\text{crit}}$ scales with $\varepsilon$), so we have

$$\gamma_c^{\text{crit}} \sim v_{\text{crit}} + \frac{\sqrt{\pi}}{2} \varepsilon \operatorname{erfi} v_{\text{crit}} = v_{\text{crit}} \left( 1 + \frac{\sqrt{\pi}}{2} \underbrace{\frac{\varepsilon \operatorname{erfi} v_{\text{crit}}}{v_{\text{crit}}}}_{\mathcal{O}(1)} \right) \overset{\varepsilon \to 0}{\sim} \sqrt{|\ln \varepsilon|} \tag{4.12}$$

and thus $\gamma_c^{\text{crit}} \to \infty$ for $\varepsilon \to 0$ in the Gauss-pulse case—whereas $\gamma_c^{\text{crit}}$ approaches the constant value $\pi/2$ in the Sauter-pulse case, which is a **remarkable physical difference** originating from the fact that $\tanh \tau$ has poles in the complex plane (assisting Sauter pulse) while $\operatorname{erf} \tau$ (Gauss pulse) has not.

### 4.1.1. Leading-order exponent

We assume here that the main singularity $t_{0,\text{main}}^{\star} = \tau_{\text{main}}^{\star}/\omega$ generates the leading-order term in $R_0^{\text{out}}$ (we will check the validity of this assumption in the next section), so we have $R_0^{\text{out}} \sim \exp[-2 \operatorname{Im} \varphi_0(t_{0,\text{main}}^{\star})]$ and thus expect

$$P_{e^+e^-} \sim |R_0^{\text{out}}|^2 \sim e^{-4 \operatorname{Im} \varphi_0(t_{0,\text{main}}^{\star})} \tag{4.13}$$

since $k = 0$ is the dominating momentum. This **main exponent** is the quantity we want to determine. In terms of the dimensionless quantities (4.5), the phase function [see Eq. (2.85)] for the Gauss profile reads

$$\varphi_\kappa(\tau) = \underbrace{\frac{1}{\chi \gamma_c}}_{m/\omega} \int_0^\tau \sqrt{1 + \left[ \kappa + \frac{\tau' + \sqrt{\pi} \varepsilon \operatorname{erf}(\tau')/2}{\gamma_c} \right]^2} \, d\tau', \tag{4.14}$$

where we have introduced the dimensionless quantity

$$\chi = \frac{E_{\text{strong}}}{E_{\text{crit}}^{\text{QED}}} \ll 1 \tag{4.15}$$

again, which measures the strength of the (subcritical) background field.

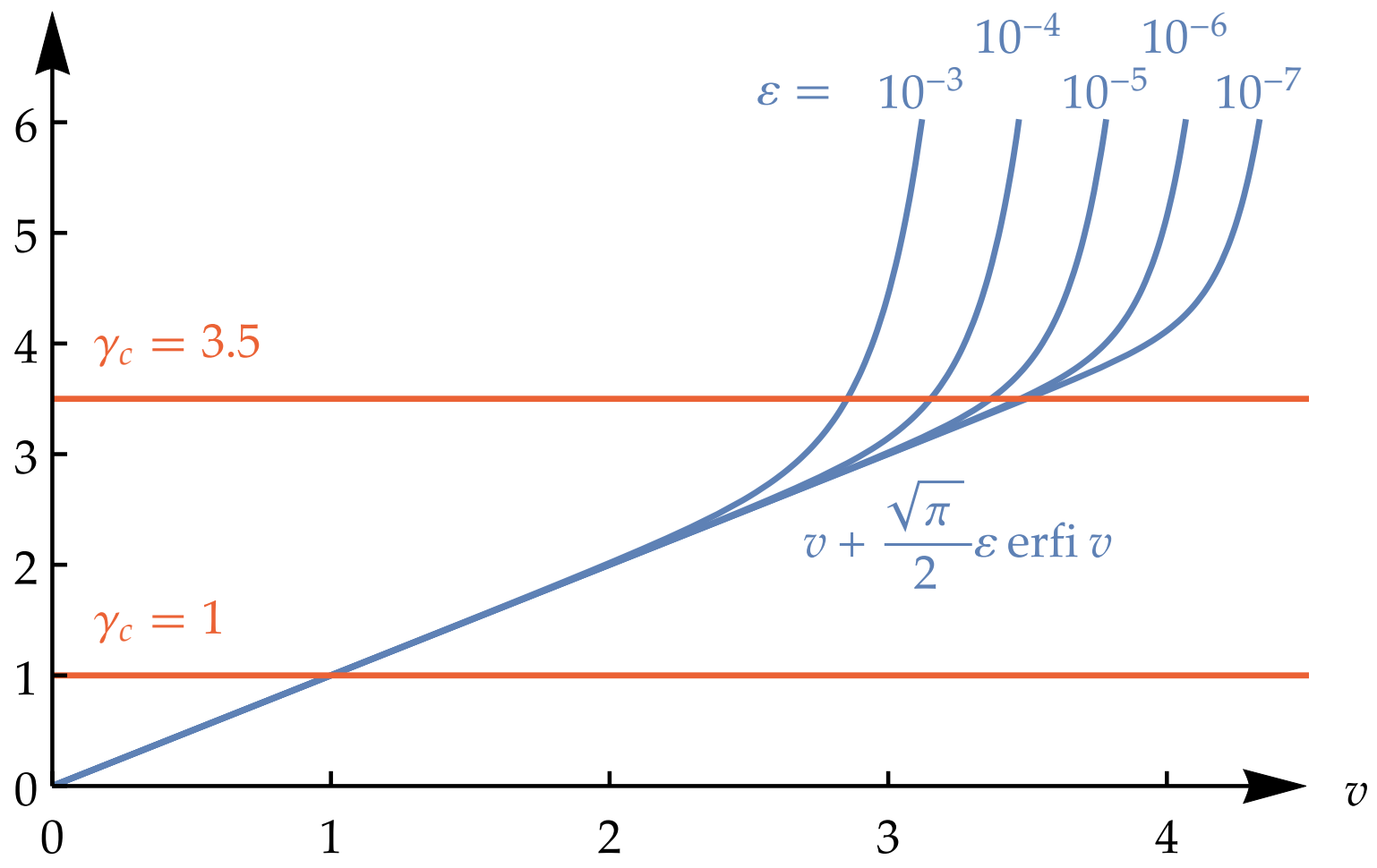

**Figure 4.2.:** Graphical solution of the singularity equation (4.6) for the main singularity $\tau^\star_{\text{main}} = \mathrm{i}v$. The left-hand side of the equation (blue plot lines) is linear for small $v$ but becomes dominated by the exponentially increasing Gauss-pulse contribution $\propto \varepsilon \operatorname{erfi} v$ above some $v_{\text{crit}}$ roughly given by Eq. (4.8). Since $v_{\text{crit}} \sim \sqrt{|\ln \varepsilon|}$ for small $\varepsilon$ [see Eq. (4.11)], the threshold for dynamical assistance, $\gamma_c^{\text{crit}}$, depends on $\varepsilon$ here, even in the limit $\varepsilon \to 0$ (in contrast to the Sauter-pulse case in Fig. 2.8 on page 76). The red $\gamma_c = 1$ line intersects all blue plot lines over their respective linear domains (i.e., where the contribution from the Gauss pulse is negligible), so this value of $\gamma_c$ is subcritical (no dynamical assistance; pure tunneling due to $E_{\text{strong}}$) for all blue graphs in this plot. The higher value, $\gamma_c = 3.5$, still looks subcritical for $\varepsilon = 10^{-6}$ and $10^{-7}$ in this plot, but it clearly intersects the $\varepsilon = 10^{-3}$ graph within its nonlinear domain, which means that the Gauss pulse corresponding to these values of $\varepsilon$ and $\gamma_c$ would assist tunneling in the background field $E_{\text{strong}}$ significantly. The exact value of $\gamma_c^{\text{crit}}$ for every $\varepsilon$, however, is a matter of definition (we will do this in Sec. 4.1.1).

**Analytical approximation of the exponent**
Unfortunately, we cannot find an exact expression for $\tau^\star_{\text{main}} = \mathrm{i}v$ since Eq. (4.6) is a transcendental equation, and the integral in the phase function (4.14) is also too difficult to solve. However, we may expand the main exponent (4.13) in **powers of the small quantity** $\varepsilon$ (for fixed values of $\chi$ and $\gamma_c$), which measures the amplitude of the assisting Gauss pulse. For $\varepsilon = 0$, the Gauss pulse vanishes, and we get $\tau^\star_{\text{main}}(\varepsilon = 0) = \gamma_c$ and the ordinary Sauter–Schwinger exponent $-\pi E^{\text{QED}}_{\text{crit}}/E_{\text{strong}} = -\pi/\chi$ [cf. Eq. (2.117)–(2.118)]. As we increase $\varepsilon$, the main singularity $\tau^\star_{\text{main}}(\varepsilon)$ starts to move, and thus the exponent changes. For sufficiently small $\varepsilon$, this change is described well by the linear order in $\varepsilon$. Hence, let us calculate this **first-order correction of the Sauter–Schwinger exponent** due to the additional Gauss pulse.

We start with the value of the phase function at $\tau^\star_{\text{main}}(\varepsilon)$ (for the dominating momentum $\kappa = 0$):

$$\varphi_0(\tau^\star_{\text{main}}(\varepsilon)) = \frac{\mathrm{i}\pi}{4\chi} + \frac{1}{\chi\gamma_c}\frac{\mathrm{d}}{\mathrm{d}\varepsilon}\int_0^{\tau^\star_{\text{main}}(\varepsilon)} \sqrt{1 + \left[\frac{\tau' + \sqrt{\pi}\varepsilon\,\mathrm{erf}(\tau')/2}{\gamma_c}\right]^2}\,\mathrm{d}\tau'\Bigg|_{\varepsilon=0} \varepsilon + \mathcal{O}(\varepsilon^2). \quad (4.16)$$

The constant term ($\propto \varepsilon^0$) corresponds to the Sauter–Schwinger exponent. Note that we have to apply the **Leibniz integral rule** to calculate the integral appearing in the first-order term because the upper limit depends on $\varepsilon$ as well, not just the integrand. The general rule reads

$$\frac{\mathrm{d}}{\mathrm{d}\varepsilon}\int_{a(\varepsilon)}^{b(\varepsilon)} f(x,\varepsilon)\,\mathrm{d}x = \underbrace{f(b(\varepsilon),\varepsilon)}_{=0\text{ above}}\frac{\mathrm{d}b(\varepsilon)}{\mathrm{d}\varepsilon} - f(a(\varepsilon),\varepsilon)\underbrace{\frac{\mathrm{d}a(\varepsilon)}{\mathrm{d}\varepsilon}}_{=0\text{ above}} + \int_{a(\varepsilon)}^{b(\varepsilon)} \frac{\partial f(x,\varepsilon)}{\partial\varepsilon}\,\mathrm{d}x. \quad (4.17)$$

When applied to the above integral in Eq. (4.16), the first term vanishes since $\tau^\star_{\text{main}}(\varepsilon)$ inserted into the integrand yields exactly zero (because the singularities are precisely the zeros of the square root). The lower integral limit is constant, so the second term does also vanish, and we consequently get

$$\begin{aligned}\varphi_0(\tau^\star_{\text{main}}(\varepsilon)) &= \frac{\mathrm{i}\pi}{4\chi} + \frac{1}{\chi\gamma_c}\int_0^{\tau^\star_{\text{main}}(0)} \frac{2\frac{\tau'}{\gamma_c}\frac{\sqrt{\pi}}{2\gamma_c}\,\mathrm{erf}\,\tau'}{2\sqrt{1+(\tau'/\gamma_c)^2}}\,\mathrm{d}\tau'\,\varepsilon + \mathcal{O}(\varepsilon^2)\\ &= \frac{\mathrm{i}\pi}{4\chi} + \frac{\sqrt{\pi}}{2\chi\gamma_c^3}\int_0^{\mathrm{i}\gamma_c} \frac{\tau'\,\mathrm{erf}\,\tau'}{\sqrt{1+(\tau'/\gamma_c)^2}}\,\mathrm{d}\tau'\,\varepsilon + \mathcal{O}(\varepsilon^2)\end{aligned}$$

$$= \frac{\mathrm{i}\pi}{4\chi} - \frac{\sqrt{\pi}}{2\chi\gamma_c}\int_0^1 \frac{\xi\,\mathrm{erf}(\mathrm{i}\gamma_c\xi)}{\sqrt{1-\xi^2}}\,\mathrm{d}\xi\,\varepsilon + \mathcal{O}(\varepsilon^2)$$

$$= \frac{\mathrm{i}\pi}{4\chi} - \frac{\mathrm{i}\pi}{4\chi}\mathrm{e}^{\gamma_c^2/2}\left[I_0\left(\frac{\gamma_c^2}{2}\right) - I_1\left(\frac{\gamma_c^2}{2}\right)\right]\varepsilon + \mathcal{O}(\varepsilon^2), \tag{4.18}$$

**Result**

where $I_n(x)$ denotes a modified Bessel function of the first kind. The **main exponent** in Eq. (4.13) for the Gauss profile thus reads ($\chi = E_{\text{strong}}/E_{\text{crit}}^{\text{QED}}$)

$$-4\,\mathrm{Im}\,\varphi_0(t^\star_{0,\text{main}})$$
$$= -\frac{\pi E_{\text{crit}}^{\text{QED}}}{E_{\text{strong}}}\left\{1 - \underbrace{\mathrm{e}^{\gamma_c^2/2}\left[I_0\left(\frac{\gamma_c^2}{2}\right) - I_1\left(\frac{\gamma_c^2}{2}\right)\right]}_{\text{factor in marginal plot}}\varepsilon + \mathcal{O}(\varepsilon^2)\right\}. \tag{4.19}$$

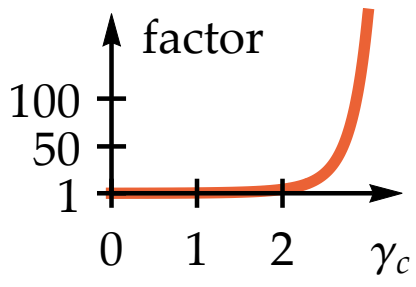

This result is confirmed in Ref. [1] via the worldline-instanton method. Neglecting the terms of order $\mathcal{O}(\varepsilon^2)$ should approximate the actual exponent well **as long as the first-order correction is small**; that is,

**Validity condition**

$$1 \gg \varepsilon\mathrm{e}^{\gamma_c^2/2}\left[I_0\left(\frac{\gamma_c^2}{2}\right) - I_1\left(\frac{\gamma_c^2}{2}\right)\right] \overset{\gamma_c\to\infty}{\sim} \frac{\varepsilon\mathrm{e}^{\gamma_c^2}}{\sqrt{\pi}\gamma_c^3}, \tag{4.20}$$

where we have used the asymptotic expansions [108]

$$I_0(x) \sim \frac{\mathrm{e}^x}{\sqrt{2\pi x}}\left[1 + \frac{1}{8x} + \mathcal{O}(x^{-2})\right] \quad \text{and} \tag{4.21}$$

$$I_1(x) \sim \frac{\mathrm{e}^x}{\sqrt{2\pi x}}\left[1 - \frac{3}{8x} + \mathcal{O}(x^{-2})\right]. \tag{4.22}$$

### Definition of the critical combined Keldysh parameter

The main exponent facilitates a precise definition of the critical threshold for dynamical assistance: given an $\varepsilon \ll 1$, say $\gamma_c^{\text{crit}}(\varepsilon)$ is the value of the combined Keldysh parameter for which the first-order correction in the curly brackets in Eq. (4.19) measures $1/100$ (i.e., the **Sauter–Schwinger exponent is lowered by 1%**):

$$\mathrm{e}^{[\gamma_c^{\text{crit}}(\varepsilon)]/2}\left[I_0\left(\frac{[\gamma_c^{\text{crit}}(\varepsilon)]^2}{2}\right) - I_1\left(\frac{[\gamma_c^{\text{crit}}(\varepsilon)]^2}{2}\right)\right]\varepsilon \overset{!}{=} 0.01. \tag{4.23}$$

In the limit $\varepsilon \to 0$, when $\gamma_c^{\text{crit}}(\varepsilon)$ approaches infinity, we may infer from the scaling law in Eq. (4.20) that

$$\gamma_c^{\text{crit}}(\varepsilon) \overset{\varepsilon\to 0}{\sim} \sqrt{|\ln\varepsilon|} \tag{4.24}$$

in accordance with Eq. (4.12). The approximation of the main exponent in Eq. (4.19) should be good up to the critical point (and maybe even a little bit above) defined by Eq. (4.23).

**Enhancement at the threshold**

Note that the reduction of the Sauter–Schwinger exponent by 1% can have a significant effect on the pair-creation yield: for $E_{\text{strong}} = E_{\text{crit}}^{\text{QED}}/10$, this reduction enhances the expected number of pairs per unit volume (approximately) by

$$\mathcal{N}_{e^+e^-} \sim e^{-\pi E_{\text{crit}}^{\text{QED}}/E_{\text{strong}}(1-0.01)} = e^{-\pi E_{\text{crit}}^{\text{QED}}/E_{\text{strong}}} \underbrace{e^{\pi/10}}_{\approx 1.37}, \tag{4.25}$$

that is, by 37%. If the background field is weaker, the enhancement will be even stronger: for $E_{\text{strong}} = E_{\text{crit}}^{\text{QED}}/100$, we get

$$\mathcal{N}_{e^+e^-} \sim e^{-\pi E_{\text{crit}}^{\text{QED}}/E_{\text{strong}}(1-0.01)} = e^{-\pi E_{\text{crit}}^{\text{QED}}/E_{\text{strong}}} \underbrace{e^{\pi}}_{\approx 23}. \tag{4.26}$$

### Critical pulse amplitude

Unfortunately, we can only calculate $\gamma_c^{\text{crit}}(\varepsilon)$ numerically since Eq. (4.23) cannot be solved for $\gamma_c^{\text{crit}}(\varepsilon)$ analytically. However, we can easily calculate the inverse relation, that is, the critical $\varepsilon$ for a given $\gamma_c$:

$$\varepsilon_{\text{crit}}(\gamma_c) = \left\{100 e^{\gamma_c^2/2}\left[I_0\left(\frac{\gamma_c^2}{2}\right) - I_1\left(\frac{\gamma_c^2}{2}\right)\right]\right\}^{-1}. \tag{4.27}$$

This quantity is plotted in Fig. 4.3.

### Comparison with the numerically calculated main exponent

Our analytical approximation of the main exponent in Eq. (4.19) is plotted in Fig. 4.4, together with the corresponding numerically calculated values. We see that our approximation coincides well with the numerical results up to the critical threshold, at which the tunneling exponent is lowered by 1%. Above the threshold, the analytical approximation drops significantly faster than the numerical results. This deviation occurs because the validity condition (4.20) is no longer satisfied well above the threshold.

## 4.2. Additional singularities

After studying the main singularity (which lies on the imaginary $t$ axis for $k = 0$) and the corresponding exponent in $R_0^{\text{out}}$ in the previous section, we now consider the contributions from the additional singularities. These are the solutions of the singularity equation (4.4) which vanish to infinity in the limit $\varepsilon \to 0$ (only the main singularity remains finite in this limit)[2]. Our main intention in this section is to show that the contributions from these additional singularities will be negligible if the Gauss pulse is not too far above the critical threshold for dynamical assistance. We focus on the dominating momentum $k = \kappa = 0$ again.

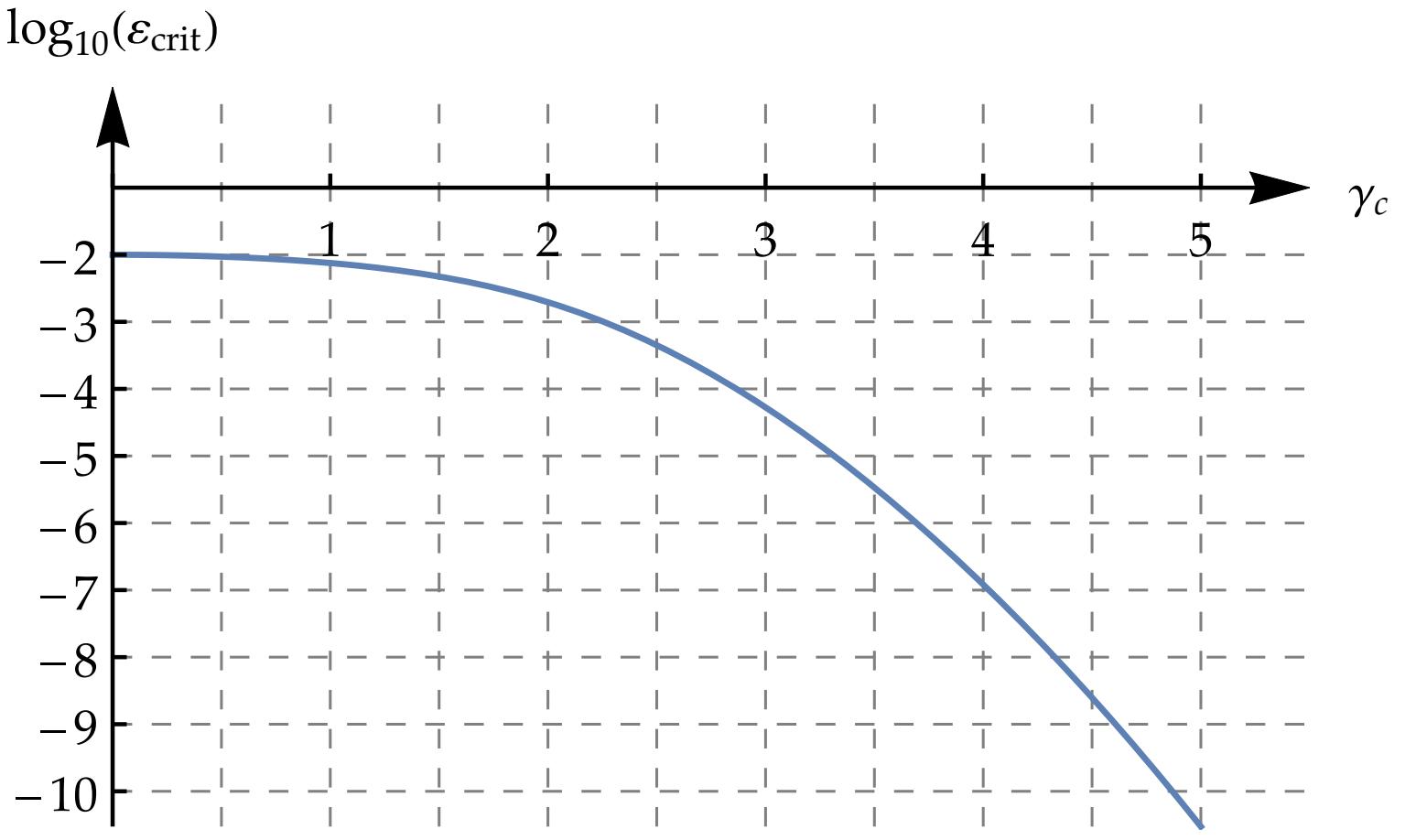

**Figure 4.3.:** Critical value of $\varepsilon$ according to the definition in Eq. (4.27) as a function of $\gamma_c$. At the critical threshold, the Sauter–Schwinger exponent $-\pi E_{\mathrm{crit}}^{\mathrm{QED}}/E_{\mathrm{strong}}$ is lowered by 1% due to dynamical assistance by the Gauss pulse.

### 4.2.1. Asymptotic solution of the singularity equation for low-intensity Gauss pulses

Say we consider an additional-singularity solution $\tau_{\mathrm{add}}^{\star} = r\exp(\mathrm{i}\vartheta)$ to the singularity equation (4.4) in the limit $\varepsilon \to 0$, so $r \to \infty$. For $\kappa = 0$, the singularity equation is invariant under $\tau^{\star} \to -\tau^{\star}$ and $\tau^{\star} \to (\tau^{\star})^{*}$ (complex conjugation). We may therefore concentrate on additional singularities in the **first quadrant** ($\mathrm{Re}\,\tau_{\mathrm{add}}^{\star} > 0$ and $\mathrm{Im}\,\tau_{\mathrm{add}}^{\star} > 0$), so we assume $\vartheta \in (0, \pi/2)$ here. Later, we can simply mirror these additional singularities into the second quadrant ($\mathrm{Re}\,\tau_{\mathrm{add}}^{\star} < 0$).

[2]Note that all additional singularities of the Gauss profile must diverge in the limit $\varepsilon \to 0$ because the error function in the singularity equation is finite everywhere. This is different from the Sauter-pulse case since the tanh function appearing in the corresponding singularity equation (2.131) has poles in the complex plane, which coincide with the additional singularities in the limit $\varepsilon \to 0$ (cf. Fig. 2.8 on page 76); that is, the additional singularities do *not* approach infinity for $\varepsilon \to 0$ in the Sauter-pulse case.

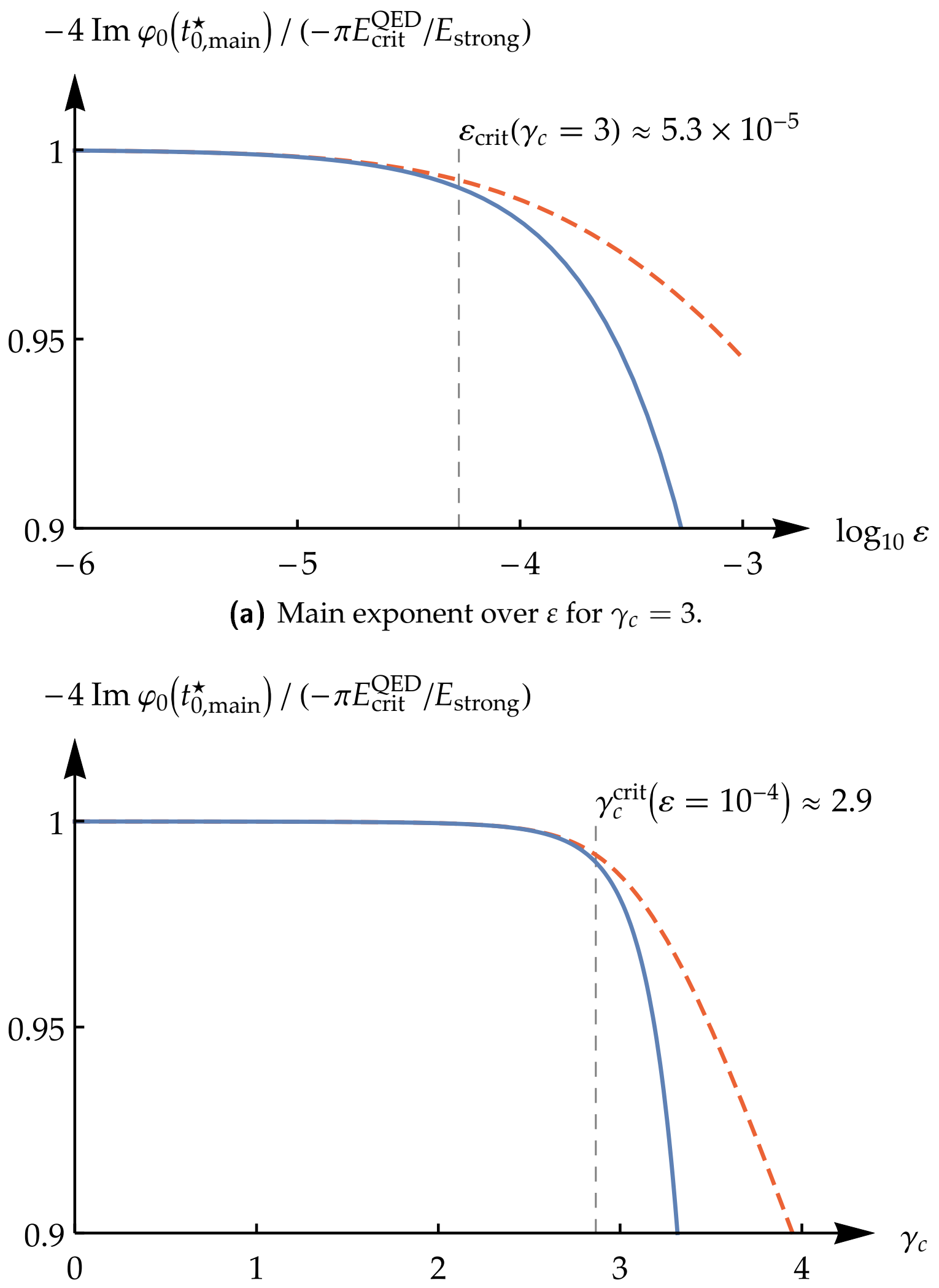

**(a)** Main exponent over $\varepsilon$ for $\gamma_c = 3$.

**(b)** Main exponent over $\gamma_c$ for $\varepsilon = 10^{-4}$.

**Figure 4.4.:** Main exponent in the pair-creation probability for the Gauss profile, in units of the pure Sauter–Schwinger exponent $-\pi E_{\mathrm{crit}}^{\mathrm{QED}}/E_{\mathrm{strong}}$. The blue, solid line is the analytical approximation in Eq. (4.19), and the red, dashed line is the numerical result [calculated by solving the singularity equation (4.6) and the integral in the phase function (4.14) numerically]. The critical values are given by Eqs. (4.23) and (4.27), respectively (the former was solved numerically).

**Asymptotic singularity equation**

Since $r$ is large for sufficiently small $\varepsilon$, we may insert the asymptotic term [108]

$$\operatorname{erf}\tau^\star_{\text{add}} \overset{r\to\infty}{\sim} 1-\frac{\mathrm{e}^{-(\tau^\star_{\text{add}})^2}}{\sqrt{\pi}\tau^\star_{\text{add}}} \tag{4.28}$$

(note that this expansion is valid for all $\vartheta$ we consider) into the singularity equation (4.4) and obtain

$$\begin{aligned}\pm\mathrm{i}\gamma_c &\approx r\mathrm{e}^{\mathrm{i}\vartheta}+\frac{\sqrt{\pi}}{2}\varepsilon\left[1-\frac{\mathrm{e}^{-r^2\exp(2\mathrm{i}\vartheta)}}{\sqrt{\pi}r\mathrm{e}^{\mathrm{i}\vartheta}}\right]\\ &= r\mathrm{e}^{\mathrm{i}\vartheta}+\frac{\sqrt{\pi}}{2}\varepsilon\left[1-\frac{\mathrm{e}^{-r^2\cos(2\vartheta)}\mathrm{e}^{-\mathrm{i}r^2\sin(2\vartheta)}}{\sqrt{\pi}r\mathrm{e}^{\mathrm{i}\vartheta}}\right].\end{aligned} \tag{4.29}$$

Since $-\cos(2\vartheta)\leq 0$ for $\vartheta\leq\pi/4$, the term in square brackets vanishes for these $\vartheta$ in the simultaneous limits $\varepsilon\to 0$ and $r\to\infty$, and therefore we end up with just the main singularity $\mathrm{i}\gamma_c$ in this case, which is not a consistent solution here. Hence, each additional singularity (in the first quadrant) must have a corresponding $\vartheta\in(\pi/4,\pi/2)$ asymptotically, which leads to $\cos(2\vartheta)<0$, so $\exp[-r^2\cos(2\vartheta)]$ can compensate the smallness of $\varepsilon$ for these $\vartheta$. Neglecting the terms $\pm\mathrm{i}\gamma_c$[3] and $\sqrt{\pi}\varepsilon/2$ in Eq. (4.29), which become negligible in the limit $r\to\infty$, the asymptotic singularity equation can be cast into the form

$$\begin{aligned} & r\mathrm{e}^{\mathrm{i}\vartheta}\approx\frac{\varepsilon}{2}\frac{\mathrm{e}^{-r^2\cos(2\vartheta)}\mathrm{e}^{-\mathrm{i}r^2\sin(2\vartheta)}}{r\mathrm{e}^{\mathrm{i}\vartheta}}\\ \Leftrightarrow\quad & \underbrace{\frac{2r^2}{\varepsilon}\mathrm{e}^{r^2\cos(2\vartheta)}}_{\text{modulus}}\underbrace{\mathrm{e}^{\mathrm{i}\,[r^2\sin(2\vartheta)+2\vartheta]}}_{\text{phase}}\approx 1.\end{aligned} \tag{4.30}$$

The modulus on both sides must be the same. After taking the logarithm, we thus get

$$\underbrace{r^2\cos(2\vartheta)}_{\to-\infty\text{ for }r\to\infty}\approx\underbrace{\ln\varepsilon}_{\to-\infty\text{ for }\varepsilon\to 0}-\underbrace{\ln(2r^2)}_{\ll r^2\text{ for }r\to\infty}, \tag{4.31}$$

**Asymptotic $r$**

which can be approximated by

$$r^2\underbrace{\cos(2\vartheta)}_{<0}\approx\underbrace{\ln\varepsilon}_{<0}\quad\Rightarrow\quad r\approx\sqrt{\left|\frac{\ln\varepsilon}{\cos(2\vartheta)}\right|}. \tag{4.32}$$

[3]Note that the fact that $\pm\mathrm{i}\gamma_c$ can be neglected for large $r$ implies that each $+\mathrm{i}\gamma_c$ solution approaches the corresponding $-\mathrm{i}\gamma_c$ solution in the limit $r\to\infty$. Keep in mind that each approximate solution $\tau^\star_{\text{add}}$ we derive in the following thus corresponds to a *pair* ($\pm$ cases) of actual solutions to the exact singularity equation (4.4).

The angle in the complex plane must also be the same on both sides of Eq. (4.30), so we demand

$$r^2 \underbrace{\sin(2\vartheta)}_{>0} + \underbrace{2\vartheta}_{\in(\pi/2,\pi)\ \Rightarrow\ n\in\mathbb{N}\ (\text{not}\ \mathbb{Z})} \overset{!}{\approx} 2\pi n \qquad \text{with} \qquad n \in \mathbb{N}. \tag{4.33}$$

**Implicit $\vartheta$**

By means of Eq. (4.32), this equation becomes

$$-\,|\ln\varepsilon|\tan(2\vartheta) + 2\vartheta \approx 2\pi n. \tag{4.34}$$

We may solve this transcendental equation numerically for a given $n$. However, in order to derive an analytical approximation for large $|\ln\varepsilon|$, we substitute $2\vartheta$ by $\vartheta' = \pi - 2\vartheta \in (0, \pi/2)$ for the moment. Due to the periodicity of the tangent function, we get

$$-|\ln\varepsilon|\tan(\pi - \vartheta') + \pi - \vartheta' \approx 2\pi n$$

$$\Leftrightarrow \qquad |\ln\varepsilon|\tan\vartheta' - \vartheta' \approx \pi\,(2n-1). \tag{4.35}$$

**Approximated $\vartheta$**

In the limit $\varepsilon \to 0$, the factor $|\ln\varepsilon|$ is large [more precisely: if $|\ln\varepsilon| \gg (2n-1)\pi$ for a fixed $n$], so small values of $\vartheta'$ on the left-hand side of this equation are sufficient to "reach" the number $(2n-1)\pi$ on the right-hand side. For these small $\vartheta'$, we may approximate $\tan\vartheta' \approx \vartheta'$, and thus find a corresponding phase angle $\vartheta = (\pi - \vartheta')/2$ after resubstitution:

$$\vartheta' \approx \pi\frac{2n-1}{|\ln\varepsilon| - 1}$$

$$\Leftrightarrow \qquad \vartheta \approx \frac{\pi}{2}\left(1 - \frac{2n-1}{|\ln\varepsilon| - 1}\right) \qquad \text{for each} \qquad n \in \mathbb{N}. \tag{4.36}$$

**Additional singularities**

In conclusion, we get one solution of the form

$$\tau^\star_{\text{asymp},n} = r_n e^{i\vartheta_n} = \sqrt{\left|\frac{\ln\varepsilon}{\cos(2\vartheta_n)}\right|}\,e^{i\vartheta_n} \qquad \text{for each} \qquad n \in \mathbb{N} \tag{4.37}$$

from the asymptotic form (4.30) of the singularity equation (4.4) for $k = 0$, with $\vartheta_n \in (\pi/4, \pi/2)$ given implicitly by Eq. (4.34) or approximately by Eq. (4.36). Remember that we have neglected the term $\pm i\gamma_c$ which appears in the exact singularity equation during the derivation of this asymptotic approximation—as a consequence, each $\tau^\star_{\text{asymp},n}$ **corresponds to two different additional singularities** ($\pm$ cases).

**Distribution of the additional singularities in the complex plane**

Each asymptotic solution (4.37) satisfies

$$\begin{aligned} \mathrm{Re}^2\,\tau^\star_{\mathrm{asymp},n} - \mathrm{Im}^2\,\tau^\star_{\mathrm{asymp},n} &= r_n^2\,(\cos^2\vartheta_n - \sin^2\vartheta_n) \\ &= r_n^2\cos(2\vartheta_n) \\ &= -|\ln\varepsilon| \\ \Leftrightarrow \qquad \underbrace{\mathrm{Im}\,\tau^\star_{\mathrm{asymp},n}}_{>0} &= \sqrt{|\ln\varepsilon| + \mathrm{Re}^2\,\tau^\star_{\mathrm{asymp},n}}\,, \end{aligned} \tag{4.38}$$

so the additional singularities **lie on a hyperbola** asymptotically (i.e., for $\varepsilon \to 0$)[4]. This hyperbola intersects the imaginary $\tau$ axis at $\mathrm{i}\sqrt{|\ln\varepsilon|}$, that is, far above the main singularity $\tau^\star_{\mathrm{main}} \approx \mathrm{i}\gamma_c$ for subcritical $\gamma_c < \gamma_c^{\mathrm{crit}}(\varepsilon) \sim \sqrt{|\ln\varepsilon|}$.

### 4.2.2. Comparison with numerically calculated singularities

**Moving singularities for varying $\varepsilon$**

The movement of the (numerically calculated) singularities $\tau^\star$ for the dominating momentum $\kappa = 0$ is depicted in Fig. 4.5 for varying $\varepsilon$, together with the asymptotic approximations (4.37) of the additional singularities. The agreement between the numerical and the asymptotic results is not perfect because the asymptotic approximation requires $|\ln\varepsilon|$ to be a very large number. However, values of $\varepsilon$ which are too tiny are probably not interesting from an experimental point of view. For example, if we set the background field to $E_{\mathrm{strong}} = E_{\mathrm{crit}}^{\mathrm{QED}}/100$, we have $E_{\mathrm{strong}} \approx 10^{16}\,\mathrm{V/m}$, and thus $\varepsilon = 10^{-15}$ (which only yields $|\ln\varepsilon| \approx 34.5$) corresponds to $E_{\mathrm{weak}} \approx 10\,\mathrm{V/m}$, which is a quite small pulse amplitude. We see in Fig. 4.5 that the main singularity (on the imaginary axis) remains approximately fixed at the "tunneling position" $\mathrm{i}\gamma_c$ as long as $\varepsilon < \varepsilon_{\mathrm{crit}}(\gamma_c)$. The additional singularities lie far above the main singularity in the complex plane for small $\varepsilon$. As $\varepsilon$ grows, the additional singularities approach the real axis. The movement of the main singularity (towards the real axis) approximately begins when the additional singularities get close. The main singularity is always the one closest to the real axis. This fact indicates that its contribution to $R_0^{\mathrm{out}}$ (we only consider the exponent) always dominates, even above the critical threshold $\varepsilon > \varepsilon_{\mathrm{crit}}(\gamma_c)$. We will check this assumption below.

**Moving singularities for varying $\gamma_c$**

In the next figure 4.6, we see the moving singularities for varying $\gamma_c$. In the subcritical regime $\gamma_c < \gamma_c^{\mathrm{crit}}(\varepsilon)$, the main singularity moves like $\tau^\star_{\mathrm{main}} \approx \mathrm{i}\gamma_c$ (pure Sauter–Schwinger case), and the additional singularities lie much deeper in the complex plane. For each pair of additional singularities ($\pm i\gamma_c$ solutions of the singularity equation), the two respective singularities lie very close to each other in the subcritical regime. As a consequence, the sum of their contributions [residues (3.25)] to $R_0^{\text{out}}$ becomes small since each pair is made up of one $+$ and one $-$ solution [if both singularities merge, which happens in the limit $\gamma_c \to 0$, their contributions cancel each other out completely according to Eq. (3.25)]. The threshold for dynamical assistance (approximately) coincides with the moment when the main singularity is at the same height as the (closest) additional singularities. Then, the movement of the main singularity slows down exponentially, and it looks like the main singularity "pushes" the additional singularities upwards. The distance between two additional singularities making up a pair starts to grow when this "pushing" process begins, which means that the mutual cancellation of the additional singularities decreases. Hence, the relative contribution from the additional singularities to $R_0^{\text{out}}$ seems to be the higher the greater $\gamma_c$ is.

[4] Note that this derivation is not based on the approximation made in Eq. (4.36) [linearization of the tangent function in Eq. (4.34)].

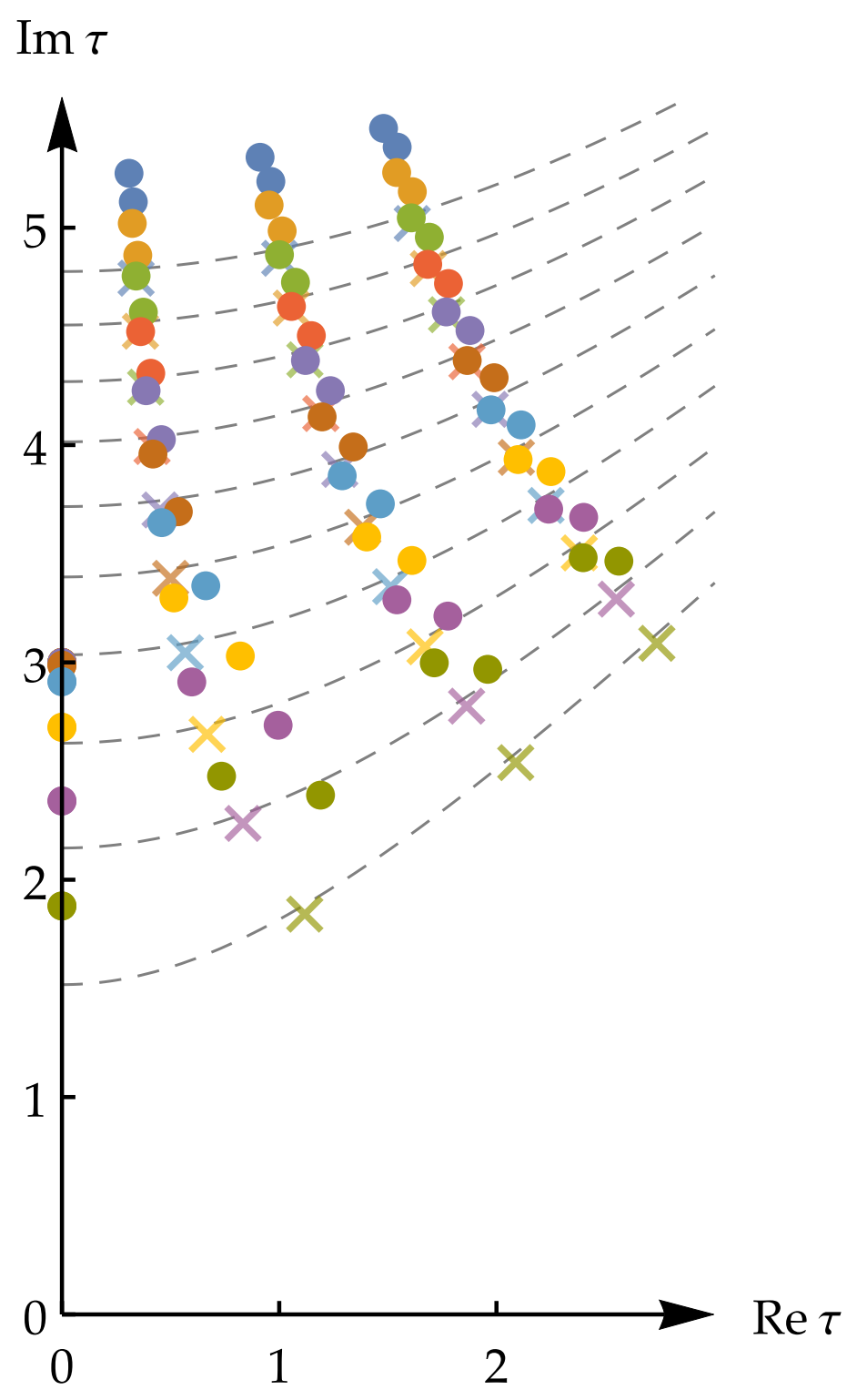

**Figure 4.5.:** Numerically calculated solutions $\tau^\star$ (colored dots) of the singularity equation (4.4) in the complex $\tau$ plane, with $\kappa = 0$ (dominating momentum) and $\gamma_c = 3$ fixed, and varying $\varepsilon = 10^{-10}$ (blue), $10^{-9}$ (ocher), ..., $10^{-1}$ (olive). According to Eq. (4.27), $\varepsilon_{\text{crit}} \approx 5.3 \times 10^{-5}$ here. Only the first three "columns" of pairs of additional singularities are included in this plot. The second quadrant is just the mirror image of the first quadrant. The colored crosses are the asymptotic solutions (4.37) [which lie on one of the gray, dashed hyperbolas (4.38) for each $\varepsilon$] which approximate the additional singularities, with $n \in \{1, 2, 3\}$. The angles $\vartheta_n$ of the asymptotic singularities were calculated numerically from Eq. (4.34). Each cross corresponds to one pair of additional singularities (dots) since we neglected the term $\pm i\gamma_c$ to derive the approximated equation (4.30).

**Exponents**

Let us conclude our study of the additional singularities by comparing the numerically calculated exponent in the contribution from $\tau^\star_{\text{main}}$ to $R_0^{\text{out}}$ (which has already been plotted in Fig. 4.4) to the exponents corresponding to the contributions from the additional singularities plotted in Figs. 4.5 and 4.6. See Fig. 4.7 for the result. We see that the main singularity always generates the smallest exponent (i.e., minimal suppression), so our guess above [Eq. (4.13)] that the main singularity determines the leading-order term in $P_{e^+e^-}$ was justified (at least when considering exponents only). When $\varepsilon$ is increased for a fixed $\gamma_c$, the difference between the main exponent and the smallest exponent corresponding to an additional singularity is approximately constant, at least in the range considered in Fig. 4.7(a). The main exponent decreases for $\varepsilon > \varepsilon_{\text{crit}}$ (dynamical assistance), and the exponents of the adjacent additional singularities seem to follow. Figure 4.7(b) indicates that the relative difference between the main exponent and the exponents of the additional singularities is mainly determined by $\gamma_c$. The decrease of the main exponent begins at $\gamma_c^{\text{crit}}(\varepsilon)$.

## 4.3. Results and conclusion

Let us summarize the main results of this chapter on the dynamically assisted Sauter–Schwinger effect via a temporal Gauss pulse.

### Critical Keldysh parameter scales with the pulse amplitude

The critical value of the combined Keldysh parameter scales with the pulse amplitude in the low-amplitude limit $\varepsilon \to 0$: $\gamma_c^{\text{crit}} \sim \sqrt{|\ln \varepsilon|}$. This is a crucial

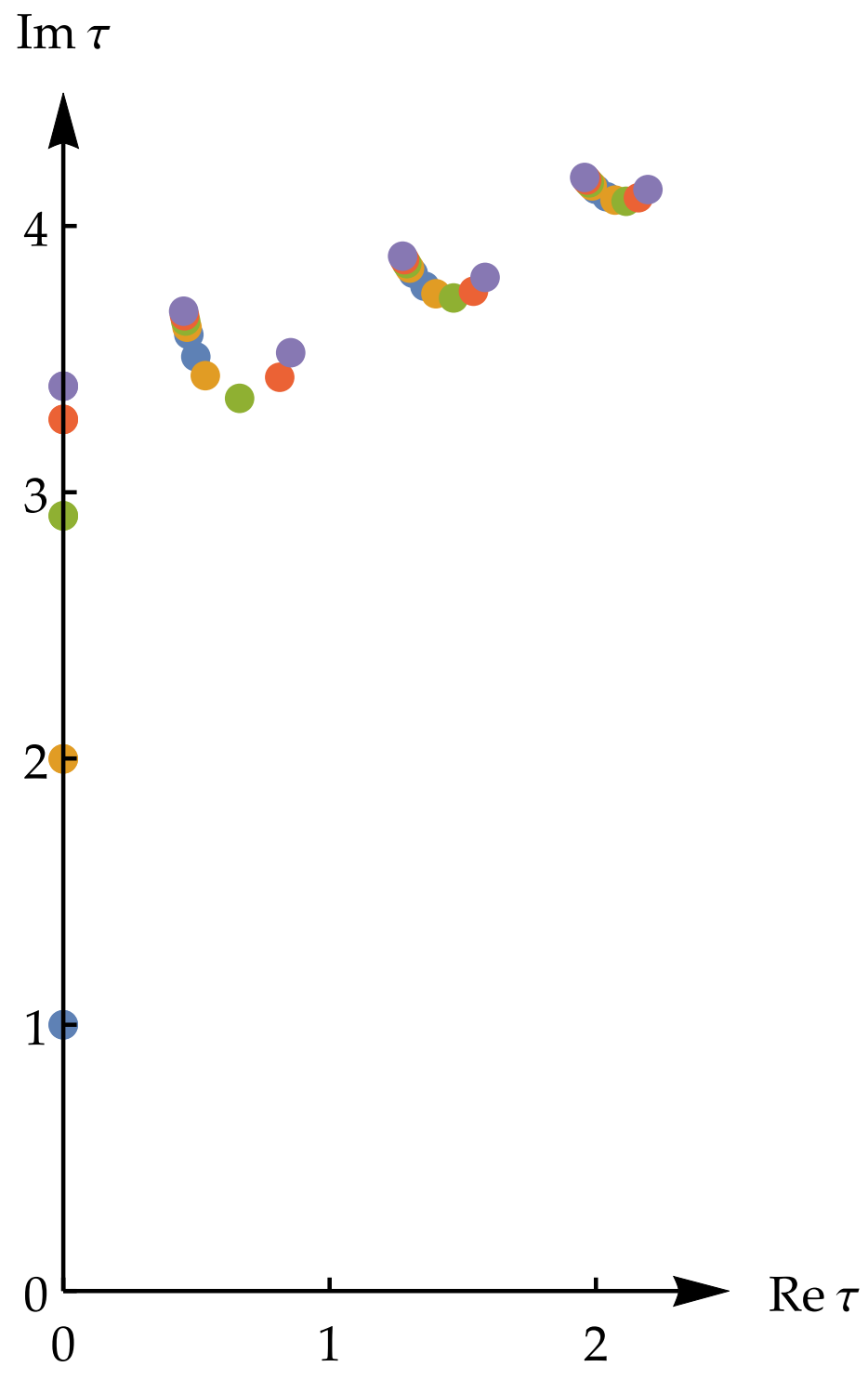

**Figure 4.6.:** Movement of the numerically calculated singularities $\tau^\star$ [colored dots; solutions to Eq. (4.4)] close to the imaginary axis in the complex $\tau$ plane as $\gamma_c$ increased from 1 (blue) to 5 (purple) in steps of 1. Again, we consider the dominating momentum $\kappa = 0$, and $\varepsilon = 10^{-4}$ is fixed, so $\gamma_c^{\text{crit}} \approx 2.9$ according to Eq. (4.23).

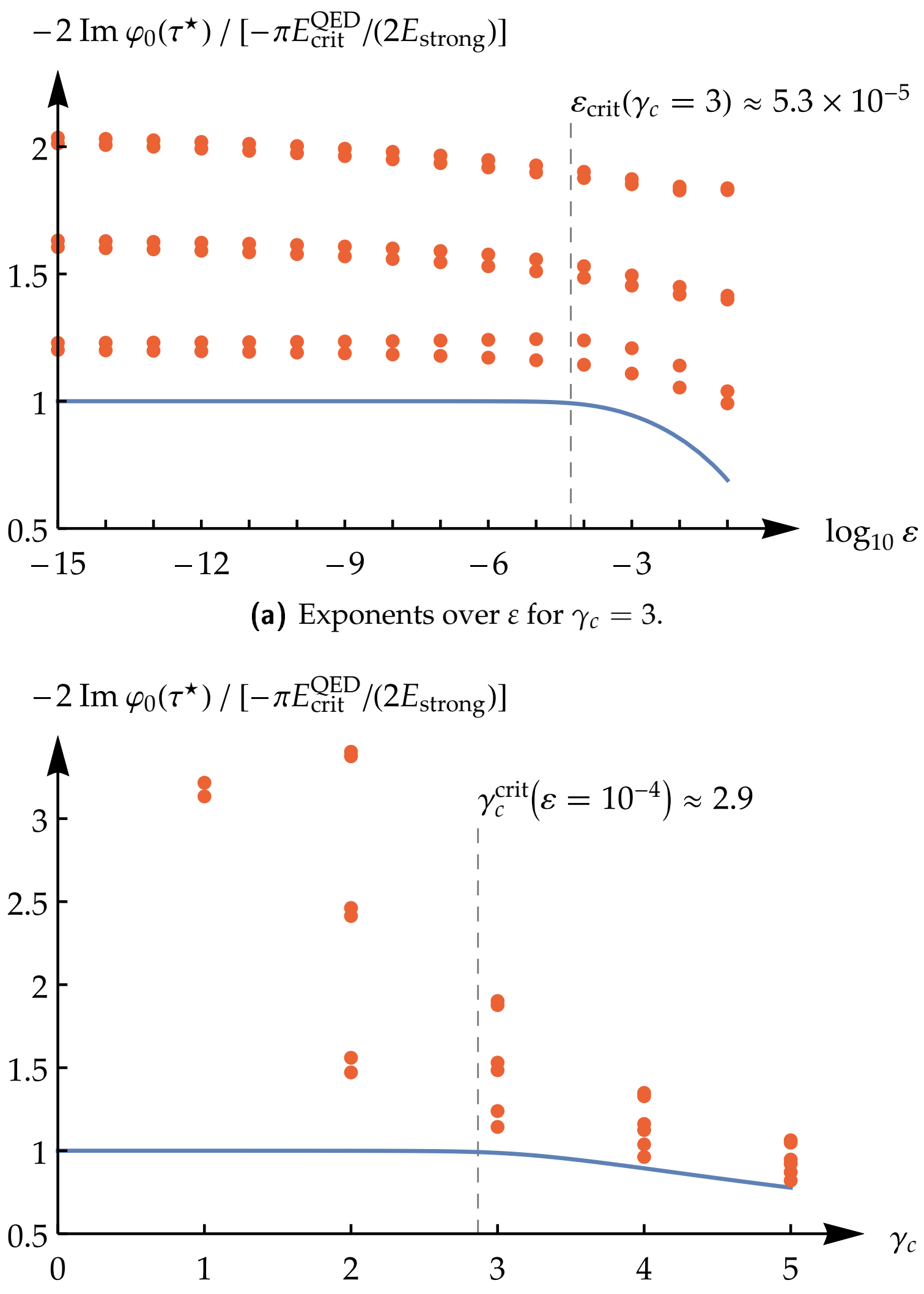

**(a)** Exponents over $\varepsilon$ for $\gamma_c = 3$.

**(b)** Exponents over $\gamma_c$ for $\varepsilon = 10^{-4}$.

**Figure 4.7.:** Numerically calculated exponents $-2\,\mathrm{Im}\,\varphi_0(\tau^\star)$ [see Eq. (4.14) for the definition of $\varphi_\kappa(\tau)$] appearing in the contributions (to $R_0^{\mathrm{out}}$) from the singularities plotted in Figs. 4.5 and 4.6 above, in units of the Sauter–Schwinger tunneling exponent. The blue plot line corresponds to the main singularity $\tau^\star_{\mathrm{main}} = \mathrm{i}v$, whose imaginary part can be found by solving the real equation (4.6) numerically. The red dots represent the exponents of the adjacent additional singularities, for the respective values of $\varepsilon$ and $\gamma_c$.

difference to the "original" dynamically assisted Sauter–Schwinger effect [67] via a temporal Sauter pulse, in which case $\gamma_c^{\text{crit}} \to \pi/2$ in the limit $\varepsilon \to 0$. Although both pulse shapes look rather similar, the structures of their corresponding vector potentials (tanh versus erf) in the complex plane are different: the (first) pole of tanh leads to $\gamma_c^{\text{crit}} \to \pi/2$, while erf is an entire function and thus $\gamma_c^{\text{crit}} \to \infty$ for $\varepsilon \to 0$. Dynamical assistance therefore **strongly depends on the shape** of the assisting field.

It has been shown recently [71] that the difference between the frequency spectra of both pulse shapes becomes important when dynamical assistance is considered as a small perturbation of the pure tunneling result in the slow background field: for the Sauter pulse, first-order perturbation theory is sufficient to describe assisted tunneling well, while the dominant correction (to the tunneling result) in the Gauss-pulse case generally comes from a higher-order perturbation term.

**Main singularity generates the dominating exponent**

The contribution from the main singularity to $R_0^{\text{out}}$ is always that with the smallest exponential suppression (see Fig. 4.7). The momentum $k = 0$ we concentrated on is preferred for symmetry reasons $[A(-t) = -A(t)]$, and we assume that it generates the dominant contribution to $P_{\mathrm{e}^+\mathrm{e}^-}$, in analogy to the Sauter-pulse case (see [62]), so the main exponent (4.19) appears in the leading-order term in $P_{\mathrm{e}^+\mathrm{e}^-}$.

We ignored all nonexponential prefactors here. It has been shown recently [41] that the typical features of dynamical assistance which are derived from the behavior of the exponents basically remain unaffected when taking also the prefactors into account.

**Critical threshold for dynamical assistance**

For a Gauss pulse, the threshold is not as sharply defined as in the Sauter-pulse case ($\gamma_c^{\text{crit}} = \pi/2$). The analytical first-order (in $\varepsilon$) expression for the main exponent in Eq. (4.19) offers a possibility for such a definition: at the threshold, the tunneling exponent is lowered by 1% (by the first-order term) according to our definition, and the resulting relation $\varepsilon_{\text{crit}}(\gamma_c)$ is plotted in Fig. 4.3. In a strong background field $E_{\text{strong}} = E_{\text{crit}}^{\text{QED}}/10$, this small decrease in the exponent enhances the pair-creation yield via tunneling by approximately 37% (we ignore the effect of the nonexponential prefactor here). For weaker background fields, the relative enhancement is even stronger.

# 5. Dynamical assistance by a harmonic oscillation

In complete analogy to the previous chapter, we study assisted tunneling in a strong, (quasi-)constant background field plus a weak harmonic oscillation,

$$E(t) = E_{\text{strong}} + E_{\text{weak}} \cos(\omega t) \tag{5.1}$$

with $\omega > 0$ and $E_{\text{weak}} \ll E_{\text{strong}}$, in this chapter. Although the Gauss pulse is qualitatively different from the oscillation on the real $t$ axis, the complex structures of the corresponding vector potentials are in fact quite similar and thus the analysis regarding dynamical assistance.

**Laser-assisted tunneling**

We consider this field profile as a model for tunneling assisted by continuous, counterpropagating laser beams. In the next part of this thesis (on the analog of Dirac theory in semiconductors), we will refer to this form of the dynamically assisted Sauter–Schwinger effect again when proposing experimental scenarios.

### Dominant momentum

The vector potential we choose to describe $E(t)$ reads

$$A(t) = E_{\text{strong}} t + \frac{E_{\text{weak}}}{\omega} \sin(\omega t). \tag{5.2}$$

This is an odd function, so we have the $k \leftrightarrow -k$ symmetry again (cf. Sec. 2.4.6.1), and thus the canonical momentum $k = 0$ is preferred. However, $E(t)$ is also invariant under $t \to t + 2\pi n/\omega$ with $n \in \mathbb{Z}$ here. The vector potential (5.2) is *not* invariant under this transformation, but $A(t)$ always appears in terms of the form $k + qA(t)$ in physical equations (the covariant operator $\hat{p}_x$ in momentum space), so any discrete time shift $t \to t + 2\pi n/\omega$ in $A(t)$ can be compensated by a shift $k \to k - 2\pi n q E_{\text{strong}}/\omega$ in momentum. However, a time shift along the real $t$ axis is irrelevant, since the physically meaningful quantity is $R_k^{\text{out}}$, which is an integral over all $t$. Two momenta $k$ and $k'$ separated by $k - k' = 2\pi n q E_{\text{strong}}/\omega$ are therefore physically equivalent, and thus there is **no single dominating momentum** here.

**Focus on $k = 0$ in most of this chapter**

From the facts that $E(t)$ has a local maximum at $t = 0$ and is an even function (like the Sauter pulse and the Gauss pulse considered before), we infer that $k = 0$ is one of the dominating momenta in the case of the oscillating profile (5.1). We will thus concentrate on $k = 0$ for simplicity in the following again, which acts as a representative for the dominating momenta $2\pi nqE_{\text{strong}}/\omega$ with $n \in \mathbb{Z}$.

## 5.1. Main singularity

The general **singularity equation** (2.112) for the vector potential (5.2) reads

$$\tau^\star + \varepsilon \sin \tau^\star = (-\kappa \pm \mathrm{i})\gamma_c, \tag{5.3}$$

**Main singularity for $k = 0$**

where we have used the dimensionless quantities (4.5) introduced in the previous chapter. For $\kappa = k = 0$, the main singularity is located on the imaginary axis again, so we insert $\tau^\star_{\text{main}} = \mathrm{i}v$ and get

$$v + \varepsilon \sinh v = \gamma_c. \tag{5.4}$$

The graphical solution of this real equation is plotted in Fig. 5.1 and looks rather similar to the corresponding plot in Fig. 4.2 on page 113 (Gauss pulse). The exponential term $\sinh v$ (coming from the oscillation) begins to dominate that left-hand side of Eq. (5.4) at

$$\frac{\varepsilon \sinh v_{\text{crit}}}{v_{\text{crit}}} = \mathcal{O}(1), \tag{5.5}$$

which roughly corresponds to the **threshold for dynamical assistance**. In the limit $\varepsilon \to 0$, we have $v_{\text{crit}} \to \infty$, so we may insert

$$\sinh v_{\text{crit}} = \frac{1}{2}\left(\mathrm{e}^{v_{\text{crit}}} - \mathrm{e}^{-v_{\text{crit}}}\right) \overset{v_{\text{crit}} \gg 1}{\approx} \frac{\mathrm{e}^{v_{\text{crit}}}}{2} \tag{5.6}$$

then. The resulting equation

$$\frac{\varepsilon \mathrm{e}^{v_{\text{crit}}}}{2v_{\text{crit}}} \overset{\varepsilon \to 0}{\sim} \mathcal{O}(1) \tag{5.7}$$

leads to

$$v_{\text{crit}} \sim |\ln \varepsilon| \tag{5.8}$$

**Scaling of $\gamma_c^{\text{crit}}$ with $\varepsilon$**

in analogy to Eq. (4.11), and thus

$$\gamma_c^{\text{crit}}(\varepsilon) \sim v_{\text{crit}} + \varepsilon \sinh v_{\text{crit}} \sim v_{\text{crit}} \overset{\varepsilon \to 0}{\sim} |\ln \varepsilon|. \tag{5.9}$$

This result resembles $\gamma_c^{\text{crit}} \sim \sqrt{|\ln \varepsilon|}$ from the Gauss-pulse case.

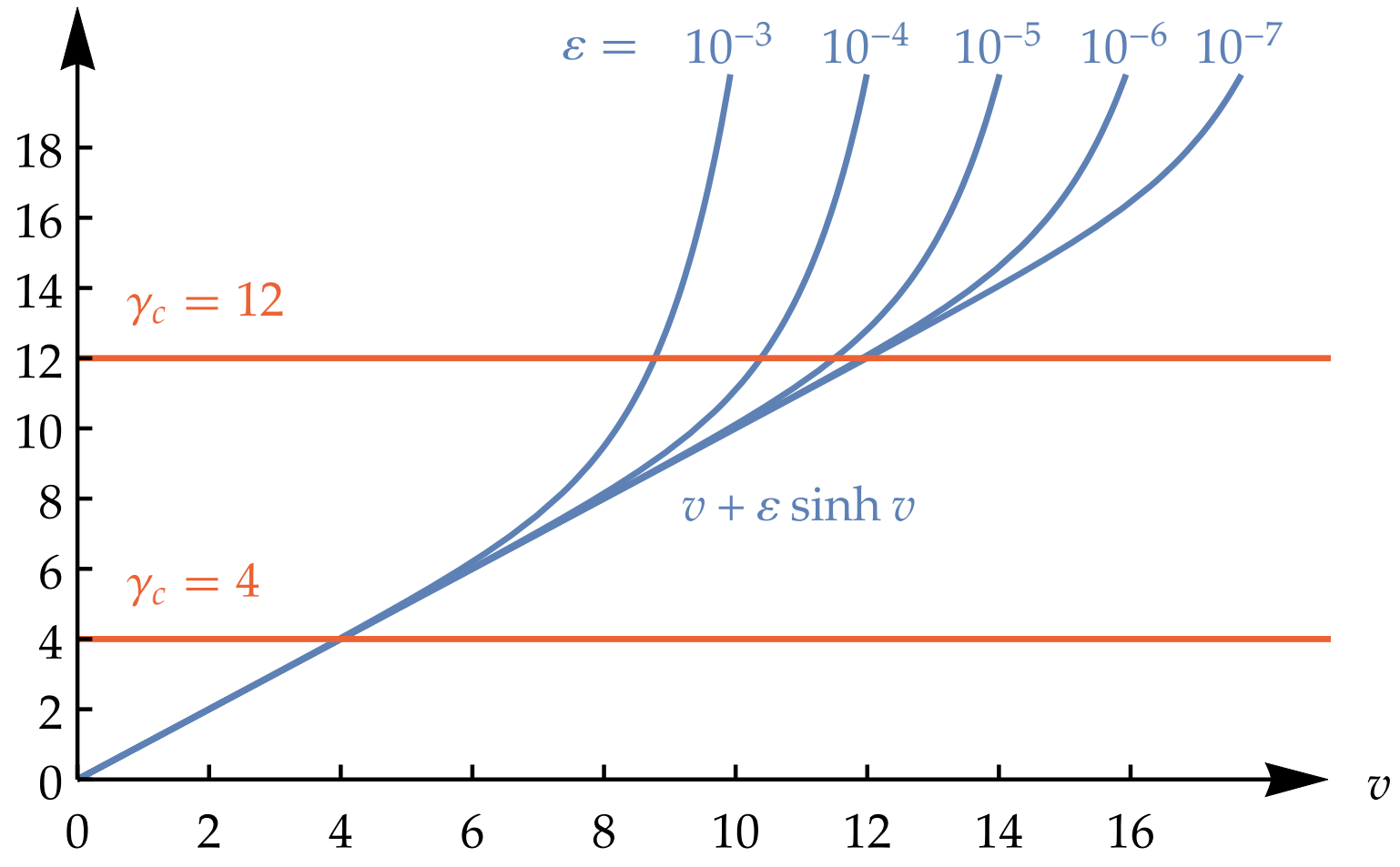

**Figure 5.1.:** Graphical solution of the singularity equation (5.4) for the main singularity $\tau^{\star}_{\text{main}} = iv$. As in the Gauss-pulse case (Fig. 4.2 on page 113), the left-hand side of Eq. (5.4) (blue plot lines) is linear for small $v$ until it becomes dominated by the exponential term $\varepsilon \sinh v$, approximately at some $v_{\text{crit}}$ roughly given by Eq. (5.5). The red $\gamma_c = 4$ line represents a subcritical value (no dynamical assistance) of the combined Keldysh parameter since it intersects all blue graphs over their respective linear domains, where the exponential contribution from the oscillation is negligible. The other value, $\gamma_c = 12$, however, looks like an overcritical value for $\varepsilon = 10^{-3}$, for example, in this plot.

### 5.1.1. Leading-order exponent

As in Sec. 4.1.1, we assume that the leading-order exponent in $P_{e^+e^-}$ is due to the exponent corresponding to $\tau^\star_{\text{main}}$ for $k = 0$ (we will check this assumption later) and calculate this main exponent up to the linear order in $\varepsilon$. The phase function (2.85) for the oscillating profile reads

$$\varphi_\kappa(\tau) = \frac{1}{\chi\gamma_c} \int_0^\tau \sqrt{1 + \left[\kappa + \frac{\tau' + \varepsilon \sin \tau'}{\gamma_c}\right]^2} \, \mathrm{d}\tau' \tag{5.10}$$

with $\chi = E_{\text{strong}}/E_{\text{crit}}^{\text{QED}}$, so we get [in analogy to Eqs. (4.16)–(4.18)]

$$\begin{aligned}
\varphi_0(\tau^\star_{\text{main}}(\varepsilon)) &= \frac{\mathrm{i}\pi}{4\chi} + \frac{1}{\chi\gamma_c}\frac{\mathrm{d}}{\mathrm{d}\varepsilon} \int_0^{\tau^\star_{\text{main}}(\varepsilon)} \sqrt{1 + \left(\frac{\tau' + \varepsilon \sin \tau'}{\gamma_c}\right)^2} \, \mathrm{d}\tau' \Bigg|_{\varepsilon=0} \varepsilon \\
&\quad + \mathcal{O}(\varepsilon^2) \\
&= \frac{\mathrm{i}\pi}{4\chi} + \frac{1}{\chi\gamma_c} \int_0^{\tau^\star_{\text{main}}(0)} \frac{2\frac{\tau'}{\gamma_c}\frac{1}{\gamma_c}\sin\tau'}{2\sqrt{1 + (\tau'/\gamma_c)^2}} \, \mathrm{d}\tau' \, \varepsilon + \mathcal{O}(\varepsilon^2) \\
&= \frac{\mathrm{i}\pi}{4\chi} + \frac{1}{\chi\gamma_c^3} \int_0^{\mathrm{i}\gamma_c} \frac{\tau' \sin\tau'}{\sqrt{1 + (\tau'/\gamma_c)^2}} \, \mathrm{d}\tau' \, \varepsilon + \mathcal{O}(\varepsilon^2) \\
&= \frac{\mathrm{i}\pi}{4\chi} - \frac{\mathrm{i}}{\chi\gamma_c} \int_0^1 \frac{\xi \sinh(\gamma_c \xi)}{\sqrt{1 - \xi^2}} \, \mathrm{d}\xi \, \varepsilon + \mathcal{O}(\varepsilon^2) \\
&= \frac{\mathrm{i}\pi}{4\chi} - \frac{\mathrm{i}\pi}{2\chi}\frac{I_1(\gamma_c)}{\gamma_c}\varepsilon + \mathcal{O}(\varepsilon^2).
\end{aligned} \tag{5.11}$$

The resulting **main exponent** is thus

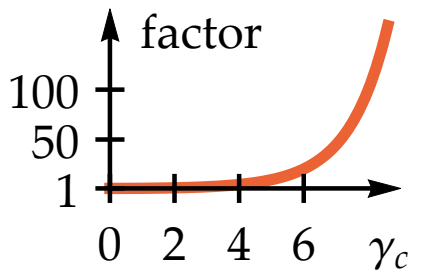

$$-4 \operatorname{Im} \varphi_0(\tau^\star_{\text{main}}) = -\frac{\pi E_{\text{crit}}^{\text{QED}}}{E_{\text{strong}}} \Bigg[1 - \underbrace{\frac{2 I_1(\gamma_c)}{\gamma_c}}_{\text{factor in marginal plot}} \varepsilon + \mathcal{O}(\varepsilon^2)\Bigg]. \tag{5.12}$$

The **validity condition** for the first-order approximation reads

$$1 \gg \frac{2 I_1(\gamma_c)}{\gamma_c}\varepsilon \overset{\gamma_c \to \infty}{\sim} \sqrt{\frac{2}{\pi}} \frac{\mathrm{e}^{\gamma_c}}{\gamma_c^{3/2}}\varepsilon, \tag{5.13}$$

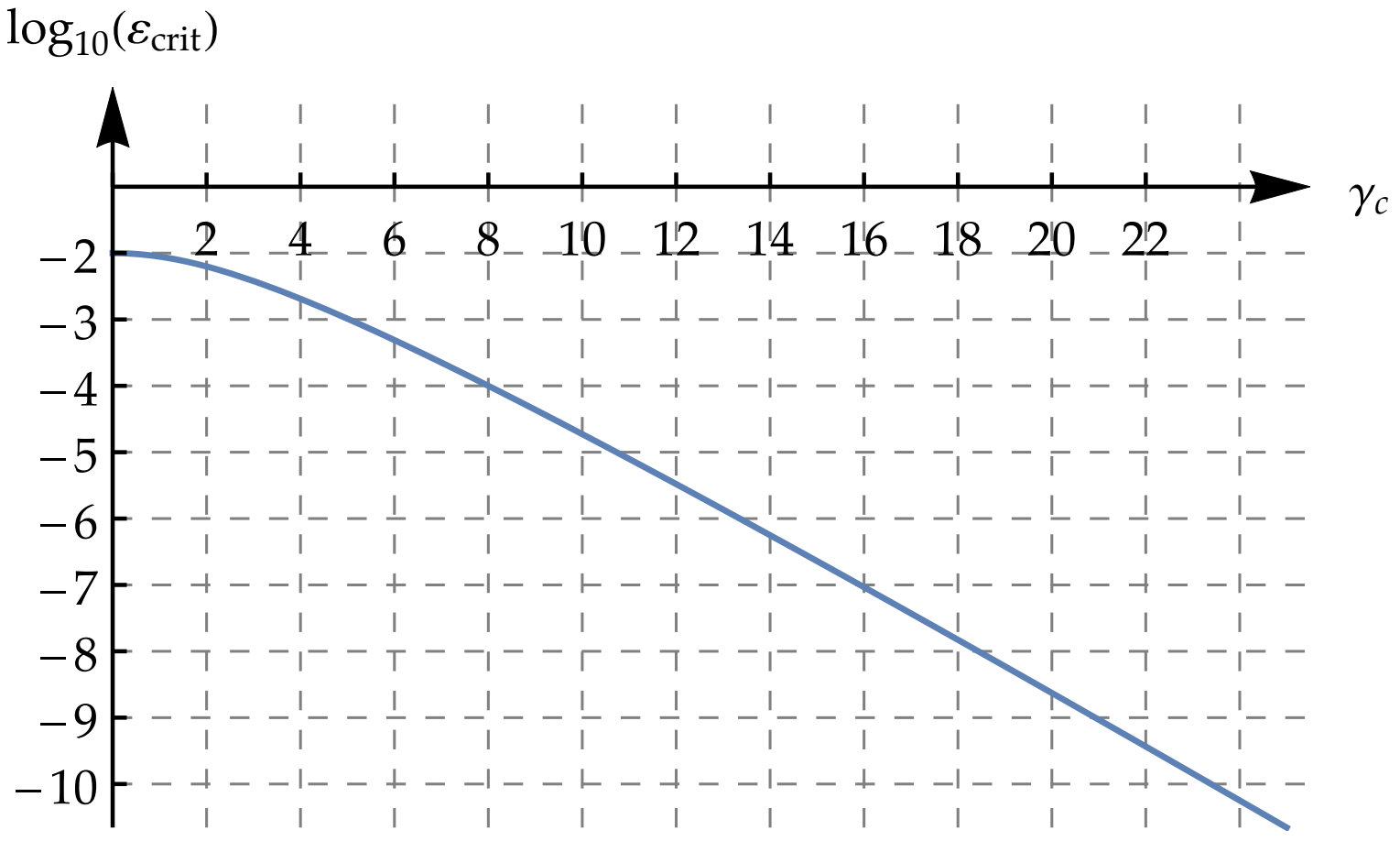

**Figure 5.2.:** Critical value of $\varepsilon$ defined in Eq. (5.15) as a function of $\gamma_c$. At the critical threshold, the oscillation lowers the Sauter–Schwinger exponent $-\pi E_{\rm crit}^{\rm QED}/E_{\rm strong}$ by 1% via dynamical assistance.

where we have inserted the leading-order asymptotic term of the Bessel function [see Eq. (4.22)].

It is shown in Ref. [1] that the worldline-instanton technique leads to the same result.

### Critical threshold for dynamical assistance

Again, we define the critical threshold as the point at which the Sauter–Schwinger tunneling exponent in Eq. (5.12) is reduced by 1% due to the linear term in $\varepsilon$. That is, when varying $\gamma_c$, $\gamma_c^{\rm crit}(\varepsilon)$ is given implicitly by

$$\frac{2\varepsilon I_1\left(\gamma_c^{\rm crit}(\varepsilon)\right)}{\gamma_c^{\rm crit}(\varepsilon)} \overset{!}{=} 0.01 \tag{5.14}$$

(this equation has to be solved numerically for $\gamma_c^{\rm crit}$), or we have

$$\varepsilon_{\rm crit}(\gamma_c) = \frac{1}{200}\frac{\gamma_c}{I_1(\gamma_c)} \tag{5.15}$$

when $\varepsilon$ is variable. The function $\varepsilon_{\rm crit}(\gamma_c)$ is plotted in Fig. 5.2.

### Comparison with the numerically calculated main exponent

The analytical approximation (5.12) of the main exponent is plotted in Fig. 5.3, together with the corresponding numerically calculated values. Again, the approximation is good up to the threshold defined by Eqs. (5.14) or (5.15).

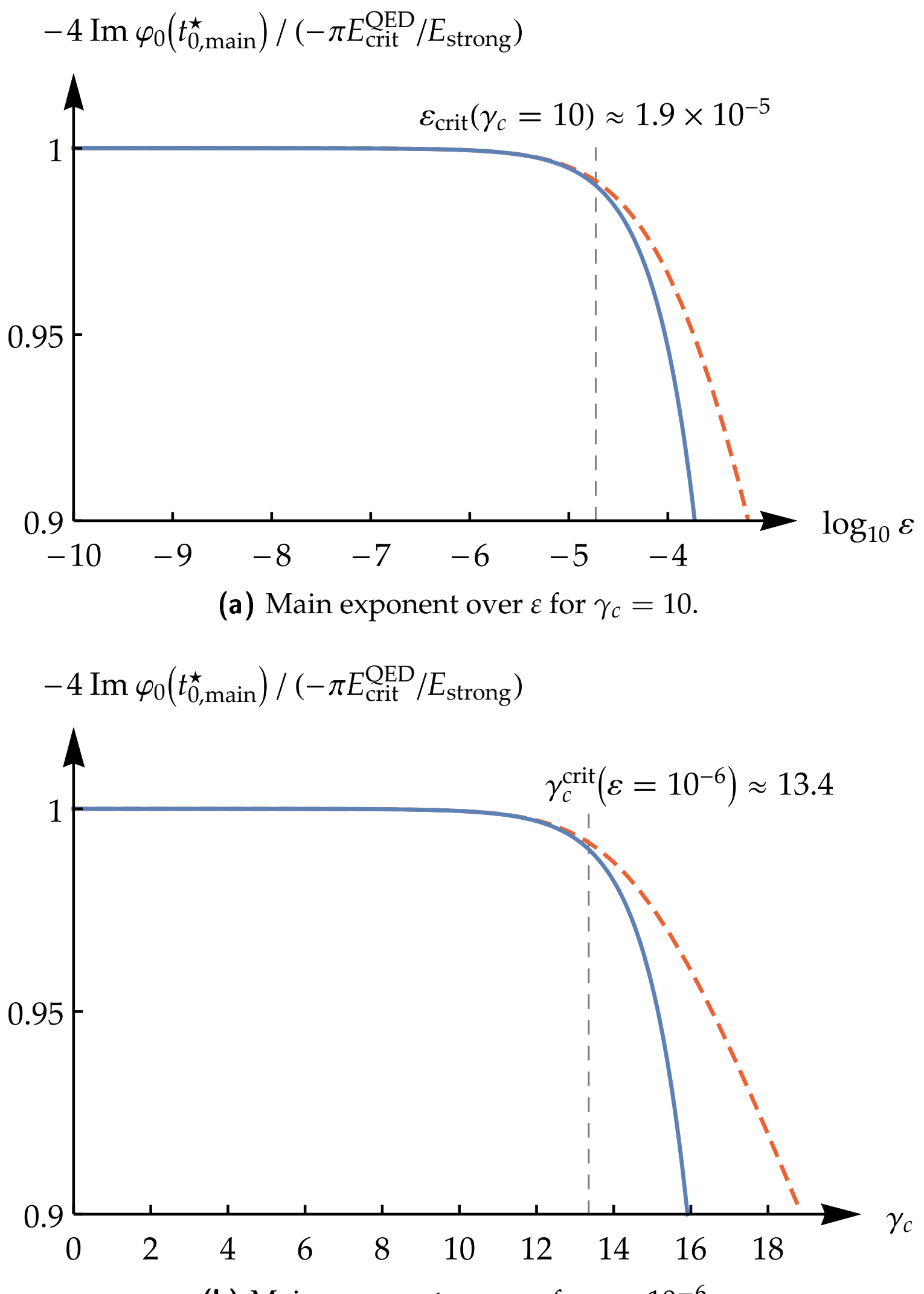

**(a)** Main exponent over $\varepsilon$ for $\gamma_c = 10$.

**(b)** Main exponent over $\gamma_c$ for $\varepsilon = 10^{-6}$.

**Figure 5.3.:** Main exponent in the pair-creation probability for the oscillating profile, in units of the pure Sauter–Schwinger exponent $-\pi E^{\mathrm{QED}}_{\mathrm{crit}}\,/\,E_{\mathrm{strong}}$. The blue, solid line is the analytical approximation in Eq. (5.12), and the red, dashed line is the numerical result [calculated by solving the singularity equation (5.4) and the integral in the phase function (5.10) numerically]. The critical values are given by Eqs. (5.14) and (5.15), respectively.

## 5.2. Additional singularities

### 5.2.1. Graphical solution of the singularity equation

We will consider the positions of the additional singularities more precisely here than in the Gauss-pulse case because the full singularity equation (5.3) is relatively easy to solve graphically. We start to solve the singularity equation by inserting

$$\tau^\star = u + \mathrm{i}v \qquad \text{with} \qquad u, v \in \mathbb{R}, \tag{5.16}$$

which gives

$$u + \mathrm{i}v + \varepsilon \underbrace{\sin(u + \mathrm{i}v)}_{\sin u \cosh v + \mathrm{i} \cos u \sinh v} = (-\kappa \pm \mathrm{i})\gamma_c, \tag{5.17}$$

so we get two real equations, one from the real part and one from the imaginary part of this equation:

$$u + \varepsilon \sin u \cosh v = -\kappa\gamma_c, \tag{5.18}$$

$$v + \varepsilon \cos u \sinh v = \pm\gamma_c. \tag{5.19}$$

Let us focus on the momentum $\kappa = 0$ again for simplicity.

**$k = 0$ in the following**

**Upper equation**

For the special case $\sin u = 0$, the upper equation (5.18) reads $u = 0$ (since $\kappa = 0$), which means that there is exactly one solution $\tau^\star = \mathrm{i}v$ in this case. However, we have already identified this solution on the imaginary axis as the main singularity in the previous section, so all *additional* singularities $\tau^\star_{\text{add}} = u + \mathrm{i}v$ must satisfy

$$\sin u \neq 0 \qquad \Leftrightarrow \qquad u \neq n\pi \quad \text{with} \quad n \in \mathbb{Z}. \tag{5.20}$$

Hence, for all additional singularities, the upper equation can be written as

$$\underbrace{\cosh v}_{\geq 1} = -\frac{1}{\varepsilon}\frac{u}{\sin u} = -\frac{|u|}{\varepsilon \sin |u|}. \tag{5.21}$$

This equation is solvable (i.e., additional singularities may exist) only if $\sin |u| < 0$, in which case we find a unique imaginary part $v > 0$ in the upper complex half-plane:

**Re $\tau^\star_{\text{add}}$ only in allowed "$\pi$ intervals"**

$$v = \operatorname{arcosh}\Bigg(\underbrace{-\frac{u}{\varepsilon \sin u}}_{\gg 1 \text{ since } -u/\sin u > 1 \text{ and } \varepsilon \ll 1}\Bigg) > 0 \quad \text{for} \quad \underbrace{|u| \in \Big((2n-1)\pi, 2n\pi\Big) \text{ with } n \in \mathbb{N}}_{\text{"}\pi\text{ intervals"}}. \tag{5.22}$$

**Asymptotic behavior of Im $\tau^\star_{\text{add}}$**

Note that since $-u/\sin u > \pi$ on each of these allowed "$\pi$ intervals"[1] and $1/\varepsilon > 1$ (or rather $\gg 1$), arcosh in Eq. (5.22) is well approximated by

$$\operatorname{arcosh} x = \ln\left(x + \sqrt{x^2 - 1}\right) \overset{x \gg 1}{\approx} \ln(2x) \tag{5.23}$$

for all relevant $u$. We can thus provide a **lower bound for the imaginary parts** $v$ of all additional singularities:

$$v \approx \ln\left(-\frac{2u}{\varepsilon \sin u}\right) = \ln 2 - \ln\varepsilon + \ln\Big(\underbrace{-\frac{u}{\sin u}}_{>1}\Big) > \ln 2 + |\ln\varepsilon|. \tag{5.24}$$

Hence, the distance of each additional singularity from the real axis diverges ($v \to \infty$) in the limit $\varepsilon \to 0$. It is also possible to obtain a $u$-dependent lower bound since $-1/\sin u \geq 1$ on the allowed $\pi$ intervals, so

$$v > \ln 2 + |\ln\varepsilon| + \ln|u|. \tag{5.25}$$

Note that this result is only meaningful over the $\pi$ intervals (where additional singularities can exist), but especially *not* between the positive and the negative $n = 1$ intervals (i.e., $|u| < \pi$), where $\ln|u|$ diverges to $-\infty$ at the imaginary axis ($u = 0$).

The result (5.25) can be compared to the hyperbolas (4.38) on which all additional singularities lie asymptotically in the Gauss-pulse case. Since $\gamma_c^{\text{crit}}(\varepsilon) \sim |\ln\varepsilon|$ for the oscillating profile, the lower bound (5.24) suggests that the main singularity will approximately remain at the position $\mathrm{i}\gamma_c$ (pure Sauter–Schwinger effect) when increasing $\varepsilon$, until the additional singularities come close (this conjecture is confirmed in Fig. 5.5 below).

### Lower equation

Since $\operatorname{Im}\tau^\star_{\text{add}} = v > 0$ (upper half-plane), $\sinh v$ in the lower equation (5.19) can be expressed via

$$\sinh v = \sqrt{\cosh^2 v - 1}, \tag{5.26}$$

**Real singularity equation**

so inserting Eq. (5.22) into the lower equation yields the **real singularity equation** for the real parts $u = \operatorname{Re}\tau^\star_{\text{add}}$ of all additional singularities:

$$\underbrace{\operatorname{arcosh}\left(-\frac{u}{\varepsilon\sin u}\right) + \varepsilon\cos u\sqrt{\left(\frac{u}{\varepsilon\sin u}\right)^2 - 1}}_{F_\varepsilon(u)} = \pm\gamma_c. \tag{5.27}$$

[1] In the first ($n = 1$) $\pi$ interval $(\pi, 2\pi)$ in the positive $u$ range, the smallest value of the nominator $u$ is $\pi$, and the maximum (absolute) value of the denominator $\sin u$ is 1—hence our simple estimate for the lower bound: $-u/\sin u > \pi$.

This equation can be **solved graphically** over the allowed $\pi$ intervals specified in Eq. (5.22). For each solution $u$, the imaginary part of the corresponding singularity is given by Eq. (5.22). See Fig. 5.4 for an example. Note that Eqs. (5.22) and (5.27) are invariant under $u \to -u$, so we may concentrate on positive $u$ in the following and just mirror the results into the negative $u$ range.

**$u \leftrightarrow -u$ symmetry for $k = 0$**

### Number of additional singularities per $\pi$ interval

Figure 5.4 suggests that the function $F_\varepsilon(u)$ in Eq. (5.27) increases strictly from $-\infty$ to $\infty$ over each allowed positive $\pi$ interval, respectively (and vice versa in the negative $u$ range). Let us check this assumption. By means of the derivative

$$\frac{\mathrm{d}}{\mathrm{d}x} \operatorname{arcosh} x = \frac{1}{\sqrt{x^2-1}} \qquad \text{for} \qquad x > 1, \tag{5.28}$$

we get

$$\begin{aligned}
\frac{\mathrm{d}}{\mathrm{d}u} F_\varepsilon(u) &= \frac{-1 + \varepsilon \cos u \frac{u}{\varepsilon \sin u}}{\sqrt{\left(\frac{u}{\varepsilon \sin u}\right)^2 - 1}} \frac{\mathrm{d}}{\mathrm{d}u} \frac{u}{\varepsilon \sin u} - \varepsilon \sin u \sqrt{\left(\frac{u}{\varepsilon \sin u}\right)^2 - 1} \\
&= \frac{-1 + u \cot u}{\varepsilon \sqrt{\left(\frac{u}{\varepsilon \sin u}\right)^2 - 1}} \frac{\sin u - u \cos u}{\sin^2 u} - \frac{\varepsilon^2 \sin u \left[\left(\frac{u}{\varepsilon \sin u}\right)^2 - 1\right]}{\varepsilon \sqrt{\left(\frac{u}{\varepsilon \sin u}\right)^2 - 1}} \\
&= \frac{\overbrace{(-1 + u \cot u)^2}^{\geq 0} + \overbrace{\varepsilon^2 \sin^2 u \left[\left(\frac{u}{\varepsilon \sin u}\right)^2 - 1\right]}^{>0}}{\underbrace{-\varepsilon \sin u \sqrt{\left(\frac{u}{\varepsilon \sin u}\right)^2 - 1}}_{>0}} > 0
\end{aligned} \tag{5.29}$$

since $\sin u < 0$ on each $\pi$ interval. The slope of $F_\varepsilon(u)$ is thus always positive on the positive $\pi$ intervals, which confirms our above assumption, so there are exactly **two different solutions** (one for the $+\gamma_c$ case and one for the $-\gamma_c$ case) to the real singularity equation (5.27) per $\pi$ interval.

In conclusion, the graphical solution method provides a systematic way to **find all additional singularities** which occur for the oscillating field profile (at least for the canonical momentum $k = 0$ we consider here).

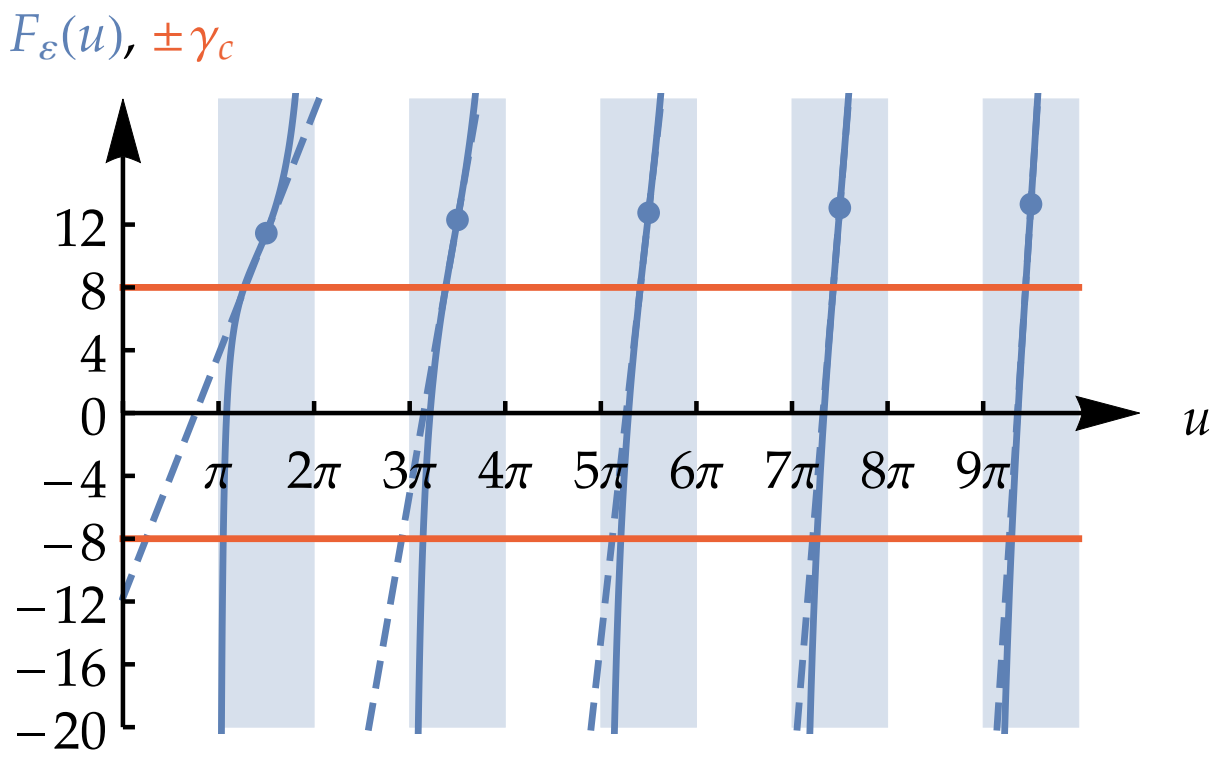

**(a)** Step 1: graphical solution of $F_\varepsilon(u) = \pm\gamma_c$.

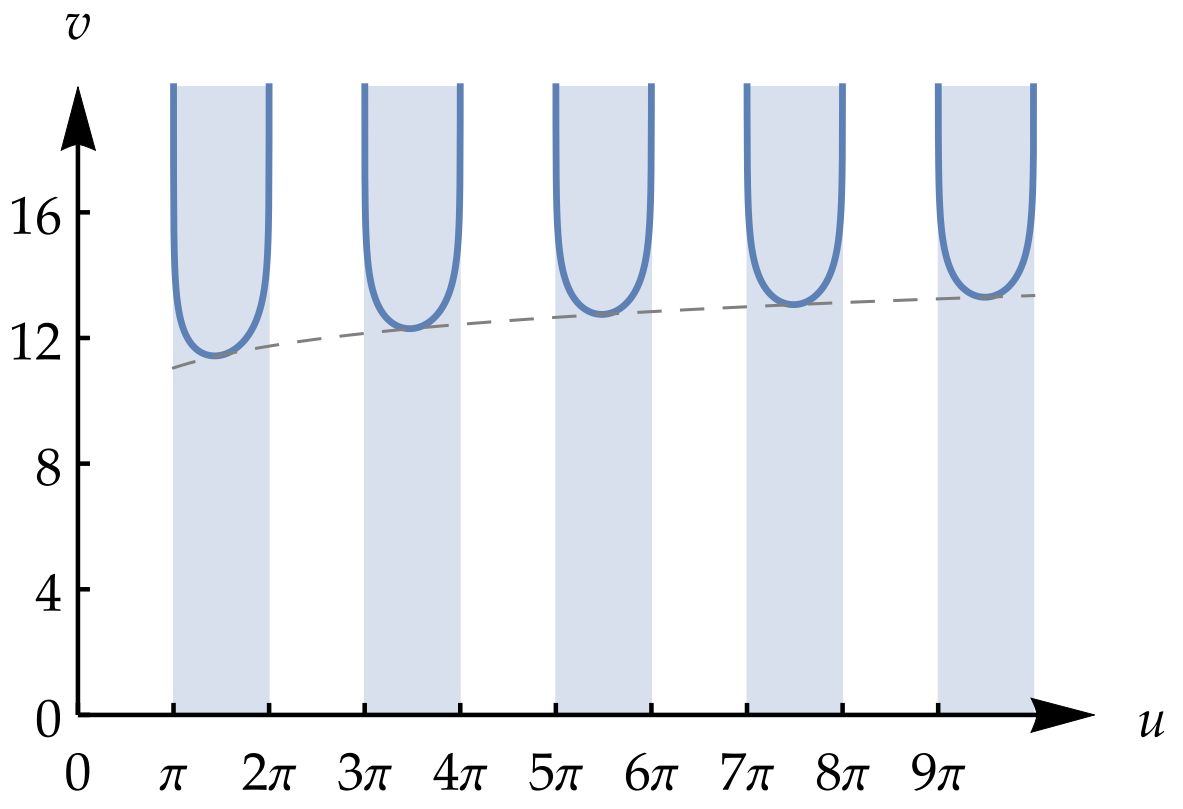

**(b)** Step 2: corresponding imaginary parts $v$.

**Figure 5.4.:** Graphical solution method which yields all additional singularities $\tau^\star_{\text{add}} = u + iv$ for $k = 0$ (the negative $u$ range is just the mirror image). The allowed "$\pi$ intervals" have a light-blue background. The parameter values in these plots are $\varepsilon = 10^{-4}$ and $\gamma_c = 8$. (a) Step 1: The function $F_\varepsilon(u)$ (blue, solid graphs) defined in Eq. (5.27) is plotted. Each position $u$ at which $F_\varepsilon$ intersects either $+\gamma_c$ or $-\gamma_c$ (horizontal, red lines) is the real part of an additional singularity. The dashed, blue lines are the tangent lines to $F_\varepsilon$ at the interval centers $u_n$ (blue points). Asymptotically (large $n$), the tangent lines approximate $F_\varepsilon$ well at the heights $\pm\gamma_c$ (see the rightmost $\pi$ interval). (b) Step 2: Equation (5.22) (blue graph) evaluated at the above solutions yields the corresponding $v$ values. The gray, dashed graph is the lower bound (5.25) for these imaginary parts.

### 5.2.2. Approximation of the real singularity equation and asymptotics for large real parts

The graphical method presented in the previous subsection is exact in the sense that it is equivalent to the complex singularity equation (5.3) for $\kappa = 0$. In this subsection, we will apply approximations in order to study the positions of the additional singularities in the complex plane further.

**Approximate real singularity equation (dependence on $\varepsilon$)**
Since $-u/(\varepsilon \sin u) \gg 1$ for $\varepsilon \ll 1$ [see Eq. (5.22)], a very good approximation of $F_\varepsilon(u)$ in the real singularity equation (5.27) is

$$\begin{aligned} F_\varepsilon(u) &= \operatorname{arcosh}\Big(-\frac{u}{\varepsilon \sin u}\Big) + \varepsilon \cos u \sqrt{\Big(\frac{u}{\varepsilon \sin u}\Big)^2 - 1} \\ &\approx \ln 2 - \ln \varepsilon + \ln\Big(-\frac{u}{\sin u}\Big) + \varepsilon \cos u \left|\frac{u}{\varepsilon \sin u}\right| \\ &= \ln 2 + |\ln \varepsilon| + \ln\Big(-\frac{u}{\sin u}\Big) - |u| \cot u, \end{aligned} \tag{5.30}$$

where we have used Eqs. (5.23)–(5.24) and $|\sin u| = -\sin u$. The **approximated real singularity equation** thus reads

$$\ln\Big(-\frac{u}{\sin u}\Big) - |u| \cot u = \pm\gamma_c - |\ln \varepsilon| - \ln 2. \tag{5.31}$$

In this form, the left-hand side is independent of the parameter $\varepsilon$, which only appears on the constant ($u$-independent), right-hand side. The effect of changing $\varepsilon$ in the graphical solution (Fig. 5.4) becomes manifest in Eq. (5.30): essentially, changing $\varepsilon$ shifts the graph of $F_\varepsilon(u)$ up or down.

**Approximation of additional singularities with large real parts**
The number of additional singularities is always infinite (for $\varepsilon > 0$) since there are infinitely many $\pi$ intervals [see Eq. (5.22)], and two different $\tau^\star_{\text{add}}$ exist per interval. For large $|u| = |\operatorname{Re} \tau^\star_{\text{add}}|$, that is, for $\pi$ intervals with large $n$, far away from the imaginary axis, we can solve the real singularity equation (5.27) approximately and thus find an asymptotic expression for the additional singularities.

We start by evaluating $F_\varepsilon(u)$ at the midpoints

$$u_n = \left(2n - \frac{1}{2}\right)\pi \qquad \text{with} \qquad n \in \mathbb{N} \tag{5.32}$$

of the allowed $\pi$ intervals (the negative $u$ range is just the mirror image, so we do not consider it here for simplicity). Since $\sin u_n = -1$ and $\cos u_n = 0$,

we get

$$F_\varepsilon(u_n) = \operatorname{arcosh}\Big(\underbrace{\frac{4n-1}{2\varepsilon}\pi}_{\gg 1}\Big) \overset{\text{Eq. (5.23)}}{\approx} \ln\left(\frac{4n-1}{\varepsilon}\pi\right) \sim \ln n, \tag{5.33}$$

so these function values increase logarithmically with $n$ (and thus with the real parts of the additional singularities). The slopes of $F_\varepsilon(u)$ at the midpoints are found by inserting $u_n$ into the derivative given in Eq. (5.29):

$$\begin{aligned}
\left.\frac{\mathrm{d}F_\varepsilon(u)}{\mathrm{d}u}\right|_{u=u_n} &= \frac{-1}{\varepsilon\sqrt{\left(\frac{u_n}{\varepsilon \sin u_n}\right)^2 - 1}}\frac{\sin u_n}{\sin^2 u_n} - \frac{\varepsilon^2 \sin u_n \left[\left(\frac{u_n}{\varepsilon \sin u_n}\right)^2 - 1\right]}{\varepsilon\sqrt{\left(\frac{u_n}{\varepsilon \sin u_n}\right)^2 - 1}} \\
&= \frac{1 + \left[(4n-1)^2 \pi^2/4 - \varepsilon^2\right]}{\sqrt{(4n-1)^2 \pi^2/4 - \varepsilon^2}} \\
&\overset{\varepsilon \ll 1}{\approx} \frac{1 + \left[(4n-1)^2 \pi^2/4\right]}{\sqrt{(4n-1)^2 \pi^2/4}} \\
&\overset{n \gg 1}{\approx} \sqrt{(4n-1)^2 \pi^2/4} \\
&= (4n-1)\frac{\pi}{2} = u_n \sim n,
\end{aligned} \tag{5.34}$$

so the slopes at the midpoints diverge linearly with $n$ [faster than the function values $F_\varepsilon(u_n) \sim \ln n$] for $\pi$ intervals far away from the imaginary axis.

Now consider the real singularity equation $F_\varepsilon(u) = \pm\gamma_c$, where $\gamma_c$ is some fixed value. For large $n \gg 1$, we may linearize $F_\varepsilon(u)$ around $u_n$ over each $\pi$ interval, respectively, in order to calculate the two corresponding solutions ($\pm\gamma_c$ cases):

$$F_\varepsilon(u) \overset{|u-u_n| \ll \pi/2}{\approx} F_\varepsilon(u_n) + \left.\frac{\mathrm{d}F_\varepsilon(u)}{\mathrm{d}u}\right|_{u=u_n} (u - u_n) \overset{!}{=} \pm\gamma_c, \tag{5.35}$$

**Approximated $\mathrm{Re}\,\tau^\star_{\mathrm{add}}$ over $\pi$ intervals with large $n$**

which yields the two approximated solutions

$$u_\pm = \underbrace{\frac{\pm\gamma_c - \overbrace{F_\varepsilon(u_n)}^{\sim \ln n}}{\left.\frac{\mathrm{d}F_\varepsilon(u)}{\mathrm{d}u}\right|_{u=u_n}}}_{\sim n} + \underbrace{u_n}_{\sim n} \approx \frac{\pm\gamma_c - \ln[(4n-1)\pi/\varepsilon]}{(4n-1)\pi/2} + \left(2n - \frac{1}{2}\right)\pi. \tag{5.36}$$

For small $n$, this approximation may be bad because the quantity

$$|u_\pm - u_n| \approx \left| \frac{\pm\gamma_c - \ln[(4n-1)\pi/\varepsilon]}{(4n-1)\pi/2} \right| \sim \frac{\ln n}{n} \tag{5.37}$$

could be even greater than $\pi/2$ (depending on the values of $\varepsilon$ and $\gamma_c$); that is, the approximated real parts $u_\pm$ of the two singularities could be outside the corresponding $\pi$ interval, which cannot happen with the exact equation $F_\varepsilon(u) = \pm\gamma_c$ [cf. Fig. 5.4(a)]. In order for the approximation to be consistent with the above linearization of $F_\varepsilon(u)$, the value (5.37) must be much smaller than $\pi/2$, which is true for sufficiently large $n$ [rightmost interval in Fig. 5.4(a)].

**Convergence of the contour integral**

The "residue" [see Eq. (3.25) for the general formula] of each singularity in the upper complex half-plane contributes to the contour integral $R_0^{\text{out}}$. This raises the question of whether the infinite series of residues converges. The residue sum of a finite number of singularities is always convergent, but we have an infinite number of additional singularities here (two over each $\pi$ interval).

However, we see in Eq. (5.37) that the two singularities with real parts $u_\pm$ over a $\pi$ interval far away from the imaginary axis ($n \gg 1$) merge at the interval center; that is, $u_\pm \to u_n$ for $n \to \infty$. Since the imaginary part of an additional singularity is a function of the corresponding real part in the case of the oscillating profile [see Eq. (5.22)], the two singularities really merge in the complex plane asymptotically. Their residues (3.25) thus coincide up to the sign since $u_\pm$ corresponds to a "plus/minus solution" of the singularity equation. As a result, the two residues cancel each other in the limit $n \to \infty$. This fact ensures the convergence of the series of residues.

Note that we have ignored the branch cuts since we assume that their contribution to $R_0^{\text{out}}$ is of the same order as that of the residues (inspired by the constant-field case; see Sec. 3.3).

### 5.2.3. Numerical results: singularity movement and exponents

**Moving singularities for varying $\varepsilon$**

The movement of the (numerically calculated) singularities $\tau^\star$ in the complex plane for varying $\varepsilon$ is depicted in Fig. 5.5. The imaginary part of the main singularity $\tau^\star_{\text{main}} = \mathrm{i}v$ is given by Eq. (5.4). The additional singularities are obtained by solving Eq. (5.27) over the allowed $\pi$ intervals, and then we insert the resulting real parts into Eq. (5.22) in order to calculate the corresponding imaginary parts. The intuitive picture conveyed by Fig. 5.5 is similar to the Gauss-pulse case (Sec. 4.2.2): as $\varepsilon$ is increased, the additional singularities approach the real axis, and $\tau^\star_{\text{main}}$ remains fixed at $\mathrm{i}\gamma_c$ (pure tunneling), until the additional singularities in the first $\pi$ intervals come close. This approximately happens when $|\ln \varepsilon| = \mathcal{O}(\gamma_c)$ since the lower bound of $\mathrm{Im}\,\tau^\star_{\text{add}}$ (dashed lines in the plot) scales with $|\ln \varepsilon|$—which is also consistent with our finding from Sec. 5.1 that $\gamma_c^{\text{crit}} \sim |\ln \varepsilon|$. Above the critical threshold, the additional singularities seem to "push" the main singularity towards the real axis (dynamical assistance). Note that additional singularities far away from the imaginary axis vanish logarithmically in the complex plane, which indicates that their contribution to $R_0^{\text{out}}$ is less important. This is confirmed (considering exponents only, again) in Fig. 5.7 below.

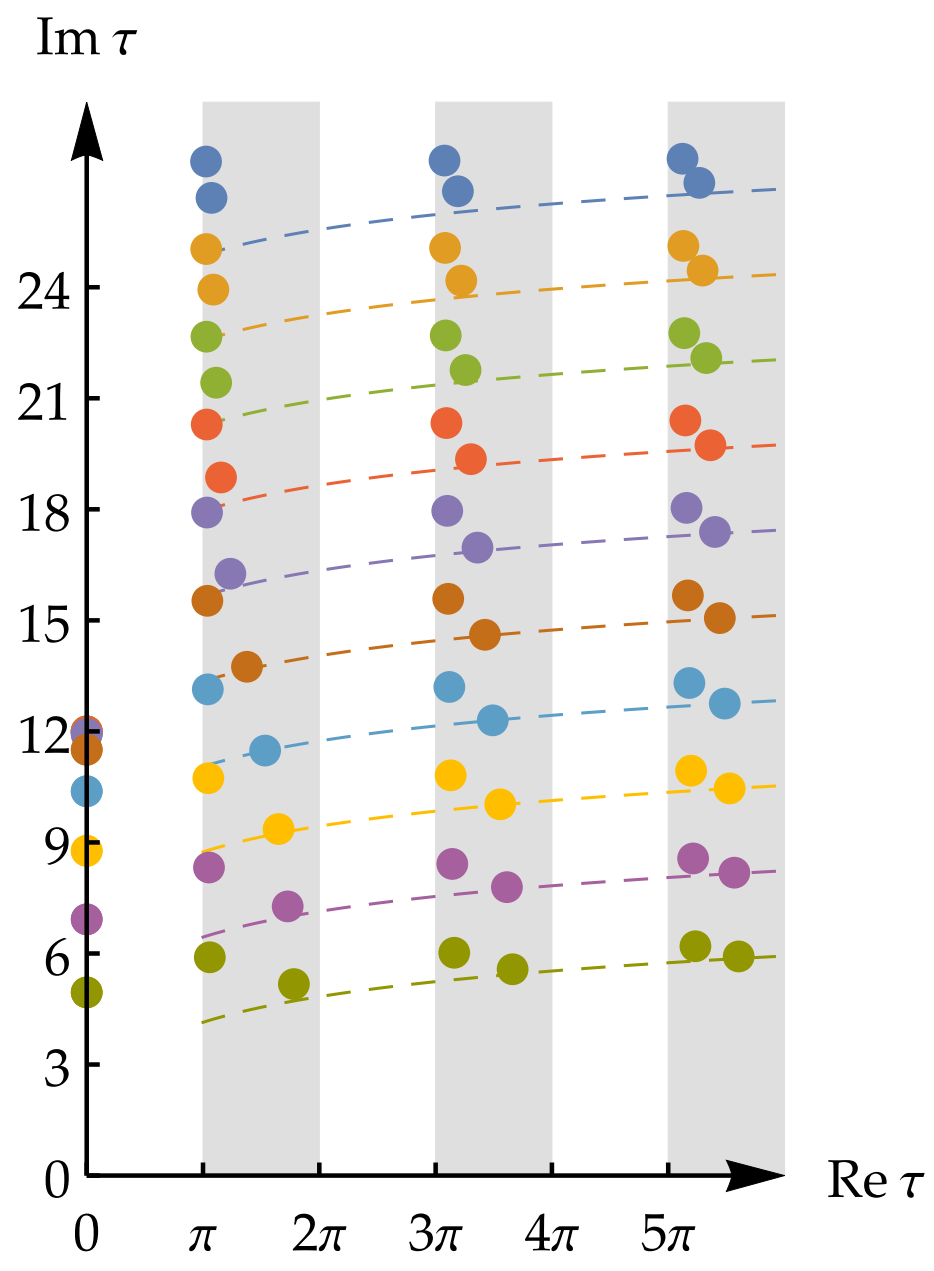

**Figure 5.5.:** Numerically calculated solutions $\tau^\star$ (colored dots) of the singularity equation (5.3) in the complex $\tau$ plane, with $\kappa = 0$ (one of the dominating momenta) and $\gamma_c = 12$ fixed, and varying $\varepsilon = 10^{-10}$ (blue), $10^{-9}$ (ocher), ..., $10^{-1}$ (olive). The second quadrant ($\mathrm{Re}\,\tau < 0$) is just the mirror image. According to Eq. (5.15), $\varepsilon_{\text{crit}} \approx 3.3 \times 10^{-6}$ here. We see the first three allowed $\pi$ intervals (gray) in this plot ($n = 1, 2, 3$), each of them containing two additional singularities. The dashed lines are the $\mathrm{Re}\,\tau$- and $\varepsilon$-dependent lower bounds (5.25) for the imaginary parts of the additional singularities within the $\pi$ intervals.

**Moving singularities for varying $\gamma_c$**

In the next plot (Fig. 5.6), we see the singularities moving as $\gamma_c$ is increased. As in the Gauss-pulse case, the main singularity moves like $\mathrm{i}\gamma_c$ below the crit-ical threshold (pure Sauter–Schwinger effect) and reaches the height of the first additional singularities approximately at $\gamma_c^{\text{crit}}(\varepsilon)$. Then, it slows down exponentially and seems to "push" the additional singularities upwards (dynamical assistance).

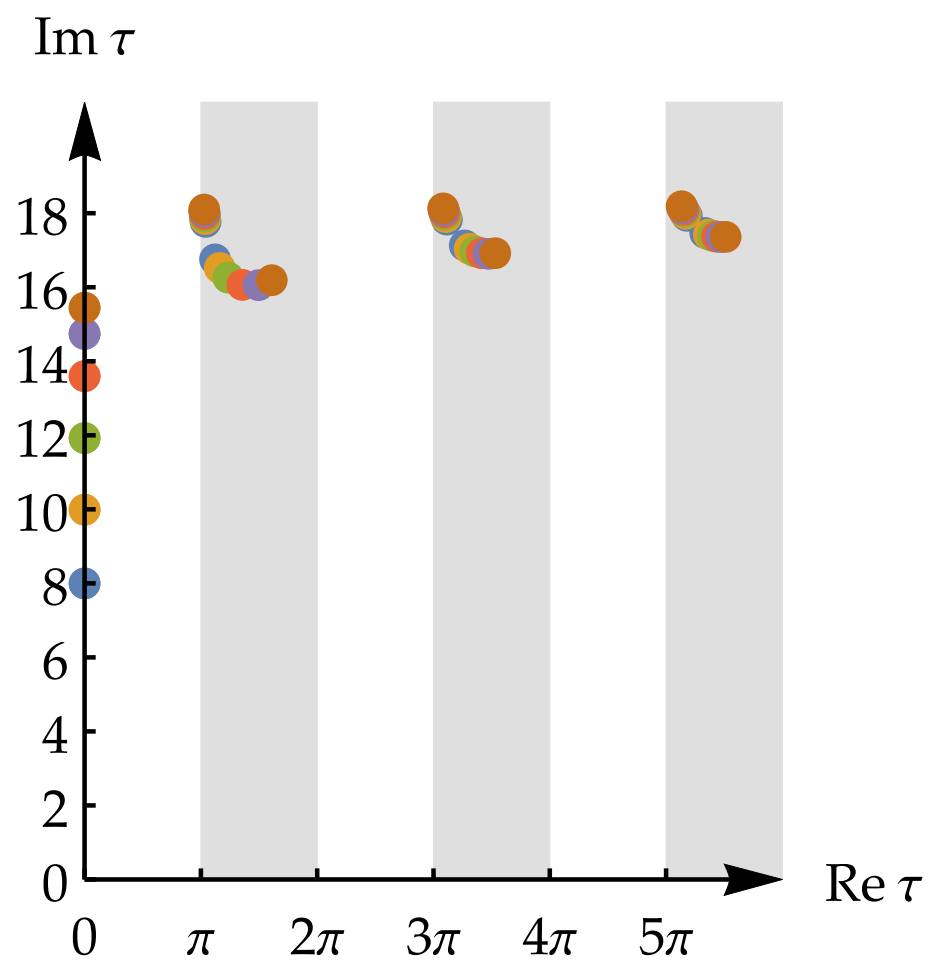

**Figure 5.6.:** Movement of the numerically calculated singularities $\tau^\star$ [colored dots; solutions to Eq. (5.3)] in the complex $\tau$ plane as $\gamma_c$ increased from 8 (blue) to 18 (brown) in steps of 2. Again, we consider $\kappa = 0$, and $\varepsilon = 10^{-6}$ is fixed, so $\gamma_c^{\text{crit}} \approx 13.4$ according to Eq. (5.14).

**Exponents** The exponents in the "residues" corresponding to these singularities (each residue contributes to $R_0^{\text{out}}$) are shown in Fig. 5.7. Again, the exponent generated by $\tau^\star_{\text{main}}$ (main exponent) is always the smallest one (least suppression), and thus we treat it as *the* exponent governing the pair-creation probability. However, the exponents associated with the additional singularities near $\tau^\star_{\text{main}}$ become comparable to the main exponent above the critical threshold, which will probably give rise to interference effects in the momentum spectrum (cf. [70, 62]).

## 5.3. Results and conclusion

The main results we found in this chapter are:

- $\gamma_c^{\text{crit}} \overset{\varepsilon\to 0}{\sim} |\ln \varepsilon|$ The critical threshold $\gamma_c^{\text{crit}}(\varepsilon)$ scales with $|\ln \varepsilon|$ in the limit $\varepsilon \to 0$. The analysis leading to this result is very similar to the Gauss-pulse case: sin and erf, the functions which appear in $A(t)$ for these field profiles, are both odd and entire functions, but $\sin x$ increases like $\exp(\text{Im}\, x)$ in the

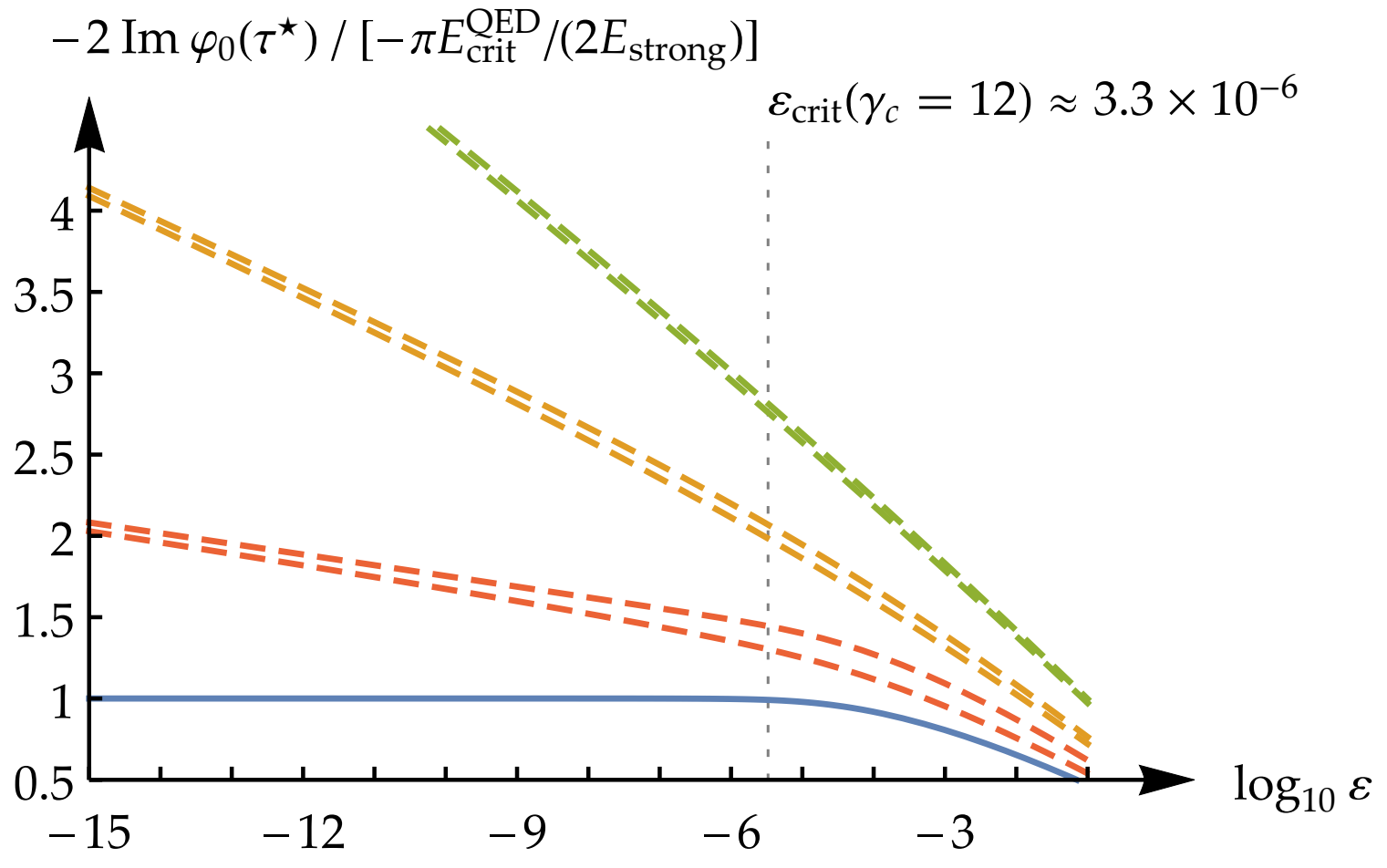

**(a)** Exponents over $\varepsilon$ for $\gamma_c = 12$.

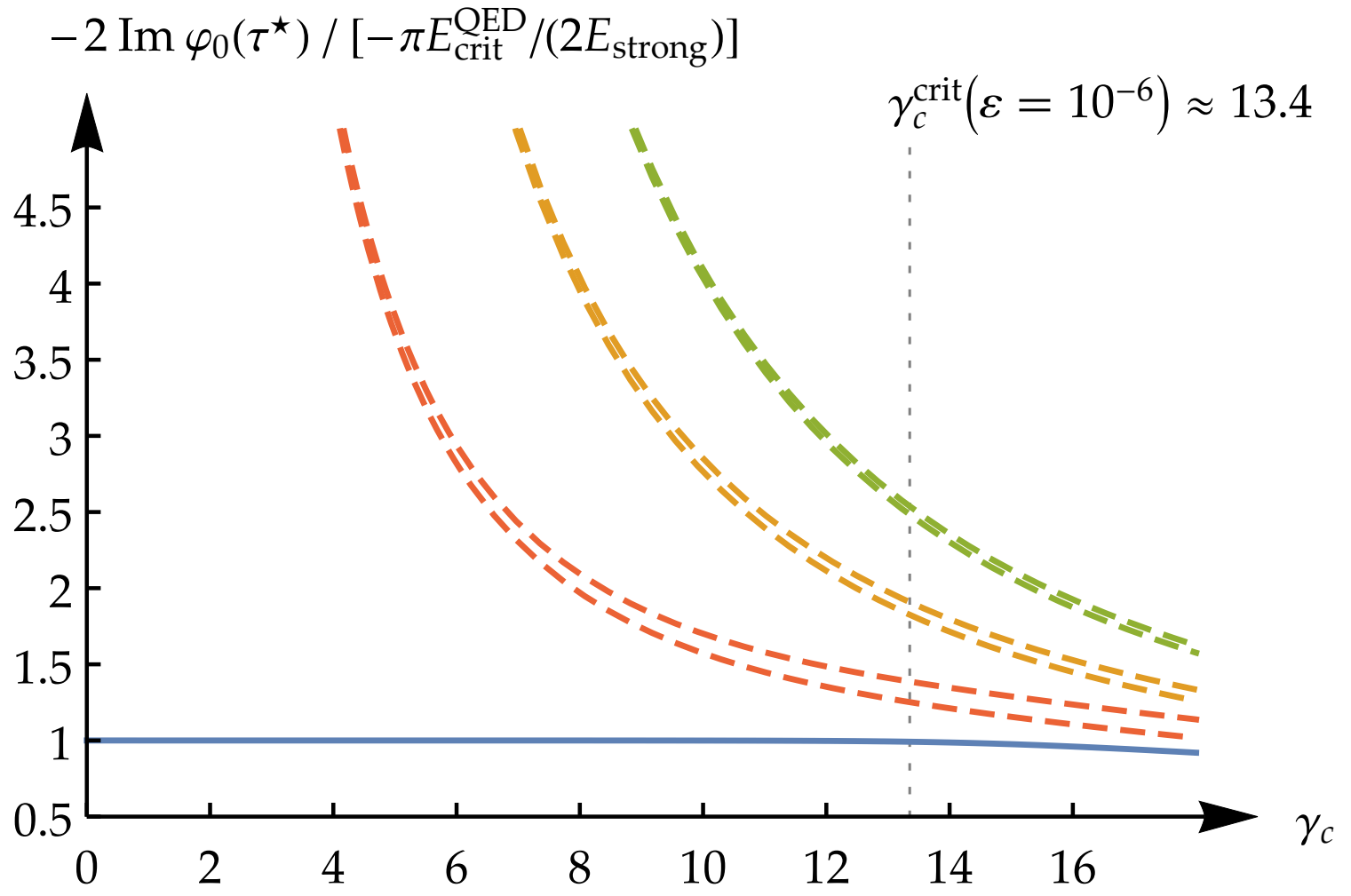

**(b)** Exponents over $\gamma_c$ for $\varepsilon = 10^{-6}$.

**Figure 5.7.:** Numerically calculated exponents $-2\,\mathrm{Im}\,\varphi_0(\tau^\star)$ [see Eq. (5.10) for the definition of $\varphi_\kappa(\tau)$] which appear in the contributions (to $R_0^{\mathrm{out}}$) from the singularities plotted in Figs. 5.5 and 5.6 above, in units of the Sauter–Schwinger tunneling exponent. The blue, solid plot line corresponds to the main singularity $\tau_{\mathrm{main}}^\star = iv$, whose imaginary part can be found by solving the real equation (5.4) numerically. The dashed plot lines represent the additional singularities in the first three $\pi$ intervals ($n = 1$: red; 2: ocher; 3: green).

upper half-plane, while the exponential increase of erf $x$ is governed by $\exp(\mathrm{Im}^2\, x)$ there (hence $\gamma_c^{\mathrm{crit}} \sim \sqrt{|\ln \varepsilon|}$ in the Gauss-pulse case).

- We defined a critical threshold for dynamical assistance in Eqs. (5.14) and (5.15); see also Fig. 5.2 on page 131. At this threshold, the Sauter–Schwinger tunneling exponent is lowered by 1% due to the first-order correction (with respect to $\varepsilon$) from the oscillation [see Eq. (5.12)].

$\gamma_c^{\mathbf{crit}}(\varepsilon)$, $\varepsilon_{\mathbf{crit}}(\gamma_c)$

- We developed a method to find all additional singularities for $k = 0$ graphically; see Fig. 5.4 on page 136. This method allows us to enumerate all additional singularities and thus provides a systematic way to calculate them numerically. Furthermore, the method helped us to understand the asymptotics of $\tau^\star_{\mathrm{add}}$ for large $\mathrm{Re}\,\tau^\star_{\mathrm{add}}$, for example (see Sec. 5.2.2).

**Graphical solution method**

- As in the Gauss-pulse case, the dominating exponent in $R_0^{\mathrm{out}}$ comes from the main singularity $\tau^\star_{\mathrm{main}}$; see Fig. 5.7. Above the critical threshold, though, the exponents associated with the nearby additional singularities take on similar values.

**Main exponent**

# Part III.

# Analog of Dirac's theory in direct-bandgap semiconductors for spacetime-dependent external fields

# 6. Basics and known results

The largeness of the Schwinger limit $E_{\text{crit}}^{\text{QED}} \approx 10^{18}\,\text{V/m}$ makes it difficult to study nonperturbative pair creation from the (spinor-)QED vacuum experimentally—so far, this effect has not been observed [54]. This fact motivates the search for analogs of this effect in other physical systems. We are talking about **quantitative analogies** in this context, which means that the essential physical laws underlying spinor QED with external, classical fields (i.e., the Dirac equation with minimally coupled fields plus the Dirac sea as the initial vacuum state) can be formally recognized in the analog system. All physical effects which are derived from these underlying laws in the original system, such as the Sauter–Schwinger effect, then have a natural counterpart in the analog system since, to quote Richard Feynman [109], **the same equations have the same solutions**. Analog systems are especially interesting if they are easy to access experimentally and if the scales in the original system have counterparts in the analog system which take on less extreme values and thus facilitate experiments. Here, $c$, $m$, $\hbar$, and $q$ appear in the Dirac equation (1.12) with external fields. Note that not all of these fundamental scales will necessarily be different in the analog system—in the semiconductor analog considered in this thesis, for example, only the vacuum speed of light $c$ and the electron rest mass $m$ have different ("effective") values $c_\star$ and $m_\star$, while the reduced Planck constant $\hbar$ and the elementary charge $q$ remain the same as in spinor QED. As we will see in this part, the **effective scales in typical, suitable semiconductors are much smaller than their QED counterparts** (i.e., $c_\star/c \ll 1$ and $m_\star/m \ll 1$), so that the analog of the Schwinger limit (1.21) reads

**Quantitative analogies**

**Effective scales**

**Analog of the Schwinger limit**

$$E_{\text{crit}}^{\text{SC}} = \frac{m_\star^2 c_\star^3}{\hbar q} = \underbrace{\left(\frac{m_\star}{m}\right)^2 \left(\frac{c_\star}{c}\right)^3}_{\ll 1 \text{ in typical semiconductors}} E_{\text{crit}}^{\text{QED}} \overset{\text{(typically)}}{\ll} E_{\text{crit}}^{\text{QED}} \tag{6.1}$$

in such a semiconductor. Hence, the field intensities required to induce the analog of nonperturbative electron–positron pair creation via strong and slow external electric fields is drastically reduced in semiconductor analogs.

There are further options to simulate spinor QED and the Sauter–Schwinger effect experimentally, for example ultracold atoms in optical lattices [110,

111, 112], trapped ions [113], or graphene [114, 115, 116, 117, 118][1].

### The analogy in semiconductors

**Dirac's theory**

Let us summarize the aspects of Dirac's theory which are important for the description of the strong-field QED phenomena considered in this work: As explained in Sec. 2.2, we **ignore the interaction between created electrons and positrons**[2] and their impact on the field which created them (i.e., we **ignore backreaction**) throughout this thesis. We may thus work within the framework of first quantization (classical, single-particle Dirac equation) with prescribed external fields. In this picture, particles are not actually created, and there is only one sort of particles (electrons). However, we need to distinguish between occupied states in the upper relativistic continuum, which are interpreted as *real* electrons, and unoccupied electron states in the lower continuum ("holes"), which represent real positrons. The vacuum state is defined as the absence of any real particles, so the upper continuum is completely unoccupied in this state, while all states in the lower continuum must be occupied by electrons (Dirac sea). Dirac-sea electrons may be excited into the upper continuum (electron–positron pair creation) under the influence of the external field. For example, the pure Sauter–Schwinger effect (constant external $\boldsymbol{E}$ field) can be interpreted as a tunneling transition between the two energy continua according to this picture (see Fig. 2.2 on page 48).

**Semiconductor analog**

Another well-known physical system in which electron energies are confined to certain **energy bands** are crystalline solids. In simple words, the atomic/molecular orbitals (electron states with discrete energies) of the constituents overlap in the crystal, leading to the formation of energy bands [122].

**Assumption: perfect crystalline structure**

In the limit of an **infinitely large crystal**, each band is a finite interval of continuously distributed allowed electron energies. In the ground state (low temperatures), all electron states up to the Fermi energy are occupied by electrons, while all energy states above this level are empty. The highest energy band containing electrons in the ground state is called the **valence band**. If the number of valence electrons of the constituents leads to a completely filled valence band in the crystal and if the next higher band (**conduction band**) is separated from the valence band by a nonzero **energy gap/bandgap** $\mathcal{E}_g$, the crystalline material will be insulating. Insulators with bandgaps up to about 3 eV (corresponding to the upper frequency edge of the visible spectrum) are commonly called **semiconductors**. Electrons can be excited into free states of higher energy (including band transitions) via external forces—for example, by absorbing an incident photon. If the photon energies provided by the external field do not exceed the bandgap too much, only transitions from the valence band to the conduction band will be likely to occur; that is, the filled bands below the valence band will be practically inert, and the bands above the conduction band will remain empty. In this situation, we may treat the semiconductor as if only the valence band and the conduction band were present (**two-band model**). Within this model, the valence band, which is completely filled with electrons in the ground state of the semiconductor, is in some way analog to the Dirac sea in spinor QED, while the conduction band resembles the positive relativistic energy continuum. There is thus at least a qualitative analogy between electron–hole pair creation due to slowly varying fields in two-band semiconductors and electron–positron pair creation in Dirac theory. (This qualitative analogy is well known; see Sec. 6.2 below for an overview.)

[1] Note, however, that the bandgap in ordinary graphene [119, 120] is zero, and thus this material is only appropriate to simulate the creation of *massless* Dirac fermions in 2+1 spacetime dimensions. Generating a nonzero bandgap in graphene is possible by using appropriate substrates for a layer of graphene [121], for example, which could then be used as an analog of spinor QED with massive particles [118].

[2] We will provide a simple estimate on why this approximation is justified later in Sec. 8.4.1.

**Approximation: two-band model**

The main objective in this part is to show that this is a true *quantitative* analogy, and we will work out the assumptions which are required to draw this analogy. Under these assumptions, a suitable semiconductor could thus be used to simulate nonperturbative electron–positron pair creation in the laboratory.

**Goals in this thesis**

## 6.1. Bloch band structures in semiconductors

While the existence of two distinct energy continua in Dirac theory is a consequence of the relativistic invariance of the Dirac equation, the physics in semiconductors at low temperatures and with slowly varying external fields (photon energies up to the bandgap, i.e., a few electronvolts) is described well by nonrelativistic quantum mechanics. This theory is usually associated with only one (positive) energy branch since the dispersion relation of a free electron is $\mathcal{E} = (\hbar \boldsymbol{k})^2/(2m)$ according to the Schrödinger equation. However, electrons within a solid "feel" a potential created by the atomic nuclei/ion cores (i.e., the electrons are not free), and this potential has the same spatial periodicity as the crystal lattice. According to the Bloch theorem, it is this **periodic crystal potential** $V(\boldsymbol{r})$ which leads to the formation of electronic energy bands in crystals [122]. Throughout this thesis, we make the following assumptions about our semiconductor, which allows us to apply this theorem:

**Semiconductor: assumptions**

- We assume that vibrations of the ion cores (**phonons**) can be neglected, so $V(\boldsymbol{r})$ should have a unique lattice symmetry and should be time independent.

- Electron–electron **interactions** are neglected[3]; each electron is subject to the same crystal potential. This assumption allows us to work in the one-electron picture.
- Any **backreaction** of the electrons, which are much lighter particles than the ion cores, on the crystal lattice is neglected.

Furthermore, we add the assumption that

- the semiconductor should be in its ground state initially (filled valence band, empty conduction band), which requires a **temperature** $T$ far below the bandgap ($k_B T \ll \mathcal{E}_g$, where $k_B$ is the Boltzmann constant). This is in accordance with the assumption above that phonons can be neglected.

**Neglect electron spin**

In the absence of any external field, the wave functions of the Bloch electrons within the semiconductor are then given by the solutions of the nonrelativistic, stationary Schrödinger equation (we **ignore spin effects** here, which usually play a minor role in the tunneling pair-creation phenomena we are interested in [97, 34])

$$\left[-\frac{\hbar^2\boldsymbol{\nabla}^2}{2m} + V(\boldsymbol{r})\right] f_n(\boldsymbol{K},\boldsymbol{r}) = \mathcal{E}_n(\boldsymbol{K}) f_n(\boldsymbol{K},\boldsymbol{r}). \tag{6.2}$$

**Bloch theorem**

The Bloch theorem states that the **Bloch (wave) functions** have the form

$$f_n(\boldsymbol{K},\boldsymbol{r}) = \mathrm{e}^{\mathrm{i}\boldsymbol{K}\cdot\boldsymbol{r}} u_n(\boldsymbol{K},\boldsymbol{r}), \tag{6.3}$$

where the functions $u_n(\boldsymbol{K},\boldsymbol{r})$, which we call **Bloch factors** here, have the same spatial periodicity as the crystal potential $V(\boldsymbol{r})$. The Bloch states are indexed by the **quasimomentum**[4] or **crystal momentum** $\boldsymbol{K}$ and the **band index** $n \in \mathbb{N}$. We assume the energy bands to be indexed in ascending order, i.e., $\mathcal{E}_1(\boldsymbol{K}) \leq \mathcal{E}_2(\boldsymbol{K}) \leq \ldots \; \forall \boldsymbol{K}$. The quantity $\boldsymbol{K}$ is called quasimomentum because it shares some properties with the ordinary momentum of an electron. Consider the Bloch-wave form (6.3), for example: except for the cell-periodic Bloch factors, the Bloch functions have the form of free-electron eigenstates $\mathrm{e}^{\mathrm{i}\boldsymbol{K}\cdot\boldsymbol{r}}$, in which case $\hbar\boldsymbol{K}$ would be the associated momentum. However, as a consequence of the lattice periodicity of $V(\boldsymbol{r})$, the quasimomentum $\boldsymbol{K}+\boldsymbol{G}$, where $\boldsymbol{G}$ is an arbitrary reciprocal-lattice vector, is equivalent to $\boldsymbol{K}$ in the sense that these two quasimomenta refer to the same Bloch state (for a fixed $n$)—a fact which distinguishes $\boldsymbol{K}$ from an "ordinary" electron wave vector $\boldsymbol{k}$. The

**$\boldsymbol{K}$ lies in the first Brillouin zone**

[3] We will present a simple argument which supports this approximation later in Sec. 8.4.1.

[4] Note that these names should be understood modulo $\hbar$.

quasimomentum of a Bloch state can thus always be chosen to lie within the **first Brillouin zone** in $\boldsymbol{K}$ space, and we will adopt this convention throughout this thesis (reduced zone scheme).

In summary, each Bloch band within a semiconductor is associated with a function $\mathcal{E}_n(\boldsymbol{K})$ which is defined over the first Brillouin zone and is a continuous function of the quasimomentum. This **band structure** can be formally compared to the relativistic energy–momentum relation (dispersion relation) $\mathcal{E}_\pm(\boldsymbol{k}) = \pm\sqrt{m^2c^4 + (c\hbar\boldsymbol{k})^2}$ of a free electron in Dirac theory.

### 6.1.1. Analogy in 1+1 spacetime dimensions

**Dirac theory**

The relativistic "band structure" for free electrons in one-dimensional space is plotted in Fig. 6.1(a). The minimal difference between both branches of the energy–momentum relation measures $2mc^2$ (**mass gap**) and is located at $k = 0$ in momentum space.

**Semiconductor**

In the semiconductor, the band structure depends on the concrete shape of $V(x)$. There is only one type of lattice symmetry in one spatial dimension: the crystal potential must satisfy

$$V(x) = V(x + \ell) \quad \forall x \tag{6.4}$$

with the **lattice constant** $\ell > 0$. The first Brillouin zone then corresponds to

$$K \in \left(-\frac{\pi}{\ell}, \frac{\pi}{\ell}\right]. \tag{6.5}$$

We are only interested in the valence band (band index $-$) and the next higher band, the conduction band ($+$), because of the two-band approximation. Such a two-band Bloch structure is plotted in Fig. 6.1(b). Although this plot is only an example, we assume that the band structures of the semiconductors considered in this thesis have the following properties, which are also visualized in this plot:

**Semiconductor: further assumptions**

- The minimal difference between the conduction band and the valence band (i.e., the bandgap $\mathcal{E}_g$) should be located at a reciprocal-space position $K$ where the valence band $\mathcal{E}_-(K)$ has a maximum and the conduction band $\mathcal{E}_+(K)$ has a minimum; that is, we consider only semiconductors with a **direct bandgap**. According to this well-known definition, the mass gap in Dirac theory [see Fig. 6.1(a)] is also a "direct bandgap", and it seems plausible that a semiconductor should share this generic property in order to serve as an analog for Dirac's theory since the majority of excitation processes via low-energy effects usually occur near the gap.

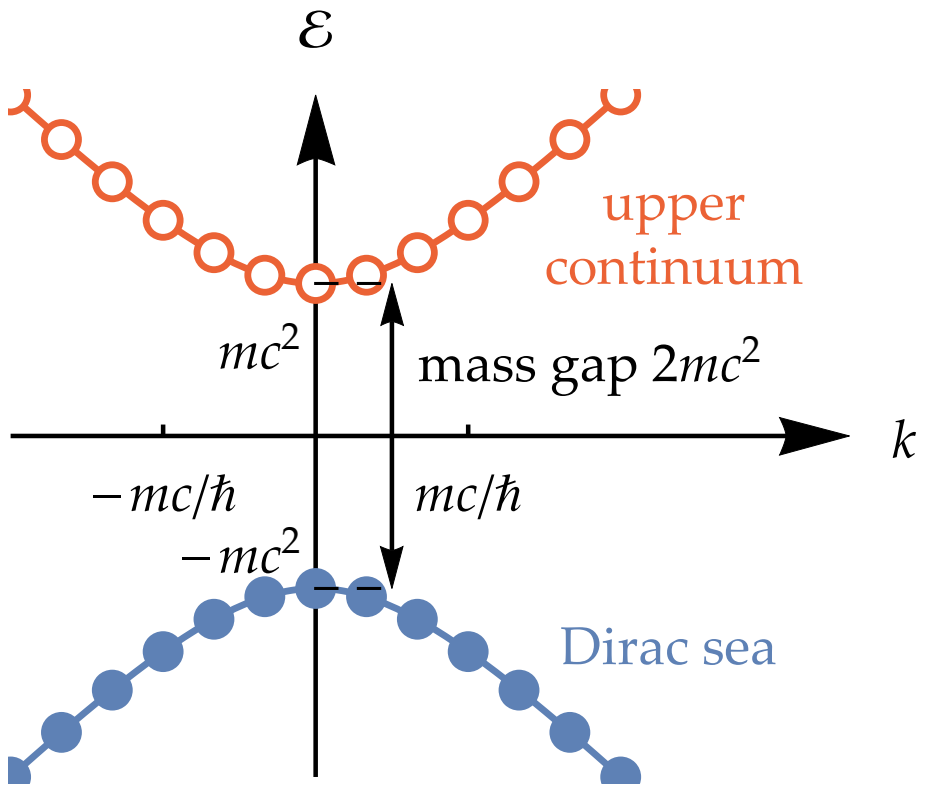

**(a)** Dirac theory

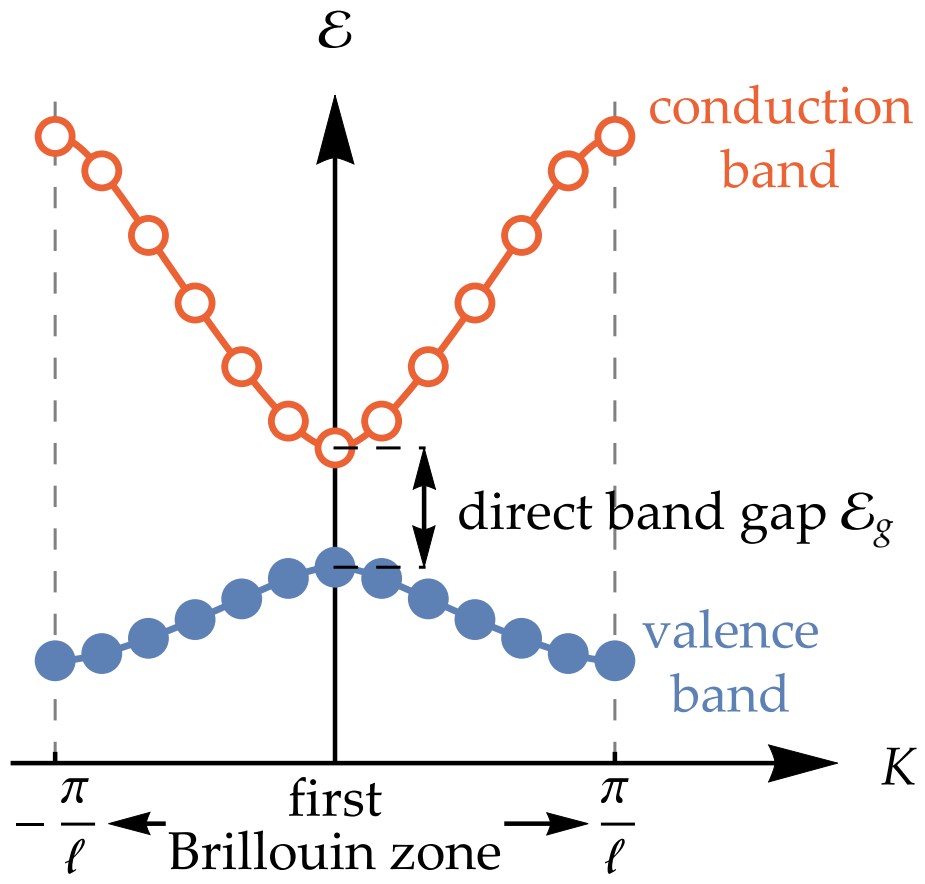

**(b)** Semiconductor analog

**Figure 6.1.:** Comparison between the electronic band structures in reciprocal space for the two systems under consideration in 1+1 spacetime dimensions. Both plots illustrate the respective ground state; filled/empty circles indicate occupied/free electron states. (a) Dirac theory: the two branches $\mathcal{E}_\pm(k) = \pm\sqrt{m^2c^4 + (c\hbar k)^2}$ of the relativistic energy–momentum relation. (b) Semiconductor analog: two-band structure $\mathcal{E}_\pm(K)$ with a direct bandgap at the zone center ($\Gamma$ point).

- The bandgap should be located at the **center of the Brillouin zone** ($K = 0$), often denoted by $\Gamma$ in band-structure plots. Although this assumption is probably *not* required to derive the analogy, we make it here for convenience.

A suitable semiconductor meeting these assumptions is **gallium arsenide (GaAs)** [122], which will be considered as a promising candidate for the semiconductor analog throughout this part. **Gallium arsenide**

### 6.1.2. Problems in 3+1 spacetime dimensions

Unfortunately, a real probe of GaAs, for example, is of course a three-dimensional object, and for this reason the band structure is in fact more complex than the simple 1+1-dimensional model depicted in Fig. 6.1(b). Drawing the analogy to Dirac theory thus becomes more difficult (or, requires more approximations) mainly for the following two reasons:

- **Anisotropy:** The shape of the dispersion curves $\mathcal{E}_\pm(\boldsymbol{K})$ in the semiconductor usually depends on the direction in reciprocal space along which the curves are considered. That is, even if the plot in Fig. 6.1(b) is valid for a given semiconductor and for a given spatial direction, the band plot will probably look different when considered for another direction (although the position of the bandgap at the zone center is universal). Dirac theory, in contrast, is isotropic because $\mathcal{E}_\pm(\boldsymbol{k}) = \pm\sqrt{m^2c^4 + (c\hbar\boldsymbol{k})^2}$ depends on $|\boldsymbol{k}|$ only, not on the direction of $\boldsymbol{k}$, and the physical constants $m$, $c$, and $\hbar$ are scalar.

- **Multiple valence bands:** A Bloch electron moving through a crystal "sees" the positive charges of the nuclei/ion cores. These charges give rise to a magnetic field in the rest frame of the Bloch electron, which couples to its spin (**spin–orbit interaction**) [122]. The resulting splitting of energy levels (which would be degenerate without this effect) leads to the existence of multiple valence bands in crystals[5]. Due to the absence of positive charges in the Dirac vacuum, this effect has no counterpart in Dirac theory [123]—there is only one lower relativistic energy continuum $\mathcal{E}_-(\boldsymbol{k})$.

We will discuss these problems with particular regard to the band structure of GaAs in the following, but there are many other direct-bandgap semiconductors with a similar band structure.

[5] We are describing the effects of spin–orbit interaction in III–V semiconductors such as GaAs here, the type of semiconductors we are primarily interested in.

GaAs belongs to the group of III–V compound semiconductors and has a Zincblende structure, which means that the ion positions are the same as in the diamond lattice, but the gallium ions and the arsenide ions are arranged alternatingly at the lattice positions. The electronic band structure within this lattice structure has been studied in [124, 125] via **$\boldsymbol{K} \cdot \boldsymbol{p}$ perturbation theory**. This approach for calculating band structures can be understood by considering the Schrödinger equation for the lattice-periodic Bloch factors for a given quasimomentum $\boldsymbol{K}$ [we insert the Bloch-wave form (6.3) into Eq. (6.2)]:

**$\boldsymbol{K} \cdot \boldsymbol{p}$ perturbation theory**

$$\left[-\frac{\hbar^2\boldsymbol{\nabla}^2}{2m} + \underbrace{\frac{\hbar}{m}\boldsymbol{K}\cdot(-\mathrm{i}\hbar\boldsymbol{\nabla})}_{\text{perturbation}} + \frac{\hbar^2\boldsymbol{K}^2}{2m} + V(\boldsymbol{r})\right] u_n(\boldsymbol{K},\boldsymbol{r}) = \mathcal{E}_n(\boldsymbol{K})u_n(\boldsymbol{K},\boldsymbol{r}). \quad (6.6)$$

Suppose we know all Bloch factors and energies at the bandgap (here: $\boldsymbol{K} = 0$); these can be obtained from measurements or computer simulations, for example, if a direct calculation is not possible. These Bloch factors $u_n(0,\boldsymbol{r})$ form a complete set of basis functions for arbitrary lattice-periodic functions and are thus also appropriate to express the Bloch factors at other $\boldsymbol{K}$ values. For small $\boldsymbol{K}$ vectors close to the bandgap (i.e., if $|\boldsymbol{K}|$ is much smaller than the size of the Brillouin zone), the term containing $\boldsymbol{K}\cdot\hat{\boldsymbol{p}}$ in Eq. (6.6) may be treated as a small perturbation (hence the name), and time-independent perturbation theory may be applied to calculate $\mathcal{E}_n(\boldsymbol{K})$ and $u_n(\boldsymbol{K},\boldsymbol{r})$ from $\mathcal{E}_n(0)$ and $u_n(0,\boldsymbol{r})$ with sufficient precision (see [122] for more details). Since the influence of the $n'$th band on the $n$th band is suppressed by the factor $1/|\mathcal{E}_n(0) - \mathcal{E}_{n'}(0)|$ according to perturbation theory, it is often practically sufficient to include only corrections from adjacent energy bands in such perturbational approaches. Applying the two-band approximation in this context means that we only consider the valence band and the conduction band and their mutual influence via $\boldsymbol{K}\cdot\boldsymbol{p}$ perturbation theory.

**Two-band approximation**

**Band structure in GaAs**

The band structure close to the bandgap in GaAs is depicted in Fig. 6.2 [122]. All bands can be approximated by parabolas around $\boldsymbol{K} = 0$ because they have an extremum there. For a given direction (say $K_x$, as in Fig. 6.2), each band thus has the form

$$\mathcal{E}(K_x) = \text{const.} \pm \frac{\hbar^2 K_x^2}{2m_\star} \quad (6.7)$$

near the gap, where $m_\star$ is the **effective electron/hole mass**[6] associated with that band (and direction!). There is one (nondegenerate) conduction band, which originates from s-like molecular orbitals. It is isotropic around the gap. This part of the band structure is approximated well by the conduction band in the 1+1-dimensional model in Fig. 6.1(b). The p-like valence orbitals, in contrast, form three different valence bands due to spin–orbit interaction. The upper two ones, the **light-hole band** and the **heavy-hole band** are both separated by $\mathcal{E}_g$ from the conduction band, respectively, and they are degenerate at the gap. The corresponding effective masses are $m_\star^{\mathrm{lh}} \approx 0.076m$ and $m_\star^{\mathrm{hh}} \approx 0.5m$ in GaAs [126]. As we will derive later, these effective masses play the role of the ordinary electron mass $m$ in the semiconductor analog. Since a higher mass suppresses pair creation via the Sauter–Schwinger effect exponentially (because $E_{\mathrm{crit}}^{\mathrm{QED}} \propto m^2$), we **ignore the heavy-hole band** in our semiconductor analog. Furthermore, the heavy-hole band does not couple to the conduction band according to first-order $\boldsymbol{K} \cdot \boldsymbol{p}$ perturbation theory [125, 127], which hints that this is a good approximation.

[6]Symbols denoting effective masses always represent positive values here.

**Neglect heavy-hole band**

The third valence band is the **split-off band**. In GaAs, it lies $\Delta \approx 0.34\,\mathrm{eV} \approx 0.24\mathcal{E}_g$ below the upper two valence bands, and the corresponding effective mass, $m_\star^{\mathrm{so}} \approx 0.15m$, is approximately twice as large as the light-hole mass [122]. For these two reasons, we expect this band to contribute much less to tunneling pair creation than the light-hole band. However, the split-off band couples to the conduction band in first-order $\boldsymbol{K} \cdot \boldsymbol{p}$ perturbation theory, and thus it was suggested that it might contribute appreciably to tunneling currents in GaAs [128, 127, 129]. This question was addressed in [130]; it was found that the exponent in the factor $\exp(-\pi E_{\mathrm{crit}}^{\mathrm{GaAs}}/|\boldsymbol{E}|)$ appearing in the tunneling pair-creation rate in GaAs (for a constant external $\boldsymbol{E}$ field, i.e., the analog of the Sauter–Schwinger effect) is reduced by the factor

**Correction due to the split-off band**

$$\sqrt{\frac{(5+4\alpha)(1+2\alpha)}{(2+2\alpha)(3+4\alpha)}} \approx 0.95 \quad \text{for} \quad \alpha = \frac{\Delta}{\mathcal{E}_g} \approx 0.24 \quad \text{in GaAs} \qquad (6.8)$$

when tunneling from the split-off band is taken into account. Hence, the incorporation of the **split-off band may lower the equivalent of the Schwinger limit** in GaAs by about 5%. This correction is useful to keep in mind (e.g., in order to interpret experimental results better), but the light-hole band seems to be crucial in the context of tunneling pair creation [131], so we will **ignore the split-off band in our two-band semiconductor analog** since it has no counterpart in Dirac theory.

**Neglect split-off band**

All in all, we will base our semiconductor analog of Dirac's theory in external fields on the 1+1-dimensional two-band structure shown in Fig. 6.1(b) with the valence band corresponding to the light-hole band, which is fortunately only slightly anisotropic in semiconductors with a Zincblende structure like GaAs [122].

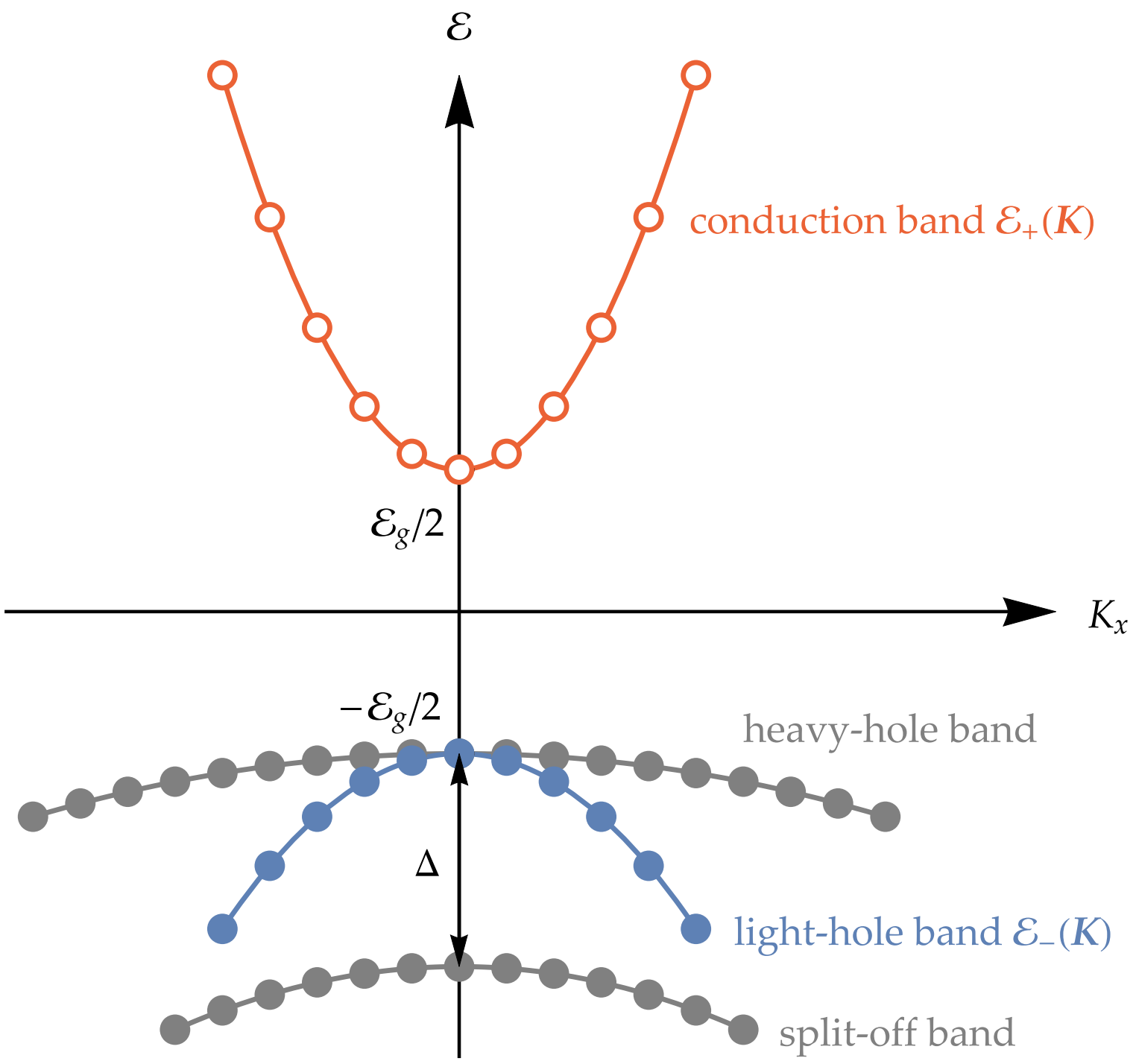

**Figure 6.2.:** Schematic band structure of GaAs in the vicinity of the bandgap (at $\boldsymbol{K} = 0$, the $\Gamma$ point) along the $K_x$ direction (arbitrary choice) in reciprocal space. Empty/filled circles indicate free/occupied electron states; this plot depicts the ground state. In our semiconductor analog of Dirac theory, the conduction band $\mathcal{E}_+(\boldsymbol{K})$ corresponds to the upper relativistic energy continuum, while the light-hole band $\mathcal{E}_-(\boldsymbol{K})$ corresponds to the Dirac sea. The heavy-hole band and the split-off band are ignored (assumed to be inert) in our analogy.

## 6.2. Known results concerning the analogy

When a direct-bandgap semiconductor (in its ground state) is exposed to a constant external $\boldsymbol{E}$ field, the energy continua representing the valence band and the conduction band become tilted in space just like the relativistic continua in Fig. 2.2 on page 48 (tunneling picture of the Sauter–Schwinger effect). This tilt of energy levels renders interband tunneling of electrons possible, which leads to an electric current—that is, tunneling is one mechanism of electrical breakdown in insulating crystals. This effect has been described for two-band crystals in [132, 133, 134] and is therefore referred to as **Landau–Zener tunneling**. The theory was further refined in [135, 136, 137, 138, 128, 139, 127, 140]. The tunneling picture immediately suggests the analogy between Landau–Zener tunneling and the Sauter–Schwinger effect, which has also been discussed in the literature; see, e.g., [141, 142, 143, 144, 115].

**Constant $\boldsymbol{E}$ field**

When applying Kane's model [125] in 1+1 spacetime dimensions to derive the electronic band structure in a direct-bandgap semiconductor close to the gap (using various approximations), only taking into account one conduction band and one valence band (both nondegenerate), one finds that the Bloch electrons in these bands obey an effective Dirac equation. This **relativistic analogy** was derived and/or discussed for constant external fields in [145, 104, 97, 123, 146, 142, 147, 148], for example. The difference between Dirac theory and the semiconductor analog are two scale substitutions, which can be understood intuitively via the following considerations (see, e.g., [128, 104, 96, 97, 146]):

**Relativistic analogy**

- The **mass gap** $2mc^2$ in Dirac theory corresponds to the **bandgap** $\mathcal{E}_g$ in the semiconductor, which is an obvious analogy in regard of the band diagrams in Fig. 6.1 on page 152;

$$2mc^2 \leftrightarrow \mathcal{E}_g. \tag{6.9}$$

- In Dirac theory, the inertial mass of electrons (and positrons) is the **electron rest mass** $m$ (for nonrelativistic velocities). In semiconductors, charge carriers close to the bandgap will change their quasimomentum $\boldsymbol{K}$ in the presence of an external electric field according to a different, **effective mass** $m_\star$, so we have the correspondence

$$m \leftrightarrow m_\star. \tag{6.10}$$

  In Ref. [96], for example, the authors assume that electrons in the conduction band have the same effective mass $m_{\star,e}$ as (light) holes in the

**Scale substitutions**

valence band ($m_{\star,h}$), in which case $m_\star = m_{\star,e} = m_{\star,h}$. It was found in Ref. [128], however, that the analogy does also work for $m_{\star,e} \neq m_{\star,h}$ (which is usually true) in the case of constant external electric fields. Then, *the* effective mass in the sense of Eq. (6.10) is given by the harmonic mean

$$m_\star = \frac{2}{\frac{1}{m_{\star,e}} + \frac{1}{m_{\star,h}}}, \tag{6.11}$$

which is closely related to the **reduced mass** known from two-body problems in Newtonian mechanics.

- The **vacuum speed of light** $c$ is a characteristic velocity in Dirac theory due to its relativistic nature. Formally, $c$ can be obtained from the mass gap and the electron mass by taking the square root of the mass gap over twice the electron mass: $\sqrt{2mc^2/(2m)} = c$. In the semiconductor, we may use the same relation to derive the equivalent of $c$: according to the above scale substitutions, we have to set $2mc^2$ to $\mathcal{E}_g$ and $m$ in the denominator to $m_\star$ in the semiconductor analog, thus obtaining the **effective speed of light**;

$$c \leftrightarrow c_\star = \sqrt{\frac{\mathcal{E}_g}{2m_\star}}. \tag{6.12}$$

**Analog of the Sauter–Schwinger effect**

In conclusion, it is well known that for *constant* external fields the differences between Dirac theory and a direct-bandgap two-band semiconductor are merely the scale substitutions $m \leftrightarrow m_\star$ and $c \leftrightarrow c_\star$. In the case of a constant $\boldsymbol{E}$ field, we just have to substitute these scales in the Sauter–Schwinger pair-creation rate (2.5) and get

$$\begin{aligned} \dot{\mathcal{N}}_{\mathrm{e^-\text{–}hole}} &= \frac{q^2|\boldsymbol{E}|^2}{4\pi^3\hbar^2 c_\star} \exp\left(-\frac{\pi m_\star^2 c_\star^3}{\hbar q|\boldsymbol{E}|}\right) \\ &= \frac{q^2|\boldsymbol{E}|^2}{2\sqrt{2}\pi^3\hbar^2}\sqrt{\frac{m_\star}{\mathcal{E}_g}} \exp\left(-\frac{\pi\sqrt{m_\star \mathcal{E}_g^3}}{2\sqrt{2}\hbar q|\boldsymbol{E}|}\right), \end{aligned} \tag{6.13}$$

the expected Landau–Zener electron–hole pair-creation rate (per unit volume) in the semiconductor. Almost the same expression was found in [128], the only difference being a little, constant deviation in the prefactor: $1/(4\pi^3)$ in Eq. (6.13) versus $1/(36\pi)$ in [128][7]. By considering the argument of the exponential function in Eq. (6.13), we can read off the **equivalent of the Schwinger**

[7] The reason for this discrepancy probably lies in the (semiclassical) JWKB approximation used in [128]: Remember that we studied QED pair creation by a constant electric field in

**limit**, which is valid in the semiconductor analog [see also Eq. (6.1)]:

$$E_{\text{crit}}^{\text{QED}} \leftrightarrow E_{\text{crit}}^{\text{SC}} = \frac{m_\star^2 c_\star^3}{\hbar q} = \frac{\sqrt{m_\star \mathcal{E}_g^3}}{2\sqrt{2}\hbar q}. \tag{6.14}$$

The expression $\exp(-\pi E_{\text{crit}}^{\text{SC}}/|\boldsymbol{E}|)$ with this analog of the Schwinger limit (although not referred to as such) can be found in many papers dealing with electron–hole pair creation in constant electric fields [149, 128, 127, 139, 52, 140, 97, 96, 142].

**Goals in this thesis**

The goal in the following chapters is to derive the **quantitative analogy** between Dirac theory and electrons in a two-band semiconductor with a direct bandgap from scratch, carefully keeping track of further required approximations, and allowing also for **time- and spacetime-dependent external fields**. Our basis will be the semiconductor model consisting of two nondegenerate energy bands in 1+1 spacetime dimensions [see Fig. 6.1(b) on page 152], which has been explained in this chapter. We will then do first steps to generalize the analogy to 2+1 spacetime dimensions in the case of **crossed, constant electric and magnetic fields** (see also [104, 96, 97]), which is the simplest field profile including electric and magnetic components in 2+1 dimensions.

Furthermore, we will apply the gathered knowledge about the analogy by proposing ways to **simulate various mechanisms of dynamical assistance for the Sauter–Schwinger effect** (such as dynamical assistance in inhomogeneous fields [90]) via semiconductor analogs in GaAs. Such analogs could help us to observe and to verify these mechanisms in the laboratory, which in turn could pave the way towards the observation of the Sauter–Schwinger effect.

Most of the results presented in this part have been published in the article [2].

Ch. 3 by means of the linearized Riccati equation, which also involves a form of the JWKB approximation. The prefactor in the resulting pair-creation probability deviated by a factor of $\pi^2/9$ from the expected result [in the limit $E \to 0$; see Sec. 3.3 and Eq. (3.44)]—which exactly coincides with $1/(36\pi)$ divided by $1/(4\pi^3)$.

# 7. Analogy in 1+1 spacetime dimensions for time-dependent electric fields

We start to study the quantitative analogy between Dirac theory and the two-band semiconductor model we have developed in the previous chapter for a **time-dependent, homogeneous** external electric field $E(t)$. We choose the **temporal gauge** with a purely time-dependent vector potential (which only has one component in 1+1 spacetime dimensions):

$$E(t) = \dot{A}(t). \tag{7.1}$$

## 7.1. Many-body Hamiltonians

We consider the many-body Hamiltonians ("second quantization") of both systems here and compare them with each other in order to find out in how far they are equivalent (except for different scales).

### 7.1.1. Dirac theory

Let us start with the Hamiltonian $\hat{H}_D(t)$ of the quantized Dirac field coupled to the prescribed external electric field. The general **real-space expression** for this operator reads

$$\hat{H}_D(t) = \int_{-\infty}^{\infty} \underline{\hat{\Psi}}^\dagger(t,x)\hat{H}_D^{\text{one}}(t,x)\underline{\hat{\Psi}}(t,x)\,\mathrm{d}x, \tag{7.2}$$

where $\underline{\hat{\Psi}}$ is the field operator of the quantized Dirac field. It has two components in 1+1-dimensional spacetime, which obey the canonical anticommutation relations (1.10). The $2 \times 2$ matrix $\hat{H}_D^{\text{one}}$, a matrix of operators acting on the field operator, is the same single-particle Hamilton operator which appears in the classical Dirac equation [cf. Eq. (2.77)]:

$$\hat{H}_D^{\text{one}}(t) = \begin{pmatrix} mc^2 & -\mathrm{i}c\hbar\partial_x + cqA(t) \\ -\mathrm{i}c\hbar\partial_x + cqA(t) & -mc^2 \end{pmatrix}. \tag{7.3}$$

**Transformation to $k$ space**

Since the analogy between Dirac theory and our two-band semiconductor model is most obvious in (crystal) momentum space (remember the band structures plotted in Fig. 6.1 on page 152), we transform to the $k$-space representation of $\hat{H}_D$ by inserting the Fourier-space expression

$$\underline{\hat{\Psi}}(t,x) = \frac{1}{\sqrt{2\pi}} \int_{-\infty}^{\infty} \underline{\hat{\tilde{\Psi}}}(t,k)\, \mathrm{e}^{\mathrm{i}kx}\, \mathrm{d}k \tag{7.4}$$

of the field operator into Eq. (7.2). This yields

$$\begin{aligned}
\hat{H}_D(t) &= \frac{1}{2\pi} \int_{-\infty}^{\infty} \int_{-\infty}^{\infty} \underline{\hat{\tilde{\Psi}}}^{\dagger}(t,k)\, \mathrm{e}^{-\mathrm{i}kx}\, \mathrm{d}k \int_{-\infty}^{\infty} \overbrace{\begin{pmatrix} mc^2 & c\hbar k' + cqA(t) \\ c\hbar k' + cqA(t) & -mc^2 \end{pmatrix}}^{\widetilde{H}_D^{\mathrm{one}}(t,k')} \\
&\qquad \cdot \underline{\hat{\tilde{\Psi}}}(t,k')\, \mathrm{e}^{\mathrm{i}k'x}\, \mathrm{d}k'\, \mathrm{d}x \\
&= \int_{-\infty}^{\infty} \int_{-\infty}^{\infty} \underline{\hat{\tilde{\Psi}}}^{\dagger}(t,k) \widetilde{H}_D^{\mathrm{one}}(t,k') \underline{\hat{\tilde{\Psi}}}(t,k')\, \underbrace{\frac{1}{2\pi} \int_{-\infty}^{\infty} \mathrm{e}^{\mathrm{i}(k'-k)x}\, \mathrm{d}x}_{\delta(k'-k)}\, \mathrm{d}k'\, \mathrm{d}k \\
&= \int_{-\infty}^{\infty} \underline{\hat{\tilde{\Psi}}}^{\dagger}(t,k) \widetilde{H}_D^{\mathrm{one}}(t,k) \underline{\hat{\tilde{\Psi}}}(t,k)\, \mathrm{d}k.
\end{aligned} \tag{7.5}$$

**Diagonalization of $\widetilde{H}_D^{\mathrm{one}}$**

We now diagonalize the single-particle Hamiltonian in $k$ space, $\widetilde{H}_D^{\mathrm{one}}(t,k)$, which is a simple $2 \times 2$ matrix. This will allow us to see the analogy between $\hat{H}_D(t)$ and the semiconductor Hamiltonian, which we will cast into the same form, more easily. We already know the diagonal form of the matrix $\widetilde{H}_D^{\mathrm{one}}$ from Sec. 2.4.3 [see Eqs. (2.80)–(2.84)], so let us just summarize the results here: we have

$$O_D(t,k) \cdot \widetilde{H}_D^{\mathrm{one}}(t,k) \cdot O_D^{\intercal}(t,k) = \begin{pmatrix} \mathcal{E}_D(t,k) & 0 \\ 0 & -\mathcal{E}_D(t,k) \end{pmatrix} \tag{7.6}$$

with the **instantaneous energy eigenvalues** $\pm\mathcal{E}_D(t,k)$, where

$$\mathcal{E}_D(t,k) = \sqrt{m^2c^4 + c^2[\hbar k + qA(t)]^2}, \tag{7.7}$$

and the orthogonal matrix

$$O_D(t,k) = \frac{1}{\sqrt{1 + d_D^2(t,k)}} \begin{pmatrix} 1 & d_D(t,k) \\ -d_D(t,k) & 1 \end{pmatrix} \tag{7.8}$$

(so $O_D \cdot O_D^\intercal = O_D^\intercal \cdot O_D = \mathbb{1}$) with the abbreviation

$$d_D(t,k) = c\frac{\hbar k + qA(t)}{mc^2 + \mathcal{E}_D(t,k)}. \tag{7.9}$$

Hence, by transforming the standard $k$-space field operator $\hat{\underline{\tilde{\Psi}}}(t,k)$ to the instantaneous energy eigenbasis via

$$\hat{\underline{\Upsilon}}(t,k) = O_D(t,k)\hat{\underline{\tilde{\Psi}}}(t,k), \tag{7.10}$$

the **single-particle Hamiltonian is diagonalized**, so we get

$$\hat{H}_D(t) = \int\limits_{-\infty}^{\infty} \hat{\underline{\Upsilon}}^\dagger(t,k) \begin{pmatrix} \mathcal{E}_D(t,k) & 0 \\ 0 & -\mathcal{E}_D(t,k) \end{pmatrix} \hat{\underline{\Upsilon}}(t,k)\, \mathrm{d}k \tag{7.11}$$

from Eq. (7.5). Note that the "rotation" (7.10) of the spinor field operator preserves the canonical anticommutation relations (1.10) and is thus a **Bogoliubov transformation**. In the new particle basis, the upper (lower) component of $\hat{\underline{\Upsilon}}$ annihilates particles in the positive (negative), instantaneous energy continuum.

### 7.1.2. Semiconductor

In accordance to our model introduced in Sec. 6.1, we only take into account the (movable) **Bloch electrons** in the semiconductor and describe them as excitations of the quantized Schrödinger field (**nonrelativistic motion**, **neglect of spin**). The corresponding many-body Hamiltonian (which includes all Bloch electrons) has the same form as in the Dirac case [Eq. (7.2)], but the field operator $\hat{\psi}(t,x)$ is scalar (no spin) and the single-particle Hamiltonian is the nonrelativistic Hamilton operator in the external electric field and the lattice-periodic crystal potential $V(x) = V(x+\ell)$:

$$\begin{aligned} &\hat{H}_S^{\mathrm{full}}(t) \\ &= \int\limits_{-\infty}^{\infty} \hat{\psi}^\dagger(t,x) \left\{ \frac{[-\mathrm{i}\hbar\partial_x + qA(t)]^2}{2m} + V(x) \right\} \hat{\psi}(t,x)\, \mathrm{d}x \\ &= \int\limits_{-\infty}^{\infty} \hat{\psi}^\dagger(t,x) \left\{ -\frac{\hbar^2\partial_x^2}{2m} + V(x) + \frac{q^2A^2(t)}{2m} + \frac{qA(t)}{m}(-\mathrm{i}\hbar\partial_x) \right\} \hat{\psi}(t,x)\, \mathrm{d}x. \end{aligned} \tag{7.12}$$

**$A^2$ term not relevant**

Note that we may absorb the purely time-dependent $A^2$ term into the potential energy, which would leave the resulting force (partial derivative with respect to $x$) invariant, so this term can simply be ignored without any physical consequences. We thus redefine

$$\hat{H}_S^{\text{full}}(t) = \int_{-\infty}^{\infty} \hat{\psi}^\dagger(t,x) \left[ -\frac{\hbar^2 \partial_x^2}{2m} + V(x) + \frac{qA(t)}{m}(-\mathrm{i}\hbar\partial_x) \right] \hat{\psi}(t,x)\,\mathrm{d}x. \quad (7.13)$$

**Transformation to $K$ space (crystal momentum)**

The analog of plane waves in the vacuum are **Bloch waves** $f_n(K,x)$ [see Eqs. (6.2) and (6.3)] in the periodic crystal potential. Hence, instead of inserting the spatial Fourier transform of $\hat{\psi}$, we expand the field operator in terms of these Bloch waves (at each instant $t$):

$$\hat{\psi}(t,x) = \sum_{n=1}^{\infty} \int_{-\pi/\ell}^{\pi/\ell} f_n(K,x)\hat{a}_n(t,K)\,\mathrm{d}K. \quad (7.14)$$

This Bloch-wave expansion allows us to transform $\hat{H}_S^{\text{full}}$ to the so-called "crystal-momentum representation"; see, e.g., [136, 137, 150, 151]. The operators $\hat{a}_n(t,K)$ play the role of expansion "coefficients" in this context. According to the Bloch theorem, the Bloch states form a complete set of basis states, provided that we include all energy bands (all $n \in \mathbb{N}$) and the entire Brillouin zone $K \in (-\pi/\ell, \pi/\ell]$, so Eq. (7.14) is just a basis transformation and does not involve any approximation. In the following, we assume throughout that the **Bloch states are orthonormalized** according to the relation

**Bra–ket notation**

$$\langle n,K|n',K'\rangle = \int_{-\infty}^{\infty} f_n^*(K,x) f_{n'}(K',x)\,\mathrm{d}x \overset{!}{=} \delta_{nn'}\delta(K'-K), \quad (7.15)$$

where we have also introduced the bra–ket notation for Bloch states (a superscript $*$ denotes complex conjugation). Then, Eq. (7.14) can be inverted to yield

$$\hat{a}_n(t,K) = \int_{-\infty}^{\infty} f_n^*(K,x)\hat{\psi}(t,x)\,\mathrm{d}x \;\Rightarrow\; \hat{a}_n^\dagger(t,K) = \int_{-\infty}^{\infty} f_n(K,x)\hat{\psi}^\dagger(t,x)\,\mathrm{d}x. \quad (7.16)$$

Using the canonical anticommutation relations (1.10) which the (one-component) Bloch-electron field operator $\hat{\psi}$ satisfies per assumption, we calculate the anticommutators of the $\hat{a}$ operators:

$$\begin{aligned}
\{\hat{a}_n(t,K), \hat{a}_{n'}(t,K')\} &= \int_{-\infty}^{\infty} f_n^*(K,x)\hat{\psi}(t,x)\,\mathrm{d}x \int_{-\infty}^{\infty} f_{n'}^*(K',x')\hat{\psi}(t,x')\,\mathrm{d}x' \\
&\quad + \int_{-\infty}^{\infty} f_{n'}^*(K',x')\hat{\psi}(t,x')\,\mathrm{d}x' \int_{-\infty}^{\infty} f_n^*(K,x)\hat{\psi}(t,x)\,\mathrm{d}x \\
&= \int_{-\infty}^{\infty}\int_{-\infty}^{\infty} f_n^*(K,x) f_{n'}^*(K',x') \underbrace{\{\hat{\psi}(t,x), \hat{\psi}(t,x')\}}_{0}\,\mathrm{d}x'\,\mathrm{d}x \\
&= 0
\end{aligned} \tag{7.17}$$

and, in total analogy,

$$\{\hat{a}_n^\dagger(t,K), \hat{a}_{n'}^\dagger(t,K')\} = 0 \tag{7.18}$$

and

$$\begin{aligned}
\{\hat{a}_n(t,K), \hat{a}_{n'}^\dagger(t,K')\} &= \int_{-\infty}^{\infty} f_n^*(K,x)\hat{\psi}(t,x)\,\mathrm{d}x \int_{-\infty}^{\infty} f_{n'}(K',x')\hat{\psi}^\dagger(t,x')\,\mathrm{d}x' \\
&\quad + \int_{-\infty}^{\infty} f_{n'}(K',x')\hat{\psi}^\dagger(t,x')\,\mathrm{d}x' \int_{-\infty}^{\infty} f_n^*(K,x)\hat{\psi}(t,x)\,\mathrm{d}x \\
&= \int_{-\infty}^{\infty}\int_{-\infty}^{\infty} f_n^*(K,x) f_{n'}(K',x') \underbrace{\{\hat{\psi}(t,x), \hat{\psi}^\dagger(t,x')\}}_{\delta(x'-x)}\,\mathrm{d}x'\,\mathrm{d}x \\
&= \int_{-\infty}^{\infty} f_n^*(K,x) f_{n'}(K',x)\,\mathrm{d}x \\
&= \delta_{nn'}\delta(K'-K).
\end{aligned} \tag{7.19}$$

The $\hat{a}$ ($\hat{a}^\dagger$) operators are thus instantaneous **annihilation (creation) operators** for Bloch electrons.

By means of the orthonormality relation (7.15) and the energy-eigenvalue equation (6.2) which the Bloch states solve, inserting the Bloch-wave expan-sion (7.14) into the semiconductor Hamiltonian (7.13) yields

$$\begin{aligned}
&\hat{H}_S^{\text{full}}(t) \\
&= \int_{-\infty}^{\infty} \sum_{n=1}^{\infty} \int_{-\pi/\ell}^{\pi/\ell} f_n^*(K,x)\hat{a}_n^\dagger(t,K)\,\mathrm{d}K \left[-\frac{\hbar^2\partial_x^2}{2m} + V(x) + \frac{qA(t)}{m}(-\mathrm{i}\hbar\partial_x)\right] \\
&\qquad\qquad \times \sum_{n'=1}^{\infty} \int_{-\pi/\ell}^{\pi/\ell} f_{n'}(K',x)\hat{a}_{n'}(t,K')\,\mathrm{d}K'\,\mathrm{d}x \\
&= \sum_{n=1}^{\infty}\sum_{n'=1}^{\infty} \int_{-\pi/\ell}^{\pi/\ell}\int_{-\pi/\ell}^{\pi/\ell} \hat{a}_n^\dagger(t,K)\hat{a}_{n'}(t,K') \int_{-\infty}^{\infty} f_n^*(K,x) \\
&\qquad\qquad \times \left[\mathcal{E}_{n'}(K') + \frac{qA(t)}{m}(-\mathrm{i}\hbar\partial_x)\right] f_{n'}(K',x)\,\mathrm{d}x\,\mathrm{d}K'\,\mathrm{d}K \\
&= \sum_{n=1}^{\infty}\sum_{n'=1}^{\infty} \int_{-\pi/\ell}^{\pi/\ell}\int_{-\pi/\ell}^{\pi/\ell} \hat{a}_n^\dagger(t,K)\hat{a}_{n'}(t,K') \left[\mathcal{E}_{n'}(K')\delta_{nn'}\delta(K'-K)\right. \\
&\qquad\qquad \left. + \frac{qA(t)}{m}\langle n,K|\hat{p}_x|n',K'\rangle\right] \mathrm{d}K'\,\mathrm{d}K \\
&= \sum_{n=1}^{\infty} \int_{-\pi/\ell}^{\pi/\ell} \mathcal{E}_n(K)\hat{a}_n^\dagger(t,K)\hat{a}_n(t,K)\,\mathrm{d}K \\
&\quad + \frac{qA(t)}{m}\sum_{n=1}^{\infty}\sum_{n'=1}^{\infty} \int_{-\pi/\ell}^{\pi/\ell}\int_{-\pi/\ell}^{\pi/\ell} \langle n,K|\hat{p}_x|n',K'\rangle\,\hat{a}_n^\dagger(t,K)\hat{a}_{n'}(t,K')\,\mathrm{d}K'\,\mathrm{d}K. \qquad (7.20)
\end{aligned}$$

#### 7.1.2.1. Momentum matrix elements in the Bloch-wave basis

In order to calculate $\langle n,K|\hat{p}_x|n',K'\rangle$ in Eq. (7.20), we first derive a general formula which is very useful for many computations in the Bloch-wave basis (cf., e.g., Ref. [151]).

**General formula** Be $g(x) = g(x+\ell)$ a lattice-periodic function. We can thus write it as a complex Fourier series

$$g(x) = \sum_{j=-\infty}^{\infty} \tilde{g}_j \mathrm{e}^{2\pi\mathrm{i}jx/\ell} \qquad (7.21)$$

with coefficients

$$\tilde{g}_j = \frac{1}{\ell} \int\limits_0^{\ell} g(x) \mathrm{e}^{-2\pi \mathrm{i} j x/\ell} \, \mathrm{d}x. \tag{7.22}$$

Be furthermore $k \in (-2\pi/\ell, 2\pi/\ell)$. Now, let us calculate the integral

$$\begin{aligned} \int\limits_{-\infty}^{\infty} g(x) \mathrm{e}^{\mathrm{i}kx} \, \mathrm{d}x &= \sum_{j=-\infty}^{\infty} \tilde{g}_j \int\limits_{-\infty}^{\infty} \mathrm{e}^{\mathrm{i}(2\pi j/\ell + k)x} \, \mathrm{d}x \\ &= 2\pi \sum_{j=-\infty}^{\infty} \tilde{g}_j \, \delta\left(k + \frac{2\pi j}{\ell}\right). \end{aligned} \tag{7.23}$$

The argument of the delta distribution is always nonzero for $j \neq 0$ since $|k| < 2\pi/\ell$, so $j = 0$ is the only term that can contribute to the integral[1] (cf. Ref. [152]):

$$\int\limits_{-\infty}^{\infty} g(x) \mathrm{e}^{\mathrm{i}kx} \, \mathrm{d}x = 2\pi \tilde{g}_0 \, \delta(k) = \frac{2\pi}{\ell} \int\limits_0^{\ell} g(x) \, \mathrm{d}x \, \delta(k). \tag{7.24}$$

**Momentum matrix element $\langle n, K | \hat{p}_x | n', K' \rangle$**

This formula is useful when evaluating a momentum matrix element in the Bloch-wave basis. After inserting the general Bloch-wave form (6.3), we get

$$\begin{aligned} &\langle n, K | \hat{p}_x | n', K' \rangle \\ &= \int\limits_{-\infty}^{\infty} \mathrm{e}^{-\mathrm{i}Kx} u_n^*(K, x) (-\mathrm{i}\hbar \partial_x) \mathrm{e}^{\mathrm{i}K'x} u_{n'}(K', x) \, \mathrm{d}x \\ &= \int\limits_{-\infty}^{\infty} \mathrm{e}^{\mathrm{i}(K'-K)x} \Bigg[ \hbar K' \underbrace{u_n^*(K, x) u_{n'}(K', x)}_{\ell \,\text{periodic}} - \mathrm{i}\hbar \underbrace{u_n^*(K, x) \frac{\partial u_{n'}(K', x)}{\partial x}}_{\ell \,\text{periodic}} \Bigg] \mathrm{d}x. \end{aligned} \tag{7.25}$$

The difference $K' - K$ in the exponential function satisfies $|K' - K| < 2\pi/\ell$ since both crystal momenta are elements of the first Brillouin zone $(-\pi/\ell, \pi/\ell]$. The terms within the square brackets in the above equation are lattice periodic (with respect to the $x$ argument) because *all* Bloch factors

[1] According to Ref. [151], this statement is not true for $|k|$ approaching $2\pi/\ell$—however, the physical results derived in this and the following chapters will not be sensitive to this detail since we will focus on long-wavelength processes (corresponding to $|k|$ much smaller than $2\pi/\ell$) anyway.

are lattice periodic. Hence, we may apply the general formula (7.24), which gives the well-known result

$$\begin{aligned}
&\langle n,K|\hat{p}_x|n',K'\rangle \\
&= \frac{2\pi}{\ell}\int_0^\ell \hbar K u_n^*(K,x)u_{n'}(K,x) - \mathrm{i}\hbar u_n^*(K,x)\frac{\partial u_{n'}(K,x)}{\partial x}\,\mathrm{d}x\,\delta(K'-K) \\
&= \Big(\hbar K\,\langle n,K|n',K\rangle_{\text{cell}} + \langle n,K|\hat{p}_x|n',K\rangle_{\text{cell}}\Big)\delta(K'-K)
\end{aligned} \tag{7.26}$$

(cf. Ref. [153]), where we have introduced the **unit-cell-product notation**

$$\langle n,K|n',K'\rangle_{\text{cell}} = \frac{2\pi}{\ell}\int_0^\ell u_n^*(K,x)u_{n'}(K',x)\,\mathrm{d}x. \tag{7.27}$$

Note that we were allowed to set $K' = K$ for all Bloch factors after the first line in Eq. (7.26) because the delta distribution $\delta(K'-K)$ makes the result vanish in any other case anyway.

**Orthonormality of the Bloch factors for fixed $K$**

We may also apply formula (7.24) to the Bloch-wave orthonormality relation (7.15):

$$\begin{aligned}
\delta_{nn'}\delta(K'-K) &\overset{!}{=} \int_{-\infty}^{\infty} f_n^*(K,x)f_{n'}(K',x)\,\mathrm{d}x \\
&= \int_{-\infty}^{\infty} \mathrm{e}^{\mathrm{i}(K'-K)x}u_n^*(K,x)u_{n'}(K',x)\,\mathrm{d}x \\
&\overset{\text{Eq. (7.24)}}{=} \frac{2\pi}{\ell}\int_0^\ell u_n^*(K,x)u_{n'}(K,x)\,\mathrm{d}x\,\delta(K'-K) \\
&= \langle n,K|n',K\rangle_{\text{cell}}\,\delta(K'-K).
\end{aligned} \tag{7.28}$$

Our Bloch factors are thus always orthonormal on a unit cell at a fixed $K$:

$$\langle n,K|n',K\rangle_{\text{cell}} = \frac{2\pi}{\ell}\int_0^\ell u_n^*(K,x)u_{n'}(K,x)\,\mathrm{d}x \overset{!}{=} \delta_{nn'}. \tag{7.29}$$

**Result**

Insertion into Eq. (7.26) yields the formula for a **general momentum matrix element**:

$$\langle n,K|\hat{p}_x|n',K'\rangle = \Big(\hbar K\delta_{nn'} + \langle n,K|\hat{p}_x|n',K\rangle_{\text{cell}}\Big)\delta(K'-K). \tag{7.30}$$

#### 7.1.2.2. Diagonal elements of the momentum matrix: group velocities

Equation (7.30) shows that the momentum matrix in the Bloch-wave basis is always diagonal with respect to the crystal momenta; however, it is generally not diagonal with respect to the band indices (which would imply that *no* electron transitions were possible in the external field). The diagonal elements $\langle n, K|\hat{p}_x|n, K\rangle$ have a special physical meaning: they are related to the **expectation values of momentum in the Bloch states**. Due to the delta distribution $\delta(K' - K)$ in our Bloch-wave orthonormalization, however, these diagonal elements diverge, so we consider the distribution

$$\langle n, K|\hat{p}_x|n, K'\rangle = \Big(\hbar K + \langle n, K|\hat{p}_x|n, K\rangle_{\text{cell}}\Big)\delta(K' - K) \tag{7.31}$$

[a special case of Eq. (7.30)] instead.

**Calculation of $\langle n, K|\hat{p}_x|n, K\rangle_{\text{cell}}$**

It is well known [154] that the expectation value of the momentum in a Bloch state $|n, K\rangle$ coincides with the electron mass $m$ times the group velocity

$$v_n^{\text{gr}}(K) = \frac{1}{\hbar}\frac{\partial \mathcal{E}_n(K)}{\partial K}, \tag{7.32}$$

which is the speed a wave packet centered around $K$ propagates at. This gives us a hint on how to find a simple expression for $\langle n, K|\hat{p}_x|n, K\rangle_{\text{cell}}$ appearing in Eq. (7.31): Since

$$\hat{p}_x f_n(K, x) = -\mathrm{i}\hbar\partial_x \mathrm{e}^{\mathrm{i}Kx} u_n(K, x) = \mathrm{e}^{\mathrm{i}Kx}(\hat{p}_x + \hbar K)u_n(K, x) \tag{7.33}$$

and the Bloch waves satisfy the stationary Schrödinger equation (6.2), the Bloch factors solve the equation

$$\left[\frac{\hbar^2}{2m}(-\mathrm{i}\partial_x + K)^2 + V(x) - \mathcal{E}_n(K)\right] u_n(K, x) = 0. \tag{7.34}$$

The partial derivative with respect to $K$ of this equation yields

$$\begin{aligned}\left[\frac{\hbar^2}{m}(-\mathrm{i}\partial_x + K) - \frac{\partial \mathcal{E}_n(K)}{\partial K}\right] u_n(K, x)&\\ = -\left[\frac{\hbar^2}{2m}(-\mathrm{i}\partial_x + K)^2 + V(x) - \mathcal{E}_n(K)\right]&\frac{\partial u_n(K, x)}{\partial K}.\end{aligned} \tag{7.35}$$

Now we "project this equation onto $u_n(K, x)$" by applying the operator $(2\pi/\ell)\int_0^\ell u_n^*(K, x)\ldots \mathrm{d}x$ to both sides. Via integration by parts (IBP), the re-sulting right-hand side becomes

$$
\begin{aligned}
&-\frac{2\pi}{\ell}\int_0^\ell u_n^*(K,x)\left[\frac{\hbar^2}{2m}(-\mathrm{i}\partial_x+K)^2+V(x)-\mathcal{E}_n(K)\right]\frac{\partial u_n(K,x)}{\partial K}\,\mathrm{d}x\\
&\overset{\text{(IBP)}}{=}-\frac{2\pi}{\ell}\int_0^\ell \frac{\partial u_n(K,x)}{\partial K}\underbrace{\left[\frac{\hbar^2}{2m}(\mathrm{i}\partial_x+K)^2+V(x)-\mathcal{E}_n(K)\right]u_n^*(K,x)}_{=0\,\text{[complex conjugate of Eq. (7.34)]}}\,\mathrm{d}x\\
&=0.
\end{aligned}
\tag{7.36}
$$

Note that the boundary terms associated with the integration by parts cancel due to the $\ell$ periodicity of all $x$-dependent functions which are involved here. Using the unit-cell bra–ket notation and the Bloch-factor orthonormality from Eq. (7.29), the projected Eq. (7.35) consequently reads

$$
\begin{aligned}
0&=\left\langle n,K\left|\frac{\hbar^2}{m}(-\mathrm{i}\partial_x+K)-\frac{\partial\mathcal{E}_n(K)}{\partial K}\right|n,K\right\rangle_{\text{cell}}\\
&=\frac{\hbar}{m}\langle n,K|\hat{p}_x|n,K\rangle_{\text{cell}}+\frac{\hbar^2K}{m}-\hbar v_n^{\text{gr}}(K),
\end{aligned}
\tag{7.37}
$$

and thus

$$
\hbar K+\langle n,K|\hat{p}_x|n,K\rangle_{\text{cell}}=mv_n^{\text{gr}}(K).
\tag{7.38}
$$

**Result**

Inserting this equation into Eq. (7.31) yields our end result for the diagonal elements of the momentum matrix:

$$
\langle n,K|\hat{p}_x|n,K'\rangle=mv_n^{\text{gr}}(K)\,\delta(K'-K),
\tag{7.39}
$$

(cf., e.g., Ref. [148]).

**Resulting $K$-space expression for $\hat{H}_S^{\text{full}}(t)$**

Gathering our knowledge [Eqs. (7.30) and (7.39)] about the momentum matrix elements, we can write the crystal-momentum representation of the full semiconductor Hamiltonian (7.20) as

$$
\begin{aligned}
\hat{H}_S^{\text{full}}(t)=\sum_{n=1}^{\infty}\int_{-\pi/\ell}^{\pi/\ell}&\left[\mathcal{E}_n(K)+qA(t)v_n^{\text{gr}}(K)\right]\hat{a}_n^\dagger(t,K)\hat{a}_n(t,K)\\
&+\frac{qA(t)}{m}\sum_{n'\in\mathbb{N}}^{n'\neq n}\langle n,K|\hat{p}_x|n',K\rangle_{\text{cell}}\,\hat{a}_n^\dagger(t,K)\hat{a}_{n'}(t,K)\,\mathrm{d}K.
\end{aligned}
\tag{7.40}
$$

#### 7.1.2.3. Two-band model

As explained in Ch. 6, we neglect the contributions from all energy bands except for the valence band (we denote its band index by "$-$") and the next higher band, the conduction band ($+$), in our semiconductor analog; that is, we effectively set $\hat{a}_n(t,K) = 0$ for "$n \notin \{+,-\}$". The resulting **two-band version** of the full Hamiltonian (7.40) reads

$$\hat{H}_S(t) = \int_{-\pi/\ell}^{\pi/\ell} \underline{\hat{a}}^\dagger(t,K) \begin{pmatrix} \mathcal{E}_+(K) + qA(t)v_+^{\mathrm{gr}}(K) & qA(t)\varkappa^*(K) \\ qA(t)\varkappa(K) & \mathcal{E}_-(K) + qA(t)v_-^{\mathrm{gr}}(K) \end{pmatrix} \underline{\hat{a}}(t,K)\,\mathrm{d}K \quad (7.41)$$

in matrix notation, with the analog

$$\underline{\hat{a}}(t,K) = \begin{pmatrix} \hat{a}_+(t,K) \\ \hat{a}_-(t,K) \end{pmatrix} \quad (7.42)$$

of the two-component Dirac spinor in $k$ space and the abbreviation

$$\varkappa(K) = \frac{\langle -,K|\hat{p}_x|+,K\rangle_{\mathrm{cell}}}{m}, \quad (7.43)$$

which quantifies the single off-diagonal element of the momentum matrix[2] appearing in the two-band model. This is the Hamiltonian we want to compare to the Dirac Hamiltonian $\hat{H}_D$.

**Making the diagonal elements in $\hat{H}_S$ symmetric**

But first, we will bring the matrix in $\hat{H}_S$ above into a diagonal form. Note that the eigenvalues of this matrix will not be symmetric around zero [in contrast to the eigenvalues $\pm\mathcal{E}_D(t,k)$ in the Dirac case; see Eq. (7.11)] in general; however, by means of the band-difference quantities

$$\Delta\mathcal{E}(K) = \mathcal{E}_+(K) - \mathcal{E}_-(K) > 0 \quad \forall K \quad \text{(no band crossing)} \quad (7.44)$$

and

$$\Delta v^{\mathrm{gr}}(K) = v_+^{\mathrm{gr}}(K) - v_-^{\mathrm{gr}}(K) = \frac{1}{\hbar}\frac{\partial \Delta\mathcal{E}(K)}{\partial K}, \quad (7.45)$$

[2]The off-diagonal elements are also referred to as "optical matrix elements" in the literature.

we can write Eq. (7.41) as

$$\hat{H}_S(t) = \int_{-\pi/\ell}^{\pi/\ell} \underline{\hat{a}}^\dagger(t,K) \begin{pmatrix} \frac{\Delta\mathcal{E}(K)}{2} + qA(t)\frac{\Delta v^{\mathrm{gr}}(K)}{2} & qA(t)\varkappa^*(K) \\ qA(t)\varkappa(K) & -\frac{\Delta\mathcal{E}(K)}{2} - qA(t)\frac{\Delta v^{\mathrm{gr}}(K)}{2} \end{pmatrix} \underline{\hat{a}}(t,K)\,\mathrm{d}K$$
$$+ \int_{-\pi/\ell}^{\pi/\ell} \left[\frac{\mathcal{E}_+(K)+\mathcal{E}_-(K)}{2} + qA(t)\frac{v_+^{\mathrm{gr}}(K)+v_-^{\mathrm{gr}}(K)}{2}\right] \underline{\hat{a}}^\dagger(t,K)\underline{\hat{a}}(t,K)\,\mathrm{d}K. \quad (7.46)$$

The diagonal elements of the matrix in the upper line are now symmetric around zero, so its eigenvalues will also be. Now consider the additional term in the lower line: The operator

$$\underline{\hat{a}}^\dagger(t,K)\underline{\hat{a}}(t,K) = \hat{a}_+^\dagger(t,K)\hat{a}_+(t,K) + \hat{a}_-^\dagger(t,K)\hat{a}_-(t,K) \quad (7.47)$$

**$K$ is conserved**

counts the total number of electrons in the valence band and in the conduction band which have a crystal momentum of $\hbar K$ at the time $t$. Since the (general) semiconductor Hamiltonian in Eq. (7.40) has the form $\hat{H}_S^{\mathrm{full}}(t) = \int \hat{\mathcal{H}}_S^{\mathrm{full}}(t,K)\,\mathrm{d}K$, this operator can only perform particle transitions at constant $K$, so $K$ is a **conserved quantity** for every electron[3]. Our initial state ($t \to -\infty$) is always the ground state, in which every valence-band state is occupied by one electron and every conduction-band state is empty. Within the two-band approximation, only transitions between these two bands are possible. Since $K$ is conserved, there is always exactly one electron per $K$, which must be either located in the valence band or in the conduction band at any time, so

$$\underline{\hat{a}}^\dagger(t,K)\underline{\hat{a}}(t,K) \overset{\text{(two-band model)}}{=} 1. \quad (7.48)$$

As a consequence, the lower line in Eq. (7.46) merely yields a time-dependent, additive constant in $\hat{H}_S(t)$, which effectively causes an insignificant phase transformation of the quantum-field state vector. We thus **redefine the semiconductor Hamiltonian** again:

$$\hat{H}_S(t) = \int_{-\pi/\ell}^{\pi/\ell} \underline{\hat{a}}^\dagger(t,K) \begin{pmatrix} \frac{\Delta\mathcal{E}(K)}{2} + qA(t)\frac{\Delta v^{\mathrm{gr}}(K)}{2} & qA(t)\varkappa^*(K) \\ qA(t)\varkappa(K) & -\frac{\Delta\mathcal{E}(K)}{2} - qA(t)\frac{\Delta v^{\mathrm{gr}}(K)}{2} \end{pmatrix} \underline{\hat{a}}(t,K)\,\mathrm{d}K. \quad (7.49)$$

**Diagonalization of the matrix**

Now, we can diagonalize the $2 \times 2$ matrix in $\hat{H}_S$ in analogy to the Dirac

[3]In total analogy, the $k$-space Dirac Hamiltonian in Eq. (7.11) has the form $\hat{H}_D(t) = \int \hat{\mathcal{H}}_D(t,k)\,\mathrm{d}k$, which means that the canonical wave vector $k$ is conserved in the Dirac case.

case [see Eqs. (7.5)–(7.11)]. Its instantaneous energy eigenvalues are $\pm\mathcal{E}_S(t,K)$ with

$$\mathcal{E}_S(t,K) = \sqrt{\left[\frac{\Delta\mathcal{E}(K) + qA(t)\Delta v^{\mathrm{gr}}(K)}{2}\right]^2 + q^2A^2(t)|\varkappa(K)|^2}. \tag{7.50}$$

The unitary matrix

$$O_S(t,K) = \frac{1}{\sqrt{1+|d_S(t,K)|^2}}\begin{pmatrix} 1 & d_S^*(t,K) \\ -d_S(t,K) & 1 \end{pmatrix} \tag{7.51}$$

with

$$d_S(t,K) = \frac{qA(t)\varkappa(K)}{\left[\Delta\mathcal{E}(K) + qA(t)\Delta v^{\mathrm{gr}}(K)\right]/2 + \mathcal{E}_S(t,K)} \tag{7.52}$$

"rotates" $\underline{\hat{a}}(t,K)$ to the instantaneous energy eigenbasis via

$$\underline{\hat{b}}(t,K) = O_S(t,K)\underline{\hat{a}}(t,K), \tag{7.53}$$

so the **diagonalized *K*-space expression for the two-band semiconductor Hamiltonian** reads **Result**

$$\hat{H}_S(t) = \int\limits_{-\pi/\ell}^{\pi/\ell} \underline{\hat{b}}^\dagger(t,K)\begin{pmatrix} \mathcal{E}_S(t,K) & 0 \\ 0 & -\mathcal{E}_S(t,K) \end{pmatrix}\underline{\hat{b}}(t,K)\,\mathrm{d}K. \tag{7.54}$$

## 7.2. Quantitative analogy between the Hamiltonians

In this section, our goal is to point out in how far the two-band semiconductor Hamiltonian $\hat{H}_S(t)$ in Eq. (7.54) is analog to the Dirac Hamiltonian $\hat{H}_D(t)$ in Eq. (7.11). Both Hamiltonians have the same overall form; the only differences are the eigenvalues of the $2\times 2$ matrices [i.e., $\boldsymbol{\mathcal{E}_S(t,K)}$ **versus** $\boldsymbol{\mathcal{E}_D(t,k)}$] and the ranges of the (quasi-)wave vectors, i.e., $\boldsymbol{K \in (-\pi/\ell, \pi/\ell]}$ **versus** $\boldsymbol{k} \in \mathbb{R}$.

Since each electron in the semiconductor (Dirac) case is associated with a unique, conserved value of $K$ ($k$), it makes sense to compare each $K$ mode in the semiconductor to a corresponding $k$ in Dirac theory, respectively. So, let us put the question like this: Say we have a fixed external electric field [i.e., $A(t)$ is fixed], and we consider a particular Bloch electron in the semiconductor with the canonical crystal momentum $K \in (-\pi/\ell, \pi/\ell]$. Is this electron suitable to mimic a Dirac electron with a certain canonical wave vector $k$ [for the same external $A(t)$]? (Note that $k$ may depend on the chosen $K$ and material constants of the semiconductor.) The answer to this question is yes if $\mathcal{E}_S(t,K)$ has the same functional form as $\mathcal{E}_D(t,k)$ *for all times* $t$. The radicand in $\mathcal{E}_S(t,K)$ [Eq. (7.50)],

$$\mathcal{E}_S^2(t,K) = \frac{\Delta\mathcal{E}^2(K)}{4} + \frac{\Delta\mathcal{E}(K)\Delta v^{\mathrm{gr}}(K)}{2} qA(t) + \left\{ |\varkappa(K)|^2 + \frac{[\Delta v^{\mathrm{gr}}(K)]^2}{4} \right\} q^2 A^2(t), \quad (7.55)$$

must thus be formally equivalent to the radicand in $\mathcal{E}_D(t,k)$ [Eq. (7.7)],

$$\mathcal{E}_D^2(t,k) = m^2c^4 + c^2\hbar^2k^2 + 2c^2\hbar k q A(t) + c^2 q^2 A^2(t), \quad (7.56)$$

*for each power* of the arbitrary time-dependent function $A(t)$.

**Equating coefficients: $q^2A^2(t)$ terms,**

Let us start with the $q^2A^2(t)$ terms in the above two equations: both terms are equivalent, but the vacuum speed of light is substituted by a material- and $K$-dependent **effective speed of light**

$$c_\star(K) = \sqrt{|\varkappa(K)|^2 + \frac{[\Delta v^{\mathrm{gr}}(K)]^2}{4}} \quad (7.57)$$

in the semiconductor.

Note that we could also define an effective electron charge $q_\star(K)$ instead, or an effective vector potential—concepts known from the simulation of the Sauter–Schwinger effect via ultracold atoms trapped in optical lattices [110, 111]. However, since the effective speed of light is an established concept in semiconductor physics [128, 104, 96, 97, 146], we go for this option here. Furthermore, the resulting effective mass which we will define below is then closely related to the usual definition of effective masses in solid-state physics (we will discuss this point later).

**$qA(t)$ terms,**

The coefficient of the linear term in the Dirac case (7.56), $2c^2\hbar k$, will thus take on the value $2c_\star^2(K)\hbar k$ when simulated in the semiconductor. The assumption that this term equals the coefficient of the linear term in Eq. (7.55), $\Delta\mathcal{E}(K)\Delta v^{\mathrm{gr}}(K)/2$, yields the equation

$$k(K) = \frac{\Delta\mathcal{E}(K)\Delta v^{\mathrm{gr}}(K)}{4\hbar c_\star^2(K)}. \quad (7.58)$$

This formula determines the canonical wave vector of the Dirac electron which can be simulated by the particular Bloch-electron state corresponding to $K$.

**and constant terms.**

Finally, the constant term $\Delta\mathcal{E}^2(K)/4$ in the semiconductor case formally equals its counterpart $m_\star^2(K)c_\star^4(K) + c_\star^2(K)\hbar^2 k^2(K)$ in Dirac theory if we substitute an **effective electron rest mass** $m_\star(K)$ which is given by

$$\begin{aligned} m_\star^2(K)c_\star^4(K) &\stackrel{!}{=} \frac{\Delta\mathcal{E}^2(K)}{4} - c_\star^2(K)\hbar^2 k^2(K) \\ &= \frac{\Delta\mathcal{E}^2(K)}{4} - \frac{\Delta\mathcal{E}^2(K)[\Delta v^{\mathrm{gr}}(K)]^2}{16c_\star^2(K)} \\ &= \frac{\Delta\mathcal{E}^2(K)}{4}\left\{1 - \frac{[\Delta v^{\mathrm{gr}}(K)]^2}{4c_\star^2(K)}\right\} \\ &= \frac{\Delta\mathcal{E}^2(K)}{4}\frac{4|\varkappa(K)|^2 + [\Delta v^{\mathrm{gr}}(K)]^2 - [\Delta v^{\mathrm{gr}}(K)]^2}{4c_\star^2(K)} \\ &= \frac{\Delta\mathcal{E}^2(K)|\varkappa(K)|^2}{4c_\star^2(K)}, \end{aligned} \tag{7.59}$$

and therefore

$$m_\star(K) = \frac{\Delta\mathcal{E}(K)|\varkappa(K)|}{2c_\star^3(K)}. \tag{7.60}$$

**Analogy in the entire Brillouin zone**

In summary, each Bloch electron in the two-band semiconductor model, which has an associated $K$ in the first Brillouin zone, is suitable to simulate a state $k(K)$ in Dirac theory, with the effective physical constants $m_\star(K)$ and $c_\star(K)$. The functional forms of these quantities depend on the specific crystal potential of the considered semiconductor. This is the **general result** of the present chapter concerning the analogy in time-dependent electric fields.

### 7.2.1. Simulating nonperturbative pair creation: analogy for long-wavelength modes

The general form of the analogy for time-dependent electric fields is not entirely satisfactory because the effective constants $m_\star(K)$ and $c_\star(K)$ depend on the Bloch state considered. Let us thus be specific about what we intend to simulate in our semiconductor analog in the following: We are interested in **(assisted) tunneling pair creation** in QED. This process is associated with long wavelengths because it happens between classical turning points (points of zero momentum) in the tunneling picture. Long wavelengths correspond to small wave vectors with $|k| \ll mc/\hbar$, which constitute a small region around the mass gap where the energy curves $\mathcal{E}_\pm(k)$ [see Fig. 6.1(a) on page 152] are approximately parabolic. By analogy, tunneling in a semiconductor from the valence band to the conduction band will also most likely occur close to the (direct) bandgap, where the energy difference between the bands is minimal. Since the bandgap is located at the zone center here per assumption, this process is associated with small crystal momenta which satisfy $|K| \ll \pi/\ell$ [see

Fig. 6.1(b)]. Hence, in order to simulate the **pair-creation process** in the semiconductor, the quantitative analogy between $\hat{H}_D$ and $\hat{H}_S$ must only be good for **long wavelengths/small (quasi-)wave vectors**.

**Implications of the temporal gauge**

Note that, as a consequence of choosing the temporal gauge $E(t) = \dot{A}(t)$ in this chapter, the evolution of each Dirac (Bloch) electron is described on the basis of plane (Bloch) waves for an individual, canonical value of $k$ ($K$), respectively. This picture must of course lead to the same physical results as other gauges, but remember that we did only include the valence band and the conduction band in our semiconductor Hamiltonian $\hat{H}_S$ (two-band approximation). In order to describe all possible dynamics of a Bloch electron using Bloch wave functions for a single $K$ only, a complete basis of Bloch states is required, which must include *all* energy bands. However, our main focus here are electron tunneling transitions (which are predominantly long-wavelength processes as explained above) between the valence band and the conduction band, and thus the long-wavelength modes (small $K$) in our two-band model $\hat{H}_S$ are the most suitable choice to describe these tunneling transitions. In conclusion, if the quantitative analogy between $\hat{H}_D$ and $\hat{H}_S$ holds for small (quasi-)momenta—$|k| \ll mc/\hbar$ on the one hand and $|K| \ll \pi/\ell$ on the other hand—then the semiconductor analog should be appropriate to simulate (assisted) nonperturbative pair-*creation* processes well.

**Post-creation trajectories**

However, the trajectories of the constituents of created pairs (formed at small $k$'s or $K$'s), which are then accelerated by the external field $E(t)$, will *not* be the same! In Dirac theory, electrons and positrons will be accelerated in accordance with the relativistic energy–momentum relation. In contrast to that, the dispersion relation of Bloch electrons in the semiconductor is usually more complicated and depends on the concrete crystal potential, so the **semiconductor analog is not suited to simulate the post-creation trajectories** of QED particles in general.

### Effective constants in the infinite-wavelength limit

Let us start with the central mode in the Brillouin zone, $K = 0$, which corresponds to infinite wavelengths. Since our semiconductor has a direct bandgap at the zone center per assumption [see Fig. 6.1(b) on page 152], the energy difference at $K = 0$ equals the bandgap $\mathcal{E}_g$ and the group velocities in both bands vanish because the $K$ derivatives of $\mathcal{E}_+(K)$ and $\mathcal{E}_-(K)$ are zero at $K = 0$, so

$$\Delta\mathcal{E}(0) = \mathcal{E}_g > 0 \qquad \text{and} \qquad \Delta v^{\mathrm{gr}}(0) = 0. \tag{7.61}$$

The off-diagonal element $\varkappa(K)$ is typically nonzero at the gap, and we choose the global phases of the Bloch bands in a way that

$$\varkappa(0) = \varkappa_0 > 0 \tag{7.62}$$

in the following for convenience [in general, $\varkappa(K)$ is a complex quantity]. Inserting these material constants into Eqs. (7.57), (7.58), and (7.60) then yields the analogies

**Analogy $K = 0 \leftrightarrow k = 0$**

$$\begin{aligned} K &= 0 && \leftrightarrow \quad k = 0, \\ c_\star(0) &= \varkappa_0 && \leftrightarrow \quad c, \\ m_\star(0) &= \frac{\mathcal{E}_g}{2\varkappa_0^2} && \leftrightarrow \quad m \end{aligned} \tag{7.63}$$

between the two-band semiconductor and Dirac theory. There is thus a quantitative analogy between the infinite-wavelength mode $K = 0$ and its QED counterpart $k = 0$. Note that the corresponding effective quantities satisfy

$$2m_\star(0)c_\star^2(0) = \Delta\mathcal{E}(0) = \mathcal{E}_g \qquad \Leftrightarrow \qquad c_\star(0) = \sqrt{\frac{\mathcal{E}_g}{2m_\star(0)}}; \tag{7.64}$$

that is, the role of the mass gap $2mc^2$ is taken by the bandgap in the semiconductor analog, which seems like a natural analogy. These formulas were also found in Refs. [128, 104, 96, 97, 146], for example. Our results here, however, show that the relations (7.64) are only correct for $K = 0$ but not for all $K$ in general [cf. Eq. (7.59)].

### Effective constants for long, finite wavelengths

As the next step, we study how the effective constants change in the vicinity of $K = 0$. We do this by evaluating the derivatives of $c_\star(K)$, $m_\star(K)$, and $k(K)$ with respect to $K$ at $K = 0$.

For the effective speed of light [Eq. (7.57)] in the semiconductor, we get

**Exact derivatives**

$$\begin{aligned} \left.\frac{\mathrm{d}c_\star(K)}{\mathrm{d}K}\right|_{K=0} &= \left.\frac{\left|\varkappa(K)\frac{\mathrm{d}\varkappa(K)}{\mathrm{d}K}\right| + \frac{\Delta v^{\mathrm{gr}}(K)}{4}\frac{\mathrm{d}\Delta v^{\mathrm{gr}}(K)}{\mathrm{d}K}}{c_\star(K)}\right|_{K=0} \\ &= \left|\frac{\mathrm{d}\varkappa(K)}{\mathrm{d}K}\right|_{K=0} \end{aligned} \tag{7.65}$$

since $\Delta v^{\mathrm{gr}}(0) = 0$ and $c_\star(0) = \varkappa_0$. For the same reasons and because of $\mathrm{d}\Delta\mathcal{E}(K)/\mathrm{d}K = \Delta v^{\mathrm{gr}}(K)$, the result for the effective electron mass (7.60) reads

$$\begin{aligned} \left.\frac{\mathrm{d}m_\star(K)}{\mathrm{d}K}\right|_{K=0} &= \left.\frac{\left[\Delta v^{\mathrm{gr}}(K)|\varkappa(K)| + \Delta\mathcal{E}(K)\left|\frac{\mathrm{d}\varkappa(K)}{\mathrm{d}K}\right|\right]c_\star^3(K)}{2c_\star^6(K)}\right|_{K=0} \\ &\quad - \left.\frac{3\Delta\mathcal{E}(K)|\varkappa(K)|c_\star^2(K)\frac{\mathrm{d}c_\star(K)}{\mathrm{d}K}}{2c_\star^6(K)}\right|_{K=0} \end{aligned}$$

$$= \frac{\mathcal{E}_g \left|\frac{\mathrm{d}\varkappa(K)}{\mathrm{d}K}\right|_{K=0} \varkappa_0 - 3\mathcal{E}_g \varkappa_0 \left|\frac{\mathrm{d}\varkappa(K)}{\mathrm{d}K}\right|_{K=0}}{2\varkappa_0^4}$$
$$= -\frac{\mathcal{E}_g}{\varkappa_0^3} \left|\frac{\mathrm{d}\varkappa(K)}{\mathrm{d}K}\right|_{K=0}. \tag{7.66}$$

Finally, the canonical Dirac-electron wave vector $k(K)$ [see Eq. (7.58)] corresponding to the mode $K$ in the semiconductor changes near $K = 0$ according to

$$\left.\frac{\mathrm{d}k(K)}{\mathrm{d}K}\right|_{K=0} = \left.\frac{\left\{[\Delta v^{\mathrm{gr}}(K)]^2 + \Delta\mathcal{E}(K)\frac{\mathrm{d}\Delta v^{\mathrm{gr}}(K)}{\mathrm{d}K}\right\} c_\star^2(K)}{4\hbar c_\star^4(K)}\right|_{K=0}$$
$$- \left.\frac{2\Delta\mathcal{E}(K)\Delta v^{\mathrm{gr}}(K) c_\star(K)\frac{\mathrm{d}c_\star(K)}{\mathrm{d}K}}{4\hbar c_\star^4(K)}\right|_{K=0}$$
$$= \frac{\mathcal{E}_g}{4\hbar\varkappa_0^2} \left.\frac{\mathrm{d}\Delta v^{\mathrm{gr}}(K)}{\mathrm{d}K}\right|_{K=0}. \tag{7.67}$$

**$K \cdot p$ perturbation theory**

Hence, in order to evaluate these derivatives, we need to determine the first $K$ derivatives of $\varkappa(K)$ and $\Delta v^{\mathrm{gr}}(K)$ at $K = 0$. One way to do this is via $\boldsymbol{K} \cdot \boldsymbol{p}$ perturbation theory: According to this application of (nondegenerate) time-independent perturbation theory to the Schrödinger-like equation (7.34) the Bloch factors solve, we can express $u_n(K, x)$ for any $K \neq 0$ as a power series in $K$ with only the Bloch factors $u_{n'}(0, x)$ at the zone center appearing in the coefficients. Up to the first order, the general result of this perturbational approach reads [122]

$$u_n(K, x) = u_n(0, x) + \frac{\hbar K}{m} \sum_{n' \in \mathbb{N}}^{n' \neq n} \frac{\langle n', 0|\hat{p}_x|n, 0\rangle_{\mathrm{cell}}}{\mathcal{E}_n(0) - \mathcal{E}_{n'}(0)} u_{n'}(0, x) + \mathcal{O}(K^2). \tag{7.68}$$

Note that the resulting Bloch factors at $K \neq 0$ are *not* normalized on a unit cell [assuming that the Bloch factors at $K = 0$ *are* normalized according to Eq. (7.29)]. However, Eq. (7.68) yields

$$\langle n, K|n, K\rangle_{\mathrm{cell}} = \langle n, 0|n, 0\rangle_{\mathrm{cell}} + \mathcal{O}(K^2) = 1 + \mathcal{O}(K^2), \tag{7.69}$$

so normalizing the result (7.68) by multiplying it with a factor of the form

$$\frac{1}{\sqrt{1 + \mathcal{O}(K^2)}} = 1 + \mathcal{O}(K^2) \tag{7.70}$$

will not change the $K^0$ and the $K^1$ terms in Eq. (7.68).

**Two-band approximation**

According to Eq. (7.68), every energy band (at $K = 0$) can contribute to $u_n(K, x)$ with $K \neq 0$ in principle; however, the first-order correction is suppressed by the band-energy difference, so the main corrections will come from bands adjacent to $n$. We will stick to the **two-band approximation** here, which means that the $n'$ sum in Eq. (7.68) is taken to run over "$\{+, -\}$" only instead of $\mathbb{N}$. This approximation yields the following Bloch factors in the conduction band and the valence band:

$$\begin{aligned} u_\pm(K,x) &= u_\pm(0,x) + \frac{\hbar K}{m} \frac{\langle \mp, 0|\hat{p}_x|\pm, 0\rangle_{\text{cell}}}{\mathcal{E}_\pm(0) - \mathcal{E}_\mp(0)} u_\mp(0,x) + \mathcal{O}(K^2) \\ &= u_\pm(0,x) \pm \frac{\varkappa_0 \hbar K}{\mathcal{E}_g} u_\mp(0,x) + \mathcal{O}(K^2). \end{aligned} \tag{7.71}$$

Taking into account the first-order term in $K$ only should provide a good approximation as long as the correction is small; that is,

$$\left| \frac{\varkappa_0 \hbar K}{\mathcal{E}_g} \right| \ll 1 \qquad \Leftrightarrow \qquad |K| \ll \frac{\mathcal{E}_g}{\hbar \varkappa_0}. \tag{7.72}$$

**$\varkappa(K)$ in the two-band model**

The Bloch-factor expansion (7.71) allows us to calculate $\varkappa(K)$ up to the first order in $K$. Using the orthonormality (7.29) of the Bloch factors at $K = 0$ again, we get

$$\begin{aligned} \varkappa(K) &= \frac{1}{m} \langle -, K|\hat{p}_x|+, K\rangle_{\text{cell}} \\ &= \frac{1}{m} \langle -, 0|\hat{p}_x|+, 0\rangle_{\text{cell}} \\ &\quad + \frac{\varkappa_0 \hbar K}{m \mathcal{E}_g} \Big( \underbrace{\langle -, 0|\hat{p}_x|-, 0\rangle_{\text{cell}}}_{= m v_-^{\text{gr}}(0) = 0 \text{ [see Eq. (7.38)]}} - \underbrace{\langle +, 0|\hat{p}_x|+, 0\rangle_{\text{cell}}}_{= m v_+^{\text{gr}}(0) = 0} \Big) + \mathcal{O}(K^2) \\ &= \varkappa_0 + \mathcal{O}(K^2), \end{aligned} \tag{7.73}$$

**Results**

and thus $\mathrm{d}\varkappa(K)/\mathrm{d}K$ vanishes at $K = 0$, which means that **$c_\star$ and $m_\star$ are approximately constant near $K = 0$** according to $\boldsymbol{K} \cdot \boldsymbol{p}$ perturbation theory within the two-band approximation:

$$\left. \frac{\mathrm{d}c_\star(K)}{\mathrm{d}K} \right|_{K=0} = \left. \frac{\mathrm{d}m_\star(K)}{\mathrm{d}K} \right|_{K=0} \overset{\text{(two-band model)}}{=} 0. \tag{7.74}$$

Note that the incorporation of more energy bands into the perturbational calculation [i.e., taking the full $n'$ sum in Eq. (7.68)] will produce a nonvanishing first-order $K$ term in $\varkappa(K)$ in general; that is, the effective constants $m_\star$ and $c_\star$ will probably not be perfectly constant around the zone center in a real semiconductor.

**$\Delta v^{\mathrm{gr}}(K)$ in the two-band model**

The perturbational approach is also appropriate to calculate the $K$ dependence of the **band energies**. The general expansion around $K = 0$ reads [122]

$$\mathcal{E}_n(K) = \mathcal{E}_n(0) + \frac{\hbar^2 K^2}{2m}\left[1 + \frac{2}{m}\sum_{n'\in\mathbb{N}}^{n'\neq n}\frac{|\langle n',0|\hat{p}_x|n,0\rangle_{\mathrm{cell}}|^2}{\mathcal{E}_n(0)-\mathcal{E}_{n'}(0)}\right] + \mathcal{O}(K^3). \quad (7.75)$$

In the **two-band model**, this expression becomes

$$\begin{aligned}\mathcal{E}_\pm(K) &= \mathcal{E}_\pm(0) + \frac{\hbar^2 K^2}{2m}\left[1 + \frac{2}{m}\frac{|\langle\mp,0|\hat{p}_x|\pm,0\rangle_{\mathrm{cell}}|^2}{\mathcal{E}_\pm(0)-\mathcal{E}_\mp(0)}\right] + \mathcal{O}(K^3)\\ &= \mathcal{E}_\pm(0) + \frac{\hbar^2 K^2}{2m}\left[1 \pm \frac{2m\varkappa_0^2}{\mathcal{E}_g}\right] + \mathcal{O}(K^3). \end{aligned} \quad (7.76)$$

The quadratic approximation of $\mathcal{E}_\pm(K)$ should be good as long as the quadratic $K$ term is much smaller than $\mathcal{E}_g/2$ since we may set $\mathcal{E}_\pm(0)$ to $\pm\mathcal{E}_g/2$ without loss of generality:

$$\frac{\hbar^2 K^2}{2m}\left|1 \pm \frac{2m\varkappa_0^2}{\mathcal{E}_g}\right| \ll \frac{\mathcal{E}_g}{2}. \quad (7.77)$$

Note that

$$\frac{2m\varkappa_0^2}{\mathcal{E}_g} = \frac{2mc_\star^2(0)}{2m_\star(0)c_\star^2(0)} = \frac{m}{m_\star(0)} \gg 1 \quad (7.78)$$

is always true in typical semiconductors, so the **validity condition** (7.77) is practically equivalent to

$$\frac{\hbar^2 K^2}{2m}\frac{2m\varkappa_0^2}{\mathcal{E}_g} \ll \frac{\mathcal{E}_g}{2} \qquad \Leftrightarrow \qquad |K| \ll \frac{\mathcal{E}_g}{\sqrt{2}\hbar\varkappa_0}, \quad (7.79)$$

which is basically the same condition as in Eq. (7.72) for the Bloch factors.

From the two-band results (7.76), we obtain

$$\Delta\mathcal{E}(K) = \mathcal{E}_+(K) - \mathcal{E}_-(K) = \mathcal{E}_g + \frac{2\varkappa_0^2\hbar^2K^2}{\mathcal{E}_g} + \mathcal{O}(K^3) \quad (7.80)$$

and

$$\Delta v^{\mathrm{gr}}(K) = \frac{1}{\hbar}\frac{\mathrm{d}\Delta\mathcal{E}(K)}{\mathrm{d}K} = \frac{4\varkappa_0^2\hbar K}{\mathcal{E}_g} + \mathcal{O}(K^2). \quad (7.81)$$

**Result**

Inserting this result into Eq. (7.67) finally yields

$$\left.\frac{\mathrm{d}k(K)}{\mathrm{d}K}\right|_{K=0} \overset{\text{(two-band model)}}{=} 1, \quad (7.82)$$

which means that **$k(K)$ coincides with $K$ for long-wavelength modes** according to the two-band approximation.

**Summary and additional notes**

In this subsection, we showed that the long-wavelength modes of the semiconductor Hamiltonian are suitable to simulate the corresponding QED modes (in the sense that $k = K$). The QED constants $m$ and $c$ take on the effective, material-dependent values $m_\star(0)$ and $c_\star(0)$ given by Eq. (7.63) in the semiconductor, which are approximately universal for all long-wavelength modes. We derived these findings by applying $\boldsymbol{K} \cdot \boldsymbol{p}$ perturbation theory up to the first nonvanishing order. The validity condition for this approach is Eq. (7.72), which thus determines what we mean by "long-wavelength modes" in this context (i.e., how small $|K|$ must be). Note that the perfect analogy for these modes is only correct within the two-band approximation—taking also the coupling of other bands to the valence band and the conduction band into account would change the picture a bit. It is thus desirable to select a semiconductor with **small optical matrix elements** (ideally much smaller than $\varkappa_0$) between the valence/conduction band on the one hand and one of the adjacent energy bands on the other hand since then the **two-band approximation is good** [cf. Eq. (7.68)].

**Effective masses**

Note that the usual notion of effective masses (at the bandgap) in solid-state physics is that they describe the bending of the dispersion curves in the semiconductor according to

$$\begin{aligned} \mathcal{E}_+(K) &= \mathcal{E}_+(0) + \frac{\hbar^2 K^2}{2m_{\star,e}} + \mathcal{O}(K^3) \quad \text{and} \\ \mathcal{E}_-(K) &= \mathcal{E}_-(0) - \frac{\hbar^2 K^2}{2m_{\star,h}} + \mathcal{O}(K^3), \end{aligned} \tag{7.83}$$

where $m_{\star,e} > 0$ is the **effective electron mass** in the conduction band, and $m_{\star,h} > 0$ is the **effective (light-)hole mass** in the valence band. Comparing this definition to the full perturbational expansions (7.75) of $\mathcal{E}_\pm(K)$ yields [122]

$$\begin{aligned} \frac{1}{m_{\star,e}} &= \frac{1}{m}\left[1 + \frac{2}{m}\sum_{n\in\mathbb{N}}^{n\neq\text{“}+\text{”}} \frac{|\langle n,0|\hat{p}_x|+,0\rangle_{\text{cell}}|^2}{\mathcal{E}_+(0) - \mathcal{E}_n(0)}\right] \quad \text{and} \\ \frac{1}{m_{\star,h}} &= -\frac{1}{m}\left[1 + \frac{2}{m}\sum_{n\in\mathbb{N}}^{n\neq\text{“}-\text{”}} \frac{|\langle n,0|\hat{p}_x|-,0\rangle_{\text{cell}}|^2}{\mathcal{E}_-(0) - \mathcal{E}_n(0)}\right], \end{aligned} \tag{7.84}$$

so all energy bands at the zone center can contribute to the effective mass in a given band. If we only consider the mutual influence between the valence band and the conduction band again and neglect all other bands (**two-band** **approximation**), these relations simplify to

$$\frac{1}{m_{\star,e}} = \frac{1}{m}\left(1 + \frac{2m\varkappa_0^2}{\mathcal{E}_g}\right) \quad \text{and} \quad \frac{1}{m_{\star,h}} = -\frac{1}{m}\left(1 - \frac{2m\varkappa_0^2}{\mathcal{E}_g}\right). \tag{7.85}$$

Adding both equations and comparing it to Eq. (7.63) yields

$$\frac{1}{m_{\star,e}} + \frac{1}{m_{\star,h}} = \frac{4\varkappa_0^2}{\mathcal{E}_g} = \frac{2}{m_\star(0)}. \tag{7.86}$$

**Relation between the effective-mass concepts**

Hence, if other bands are neglected, "our" effective mass $m_\star(K)$ coincides with the **harmonic mean** (twice the reduced mass) of the two effective charge-carrier masses at the zone center:

$$m_\star(0) = \frac{\mathcal{E}_g}{2\varkappa_0^2} \overset{\text{(two-band model)}}{=} \frac{2}{m_{\star,e}^{-1} + m_{\star,h}^{-1}}, \tag{7.87}$$

so both concepts are closely connected. This result was also found in Ref. [128] for constant electric fields, and we can confirm it here for electric fields with an arbitrary time dependence. Note that $m_{\star,e}$ and $m_{\star,h}$ [i.e., the parabolic band curvatures of $\mathcal{E}_\pm(K)$ near the zone center] **do not need to be equal** in order to draw the analogy here in the purely time-dependent case.

The use of Eq. (7.87) is twofold: On the one hand, it provides a way to calculate $m_\star(0)$ approximately from $m_{\star,e}$ and $m_{\star,h}$ (quantities which can be found in standard textbooks for many typical semiconductors). On the other hand, if we know the optical matrix element $\varkappa_0$ [and thus can calculate $m_\star(0)$ directly], the degree to which Eq. (7.87) is satisfied tells us how good the two-band approximation works for the considered semiconductor because the coupling of other bands to $m_{\star,e}$ and $m_{\star,h}$ has been neglected in this equation.

## 7.3. Analog of the Sauter–Schwinger effect and dynamical assistance in GaAs

Now that we know that the two-band semiconductor analog is suitable to simulate (assisted) nonperturbative QED pair creation with effective parameters $m \to m_\star(0)$ and $c \to c_\star(0)$ (we will just write $m_\star$ and $c_\star$ in the following for brevity when referring to the effective constants at $K = 0$), we will consider some experimental scenarios with time-dependent electric fields in this section. As pointed out in the introduction of this part (Ch. 6), we focus on **gallium arsenide (GaAs)** here, which is a semiconductor with a direct bandgap at the zone center. Table 7.1 lists the crucial properties of GaAs.

| Property | Value in GaAs | | | Reference |
|---|---|---|---|---|
| Lattice constant | $\ell^{\mathrm{GaAs}}$ | $=$ | $0.565\,\mathrm{nm}$ | [126, p. 789] |
| Bandgap | $\mathcal{E}_g^{\mathrm{GaAs}}$ | $=$ | $1.5\,\mathrm{eV}$ | [126, p. 789] |
| Effective electron mass | $m_{\star,e}^{\mathrm{GaAs}}$ | $=$ | $0.063m$ | [126, p. 789] |
| Effective light-hole mass | $m_{\star,h}^{\mathrm{GaAs}}$ | $=$ | $0.076m$ | [126, p. 789] |
| Optical matrix element | $\varkappa_0^{\mathrm{GaAs}}$ | $=$ | $0.0050c$ | [122, p. 104] |
| Dielectric-breakdown field | $E_{\mathrm{BD}}^{\mathrm{GaAs}}$ | $=$ | $30\text{–}90\,\mathrm{MV/m}$ | [126, p. 790] |

**Table 7.1.:** Properties of gallium arsenide which we will use throughout this thesis. Some values were rounded for simplicity.

The resulting **effective constants** within the semiconductor analogy are calculated by Eq. (7.63):

$$m_\star^{\mathrm{GaAs}} = 0.058m \qquad \text{and} \qquad c_\star^{\mathrm{GaAs}} = 0.0050c. \tag{7.88}$$

According to the two-band approximation, these constants are valid for the long-wavelength modes in GaAs which satisfy Eq. (7.72):

$$\left|\frac{K}{\pi/\ell^{\mathrm{GaAs}}}\right| \ll \frac{\mathcal{E}_g^{\mathrm{GaAs}}\ell^{\mathrm{GaAs}}}{\pi\hbar\varkappa_0^{\mathrm{GaAs}}} = 0.27. \tag{7.89}$$

The harmonic mean of the effective charge-carrier masses,

$$\frac{2}{1/m_{\star,e}^{\mathrm{GaAs}} + 1/m_{\star,h}^{\mathrm{GaAs}}} = 0.069m, \tag{7.90}$$

deviates from $m_\star^{\mathrm{GaAs}}$ above by about 19.4%. In principle (ignoring measurement errors etc.), this deviation is caused by other bands coupling to the valence band and the conduction band [see Eqs. (7.84)–(7.87)]. As explained in Sec. 6.1.2, a major part of this deviation could come from the split-off band, which is neglected in the two-band model.

### 7.3.1. Constant electric field

A constant electric field gives rise to purely nonperturbative pair creation. The analogy between the Sauter–Schwinger effect and Landau–Zener tunneling is well known (see Sec. 6.2); we just have to substitute the Schwinger limit

$E_{\text{crit}}^{\text{QED}} \propto m^2c^3$ by

$$E_{\text{crit}}^{\text{SC}} = \frac{m_\star^2 c_\star^3}{\hbar q} \overset{\text{Eq. (7.64)}}{=} \frac{\sqrt{2m_\star}\mathcal{E}_g^{3/2}}{4\hbar q} \overset{\text{Eq. (7.63)}}{=} \frac{\mathcal{E}_g^2}{4\hbar q \varkappa_0} \tag{7.91}$$

in the semiconductor (cf. [149, 128, 127, 139, 52, 140, 97, 96, 142]). The **analog of the Schwinger limit** in GaAs thus measures

$$E_{\text{crit}}^{\text{GaAs}} = 5.65\,\frac{\text{MV}}{\text{cm}} = 5.65\times 10^8\,\frac{\text{V}}{\text{m}}, \tag{7.92}$$

a typical value [129, 131], which is 10 orders of magnitude smaller than $E_{\text{crit}}^{\text{QED}} \approx 10^{18}\,\text{V/m}$.

**Critical field versus breakdown field**

Note that $E_{\text{crit}}^{\text{GaAs}}$ is roughly one order of magnitude stronger than the dielectric-breakdown field strength $E_{\text{BD}}^{\text{GaAs}}$ found in the literature (see Table 7.1). This ratio makes sense because tunneling (one of the seeds for dielectric breakdown) is strongly suppressed far below $E_{\text{crit}}^{\text{GaAs}}$, while the tunneling current becomes huge close to the critical field, so a value of $E_{\text{BD}}^{\text{GaAs}}$ approximately one order of magnitude below $E_{\text{crit}}^{\text{GaAs}}$ seems reasonable. Compare this to the situation in QED where we expect a significant pair-creation rate for $E \approx E_{\text{crit}}^{\text{QED}}/10$ but only a very tiny (about 120 orders of magnitude smaller) tunneling current for $E \approx E_{\text{crit}}^{\text{QED}}/100$; see Fig. 2.1 on page 36.

### 7.3.2. Dynamical assistance by a temporal Sauter pulse

**In QED**

One way to assist tunneling in a constant ("background") electric field $E_{\text{strong}}$ is via an additional time-dependent Sauter pulse $E_{\text{weak}}/\cosh^2(\omega t)$ with $E_{\text{weak}} \ll E_{\text{strong}}$—the dynamically assisted Sauter–Schwinger effect [67] (introduced in Sec. 2.4.6). In QED, the combined Keldysh parameter (2.125) reaches its threshold value $\pi/2$ when the pulse's frequency-scale parameter is given by

$$\hbar\omega_{\text{crit}}^{\text{QED}} = \frac{\pi}{2}\frac{E_{\text{strong}}}{E_{\text{crit}}^{\text{QED}}}\frac{2mc^2}{2}, \tag{7.93}$$

so, for a rather strong background field $E_{\text{strong}} = E_{\text{crit}}^{\text{QED}}/10$, we get

$$\hbar\omega_{\text{crit}}^{\text{QED}} = 80\,\text{keV}, \tag{7.94}$$

a photon energy in the **hard X-ray** part of the spectrum. Since $\omega_{\text{crit}}^{\text{QED}} \propto E_{\text{strong}}$, this value will be correspondingly lower for weaker background fields.

**Analog in GaAs**

When we transfer this scenario to the semiconductor analog, we have to substitute $E_{\text{crit}}^{\text{QED}} \to E_{\text{crit}}^{\text{SC}}$ and $2mc^2 \to \mathcal{E}_g$ [see Eq. (7.64)] in Eq. (7.93) so that

$$\hbar\omega_{\text{crit}}^{\text{SC}} = \frac{\pi}{2} \frac{E_{\text{strong}}}{E_{\text{crit}}^{\text{SC}}} \frac{\mathcal{E}_g}{2}. \tag{7.95}$$

Assuming $E_{\text{strong}} = E_{\text{crit}}^{\text{GaAs}}/10$ in analogy to the QED case above, we get

$$\hbar\omega_{\text{crit}}^{\text{GaAs}} = 0.12\,\text{eV} \tag{7.96}$$

in GaAs, which lies in the **infrared** part of the spectrum.

### 7.3.3. Dynamical assistance by a harmonic oscillation

As we have studied in Ch. 5, tunneling in a constant field $E_{\text{strong}}$ can also be assisted by a sinusoidal oscillation $E_{\text{weak}} \cos(\omega t)$, which is a good model for assistance via counterpropagating laser beams. Let us consider a specific example for our GaAs analog here: The oscillation is generated by a **carbon-dioxide ($CO_2$) laser**, which has a wavelength of $10.6\,\mu$m, corresponding to a photon energy of $\hbar\omega = 0.117\,\text{eV}$.[4] If we set the background field to $E_{\text{strong}} = E_{\text{crit}}^{\text{GaAs}}/10$ again (i.e., approximately just below the breakdown field), the combined Keldysh parameter (2.125) will take on the value (remember $2mc^2 \to \mathcal{E}_g$)

$$\gamma_c = 2 \frac{E_{\text{crit}}^{\text{GaAs}}}{E_{\text{strong}}} \frac{\hbar\omega}{\mathcal{E}_g^{\text{GaAs}}} = 1.56. \tag{7.97}$$

This value is fixed since the laser frequency is fixed and we assume that $E_{\text{strong}}$ is kept constant. However, the laser intensity and thus $E_{\text{weak}}$ are easy to vary. According to Eq. (5.15), the **critical amplitude** of the oscillation is given by

$$\frac{E_{\text{weak}}^{\text{crit}}}{E_{\text{strong}}} = \frac{1}{200} \frac{\gamma_c}{I_1(\gamma_c)} = 0.0075, \tag{7.98}$$

$I_1(x)$ denoting a modified Bessel function of the first kind. Remember that we have defined the critical threshold as the point at which the magnitude of the leading-order exponent suppressing tunneling is decreased by 1% via dynamical assistance. For $E_{\text{strong}} = E_{\text{crit}}^{\text{GaAs}}/10$, this corresponds to a **37% increase of the pair-creation yield** [cf. Eq. (4.25)] when considering the exponential function only (i.e., when ignoring the minor effect [41] of the nonexponential

[4] This photon energy measures 7.8% of the bandgap $\mathcal{E}_g^{\text{GaAs}}$ and thus corresponds to an oscillation with $\hbar\omega = 80\,\text{keV}$ in QED (7.8% of the mass gap).

prefactor). The resulting critical amplitude and the corresponding intensity read[5]

$$E_{\rm weak}^{\rm crit} = 4.2\,\frac{\rm kV}{\rm cm} \qquad \Rightarrow \qquad I_{\rm crit} = \frac{c\varepsilon_0 (E_{\rm weak}^{\rm crit})^2}{2} = 23.7\,\frac{\rm kW}{\rm cm^2}. \tag{7.99}$$

**Laser-induced probe damage**

One might ask whether a high-quality probe of GaAs, which does not absorb too much energy via defects etc., will sustain this amount of incident radiation without being damaged. According to the studies [155, 156, 157], the above value of $I_{\rm crit}$ should not be problematic; Reference [157], for example, reports a threshold of the order of $10\,{\rm MW/cm^2}$ for GaAs surface damage via $CO_2$-laser radiation. Note that these studies only consider laser pulses with durations in the $10^{-8}$–$10^{-7}$ s range. Let us compare these pulse durations with the timescale associated with tunneling in the background field $E_{\rm strong}$ ("formation time" of an electron–hole pair): the analog of Eq. (2.63) for our GaAs setup,

$$t_{\rm form}^{\rm GaAs} = \frac{E_{\rm crit}^{\rm GaAs}}{E_{\rm strong}} \frac{\hbar}{m_\star^{\rm GaAs}(c_\star^{\rm GaAs})^2} = \frac{E_{\rm crit}^{\rm GaAs}}{E_{\rm strong}} \frac{2\hbar}{\mathcal{E}_g^{\rm GaAs}} = 8.8 \times 10^{-15}\,{\rm s}, \tag{7.100}$$

is more than six orders of magnitude smaller than the pulse durations considered in Ref. [157]. Hence, even these short laser pulses could be slow enough to assist tunneling effectively via a temporal oscillation $E_{\rm weak}\cos(\omega t)$ for a sufficient amount of time. Furthermore, since $I_{\rm crit}$ is almost three orders of magnitude smaller than the damage threshold given in Ref. [157], chances are that it is possible to keep the assisting laser switched on for even longer periods of time without damaging the GaAs probe.

## 7.4. Summary

Assuming a 1+1-dimensional spacetime and a purely time-dependent external field $E(t) = \dot{A}(t)$ in temporal gauge, we have shown in this chapter that the evolution of each $K$ mode in a two-band semiconductor is formally equivalent to that of a $k = k(K)$ mode in Dirac theory with effective physical constants $m \to m_\star(K)$ and $c \to c_\star(K)$ given by Eqs. (7.57), (7.58), and (7.60). The $K$ dependence of the quantities is unfavorable in the context of quantum simulation; however, when we restrict ourselves to **long-wavelength modes** only (small $|K|$), which describe **(assisted) nonperturbative pair-*creation* processes** for example, we then may assume $k = K$ and **constant effective values**

[5] If we set $E_{\rm strong} = E_{\rm crit}^{\rm QED}/10$ and $\hbar\omega = 80\,{\rm keV}$ in QED, we get the same value of $\gamma_c$ as in Eq. (7.97), and the resulting critical intensity reads $I_{\rm crit} = 1.3 \times 10^{23}\,{\rm W/cm^2}$.

| QED quantity | | Semiconductor analog |
|---|---|---|
| electron mass | $\leftrightarrow$ | effective electron mass |
| $m$ | | $m_\star = \mathcal{E}_g/(2\varkappa_0^2) \overset{\text{(GaAs)}}{=} 0.058m$ |
| speed of light | $\leftrightarrow$ | effective speed of light |
| $c$ | | $c_\star = \sqrt{\mathcal{E}_g/(2m_\star)} \overset{\text{(GaAs)}}{=} 0.0050c$ |
| mass gap | $\leftrightarrow$ | bandgap |
| $1\,\text{MeV} = 2mc^2$ | | $\mathcal{E}_g \overset{\text{(GaAs)}}{=} 1.5\,\text{eV}$ |
| Schwinger limit | $\leftrightarrow$ | critical field |
| $E_{\text{crit}}^{\text{QED}} = m^2c^3/(\hbar q)$ | | $E_{\text{crit}}^{\text{SC}} = \sqrt{2m_\star}\mathcal{E}_g^{3/2}/(4\hbar q)$ |
| $= 1.3 \times 10^{18}\,\text{V/m}$ | | $\overset{\text{(GaAs)}}{=} 5.65 \times 10^{8}\,\text{V/m}$ |
| Dynamically assisted Sauter–Schwinger effect (Sauter pulse) | | |
| $E_{\text{strong}} = E_{\text{crit}}^{\text{QED}}/10$ | $\leftrightarrow$ | $E_{\text{strong}} = E_{\text{crit}}^{\text{GaAs}}/10$ |
| $\hbar\omega_{\text{crit}}^{\text{QED}} = 80\,\text{keV}$ | $\leftrightarrow$ | $\hbar\omega_{\text{crit}}^{\text{GaAs}} = 0.12\,\text{eV}$ |
| Dynamically assisted Sauter–Schwinger effect (oscillation) | | |
| $E_{\text{strong}} = E_{\text{crit}}^{\text{QED}}/10$ | $\leftrightarrow$ | $E_{\text{strong}} = E_{\text{crit}}^{\text{GaAs}}/10$ |
| $\hbar\omega = 80\,\text{keV}$ | $\leftrightarrow$ | $\hbar\omega = 0.117\,\text{eV}$ ($CO_2$ laser) |
| $I_{\text{crit}}^{\text{QED}} = 1.3 \times 10^{23}\,\text{W/cm}^2$ | $\leftrightarrow$ | $I_{\text{crit}}^{\text{GaAs}} = 23.7\,\text{kW/cm}^2$ |

**Table 7.2.**: Comparison between various QED scales and their counterparts in the semiconductor analog.

$m_\star(0)$ and $c_\star(0)$ [see Eq. (7.63)] in the semiconductor analog. This analogy is valid by means of the **two-band approximation**. The coupling with other energy bands in the semiconductor gives rise to deviations, so a careful selection of the semiconductor is required. We focus on GaAs throughout this part, and we demonstrated the effect of the smaller scales in the semiconductor analog (in comparison to QED) on the required external field strengths and photon energies by considering some example field profiles giving rise to (assisted) nonperturbative pair production. The results of this comparison of scales between QED and the GaAs analog are summarized in Table 7.2.

# 8. Analogy in 1+1 spacetime dimensions for spacetime-dependent electric fields

The goal in this chapter is to generalize the analogy between Dirac theory and a two-band semiconductor for external electric fields depending not only on time but also on the space coordinate: $E = E(t, x)$. In order to describe such a field, we choose the **gauge**

$$E(t, x) = \partial_x \Phi(t, x) \tag{8.1}$$

with $A = 0$ throughout this chapter. It is always possible to find a suitable scalar potential because force fields in 1+1 dimensions are always conservative.

## 8.1. Many-body Hamiltonians

In this section, we will derive the many-body Hamiltonians in the present gauge (8.1) for both systems and bring them into a form which facilitates their comparison. We will proceed in analogy to the purely time-dependent case (Sec. 7.1).

### 8.1.1. Dirac theory

The Dirac Hamiltonian $\hat{H}_D(t)$ in second quantization has the same general form as in Eq. (7.2), but the single-body Hamilton operator is different due to the gauge (8.1):

$$\hat{H}_D^{\text{one}}(t, x) = \begin{pmatrix} mc^2 & -\mathrm{i}c\hbar\partial_x \\ -\mathrm{i}c\hbar\partial_x & -mc^2 \end{pmatrix} - q\Phi(t, x)\mathbb{1}. \tag{8.2}$$

In order to transform $\hat{H}_D(t)$ to the $k$-space representation, we insert the inverse **Fourier transform** (7.4) of the field operator and also that of $\Phi(t, x)$,

which yields

$$
\begin{aligned}
&\hat{H}_D(t)\\
&= \frac{1}{2\pi}\int_{-\infty}^{\infty}\int_{-\infty}^{\infty} \hat{\tilde{\underline{\Psi}}}^{\dagger}(t,k)\,\mathrm{e}^{-\mathrm{i}kx}\,\mathrm{d}k \int_{-\infty}^{\infty} \begin{pmatrix} mc^2 & c\hbar k' \\ c\hbar k' & -mc^2 \end{pmatrix} \hat{\tilde{\underline{\Psi}}}(t,k')\,\mathrm{e}^{\mathrm{i}k'x}\,\mathrm{d}k'\,\mathrm{d}x\\
&\quad - \frac{q}{\sqrt{2\pi}^3}\int_{-\infty}^{\infty}\int_{-\infty}^{\infty} \hat{\tilde{\underline{\Psi}}}^{\dagger}(t,k)\,\mathrm{e}^{-\mathrm{i}kx}\,\mathrm{d}k \int_{-\infty}^{\infty} \tilde{\Phi}(t,k'')\,\mathrm{e}^{\mathrm{i}k''x}\,\mathrm{d}k''\\
&\qquad\qquad\qquad\qquad \times \int_{-\infty}^{\infty} \hat{\tilde{\underline{\Psi}}}(t,k')\,\mathrm{e}^{\mathrm{i}k'x}\,\mathrm{d}k'\,\mathrm{d}x\\
&= \int_{-\infty}^{\infty}\int_{-\infty}^{\infty} \hat{\tilde{\underline{\Psi}}}^{\dagger}(t,k) \begin{pmatrix} mc^2 & c\hbar k' \\ c\hbar k' & -mc^2 \end{pmatrix} \hat{\tilde{\underline{\Psi}}}(t,k')\,\delta(k'-k)\,\mathrm{d}k\,\mathrm{d}k'\\
&\quad - \frac{q}{\sqrt{2\pi}}\int_{-\infty}^{\infty}\int_{-\infty}^{\infty}\int_{-\infty}^{\infty} \hat{\tilde{\underline{\Psi}}}^{\dagger}(t,k)\tilde{\Phi}(t,k'')\hat{\tilde{\underline{\Psi}}}(t,k')\,\delta(k''-k+k')\,\mathrm{d}k\,\mathrm{d}k'\,\mathrm{d}k''\\
&= \int_{-\infty}^{\infty} \hat{\tilde{\underline{\Psi}}}^{\dagger}(t,k) \begin{pmatrix} mc^2 & c\hbar k \\ c\hbar k & -mc^2 \end{pmatrix} \hat{\tilde{\underline{\Psi}}}(t,k)\,\mathrm{d}k\\
&\quad - \frac{q}{\sqrt{2\pi}}\int_{-\infty}^{\infty}\int_{-\infty}^{\infty} \hat{\tilde{\underline{\Psi}}}^{\dagger}(t,k)\tilde{\Phi}(t,k-k')\hat{\tilde{\underline{\Psi}}}(t,k')\,\mathrm{d}k\,\mathrm{d}k'. \qquad (8.3)
\end{aligned}
$$

**Matrix diagonalization**

The matrix in the first line of the resulting $\hat{H}_D$ is diagonalized by the same "rotation" $\hat{\tilde{\underline{\Psi}}}(t,k) \to \hat{\underline{\Upsilon}}(t,k)$ of the field operator as in the previous chapter, but we have to set $A(t) = 0$ when reusing the corresponding Eqs. (7.6)–(7.10) here. As a consequence, all quantities related to the diagonalization are time independent in this chapter. We thus get the **Dirac Hamiltonian**

$$
\begin{aligned}
\hat{H}_D(t) = &\int_{-\infty}^{\infty} \hat{\underline{\Upsilon}}^{\dagger}(t,k) \begin{pmatrix} \sqrt{m^2c^4+c^2\hbar^2k^2} & 0 \\ 0 & -\sqrt{m^2c^4+c^2\hbar^2k^2} \end{pmatrix} \hat{\underline{\Upsilon}}(t,k)\,\mathrm{d}k\\
&- \frac{q}{\sqrt{2\pi}}\int_{-\infty}^{\infty}\int_{-\infty}^{\infty} \hat{\underline{\Upsilon}}^{\dagger}(t,k)\tilde{\Phi}(t,k-k')M_D(k,k')\hat{\underline{\Upsilon}}(t,k')\,\mathrm{d}k\,\mathrm{d}k' \qquad (8.4)
\end{aligned}
$$

with the matrix

$$\begin{aligned} &M_D(k,k') \\ &= O_D(k) \cdot O_D^{\mathsf{T}}(k') \\ &= \frac{1}{\sqrt{\left[1+d_D^2(k)\right]\left[1+d_D^2(k')\right]}} \begin{pmatrix} 1 & d_D(k) \\ -d_D(k) & 1 \end{pmatrix} \cdot \begin{pmatrix} 1 & -d_D(k') \\ d_D(k') & 1 \end{pmatrix} \end{aligned} \tag{8.5}$$

and the abbreviation

$$d_D(k) = \frac{c\hbar k}{mc^2 + \sqrt{m^2c^4 + c^2\hbar^2 k^2}} = \frac{\hbar k/(mc)}{1 + \sqrt{1 + [\hbar k/(mc)]^2}}. \tag{8.6}$$

**Momentum $\hbar k$ not conserved**

Note that $\hbar k$ coincides with the **mechanical momentum** in the present gauge (since $A = 0$). This momentum is **not conserved** in the presence of an electric field $\partial_x \Phi$ since the term in the lower line of the Hamiltonian (8.4) ("$\Phi$ part") couples particle states with arbitrary wave vectors $k$ and $k'$. The transition amplitude is governed by the Fourier transform $\tilde{\Phi}(t, k - k')$.

### 8.1.2. Semiconductor

The derivation starts with the full (all bands) Bloch-electron Hamiltonian

$$\hat{H}_S^{\text{full}}(t) = \int_{-\infty}^{\infty} \hat{\psi}^\dagger(t,x) \left[ -\frac{\hbar^2 \partial_x^2}{2m} + V(x) - q\Phi(t,x) \right] \hat{\psi}(t,x)\, \mathrm{d}x, \tag{8.7}$$

which is just Eq. (7.12) in the present gauge. The **transformation to $K$ space** and the **two-band approximation** can be performed in one step by expanding

$$\hat{\psi}(t,x) \overset{\text{(two-band model)}}{=} \int_{-\pi/\ell}^{\pi/\ell} f_+(K,x)\hat{a}_+(t,K) + f_-(K,x)\hat{a}_-(t,K)\, \mathrm{d}K. \tag{8.8}$$

The resulting **two-band semiconductor Hamiltonian** $\hat{H}_S(t)$ is equal to its counterpart (7.41) from the previous chapter with $A(t)$ set to zero and with an additional term due to the scalar potential ("Φ part"):

$$\begin{aligned}
&\hat{H}_S(t) \\
&= \int_{-\pi/\ell}^{\pi/\ell} \underline{\hat{a}}^\dagger(t,K) \begin{pmatrix} \mathcal{E}_+(K) & 0 \\ 0 & \mathcal{E}_-(K) \end{pmatrix} \underline{\hat{a}}(t,K)\,\mathrm{d}K \\
&\quad - q \int_{-\infty}^{\infty} \int_{-\pi/\ell}^{\pi/\ell} \Big[ f_+^*(K,x)\hat{a}_+^\dagger(t,K) + f_-^*(K,x)\hat{a}_-^\dagger(t,K) \Big]\,\mathrm{d}K\,\Phi(t,x) \\
&\qquad \times \int_{-\pi/\ell}^{\pi/\ell} \big[ f_+(K',x)\hat{a}_+(t,K') + f_-(K',x)\hat{a}_-(t,K') \big]\,\mathrm{d}K'\,\mathrm{d}x \\
&= \int_{-\pi/\ell}^{\pi/\ell} \underline{\hat{a}}^\dagger(t,K) \begin{pmatrix} \mathcal{E}_+(K) & 0 \\ 0 & \mathcal{E}_-(K) \end{pmatrix} \underline{\hat{a}}(t,K)\,\mathrm{d}K \\
&\quad - q \int_{-\pi/\ell}^{\pi/\ell} \int_{-\pi/\ell}^{\pi/\ell} \underline{\hat{a}}^\dagger(t,K) M_S(t,K,K') \underline{\hat{a}}(t,K')\,\mathrm{d}K\,\mathrm{d}K'
\end{aligned} \tag{8.9}$$

with the **Φ matrix** in the Bloch-wave basis

$$M_S(t,K,K') = \begin{pmatrix} \langle +,K|\Phi(t,x)|+,K'\rangle & \langle +,K|\Phi(t,x)|-,K'\rangle \\ \langle -,K|\Phi(t,x)|+,K'\rangle & \langle -,K|\Phi(t,x)|-,K'\rangle \end{pmatrix} \tag{8.10}$$

[the bra–ket notation has been introduced in Eq. (7.15)]. In analogy to the Φ part in $\hat{H}_D$, the scalar potential manifestly couples Bloch states with different crystal momenta here, so ***K* is not conserved** in this chapter.

Note that the matrix in the Φ-independent part in $\hat{H}_S$ is already diagonal in the crystal-momentum representation, so no Bogoliubov transformation ("rotation") of $\underline{\hat{a}}$ is required here.

**Difference from the time-dependent case**

In the purely time-dependent case, we were able to **make the eigenvalues of diag $(\mathcal{E}_+(K), \mathcal{E}_-(K))$ symmetric around zero** by setting them to $\pm\Delta\mathcal{E}(K)/2$ [where $\Delta\mathcal{E}(K) = \mathcal{E}_+(K) - \mathcal{E}_-(K)$] via a suitable phase/gauge transformation; see Eqs. (7.41)–(7.49). Trying the same approach here leads us to the equation [cf. Eq. (7.46) with $A(t) = 0$]

$$\hat{H}_S(t) = (\Phi\,\text{part}) + \int_{-\pi/\ell}^{\pi/\ell} \underline{\hat{a}}^\dagger(t,K) \begin{pmatrix} \frac{\Delta\mathcal{E}(K)}{2} & 0 \\ 0 & -\frac{\Delta\mathcal{E}(K)}{2} \end{pmatrix} \underline{\hat{a}}(t,K)\,\mathrm{d}K$$
$$+ \int_{-\pi/\ell}^{\pi/\ell} \frac{\mathcal{E}_+(K) + \mathcal{E}_-(K)}{2} \Big[\underbrace{\hat{a}_+^\dagger(t,K)\hat{a}_+(t,K) + \hat{a}_-^\dagger(t,K)\hat{a}_-(t,K)}_{=\underline{\hat{a}}^\dagger(t,K)\underline{\hat{a}}(t,K)}\Big]\,\mathrm{d}K. \quad (8.11)$$

The expression $\mathcal{E}_+(K) + \mathcal{E}_-(K)$ is generally $K$ dependent, and $K$ is *not* a conserved quantity in the present gauge. Therefore, moving a particle from the Bloch state $|\pm, K\rangle$ to a state $|\pm, K'\rangle$ with $K' \neq K$ (i.e., accelerating the particle) changes the value of the expression in the lower line in Eq. (8.11), so this term has an effect on the particle dynamics and thus cannot simply be ignored. We will see in the next section that this fact has an important physical consequence for the analogy in spacetime-dependent fields. In conclusion, the Hamiltonian in Eq. (8.9) is our end result in this subsection.

## 8.2. Quantitative analogy for long-wavelength modes

As in the purely time-dependent case, our primary interest are **pair-*creation* processes via (assisted) tunneling**, which is a long-wavelength phenomenon and thus predominantly occurs at **small $\boldsymbol{k}$ or $\boldsymbol{K}$**. Hence, we will compare the actions of the Hamiltonians $\hat{H}_D$ and $\hat{H}_S$ on states with small (crystal) momenta, that is, $|k| \ll mc/\hbar$ (nonrelativistic regime) and $|K| \ll \mathcal{E}_g/(\hbar\varkappa_0)$ [where $\boldsymbol{K}\cdot\boldsymbol{p}$ perturbation theory up to the first nonvanishing order is a good approximation according to Eq. (7.72)]. Figure 6.1 on page 152 shows that the dispersion relations in both systems are parabolic for these (crystal) momenta, which facilitates the quantitative analogy.

### Analogy between the potential-independent parts of the Hamiltonians

Let us start to compare the $2 \times 2$ matrices in the $\Phi$-independent parts of the Hamiltonians [see Eqs. (8.4) and (8.9)]. We do this by expanding these matrices around the mass gap/bandgap at $k = K = 0$ up to the first nonvanishing order in $k$ or $K$, respectively. Using Eq. (7.83) and setting $\mathcal{E}_\pm(0) = \pm\mathcal{E}_g/2$ in the semiconductor case without loss of generality, we get

$$\begin{aligned} &\sqrt{m^2c^4 + c^2\hbar^2k^2}\begin{pmatrix}1 & 0\\ 0 & -1\end{pmatrix} && \begin{pmatrix}\mathcal{E}_+(K) & 0\\ 0 & \mathcal{E}_-(K)\end{pmatrix} \\ &= \left[mc^2 + \frac{\hbar^2k^2}{2m}\right]\begin{pmatrix}1 & 0\\ 0 & -1\end{pmatrix} \quad \leftrightarrow && = \begin{pmatrix}\frac{\mathcal{E}_g}{2} + \frac{\hbar^2K^2}{2m_{\star,e}} & 0\\ 0 & -\frac{\mathcal{E}_g}{2} - \frac{\hbar^2K^2}{2m_{\star,h}}\end{pmatrix} \\ &\quad + \mathcal{O}\left[\left(\frac{\hbar k}{mc}\right)^4\right] && + \mathcal{O}\left[\left(\frac{\hbar K \varkappa_0}{\mathcal{E}_g}\right)^3\right]. \end{aligned} \tag{8.12}$$

**Additional assumption in the spacetime-dependent case**

We see that the **quantitative analogy can only work if $m_{\star,e} = m_{\star,h}$** in the semiconductor. These effective masses then play the role of *the* effective electron mass within the analogy with QED:

$$m_\star = m_{\star,e} \overset{!}{=} m_{\star,h}. \tag{8.13}$$

This additional assumption is required here because it is not possible, as explained in the previous section, to make the diagonal elements $\mathcal{E}_\pm(K)$ symmetric around zero [i.e., $\pm\Delta\mathcal{E}(K)/2$] via a gauge transformation in the case of a spacetime-dependent $E$ field—**in contrast to the purely time-dependent case**, in which $m_{\star,e}$ and $m_{\star,h}$ may be different and $m_\star$ is their harmonic mean (within the two-band approximation); see Eq. (7.87).

**Example: GaAs**

Assumption (8.13) is of course not perfectly satisfied in many semiconductors. In GaAs, $m_{\star,h}^{\text{GaAs}} = 0.076m$ is 20.6% greater than $m_{\star,e}^{\text{GaAs}} = 0.063m$, which is a significant deviation. However, the deviation measures several hundred percents in other typical direct-bandgap semiconductors (see, e.g., [126]).

Let us assume $m_\star = m_{\star,e} = m_{\star,h}$ in the following. Then, **$K$ directly corresponds to $k$** for long-wavelength modes, and by defining an **effective speed of light** in the semiconductor via $m_\star c_\star^2 = \mathcal{E}_g/2$, we get the same relation

$$c_\star = \sqrt{\frac{\mathcal{E}_g}{2m_\star}} \tag{8.14}$$

as in the time-dependent case in the long-wavelength limit [see Eq. (7.64)].

### Analogy between the potential-dependent parts

**QED**

In the Dirac Hamiltonian (8.4), the spatial Fourier transform $\tilde{\Phi}$ of the scalar potential couples states with different $k$. Note that $E(t,x)$ typically only incorporates photon energies $\hbar\omega \ll 2mc^2$ in the context of assisted tunneling pair creation, so **$\tilde{\Phi}(t,k)$ is only nonvanishing for wave vectors which satisfy $|k| \ll 2mc/\hbar$** since $\omega = c|k|$ in the vacuum. A long-wavelength particle state with a small $|k'| \ll mc/\hbar$ can thus only be coupled (directly) with another long-wavelength state $|k| \ll mc/\hbar$ via $\tilde{\Phi}(t, k-k')$. For such transitions near the mass gap, the matrix $M_D(k,k')$ in Eq. (8.5) may be approximated by expanding it up to the first order in one of the small quantities around the mass gap:

$$M_D(k,k') = \begin{pmatrix} 1 & 0 \\ 0 & 1 \end{pmatrix} + \frac{\hbar(k-k')}{2mc} \begin{pmatrix} 0 & 1 \\ -1 & 0 \end{pmatrix} + \mathcal{O}\left[\left(\frac{\hbar k}{mc}\right)^2\right] + \mathcal{O}\left[\left(\frac{\hbar k'}{mc}\right)^2\right] + \mathcal{O}\left[\frac{\hbar^2 k k'}{m^2 c^2}\right]. \tag{8.15}$$

**Semiconductor**

In the two-band semiconductor Hamiltonian (8.9), $\Phi$ appears in the elements of the matrix $M_S(t,K,K')$, which are all of the form

$$\begin{aligned} \langle n, K|\Phi(t,x)|n', K'\rangle &= \int_{-\infty}^{\infty} f_n^*(K,x)\Phi(t,x) f_{n'}(K',x)\, \mathrm{d}x \\ &= \int_{-\infty}^{\infty} \mathrm{e}^{-\mathrm{i}Kx} u_n^*(K,x) \int_{-\infty}^{\infty} \frac{\tilde{\Phi}(t,k)}{\sqrt{2\pi}} \mathrm{e}^{\mathrm{i}kx}\, \mathrm{d}k\, \mathrm{e}^{\mathrm{i}K'x} u_{n'}(K',x)\, \mathrm{d}x \\ &= \int_{-\infty}^{\infty} \frac{\tilde{\Phi}(t,k)}{\sqrt{2\pi}} \int_{-\infty}^{\infty} u_n^*(K,x) u_{n'}(K',x) \mathrm{e}^{\mathrm{i}(k+K'-K)x}\, \mathrm{d}x\, \mathrm{d}k \end{aligned} \tag{8.16}$$

with $n, n' \in \mathbb{N}$. When we calculate this expression for **particle transitions near the bandgap**, we have $|K| \ll \pi/\ell$ and $|K'| \ll \pi/\ell$, but what can we say about $\tilde{\Phi}(t,k)$?

**Slowly varying potential**

In analogy to the Dirac case above, we assume that the **energies of the photons** making up the external field are all **well below the bandgap: $\hbar\omega \ll \mathcal{E}_g$**. Within the semiconductor crystal, photon frequency and wave number are related via

$$\omega = \frac{c}{n_{\text{ref}}}|k|, \tag{8.17}$$

where $n_{\text{ref}}$ denotes the refractive index (which might depend on $k$) in this context, so the Fourier components $\tilde{\Phi}(t,k)$ should vanish except for small

$$|k| \ll \frac{n_{\text{ref}}\mathcal{E}_g}{\hbar c}. \tag{8.18}$$

We may estimate the quantity on the right-hand side of this inequality by using that $\mathcal{E}_g$ is always (much) smaller than the Fermi energy in the empty lattice,

$$\mathcal{E}_F = \frac{\hbar^2(\pi/\ell)^2}{2m} \tag{8.19}$$

(i.e., the kinetic energy of a free electron at the zone boundary), according to the nearly-free electron model, which works well for GaAs [122]. We thus get

$$|k| \ll \frac{n_{\text{ref}}\mathcal{E}_g}{\hbar c} < \frac{n_{\text{ref}}\mathcal{E}_F}{\hbar c} = \frac{n_{\text{ref}}}{\hbar c}\frac{\pi^2\hbar^2}{2m\ell^2} = \frac{\pi}{\ell}\frac{n_{\text{ref}}}{4}\frac{\lambda_C}{\ell}. \tag{8.20}$$

The value of $n_{\text{ref}}/4$ is usually of order 1; in GaAs, for example, we have $n_{\text{ref}}^{\text{GaAs}} \approx 3.7$ at the bandgap (i.e., at $\hbar\omega = 1.5\,\text{eV}$) and even smaller values of $n_{\text{ref}}^{\text{GaAs}}$ for lower frequencies (see [158] and cf. [126]). Since the lattice constant (e.g., $\ell^{\text{GaAs}} = 0.565\,\text{nm}$) is always much greater than the Compton wavelength of the electron, $\lambda_C \approx 10^{-12}\,\text{m}$, we infer from the inequality (8.20) that $\tilde{\boldsymbol{\Phi}}(\boldsymbol{t},\boldsymbol{k})$ **is only nonzero for** $|\boldsymbol{k}| \ll \boldsymbol{\pi/\ell}$ in typical semiconductors if the photon energies are all well below the bandgap, which was our initial assumption. An equivalent form of this statement can be obtained by inserting $|k| = 2\pi/\lambda$ and then rearranging Eq. (8.20) into the form

$$\lambda \gg \overbrace{\frac{8}{n_{\text{ref}}}}^{\gtrsim 1} \underbrace{\frac{\ell}{\lambda_C}}_{\gg 1} \ell \overset{\text{(typically)}}{\gg} \ell. \tag{8.21}$$

That is, the external electric field is composed of wavelengths which are all much greater than the lattice constant—the corresponding **potential is thus slowly varying in space**.

**$\langle n,K|\Phi|n',K'\rangle$ for slowly varying potentials**

In conclusion, for such a slowly varying potential, we may assume $|k| \ll \pi/\ell$ when calculating Eq. (8.16) since $\tilde{\Phi}(t,k)$ is zero for greater $k$ anyway. The $x$ integral in Eq. (8.16) can be calculated by means of the formula (7.24) since the Bloch factors are $\ell$ periodic and $|k+K'-K| < 2\pi/\ell$ is certainly true near the bandgap ($|K|, |K'| \ll \pi/\ell$), so we get

$$\begin{aligned}\langle n,K|\Phi(t,x)|n',K'\rangle &= \int_{-\infty}^{\infty} \frac{\tilde{\Phi}(t,k)}{\sqrt{2\pi}}\frac{2\pi}{\ell}\int_0^{\ell} u_n^*(K,x)u_{n'}(K',x)\,\mathrm{d}x\,\delta(k+K'-K)\,\mathrm{d}k \\ &= \frac{\tilde{\Phi}(t,K-K')}{\sqrt{2\pi}}\langle n,K|n',K'\rangle_{\text{cell}}\,. \end{aligned} \tag{8.22}$$

**$K \cdot p$ perturbation theory**

The Bloch factors appearing in the unit-cell product are well approximated near the gap by first-order $\boldsymbol{K}\cdot\boldsymbol{p}$ perturbation theory. Inserting the perturbational expansions from Eq. (7.68) and using the orthonormality (7.29) of the Bloch factors at $K=0$ yields

$$\begin{aligned}
&\langle n,K|n',K'\rangle_{\text{cell}} \\
&= \Bigg\langle u_n(0,x) + \frac{\hbar K}{m}\sum_{\tilde n\in\mathbb{N}}^{\tilde n\neq n}\frac{\langle\tilde n,0|\hat p_x|n,0\rangle_{\text{cell}}}{\mathcal{E}_n(0)-\mathcal{E}_{\tilde n}(0)}u_{\tilde n}(0,x) + \mathcal{O}\left[\left(\frac{\varkappa_0\hbar K}{\mathcal{E}_g}\right)^2\right]\Bigg| \\
&\qquad u_{n'}(0,x) + \frac{\hbar K'}{m}\sum_{\tilde n'\in\mathbb{N}}^{\tilde n'\neq n'}\frac{\langle\tilde n',0|\hat p_x|n',0\rangle_{\text{cell}}}{\mathcal{E}_{n'}(0)-\mathcal{E}_{\tilde n'}(0)}u_{\tilde n'}(0,x) + \mathcal{O}\left[\left(\frac{\varkappa_0\hbar K'}{\mathcal{E}_g}\right)^2\right]\Bigg\rangle_{\text{cell}} \\
&= \delta_{nn'} + \frac{\hbar K}{m}\sum_{\tilde n\in\mathbb{N}}^{\tilde n\neq n}\frac{\langle\tilde n,0|\hat p_x|n,0\rangle_{\text{cell}}}{\mathcal{E}_n(0)-\mathcal{E}_{\tilde n}(0)}\delta_{\tilde n n'} + \frac{\hbar K'}{m}\sum_{\tilde n'\in\mathbb{N}}^{\tilde n'\neq n'}\frac{\langle\tilde n',0|\hat p_x|n',0\rangle_{\text{cell}}}{\mathcal{E}_{n'}(0)-\mathcal{E}_{\tilde n'}(0)}\delta_{\tilde n' n} \\
&\quad \underbrace{+\,\mathcal{O}\left[\left(\frac{\varkappa_0\hbar K}{\mathcal{E}_g}\right)^2\right] + \mathcal{O}\left[\left(\frac{\varkappa_0\hbar K'}{\mathcal{E}_g}\right)^2\right] + \mathcal{O}\left[\frac{\varkappa_0^2\hbar^2 KK'}{\mathcal{E}_g^2}\right]}_{\text{second order}} \\
&= \delta_{nn'} + \frac{\hbar K}{m}\frac{\langle n',0|\hat p_x|n,0\rangle_{\text{cell}}}{\mathcal{E}_n(0)-\mathcal{E}_{n'}(0)}(1-\delta_{nn'}) + \frac{\hbar K'}{m}\frac{\langle n,0|\hat p_x|n',0\rangle_{\text{cell}}}{\mathcal{E}_{n'}(0)-\mathcal{E}_n(0)}(1-\delta_{nn'}) \\
&\quad + \text{second order} \\
&= \delta_{nn'} + \frac{\hbar\,(1-\delta_{nn'})}{m\,[\mathcal{E}_n(0)-\mathcal{E}_{n'}(0)]}\Big(\langle n',0|\hat p_x|n,0\rangle_{\text{cell}}\,K - \langle n',0|\hat p_x|n,0\rangle^*_{\text{cell}}\,K'\Big) \\
&\quad + \text{second order.}
\end{aligned} \tag{8.23}$$

Note that the terms proportional to $1-\delta_{nn'}$ are meant to vanish for $n=n'$. The last two equations allow us to write the matrix (8.10) which appears in the $\Phi$ part of $\hat H_S$ as

$$\begin{aligned}
&M_S(t,K,K') \\
&= \frac{\tilde\Phi(t,K-K')}{\sqrt{2\pi}}\begin{pmatrix}\langle +,K|+,K'\rangle_{\text{cell}} & \langle +,K|-,K'\rangle_{\text{cell}} \\ \langle -,K|+,K'\rangle_{\text{cell}} & \langle -,K|-,K'\rangle_{\text{cell}}\end{pmatrix} \\
&= \frac{\tilde\Phi(t,K-K')}{\sqrt{2\pi}}\left\{\begin{pmatrix}1 & 0\\ 0 & 1\end{pmatrix} + \frac{\varkappa_0\hbar(K-K')}{\mathcal{E}_g}\begin{pmatrix}0 & 1\\ -1 & 0\end{pmatrix} + \mathcal{O}\left[\left(\frac{\varkappa_0\hbar K}{\mathcal{E}_g}\right)^2\right]\right. \\
&\qquad\qquad \left. + \mathcal{O}\left[\left(\frac{\varkappa_0\hbar K'}{\mathcal{E}_g}\right)^2\right] + \mathcal{O}\left[\frac{\varkappa_0^2\hbar^2 KK'}{\mathcal{E}_g^2}\right]\right\}
\end{aligned} \tag{8.24}$$

near the bandgap and for slowly varying potentials. Comparing this result to the Dirac-case matrix $M_D(k,k')$ in Eq. (8.15)—note that the prefactor $\tilde\Phi(t,k-k')/\sqrt{2\pi}$ does not appear in $M_D$ because it is an explicit part of the Dirac

Hamiltonian (8.4) according to our definition—we see that the **quantitative analogy will be correct** close to the bandgap/mass gap **if $2mc$ in Dirac theory corresponds to $\mathcal{E}_g/\varkappa_0$** in the semiconductor analog.

**Two-band approximation**

Let us check whether this is true. The term $2mc$ becomes $2m_\star c_\star$ in the semiconductor analog, where $m_\star = m_{\star,e} = m_{\star,h}$ according to our above assumption (8.13) and $c_\star = \sqrt{\mathcal{E}_g/(2m_\star)}$ in order to render the analogy possible for the $\Phi$-independent parts of the Hamiltonians:

$$2mc \leftrightarrow 2m_\star c_\star = 2m_\star\sqrt{\frac{\mathcal{E}_g}{2m_\star}} = \sqrt{2m_\star\mathcal{E}_g}. \tag{8.25}$$

According to the **two-band approximation**, there is a relation between $m_\star$ and $\varkappa_0$ given by Eq. (7.87), which yields

$$2mc \leftrightarrow 2m_\star c_\star \overset{\text{(two-band model)}}{=} \sqrt{2\frac{\mathcal{E}_g}{2\varkappa_0^2}\mathcal{E}_g} = \frac{\mathcal{E}_g}{\varkappa_0}, \tag{8.26}$$

**Result**

so we have **confirmed the quantitative analogy** between the long-wavelength components of $\hat{H}_D$ and $\hat{H}_S$ within the two-band approximation for a spacetime-dependent, spatially slowly varying potential $\Phi(t,x)$.

## 8.3. Dynamically assisted Sauter–Schwinger effect in GaAs with a space-dependent background field

### 8.3.1. Assisting temporal Sauter pulse

**QED**

Tunneling in QED in a strong, space-dependent background field (Sauter-pulse shape) assisted by a temporal Sauter pulse was studied in Ref. [90] via the worldline-instanton method[1]. The **electric-field profile** considered was therefore

$$E(t,x) = \frac{E_{\text{strong}}}{\cosh^2(kx)} + \frac{E_{\text{weak}}}{\cosh^2(\omega t)}. \tag{8.27}$$

The spatial width of the localized, static background field is characterized by the quantity

$$L = \frac{2\pi}{k}. \tag{8.28}$$

It was found that the critical threshold $\omega_{\text{crit}}^{\text{QED}}(k)$ for dynamical assistance decreases according to Eq. (2.145) when $k$ is increased (i.e., when the background field gets "narrower" due to an decreasing $L$), until $\omega_{\text{crit}}^{\text{QED}}$ becomes zero precisely when the **spatial Keldysh parameter**

$$\gamma_k = \frac{mc^2k}{qE_{\text{strong}}} = \frac{E_{\text{crit}}^{\text{QED}}}{E_{\text{strong}}}\frac{\hbar k}{mc} = \frac{E_{\text{crit}}^{\text{QED}}}{E_{\text{strong}}}\frac{\lambda_C^{\text{QED}}}{L} \tag{8.29}$$

attains the value 1 (no-tunneling limit).

[1] We have introduced the concepts and results from Ref. [90] in Sec. 2.5.1.

**Semiconductor analog**

Due to the quantitative analogy between the Hamiltonians $\hat{H}_D$ and $\hat{H}_S$ for long-wavelength processes, we may directly transfer these results to the two-band semiconductor analog, with $m \to m_\star$ and $c \to c_\star$ from Eqs. (8.13) and (8.14). Hence, we substitute $E_{\text{crit}}^{\text{QED}} \to E_{\text{crit}}^{\text{SC}}$ again, and the **Compton wavelength** also takes on an **effective value** within the semiconductor analogy:

$$\lambda_C^{\text{QED}} = \frac{h}{mc} \to \lambda_C^{\text{SC}} = \frac{h}{m_\star c_\star} = \underbrace{\frac{m}{m_\star}\frac{c}{c_\star}}_{\text{typically} \gg 1} \lambda_C^{\text{QED}}. \tag{8.30}$$

In GaAs, for example, we have (we use the GaAs values from Table 7.2 on page 187 in the following again)

$$\lambda_C^{\text{GaAs}} = 3437\lambda_C^{\text{QED}} = 8.34\,\text{nm}. \tag{8.31}$$

**Spatially slowly varying background field?**

The electric field must be **slowly varying in space** in order for the analogy to hold. The spatial pulse width $L$ must thus be much greater than the lattice constant $\ell$, or, equivalently, $k$ must be much smaller than the width $2\pi/\ell$ of the Brillouin zone. The minimum value of $L$ we are interested in corresponds to the no-tunneling limit $\gamma_k = 1$ (where $\omega_{\text{crit}}$ approaches zero), so

$$\frac{E_{\text{crit}}^{\text{SC}}}{E_{\text{strong}}}\frac{\lambda_C^{\text{SC}}}{L_{\text{min}}} = 1 \qquad \Leftrightarrow \qquad L_{\text{min}} = \frac{E_{\text{crit}}^{\text{SC}}}{E_{\text{strong}}}\lambda_C^{\text{SC}}. \tag{8.32}$$

A spatial Sauter pulse with this $L$ can be considered to be slowly varying if $L_{\text{min}} \gg \ell$, which is equivalent to

$$\frac{E_{\text{strong}}}{E_{\text{crit}}^{\text{SC}}} \ll \frac{\lambda_C^{\text{SC}}}{\ell} \overset{\text{(GaAs)}}{=} 14.8. \tag{8.33}$$

We see that this condition is **always satisfied in GaAs** since we assume electric field strengths well below $E_{\text{crit}}^{\text{SC}}$ throughout (as required by the worldline-instanton method, the JWKB approximation, etc.)

**Example for GaAs**

Let us thus consider a concrete example for GaAs: We assume that there is a localized, static electric field of the form $E_{\text{strong}}/\cosh^2(kx)$ in a probe of GaAs, with a maximum field strength of $E_{\text{strong}} = E_{\text{crit}}^{\text{GaAs}}/10$ at the position $x = 0$. The band bending caused by this **built-in field**—which could be induced by a suitable doping profile, for example—matches the relativistic energy continua plotted in Fig. 2.3 on page 52 (with $2mc^2 \to \mathcal{E}_g^{\text{GaAs}}$). In the limit $L \to \infty$ (so $k \to 0$), the background field becomes a constant field $E_{\text{strong}}$. We know from Sec. 7.3.2 that the temporal Sauter pulse will assist tunneling above the threshold

$$\hbar\omega_{\text{crit}}^{\text{GaAs}}(L \to \infty) = 0.12\,\text{eV} \tag{8.34}$$

in this case. For a finite $L$, however, the **critical frequency decreases** according to Eq. (2.145) (remember $m \to m_\star$ and $c \to c_\star$),

$$\begin{aligned}\hbar\omega_{\text{crit}}^{\text{SC}}(L) &= \frac{\pi}{2}\frac{\hbar q E_{\text{strong}}}{m_\star c_\star}\frac{\gamma_k\sqrt{1-\gamma_k^2}}{\arcsin\gamma_k}\\ &= \frac{\pi}{2}\frac{\hbar q E_{\text{strong}}}{m_\star c_\star}\frac{E_{\text{crit}}^{\text{SC}}}{E_{\text{strong}}}\frac{\lambda_C^{\text{SC}}}{L}\frac{\sqrt{1-\gamma_k^2}}{\arcsin\gamma_k}\\ &= \frac{\pi}{2}m_\star c_\star^2\frac{\lambda_C^{\text{SC}}}{L}\frac{\sqrt{1-\gamma_k^2}}{\arcsin\gamma_k}\\ &= \frac{\pi}{4}\mathcal{E}_g\frac{\lambda_C^{\text{SC}}}{L}\frac{\sqrt{1-\gamma_k^2}}{\arcsin\gamma_k},\end{aligned} \tag{8.35}$$

until it reaches zero at the pulse width given in Eq. (8.32), which reads

$$L_{\text{min}}^{\text{GaAs}} = 83.4\,\text{nm} \tag{8.36}$$

in this example. The dependence of $\omega_{\text{crit}}^{\text{GaAs}}$ on $L$ is plotted in Fig. 8.1.

In summary, by lowering the width $L$ of the built-in field in a semiconductor, the critical frequency scale for dynamical assistance by a temporal Sauter pulse should be reduced in a semiconductor.

#### Shape of the built-in field

Inspired by Ref. [90], we have assumed that the built-in field has the form of a spatial Sauter pulse (i.e., a tanh-shaped band bending in space) in this example. Realizing this background-field profile in a semiconductor requires a very specific doping profile. However, even if the built-in field only roughly has the form of a **generic spatial pulse** with a maximum field strength of $E_{\text{strong}}$ and a width quantified by $L$, it seems likely that some features of the dynamically assisted Sauter–Schwinger effect with a spatial Sauter pulse as background field [90] remain: the **scaling** $\omega_{\text{crit}}^{\text{SC}} \sim \sqrt{1-\gamma_k^2}$ [see Eq. (2.146)] in the no-tunneling limit $\gamma_k \nearrow 1$.

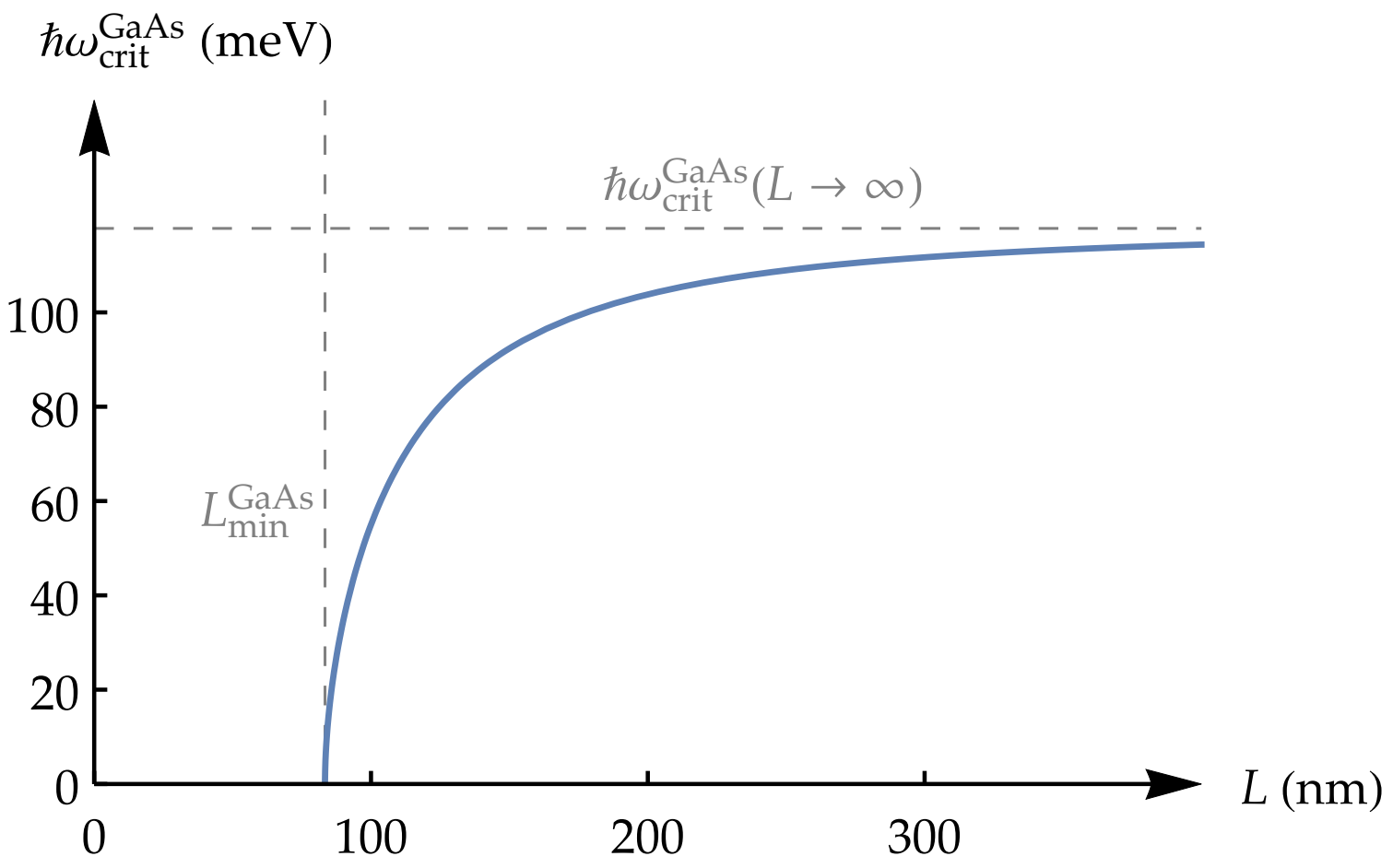

**Figure 8.1.:** Critical threshold [Eq. (8.35)] for dynamical assistance via a temporal Sauter pulse $E_{\mathrm{weak}}/\cosh^2(\omega t)$ in a GaAs sample with a localized built-in field $E_{\mathrm{strong}}/\cosh^2(kx)$, with the maximum value $E_{\mathrm{strong}} = E_{\mathrm{crit}}^{\mathrm{GaAs}}/10 = 57\,\mathrm{MV/m}$ and the width $L = 2\pi/k$, plotted as a function of $L$. The threshold approaches the value $\hbar\omega_{\mathrm{crit}}^{\mathrm{GaAs}} = 0.12\,\mathrm{eV}$ for $L \to \infty$ (constant-field limit). At $L_{\mathrm{min}}^{\mathrm{GaAs}} = 83.4\,\mathrm{nm}$, which corresponds to the no-tunneling limit for our value of $E_{\mathrm{strong}}$ here, $\omega_{\mathrm{crit}}^{\mathrm{GaAs}}$ drops to zero.

Before we give reasons for this statement, let us consider the maximal electrostatic energy the spatial Sauter pulse can provide: $q\Delta\Phi = 2qE_{\text{strong}}/k$ [see Eqs. (2.58)–(2.60)]. In terms of this quantity, the no-tunneling limit for a built-in field in a semiconductor can be expressed as $q\Delta\Phi \searrow \mathcal{E}_g$, and the corresponding scaling of $\omega_{\text{crit}}^{\text{SC}}$ becomes

$$\omega_{\text{crit}}^{\text{SC}} \sim \sqrt{1-\gamma_k^2} = \sqrt{1-\left(\frac{m_\star c_\star^2 k}{qE_{\text{strong}}}\right)^2} \propto \sqrt{\left(\frac{2qE_{\text{strong}}}{k}\right)^2 - (2m_\star c_\star^2)^2}$$
$$= \sqrt{(q\Delta\Phi)^2 - \mathcal{E}_g^2} = \sqrt{(q\Delta\Phi + \mathcal{E}_g)(q\Delta\Phi - \mathcal{E}_g)} \sim \sqrt{q\Delta\Phi - \mathcal{E}_g}. \quad (8.37)$$

This formulation makes more sense in the context of generic pulse shapes.

**Exponentially decaying built-in field for large $|x|$**

Here is why this scaling should be independent of the exact pulse shape of the background field: It seems reasonable to assume that a built-in field in a semiconductor decays exponentially far away from the maximum, which we locate at $x = 0$ without loss of generality. According to the standard depletion approximation, the built-in field vanishes identically for sufficiently large, finite $x$ [126]. This approximation is based on the assumption that the space-charge density within the depletion region is piecewise constant, including an abrupt transition at $x = 0$ from the positive to the negative region. However, this simplification is generally considered to be unrealistic [126] since a real charge density will always be somehow "smeared", and exponential "tails" at both ends of the depletion layer seem plausible. The local distribution of the mobile charge carriers is governed by the Boltzmann statistics, and, altogether, we thus expect the resulting **built-in field to also decay exponentially**.

If thermal "smearing" effects are neglected, the majority carriers cannot cross the junction due to the built-in field, which acts like a potential barrier. However, the wave functions of the majority carriers will leak into the forbidden region. This quantum effect should thus even give rise to an exponentially decaying built-in field at zero Kelvin.

**Consequences of the exponential decay**

Assuming that any pulse-shaped, space-dependent background field $E(x)$ in the semiconductor decays exponentially for large $|x|$, this field is described well by the Sauter pulse $E_{\text{strong}}/\cosh^2(kx)$ asymptotically (which also approaches zero exponentially for large $|x|$), even if $E(x)$ is not shaped like a spatial Sauter pulse near $x = 0$—we just focus on the exponential tails of the background field here. These exponential tails make the spatial turning points $x_\star^\pm$ diverge logarithmically[2] in the no-tunneling limit; cf. the band diagram in Fig. 2.3(b) on page 52. The instanton trajectories in the background field $E(x)$ alone (i.e., without the assisting, temporal pulse) coincide with those for a spatial Sauter pulse (see Fig. 2.9 on page 83) for large $|x|$ where $E(x)$ is exponentially decaying because this functional form of $E(x)$ uniquely fixes the shape of the instanton trajectories. These **instanton trajectories grow large** in the no-tunneling limit $\gamma_k \nearrow 1$, the intersection points with the imaginary-time axis diverging like $\mathcal{T}_0 \sim 1/\sqrt{1-\gamma_k^2}$ according to Eq. (2.143), which in turn accounts for the scaling $\omega_{\text{crit}}^{\text{SC}} \sim \sqrt{1-\gamma_k^2}$ in this limit and in the case of a perfect spatial Sauter pulse.

[2]See Eq. (2.144); the argument of arsinh becomes large in the no-tunneling limit $\gamma_k \nearrow 1$, in which case arsinh increases logarithmically (asymptotically).

Now, even if the shape of the actual instanton trajectories for $E(x)$ is **different near $x = 0$**, where the built-in field induced by a real p–n junction may not be formed like a spatial Sauter pulse, this **should not change the scaling $\omega_{\text{crit}}^{\text{SC}} \sim \sqrt{q\Delta\Phi - \mathcal{E}_g}$** since this scaling is a pure consequence of the growing instanton trajectories over the exponential tails of $E(x)$.

In conclusion, $\omega_{\text{crit}}^{\text{SC}}$ should be smaller in a localized background field $E(x)$ with a maximum field strength of $E_{\text{strong}}$ than in a constant background field $E_{\text{strong}}$ [in which case it is given by Eq. (7.95)], and the way $\omega_{\text{crit}}^{\text{SC}} \sim \sqrt{q\Delta\Phi - \mathcal{E}_g}$ approaches zero in the no-tunneling limit should be universal in a semiconductor analog since a realistic built-in field probably always decays exponentially.

Note that the idea that the asymptotic properties of a space-dependent electric field lead to universal scaling laws in the no-tunneling limit is not restricted to $\omega_{\text{crit}}$—it is also true for the tunneling current in the space-dependent field, for example; see Refs. [94, 95].

### 8.3.2. Assisting harmonic oscillation

Another way to assist tunneling in a localized space-dependent background field $E(x)$ with $|E(x)| \leq E_{\text{strong}}$ is via a harmonic oscillation $E_{\text{weak}} \cos(\omega t)$ (e.g., counterpropagating laser beams). For a constant background field, this field profile has been studied in Ch. 5. In this subsection, we **estimate the dependence of the critical threshold** $\varepsilon_{\text{crit}} = E_{\text{weak}}^{\text{crit}}/E_{\text{strong}}$ [given in Eq. (5.15) for a constant background field] **on the background-field width** $L$ for the case of a Sauter-pulse-shaped background field instead of a constant field, and we will apply that result to the **GaAs analog**. Our argumentation here is completely analogous to the case of an assisting temporal Sauter pulse [90] (see Sec. 2.5.1 and especially Fig. 2.9 on page 83). The crucial difference in the case of the field profile we consider here,

$$E(t,x) = \frac{E_{\text{strong}}}{\cosh^2(kx)} + E_{\text{weak}} \cos(\omega t), \tag{8.38}$$

is that the **oscillation does not establish "walls" in Euclidean spacetime** (unlike the temporal Sauter pulse), which "reflect" the instanton trajectory when hit (dynamical assistance) and leave the instanton trajectory of the spatial Sauter pulse basically unaffected in between. The reason for that difference is that the oscillation $E_{\text{weak}} \cosh(\omega \mathcal{T})$ does not have singularities in Euclidean spacetime which could act like "walls" in the limit $E_{\text{weak}}/E_{\text{strong}} \to 0$. However, we are primarily interested in the scaling of $E^{\text{crit}}_{\text{weak}}$ in the no-tunneling limit here. The approach presented in the following should yield the correct scaling in this limit.

**$E^{\text{crit}}_{\text{weak}}$ for a constant background field ($k = 0$)**

We consider the instanton trajectory in the static background Sauter pulse only (see Fig. 2.9). In the constant-field case ($k = 0$, $L \to \infty$), this is a circle around the origin in Euclidean spacetime. The maximum value the oscillating field takes on along this instanton trajectory reads $E_{\text{weak}} \cosh[\omega \mathcal{T}_0(\gamma_k)]$ and occurs on the imaginary-time axis, where the background field takes on its maximum value $E_{\text{strong}}$. The "temporal turning point" $\mathcal{T}_0(\gamma_k)$ for a general Sauter-pulse-shaped background field is given in Eq. (2.143); for a constant background field, we get the combined Keldysh parameter: $\omega \mathcal{T}_0(0) = \gamma_c = mc\omega/(qE_{\text{strong}})$. We can think of the known threshold condition (5.15) as defining that dynamical assistance sets in when the ratio of the maximum $E_{\text{weak}} \cosh[\omega \mathcal{T}_0(\gamma_k)]$ along the trajectory to the background field at the same point, $E_{\text{strong}}$, takes on a certain, critical value $\cosh[\omega \mathcal{T}_0(\gamma_k)]\varepsilon_{\text{crit}}$. We assume that $E_{\text{weak}}$ is the variable quantity here, so the threshold condition we know from Ch. 5 can be written as

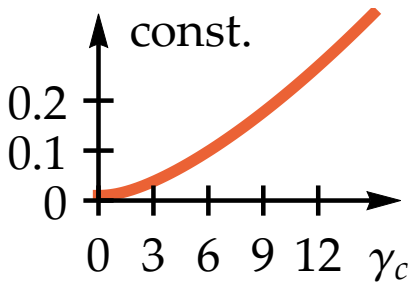

$$\frac{E^{\text{crit}}_{\text{weak}} \cosh[\omega \mathcal{T}_0(\gamma_k)]}{E_{\text{strong}}} \overset{(\text{for } k=0)}{=} \frac{\cosh \gamma_c}{200} \frac{\gamma_c}{I_1(\gamma_c)} = \text{const.} \tag{8.39}$$

Note that the actual instanton trajectory in the field (8.38) will not be a perfect circle, even below the threshold, since the oscillation $E_{\text{weak}} \cosh(\omega \mathcal{T})$ will influence its shape; however, for not too large $\gamma_c$, the ratio (8.39) is very small, so that the effect of the oscillation on the instanton trajectory can be neglected at the threshold and below.

**$E^{\text{crit}}_{\text{weak}}$ for a $k > 0$**

In order to determine $E^{\text{crit}}_{\text{weak}}$ in a Sauter-pulse-shaped background field, we simply define the ratio (8.39) to be always the same at the critical threshold for arbitrary $k > 0$ (finite $L$):

$$\frac{E^{\text{crit}}_{\text{weak}}(\gamma_k) \cosh[\omega \mathcal{T}_0(\gamma_k)]}{E_{\text{strong}}} \overset{!}{=} \frac{E^{\text{crit}}_{\text{weak}}(\gamma_k = 0) \cosh \gamma_c}{E_{\text{strong}}}, \tag{8.40}$$

which is equivalent to

$$E^{\text{crit}}_{\text{weak}}(\gamma_k) = \frac{\cosh \gamma_c}{\cosh\left[\gamma_c \arcsin(\gamma_k) / \left(\gamma_k \sqrt{1 - \gamma_k^2}\right)\right]} \underbrace{E^{\text{crit}}_{\text{weak}}(\gamma_k = 0)}_{E_{\text{strong}} \gamma_c / [200 I_1(\gamma_c)]} . \tag{8.41}$$

Note that, in the constant-field case ($\gamma_k = 0$), $E_{\mathrm{weak}}^{\mathrm{crit}}$ has been defined in Eq. (5.15) as the amplitude at which the oscillation effectively lowers the Sauter–Schwinger tunneling exponent by 1%. However, the way we have just derived Eq. (8.41) does not guarantee that this property is preserved for $\gamma_k > 0$.

By squaring Eq. (8.41), we switch to intensities:

$$I_{\mathrm{crit}}(\gamma_k) = \frac{\cosh^2 \gamma_c}{\cosh^2\left[\gamma_c \arcsin(\gamma_k)/\left(\gamma_k\sqrt{1-\gamma_k^2}\right)\right]} I_{\mathrm{crit}}(\gamma_k = 0). \tag{8.42}$$

**Scaling in the no-tunneling limit**

In the no-tunneling limit $\gamma_k \nearrow 1$, the critical intensity thus approaches zero like

$$I_{\mathrm{crit}}(\gamma_k) \overset{\gamma_k \nearrow 1}{\sim} \frac{\cosh^2 \gamma_c}{\cosh^2\left[\pi\gamma_c/\left(2\sqrt{1-\gamma_k^2}\right)\right]} I_{\mathrm{crit}}(\gamma_k = 0), \tag{8.43}$$

or, equivalently [cf. Eq. (8.37)],

$$\begin{aligned}
I_{\mathrm{crit}}(\gamma_k) \overset{q\Delta\Phi \searrow \mathcal{E}_g}{\sim} & \frac{\cosh^2 \gamma_c}{\cosh^2\left\{\pi\gamma_c/\left[2\sqrt{1-\left(\frac{\mathcal{E}_g}{q\Delta\Phi}\right)^2}\right]\right\}} I_{\mathrm{crit}}(\gamma_k = 0) \\
= & \frac{\cosh^2 \gamma_c}{\cosh^2\left[\pi\gamma_c/\left(2\sqrt{1+\frac{\mathcal{E}_g}{q\Delta\Phi}}\sqrt{1-\frac{\mathcal{E}_g}{q\Delta\Phi}}\right)\right]} I_{\mathrm{crit}}(\gamma_k = 0) \\
\sim & \frac{\cosh^2 \gamma_c}{\cosh^2\left[\pi\gamma_c/\left(2^{3/2}\sqrt{1-\frac{\mathcal{E}_g}{q\Delta\Phi}}\right)\right]} I_{\mathrm{crit}}(\gamma_k = 0) \\
\sim & \exp\left[-\frac{\pi\gamma_c}{\sqrt{2}\sqrt{1-\mathcal{E}_g/(q\Delta\Phi)}}\right].
\end{aligned} \tag{8.44}$$

As explained in the previous subsection, we expect that this scaling of the critical threshold for dynamical assistance is universal just above the tunneling threshold for arbitrary pulse-shaped background fields in a semiconductor analog.

**Example for GaAs**

Let us apply these findings to the GaAs analog. We continue with the scenario described in Sec. 7.3.3 ($E_{\mathrm{strong}} = E_{\mathrm{crit}}^{\mathrm{GaAs}}/10$, assisting $CO_2$ laser, so $\gamma_c = 1.56$), where we found $I_{\mathrm{crit}}(0) = 23.7\,\mathrm{kW/cm^2}$ for a constant background field. Assuming that the built-in field is a spatial Sauter pulse now (with the same $E_{\mathrm{strong}}$ as before), we get the plot of $I_{\mathrm{crit}}(L)$ shown in Fig. 8.2.

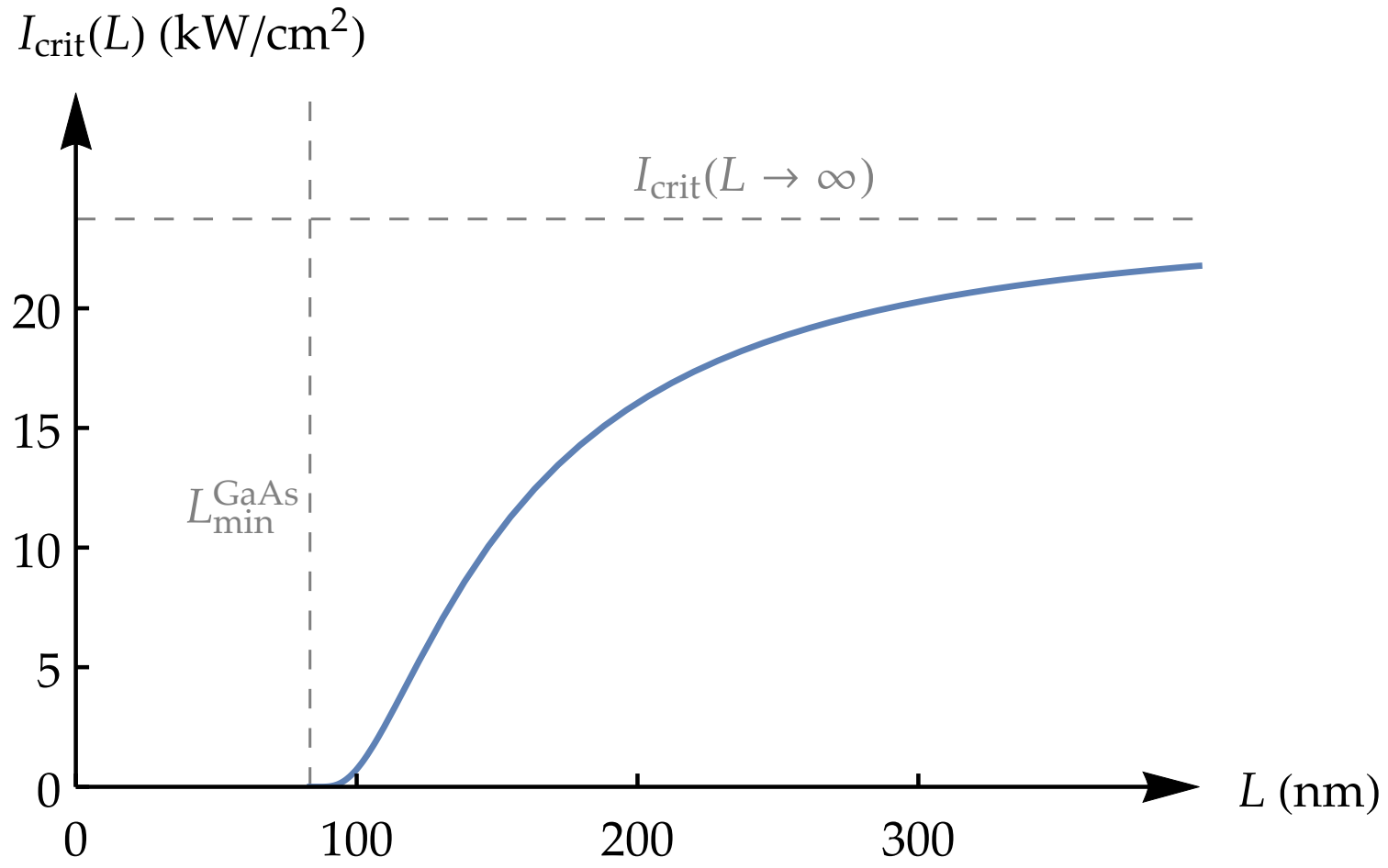

**Figure 8.2.:** Estimated threshold intensity [Eq. (8.42)] for a $CO_2$-laser-induced oscillation $E_{weak} \cos(\omega t)$ in a probe of GaAs with a Sauter-pulse-shaped built-in field $E_{strong} / \cosh^2(kx)$ with $E_{strong} = E_{crit}^{GaAs}/10 = 57\,\mathrm{MV/m}$, plotted as a function of the field width $L = 2\pi/k$. In this example, we have $\gamma_c = 1.56$ [see Eq. (7.97)] and $\gamma_k = L_{min}^{GaAs}/L$ [see Eqs. (8.29), (8.32), and (8.36)], with $L_{min}^{GaAs} = 83.4\,\mathrm{nm}$ corresponding to the no-tunneling limit, where $I_{crit}$ drops to zero. The constant-field limit known from Eq. (7.99) reads $I_{crit}(L \to \infty) = 23.7\,\mathrm{kW/cm^2}$.

## 8.4. Summary

We showed that the quantitative analogy between the long-wavelength parts of $\hat{H}_D$ and $\hat{H}_S$, respectively, continues to hold in spacetime-dependent electric fields in 1+1 dimensions under the following conditions:

1. **Two-band approximation in $\boldsymbol{K} \cdot \boldsymbol{p}$ perturbation theory:** In order to derive the relation (8.26) between $m_\star$ and $\varkappa_0$, we had to neglect the contributions from other bands to the effective mass. This approximation was also required in the purely time-dependent case [see Eqs. (7.73), (7.76), and (7.87)].

2. **Slowly varying potential:** The electric potential and the corresponding field must be approximately constant over the size of a unit cell (lattice constant $\ell$).

3. **Effective electron mass $\stackrel{!}{=}$ effective hole mass:** In a space-dependent field, there is a fundamental physical difference between electrons and holes which are accelerated differently strong and electrons and positrons which have the same rest mass $m$, hence the assumption $m_{\star,e} = m_{\star,h}$ (and thus $m_\star = m_{\star,e} = m_{\star,h}$). Note that this assumption can never be satisfied within two-band $\boldsymbol{K} \cdot \boldsymbol{p}$ perturbation theory [cf. Eq. (7.85)], so, in practice, one has to make a reasonable compromise between this assumption and the first item in this list.

We then studied the analog of the dynamically assisted Sauter–Schwinger effect in GaAs with a space-dependent, Sauter-pulse-shaped background field (e.g., a built-in field due to a p–n junction) by applying the QED results from Ref. [90] to the semiconductor analog. Such background fields can be considered to be slowly varying in GaAs up to the no-tunneling limit if they are subcritical [see Eq. (8.33)]. For an assisting temporal Sauter pulse, the threshold $\omega_{\text{crit}}^{\text{GaAs}}$ decreases in the Sauter-pulse background field as depicted in Fig. 8.1 on page 201. We argued that localized fields in semiconductors always decay exponentially, which renders the scaling $\omega_{\text{crit}}^{\text{SC}} \sim \sqrt{q\Delta\Phi - \mathcal{E}_g}$ in the no-tunneling limit $q\Delta\Phi \searrow \mathcal{E}_g$ of such fields universal (i.e., independent of the pulse shape). Finally, inspired by the worldline-instanton method used in Ref. [90], we estimated the decrease of the critical intensity $I_{\text{crit}}$ for an assisting temporal oscillation with a fixed frequency ($CO_2$ laser) in a Sauter-pulse background field in GaAs (see Fig. 8.2). The corresponding scaling in the no-tunneling limit is given in Eq. (8.44).

### 8.4.1. A remark on interactions

We have ignored the Coulomb interaction between created particles (electrons and positrons in QED, electrons and holes in the semiconductor) throughout this thesis, which allowed us to treat pair-creation problems in the one-electron picture, for example. Let us do a rough estimate (cf., e.g., Ref. [50]) in order to show that this approximation is justified for subcritical electric fields $E < E_{\text{crit}}^{\text{QED}}$ in QED and also in semiconductors.

**QED**

We consider a constant external $E$ field for simplicity. According to the tunneling picture [see Fig. 2.2 on page 48 and Eq. (2.56)], electrons and positrons created via the Sauter–Schwinger effect are separated by the tunneling length $\Delta x_\star = (E_{\text{crit}}^{\text{QED}}/E)(\lambda_C^{\text{QED}}/\pi)$ immediately after their creation (point-particle picture). There are two forces acting on these particles: the **external force**

$$F_{\text{ext}} = qE \tag{8.45}$$

accelerates the particles away from each other while the **Coulomb force**

$$F_{\text{Coul}} = \frac{1}{4\pi\varepsilon_0}\frac{q^2}{\Delta x_\star^2} = \frac{1}{16\pi}\frac{q^2 m^2 c^2}{\varepsilon_0 \hbar^2}\left(\frac{E}{E_{\text{crit}}^{\text{QED}}}\right)^2 = \frac{qE}{4}\underbrace{\frac{q^2}{4\pi\varepsilon_0\hbar c}}_{\alpha_{\text{QED}}}\frac{E}{E_{\text{crit}}^{\text{QED}}} \tag{8.46}$$

is attractive. The quantity $\alpha_{\text{QED}} \approx 1/137$ is the fine-structure constant. The ratio of both forces reads

$$\frac{F_{\text{Coul}}}{F_{\text{ext}}} = \frac{\alpha_{\text{QED}}}{4}\frac{E}{E_{\text{crit}}^{\text{QED}}} < 0.2\% \qquad \text{for} \qquad E < E_{\text{crit}}^{\text{QED}}, \tag{8.47}$$

which indicates that interactions play a minor role in tunneling pair creation in QED via subcritical electric fields.

**Semiconductor**

In the semiconductor analog, we have $E_{\text{crit}}^{\text{QED}} \to E_{\text{crit}}^{\text{SC}}$ and $c \to c_\star$, so we get an **effective fine-structure constant**

$$\alpha_{\text{QED}} \to \frac{q^2}{4\pi\varepsilon_0\hbar c_\star} = \alpha_{\text{QED}}\underbrace{\frac{c}{c_\star}}_{\text{typically} \gg 1}, \tag{8.48}$$

and thus the force ratio becomes

$$\frac{F_{\text{Coul}}}{F_{\text{ext}}} = \frac{\alpha_{\text{QED}}}{4}\frac{c}{c_\star}\frac{E}{E_{\text{crit}}^{\text{SC}}}. \tag{8.49}$$

Due to the large, additional factor $c/c_\star$, which measures approximately 198 in GaAs[3] (cf. Table 7.2 on page 187), interactions have a stronger effect in semiconductor analogs than in QED. However, considering GaAs again, if we have a strong external field $E = E_{\rm crit}^{\rm GaAs}/10$ (i.e., close to the breakdown field), the resulting ratio is still small:

**Example for GaAs**

$$\frac{F_{\rm Coul}}{F_{\rm ext}} = \frac{\alpha_{\rm QED}}{4} \frac{c}{c_\star^{\rm GaAs}} \frac{E}{E_{\rm crit}^{\rm GaAs}} = 3.6\%. \tag{8.50}$$

We take this result as a hint that neglecting interactions is a good approximation in the context of (assisted) tunneling pair creation, even in the semiconductor analog. This is consistent with Ref. [159], according to which many-body effects "generally appear as merely small bumps and wiggles on the tunneling curves."

[3]Compare this to the corresponding value $c/c_\star \approx 300$ in graphene [119], for example.

# 9. Analogy in 2+1 spacetime dimensions for crossed constant electric and magnetic fields

The goal of this final chapter is to make some first steps towards the generalization of the analogy between Dirac theory and two-band semiconductors to 2+1 spacetime dimensions. Besides GaAs, a potentially interesting analog system for this scenario is semiconducting graphene [121, 118]. Two space dimensions allow us to incorporate also the effect of **magnetic ($B$) fields** which are oriented **perpendicular** to the movement of the charge carriers (electrons, positrons, holes) and to the electric field. There is still **no spin** in 2+1 dimensions. We will only consider one specific field profile here: a **constant electric field plus a constant, effectively perpendicular magnetic field** ("crossed fields"). As explained in Sec. 2.6, it is well known that perpendicular magnetic fields reduce the pair-creation yield in QED by **effectively weakening the electric field which creates pairs** via the Sauter–Schwinger effect (see also Ref. [100]). The analog of this effect is known to occur in semiconductors [104, 97, 96, 106].

In this chapter, we will derive the two-band semiconductor Hamiltonian for crossed, constant fields in 2+1 dimensions in total analogy to the previous two chapters. However, we will not perform a detailed comparison between this Hamiltonian and the corresponding Dirac Hamiltonian, but rather rederive some of the known results concerning the analogy (see Refs. [104, 97, 96, 106]) from this semiconductor Hamiltonian as an outlook.

**General electromagnetic field in 2+1 spacetime dimensions**

Let us start with the basic formalism. In a 2+1-dimensional spacetime, we have $A_\mu = (-\Phi/c, A_x, A_y)$ and $\partial_\mu = (\partial_t/c, \partial_x, \partial_y)$ with $\mu \in \{0, 1, 2\}$, so the electromagnetic field tensor (1.23) reads

$$(F_{\mu\nu}) = \frac{1}{c}\begin{pmatrix} 0 & \partial_x\Phi + \dot{A}_x & \partial_y\Phi + \dot{A}_y \\ -\partial_x\Phi - \dot{A}_x & 0 & c\partial_x A_y - c\partial_y A_x \\ -\partial_y\Phi - \dot{A}_y & c\partial_y A_x - c\partial_x A_y & 0 \end{pmatrix}$$

$$\overset{!}{=} \frac{1}{c}\begin{pmatrix} 0 & E_x & E_y \\ -E_x & 0 & -cB \\ -E_y & cB & 0 \end{pmatrix} \tag{9.1}$$

(each component of the fields or $A_\mu$ may depend on the spacetime coordinates $t$, $x$, and $y$ in general). We can read off the components of the electromagnetic field from this tensor: the **electric field** is the two-dimensional[1] vector field

$$\vec{E}(t,\vec{r}) = \begin{pmatrix} E_x(t,\vec{r}) \\ E_y(t,\vec{r}) \end{pmatrix} = \begin{pmatrix} \partial_x \Phi(t,\vec{r}) + \dot{A}_x(t,\vec{r}) \\ \partial_y \Phi(t,\vec{r}) + \dot{A}_y(t,\vec{r}) \end{pmatrix} = \vec{\nabla}\Phi(t,\vec{r}) + \dot{\vec{A}}(t,\vec{r}), \tag{9.2}$$

and the **magnetic field**

$$B(t,\vec{r}) = \partial_y A_x(t,\vec{r}) - \partial_x A_y(t,\vec{r}) \tag{9.3}$$

is scalar in a 2+1-dimensional spacetime. Note that this scalar coincides with the $z$ component of the magnetic (vector) field in 3+1 dimensions,

$$\boldsymbol{B}(t,\boldsymbol{r}) \cdot \boldsymbol{e}_z = [-\boldsymbol{\nabla} \times \boldsymbol{A}(t,\boldsymbol{r})] \cdot \boldsymbol{e}_z = \partial_y A_x(t,\boldsymbol{r}) - \partial_x A_y(t,\boldsymbol{r}). \tag{9.4}$$

We may thus imagine the 2+1-dimensional scenario as taking place in the $x$–$y$ plane of three-dimensional space (without spin), with the field components $E_z$, $B_x$, and $B_y$ set to zero since these components could accelerate charged particles "living" in the $x$–$y$ plane away from this plane due to the Lorentz force. In this picture, the electric field (9.2) is always perpendicular to the magnetic field (9.3) ("crossed fields").

**Field profile**

Throughout this chapter, we focus on the simplest case of **constant** and (necessarily) **crossed electric and magnetic fields**, so we set, without loss of generality, $\vec{E} = E\vec{e}_x$ with $E > 0$ and $B > 0$ constant and choose the corresponding potentials

$$\Phi(x) = Ex \qquad \text{and} \qquad \vec{A}(x) = -Bx\vec{e}_y. \tag{9.5}$$

This gauge has the advantage that it depends on $x$ only, such that the **energy** and the ***y* component of the canonical momentum** are **conserved**.

## 9.1. Many-body Hamiltonians

### 9.1.1. Dirac theory

**Gamma matrices**

In 1+1 spacetime dimensions, we chose the gamma matrices $\gamma^0 = \sigma_z$ and $\gamma^1 = \mathrm{i}\sigma_y$. Hence, the third gamma matrix which is required in 2+1 dimensions can be expressed by the remaining Pauli matrix (1.7) via $\gamma^2 = -\mathrm{i}\sigma_x$ such that the Clifford algebra (1.5) is satisfied, so Dirac spinors still have only two components in 2+1 dimensions (**no spin**). The resulting components of the vector $\vec{\alpha}$ of matrices [see Eq. (1.13)] read

$$\alpha_x = \gamma^0 \cdot \gamma^1 = \begin{pmatrix} 1 & 0 \\ 0 & -1 \end{pmatrix} \cdot \begin{pmatrix} 0 & 1 \\ -1 & 0 \end{pmatrix} = \begin{pmatrix} 0 & 1 \\ 1 & 0 \end{pmatrix} \quad \text{and}$$
$$\alpha_y = \gamma^0 \cdot \gamma^2 = \begin{pmatrix} 1 & 0 \\ 0 & -1 \end{pmatrix} \cdot \begin{pmatrix} 0 & -\mathrm{i} \\ -\mathrm{i} & 0 \end{pmatrix} = \begin{pmatrix} 0 & -\mathrm{i} \\ \mathrm{i} & 0 \end{pmatrix}. \tag{9.6}$$

[1] We indicate vectors in two-dimensional space by vector arrows ($\vec{v}$) in order to distinguish them from three-dimensional vectors ($\boldsymbol{v}$), which are printed in boldface.

**Hamiltonian**

From the general Schrödinger form (1.12) of the classical Dirac equation in 3+1 dimensions, we can now read off the single-electron Hamiltonian in our potential (9.5) here:

$$\hat{H}_D^{\text{one}}(x) = \begin{pmatrix} mc^2 - qEx & -\mathrm{i}c\hbar\partial_x - \mathrm{i}c(-\mathrm{i}\hbar\partial_y - qBx) \\ -\mathrm{i}c\hbar\partial_x + \mathrm{i}c(-\mathrm{i}\hbar\partial_y - qBx) & -mc^2 - qEx \end{pmatrix}. \tag{9.7}$$

In analogy to Eq. (7.2), the real-space representation of the corresponding many-body Dirac Hamiltonian thus reads

$$\hat{H}_D(t) = \iint\limits_{\mathbb{R}^2} \underline{\hat{\Psi}}^\dagger(t,\vec{r}) \left[ \begin{pmatrix} mc^2 & -\mathrm{i}c\hbar\partial_x - c\hbar\partial_y \\ -\mathrm{i}c\hbar\partial_x + c\hbar\partial_y & -mc^2 \end{pmatrix} - q \begin{pmatrix} E & -\mathrm{i}cB \\ \mathrm{i}cB & E \end{pmatrix} x \right] \underline{\hat{\Psi}}(t,\vec{r})\, \mathrm{d}^2 r. \tag{9.8}$$

**Transformation to $k$ space**

We transform this Hamiltonian to reciprocal space by inserting the two-dimensional form of the spatial Fourier transform (7.4),

$$\underline{\hat{\Psi}}(t,\vec{r}) = \frac{1}{2\pi} \iint\limits_{\mathbb{R}^2} \underline{\hat{\tilde{\Psi}}}(t,\vec{k})\, \mathrm{e}^{\mathrm{i}\vec{k}\cdot\vec{r}}\, \mathrm{d}^2 k, \tag{9.9}$$

which yields

$$\hat{H}_D(t) = \iint\limits_{\mathbb{R}^2} \iint\limits_{\mathbb{R}^2} \underline{\hat{\tilde{\Psi}}}^\dagger(t,\vec{k}) \frac{1}{4\pi^2} \iint\limits_{\mathbb{R}^2} \left[ \begin{pmatrix} mc^2 & c\hbar k'_x - \mathrm{i}c\hbar k'_y \\ c\hbar k'_x + \mathrm{i}c\hbar k'_y & -mc^2 \end{pmatrix} - q \begin{pmatrix} E & -\mathrm{i}cB \\ \mathrm{i}cB & E \end{pmatrix} x \right] \mathrm{e}^{\mathrm{i}(\vec{k}'-\vec{k})\cdot\vec{r}}\, \mathrm{d}^2 r\, \underline{\hat{\tilde{\Psi}}}(t,\vec{k}')\, \mathrm{d}^2 k\, \mathrm{d}^2 k'. \tag{9.10}$$

This expression contains the Fourier transform of $x$ (in the distributional sense), which becomes a differential operator $\mathrm{i}\partial_{k_x}$ in $k$ space since

$$\begin{aligned}\frac{1}{4\pi^2}\iint\limits_{\mathbb{R}^2} x\mathrm{e}^{\mathrm{i}(\vec{k}'-\vec{k})\cdot\vec{r}}\,\mathrm{d}^2r &= \frac{1}{2\pi}\int\limits_{-\infty}^{\infty}\underbrace{x}_{-\mathrm{i}\partial_{k'_x}}\mathrm{e}^{\mathrm{i}(k'_x-k_x)x}\,\mathrm{d}x\;\underbrace{\frac{1}{2\pi}\int\limits_{-\infty}^{\infty}\mathrm{e}^{\mathrm{i}(k'_y-k_y)y}\,\mathrm{d}y}_{\delta(k'_y-k_y)}\\ &= -\mathrm{i}\,\frac{\partial\delta(k'_x-k_x)}{\partial k'_x}\,\delta(k'_y-k_y)\\ &= \mathrm{i}\,\underbrace{\delta(k'_x-k_x)\,\delta(k'_y-k_y)}_{\delta^{(2)}(\vec{k}'-\vec{k})}\,\partial_{k'_x}\,, \end{aligned}\tag{9.11}$$

**Result**

so the resulting **momentum-space representation of the Dirac Hamiltonian** reads

$$\begin{aligned}\hat{H}_D(t) = \iint\limits_{\mathbb{R}^2}\hat{\underline{\tilde{\Psi}}}^{\dagger}(t,\vec{k})\Bigg[&\begin{pmatrix} mc^2 & c\hbar k_x - \mathrm{i}c\hbar k_y\\ c\hbar k_x + \mathrm{i}c\hbar k_y & -mc^2\end{pmatrix}\\ &- \mathrm{i}q\begin{pmatrix} E & -\mathrm{i}cB\\ \mathrm{i}cB & E\end{pmatrix}\partial_{k_x}\Bigg]\hat{\underline{\tilde{\Psi}}}(t,\vec{k})\,\mathrm{d}^2k.\end{aligned}\tag{9.12}$$

### 9.1.2. Semiconductor

**Full Hamiltonian in real space**

In the semiconductor, we start with the **nonrelativistic Schrödinger Hamiltonian** [cf. Eq. (7.12)] in real space again, which describes all Bloch electrons. In 2+1 dimensions and for the crossed-fields profile (9.5), it reads

$$\begin{aligned}&\hat{H}_S^{\mathrm{full}}(t)\\ &= \iint\limits_{\mathbb{R}^2}\hat{\psi}^{\dagger}(t,\vec{r})\left[\frac{-\hbar^2\partial_x^2 + (-\mathrm{i}\hbar\partial_y - qBx)^2}{2m} + V(\vec{r}) - qEx\right]\hat{\psi}(t,\vec{r})\,\mathrm{d}^2r\\ &= \iint\limits_{\mathbb{R}^2}\hat{\psi}^{\dagger}(t,\vec{r})\left[-\frac{\hbar^2\vec{\nabla}^2}{2m} + V(\vec{r}) - qEx - \frac{qB}{m}\hat{p}_y x + \frac{q^2B^2}{2m}x^2\right]\hat{\psi}(t,\vec{r})\,\mathrm{d}^2r,\end{aligned}\tag{9.13}$$

where $V(\vec{r})$ denotes the periodic crystal potential in two-dimensional space (we leave the **lattice type unspecified** here) and $\hat{p}_y = -\mathrm{i}\hbar\partial_y$.

**Bloch waves in 2+1 spacetime dimensions**

#### 9.1.2.1. Transformation to crystal-momentum space

In complete analogy to the 1+1-dimensional case, we assume that the Bloch

waves

$$f_n(\vec{K},\vec{r}) = \mathrm{e}^{\mathrm{i}\vec{K}\cdot\vec{r}} u_n(\vec{K},\vec{r}), \tag{9.14}$$

which solve the eigenvalue equation

$$\left[-\frac{\hbar^2\vec{\nabla}^2}{2m} + V(\vec{r})\right] f_n(\vec{K},\vec{r}) = \mathcal{E}_n(\vec{K}) f_n(\vec{K},\vec{r}), \tag{9.15}$$

are **orthonormalized**:

$$\langle n,\vec{K}|n',\vec{K}'\rangle = \iint\limits_{\mathbb{R}^2} f_n^*(\vec{K},\vec{r}) f_{n'}(\vec{K}',\vec{r})\,\mathrm{d}^2r \stackrel{!}{=} \delta_{nn'}\delta^{(2)}(\vec{K}'-\vec{K}). \tag{9.16}$$

Note that the $\vec{K}$'s are always elements of the first (two-dimensional) Brillouin zone, whose shape depends on the lattice type.

**General theorem**

In order to derive the resulting orthonormalization of the Bloch factors $u_n(\vec{K},\vec{r})$, we use the generalized form of the well-known theorem (7.24) for more than one spatial dimension (see, e.g., Ref. [151]): We can write any lattice-periodic function $g(\vec{r})$ as

$$g(\vec{r}) = \sum_{\vec{G}} \tilde{g}_{\vec{G}} \mathrm{e}^{\mathrm{i}\vec{G}\cdot\vec{r}}, \tag{9.17}$$

where the summation runs over all reciprocal-lattice vectors $\vec{G}$, and the corresponding $\tilde{g}_{\vec{G}}$ are the (complex) coefficients describing the function in reciprocal space [analog of the Fourier coefficients (7.22) in 1+1 dimensions]. Now consider the expression

$$\begin{aligned}\iint\limits_{\mathbb{R}^2} \mathrm{e}^{\mathrm{i}(\vec{K}'-\vec{K})\cdot\vec{r}} g(\vec{r})\,\mathrm{d}^2r &= \sum_{\vec{G}} \tilde{g}_{\vec{G}} \iint\limits_{\mathbb{R}^2} \mathrm{e}^{\mathrm{i}(\vec{K}'-\vec{K}+\vec{G})\cdot\vec{r}}\,\mathrm{d}^2r \\ &= 4\pi^2 \sum_{\vec{G}} \tilde{g}_{\vec{G}}\, \delta^{(2)}(\vec{K}'-\vec{K}+\vec{G}). \end{aligned} \tag{9.18}$$

Assuming that $\vec{K}$ and $\vec{K}'$ are both elements of the first Brillouin zone (Wigner–Seitz cell in reciprocal space), $\vec{K}'-\vec{K}$ can never coincide with any reciprocal-lattice vector except for $\vec{G}=0$.[2] Together with the inversion of Eq. (9.17),

$$\tilde{g}_{\vec{G}} = \frac{1}{\mathcal{V}_{\mathrm{cell}}} \iint\limits_{\mathrm{cell}} g(\vec{r})\mathrm{e}^{-\mathrm{i}\vec{G}\cdot\vec{r}}\,\mathrm{d}^2r \tag{9.19}$$

**Resulting theorem**

(see, e.g., Refs. [152, 151, 154]), where $\mathcal{V}_{\mathrm{cell}}$ denotes the volume (here: area) of

[2] This argument may not hold for $\vec{K}$ and $\vec{K}'$ lying very close to opposite edges of the Brillouin zone [151]. However, since our main focus will be long-wavelength processes at the zone center anyway, our derivation here should be valid for that purpose (cf., e.g., Ref. [152]).

a unit cell, we thus get

$$\iint\limits_{\mathbb{R}^2} e^{i(\vec{K}'-\vec{K})\cdot\vec{r}} g(\vec{r})\, d^2r = 4\pi^2 \tilde{g}_{\vec{0}}\, \delta^{(2)}(\vec{K}'-\vec{K})$$
$$= \frac{4\pi^2}{\mathcal{V}_{\text{cell}}} \iint\limits_{\text{cell}} g(\vec{r})\, d^2r\, \delta^{(2)}(\vec{K}'-\vec{K}). \tag{9.20}$$

Compare this general result to Eq. (7.24) for 1+1 dimensions (in which case $\mathcal{V}_{\text{cell}} = \ell$).

**Bloch factors are orthonormal**

Applying this theorem to the Bloch-wave orthonormality relation (9.16) yields

$$\delta_{nn'}\delta^{(2)}(\vec{K}'-\vec{K}) \stackrel{!}{=} \iint\limits_{\mathbb{R}^2} e^{i(\vec{K}'-\vec{K})\cdot\vec{r}} \overbrace{u_n^*(\vec{K},\vec{r})u_{n'}(\vec{K}',\vec{r})}^{\text{lattice periodic}}\, d^2r$$
$$= \frac{4\pi^2}{\mathcal{V}_{\text{cell}}} \iint\limits_{\text{cell}} u_n^*(\vec{K},\vec{r})u_{n'}(\vec{K}',\vec{r})\, d^2r\, \delta^{(2)}(\vec{K}'-\vec{K})$$
$$= \langle n,\vec{K}|n',\vec{K}\rangle_{\text{cell}}\, \delta^{(2)}(\vec{K}'-\vec{K}), \tag{9.21}$$

where we have defined the **unit-cell product** between Bloch factors in 2+1 dimensions. From this equation follows the **orthonormality** of the Bloch factors for a fixed $\vec{K}$ over a unit cell:

$$\langle n,\vec{K}|n',\vec{K}\rangle_{\text{cell}} = \frac{4\pi^2}{\mathcal{V}_{\text{cell}}} \iint\limits_{\text{cell}} u_n^*(\vec{K},\vec{r})u_{n'}(\vec{K},\vec{r})\, d^2r = \delta_{nn'}. \tag{9.22}$$

### Many-body Hamiltonian in crystal-momentum space

By expanding

$$\hat{\psi}(t,\vec{r}) = \sum_{n=1}^{\infty} \iint\limits_{\text{BZ}} f_n(\vec{K},\vec{r})\hat{a}_n(t,\vec{K})\, d^2K \tag{9.23}$$

**Full Hamiltonian in crystal-momentum space**

(BZ: first Brillouin zone) in $\hat{H}_S^{\text{full}}$ in Eq. (9.13), we obtain

$$\hat{H}_S^{\text{full}}(t)$$
$$= \iint\limits_{\mathbb{R}^2} \sum_{n=1}^{\infty} \iint\limits_{\text{BZ}} f_n^*(\vec{K},\vec{r})\hat{a}_n^\dagger(t,\vec{K})\, d^2K \sum_{n'=1}^{\infty} \iint\limits_{\text{BZ}} \Big[\mathcal{E}_{n'}(\vec{K}') - qEx - \frac{qB}{m}\hat{p}_y x$$
$$+ \frac{q^2B^2}{2m}x^2\Big] f_{n'}(\vec{K}',\vec{r})\hat{a}_{n'}(t,\vec{K}')\, d^2K'\, d^2r$$

$$= \sum_{n=1}^{\infty} \sum_{n'=1}^{\infty} \iint\limits_{\mathrm{BZ}} \iint\limits_{\mathrm{BZ}} \hat{a}_n^\dagger(t,\vec{K}) \Big[ \mathcal{E}_{n'}(\vec{K}') \langle n,\vec{K}|n',\vec{K}'\rangle - qE \langle n,\vec{K}|x|n',\vec{K}'\rangle$$
$$- \frac{qB}{m} \langle n,\vec{K}|\hat{p}_y x|n',\vec{K}'\rangle + \frac{q^2B^2}{2m} \langle n,\vec{K}|x^2|n',\vec{K}'\rangle \Big] \hat{a}_{n'}(t,\vec{K}')\, \mathrm{d}^2K'\, \mathrm{d}^2K. \quad (9.24)$$

Hence, we need to calculate the matrix elements of $x$, $x^2$, and $\hat{p}_y x$ in the Bloch-wave basis.

#### 9.1.2.2. Matrix elements in the Bloch-wave basis

**Matrix elements of $x$**

These matrix elements can be calculated by expressing $x$ as a partial derivative with respect to $K_x$; see Refs. [150, 151, 153]. Note that we have to treat these matrix elements as distributions which act on the annihilation operator $\hat{a}_{n'}(t,\vec{K}')$. Using the Bloch-wave orthonormality (9.16) and the theorem (9.20), we get

$$\begin{aligned}
\langle n,\vec{K}|x|n',\vec{K}'\rangle &= -\mathrm{i} \iint\limits_{\mathbb{R}^2} f_n^*(\vec{K},\vec{r}) u_{n'}(\vec{K}',\vec{r}) \frac{\partial \mathrm{e}^{\mathrm{i}\vec{K}'\cdot\vec{r}}}{\partial K'_x}\, \mathrm{d}^2r \\
&= -\mathrm{i}\frac{\partial \langle n,\vec{K}|n',\vec{K}'\rangle}{\partial K'_x} + \mathrm{i} \iint\limits_{\mathbb{R}^2} \mathrm{e}^{\mathrm{i}(\vec{K}'-\vec{K})\cdot\vec{r}} \underbrace{u_n^*(\vec{K},\vec{r}) \frac{\partial u_{n'}(\vec{K}',\vec{r})}{\partial K'_x}}_{\text{lattice periodic}} \mathrm{d}^2r \\
&= -\mathrm{i}\delta_{nn'} \frac{\partial \delta^{(2)}(\vec{K}'-\vec{K})}{\partial K'_x} + \langle n,\vec{K}|\mathrm{i}\partial_{K_x}|n',\vec{K}\rangle_{\text{cell}}\, \delta^{(2)}(\vec{K}'-\vec{K}) \\
&= \delta^{(2)}(\vec{K}'-\vec{K}) \Big( \mathrm{i}\delta_{nn'}\partial_{K'_x} + \underbrace{\langle n,\vec{K}|\mathrm{i}\partial_{K_x}|n',\vec{K}\rangle_{\text{cell}}}_{U_{nn'}(\vec{K})} \Big),
\end{aligned} \quad (9.25)$$

where we have shifted $\partial_{K'_x}$ from the delta distribution to the annihilation operator (which we did not write explicitly here) via integration by parts in the last line.

**Unit-cell matrix elements of $\mathrm{i}\partial_{K_x}$**

The unit-cell matrix elements $U_{nn'}(\vec{K})$ defined in this equation may be calculated by "linearizing" the right-hand Bloch factor $u_{n'}(\vec{K},\vec{r})$ with respect to $\vec{K}$ around this $\vec{K}$ via (nondegenerate) first-order $\boldsymbol{K} \cdot \boldsymbol{p}$ perturbation theory [122]

[this is a generalized form of Eq. (7.68)]:

$$u_n(\vec{K}+\Delta\vec{K},\vec{r}) = u_n(\vec{K},\vec{r}) + \frac{\hbar}{m}\Delta\vec{K}\cdot\sum_{\tilde{n}\in\mathbb{N}}^{\tilde{n}\neq n}\frac{\langle\tilde{n},\vec{K}|\hat{\vec{p}}|n,\vec{K}\rangle_{\text{cell}}}{\mathcal{E}_n(\vec{K})-\mathcal{E}_{\tilde{n}}(\vec{K})}u_{\tilde{n}}(\vec{K},\vec{r}) + \mathcal{O}(\Delta\vec{K}^2). \tag{9.26}$$

Using this perturbational expansion, we find

$$U_{nn'}(\vec{K}) = \left\langle u_n(\vec{K},\vec{r}) \,\middle|\, \mathrm{i}\frac{\partial u_{n'}(\vec{K}+\Delta\vec{K},\vec{r})}{\partial\Delta K_x}\Big|_{\Delta\vec{K}=0}\right\rangle_{\text{cell}} = \frac{\mathrm{i}\hbar}{m}\frac{\langle n,\vec{K}|\hat{p}_x|n',\vec{K}\rangle_{\text{cell}}}{\mathcal{E}_{n'}(\vec{K})-\mathcal{E}_n(\vec{K})}(1-\delta_{nn'}), \tag{9.27}$$

so the interband elements of $U_{nn'}$ (i.e., $n \neq n'$) can be expressed in terms of momentum-matrix elements (the same result is derived in Ref. [136] in a more rigorous way), while the intraband elements ($n = n'$) are all zero according to this calculation[3].

**Result** Inserting Eq. (9.27) into the above matrix elements of $x$ yields

$$\langle n,\vec{K}|x|n',\vec{K}'\rangle = \delta^{(2)}(\vec{K}'-\vec{K})\begin{cases}\mathrm{i}\partial_{K'_x} & \text{if } n=n',\\ U_{nn'}(\vec{K}) & \text{if } n\neq n'.\end{cases} \tag{9.28}$$

### Matrix elements of $x^2$

Here, we make use of the completeness of the Bloch waves,

$$\mathbb{1} = \sum_{n=1}^{\infty}\iint_{\text{BZ}}|n,\vec{K}\rangle\,\langle n,\vec{K}|\;\mathrm{d}^2K, \tag{9.29}$$

and the completeness of all Bloch factors for a fixed, arbitrary $\vec{K}$ on the space of lattice-periodic functions, which we write as

$$\mathbb{1}_{\text{cell}} = \sum_{n=1}^{\infty}|n,\vec{K}\rangle\,\langle n,\vec{K}|_{\text{cell}} \qquad \forall\vec{K}. \tag{9.30}$$

[3] Note that the relative phases between the Bloch factors within the $n$th band have to be chosen in a specific way in order for this statement to hold [136, 137, 160]. A diagonal element $U_{nn}(\vec{K})$ cannot vanish identically for all $\vec{K}$ since its integral over the whole Brillouin zone ("Berry/Zak's phase") must be a nonvanishing, phase-invariant quantity; see Refs. [160, 161, 162]. However, we *can* make $U_{nn}(\vec{K})$ vanish within an arbitrary large part of the Brillouin zone via a suitable phase choice, so, as long as the electrons we consider do not perform full cycles through the Brillouin zone, we may assume $U_{nn}(\vec{K}) = 0$, which we will do in the following.

This allows us to reuse the previous result (9.25) by inserting $\mathbb{1}$:

$$\begin{aligned}
\langle n,\vec{K}|x^2|n',\vec{K}'\rangle &= \langle n,\vec{K}|x\,\mathbb{1}\,x|n',\vec{K}'\rangle \\
&= \sum_{\tilde{n}=1}^{\infty} \iint_{\mathrm{BZ}} \langle n,\vec{K}|x|\tilde{n},\tilde{\vec{K}}\rangle\,\langle \tilde{n},\tilde{\vec{K}}|x|n',\vec{K}'\rangle\;\mathrm{d}^2\tilde{K} \\
&= \sum_{\tilde{n}=1}^{\infty} \iint_{\mathrm{BZ}} \delta^{(2)}(\tilde{\vec{K}}-\vec{K})\left[\mathrm{i}\delta_{n\tilde{n}}\partial_{\tilde{K}_x} + U_{n\tilde{n}}(\vec{K})\right] \\
&\qquad\qquad \times \delta^{(2)}(\vec{K}'-\tilde{\vec{K}})\left[\mathrm{i}\delta_{\tilde{n}n'}\partial_{K'_x} + U_{\tilde{n}n'}(\tilde{\vec{K}})\right]\mathrm{d}^2\tilde{K} \\
&= \delta^{(2)}(\vec{K}'-\vec{K})\left[-\delta_{nn'}\partial^2_{K'_x} + 2U_{nn'}(\vec{K})\mathrm{i}\partial_{K'_x} + \mathrm{i}\frac{\partial U_{nn'}(\vec{K})}{\partial K_x}\right. \\
&\qquad\qquad \left. + \sum_{\tilde{n}=1}^{\infty} U_{n\tilde{n}}(\vec{K})U_{\tilde{n}n'}(\vec{K})\right]. \qquad (9.31)
\end{aligned}$$

(Note that the sum over $\tilde{n}$ yields

$$\begin{aligned}
\sum_{\tilde{n}=1}^{\infty} U_{n\tilde{n}}(\vec{K})U_{\tilde{n}n'}(\vec{K}) &= \langle n,\vec{K}|\,\mathrm{i}\partial_{K_x} \underbrace{\sum_{\tilde{n}=1}^{\infty} |\tilde{n},\vec{K}\rangle_{\mathrm{cell}}\,\langle \tilde{n},\vec{K}|}_{\mathbb{1}_{\mathrm{cell}}}\,\mathrm{i}\partial_{K_x}\,|n',\vec{K}\rangle_{\mathrm{cell}} \\
&= \langle n,\vec{K}| - \partial^2_{K_x}|n',\vec{K}\rangle_{\mathrm{cell}}, \qquad (9.32)
\end{aligned}$$

that is, the matrix elements of $-\partial^2_{K_x}$. However, we will stick to the sum representation in the following.) Since $U_{\tilde{n}n}(\vec{K}) = U^*_{n\tilde{n}}(\vec{K})$, we may also write

**Result**

$$\begin{aligned}
&\langle n,\vec{K}|x^2|n',\vec{K}'\rangle = \delta^{(2)}(\vec{K}'-\vec{K}) \\
&\times \begin{cases} -\partial^2_{K'_x} + \sum_{\tilde{n}\in\mathbb{N}}^{\tilde{n}\neq n} |U_{n\tilde{n}}(\vec{K})|^2 & \text{if } n = n', \\ 2U_{nn'}(\vec{K})\mathrm{i}\partial_{K'_x} + \mathrm{i}\frac{\partial U_{nn'}(\vec{K})}{\partial K_x} + \sum_{\tilde{n}\in\mathbb{N}}^{\tilde{n}\neq n,n'} U_{n\tilde{n}}(\vec{K})U_{\tilde{n}n'}(\vec{K}) & \text{if } n \neq n'. \end{cases} \qquad (9.33)
\end{aligned}$$

**General momentum-matrix elements**

Before we calculate the matrix elements of $\hat{p}_y x$, let us first note that Eqs. (7.30) and (7.39) derived in Secs. 7.1.2.1 and 7.1.2.2 for a 1+1-dimensional spacetime are straightforward to generalize for two or three spatial dimensions, so we have

$$\begin{aligned}
\langle n,\vec{K}|\hat{\vec{p}}|n',\vec{K}'\rangle &= \begin{pmatrix} \langle n,\vec{K}|\hat{p}_x|n',\vec{K}'\rangle \\ \langle n,\vec{K}|\hat{p}_y|n',\vec{K}'\rangle \end{pmatrix} \\
&= \left(\hbar\vec{K}\delta_{nn'} + \langle n,\vec{K}|\hat{\vec{p}}|n',\vec{K}\rangle_{\mathrm{cell}}\right)\delta^{(2)}(\vec{K}'-\vec{K}) \qquad (9.34)
\end{aligned}$$

in general here in 2+1 dimensions, and, in particular, the intraband momentum-matrix elements are related to the **group velocity** via

$$\langle n, \vec{K} | \hat{\vec{p}} | n, \vec{K}' \rangle = m \vec{v}_n^{\mathrm{gr}}(\vec{K})\, \delta^{(2)}(\vec{K}' - \vec{K}) = \frac{m}{\hbar} \frac{\partial \mathcal{E}_n(\vec{K})}{\partial \vec{K}}\, \delta^{(2)}(\vec{K}' - \vec{K}); \qquad (9.35)$$

cf., e.g., Refs. [154, 148].

**Matrix elements of $\hat{p}_y x$**

In analogy to the matrix elements of $x^2$, we calculate these matrix elements by inserting an identity (9.29) and reusing the above results (9.25), (9.34), and (9.35):

$$\begin{aligned}
&\langle n, \vec{K} | \hat{p}_y x | n', \vec{K}' \rangle \\
&= \langle n, \vec{K} | \hat{p}_y \, \mathbb{1} \, x | n', \vec{K}' \rangle \\
&= \sum_{\tilde{n}=1}^{\infty} \iint\limits_{\mathrm{BZ}} \langle n, \vec{K} | \hat{p}_y | \tilde{n}, \tilde{\vec{K}} \rangle \, \langle \tilde{n}, \tilde{\vec{K}} | x | n', \vec{K}' \rangle \; \mathrm{d}^2 \tilde{K} \\
&= \sum_{\tilde{n}=1}^{\infty} \iint\limits_{\mathrm{BZ}} \delta^{(2)}(\tilde{\vec{K}} - \vec{K}) \left[ \hbar K_y \delta_{n\tilde{n}} + \langle n, \vec{K} | \hat{p}_y | \tilde{n}, \vec{K} \rangle_{\mathrm{cell}} \right] \\
&\qquad \times \delta^{(2)}(\vec{K}' - \tilde{\vec{K}}) \left[ \mathrm{i} \delta_{\tilde{n}n'} \partial_{K'_x} + U_{\tilde{n}n'}(\tilde{\vec{K}}) \right] \mathrm{d}^2 \tilde{K} \\
&= \delta^{(2)}(\vec{K}' - \vec{K}) \left[ \left( \hbar K_y \delta_{nn'} + \langle n, \vec{K} | \hat{p}_y | n', \vec{K} \rangle_{\mathrm{cell}} \right) \mathrm{i} \partial_{K'_x} + \hbar K_y U_{nn'}(\vec{K}) \right. \\
&\qquad \left. + \sum_{\tilde{n}=1}^{\infty} \langle n, \vec{K} | \hat{p}_y | \tilde{n}, \vec{K} \rangle_{\mathrm{cell}} \, U_{\tilde{n}n'}(\vec{K}) \right]. 
\end{aligned} \qquad (9.36)$$

**Result**

Hence, the **intraband elements** ($n' = n$) are

$$\begin{aligned}
\langle n, \vec{K} | \hat{p}_y x | n, \vec{K}' \rangle &= \delta^{(2)}(\vec{K}' - \vec{K}) \\
&\quad \times \left[ m v_{n,y}^{\mathrm{gr}}(\vec{K}) \mathrm{i} \partial_{K'_x} + \sum_{\tilde{n} \in \mathbb{N}}^{\tilde{n} \neq n} \langle n, \vec{K} | \hat{p}_y | \tilde{n}, \vec{K} \rangle_{\mathrm{cell}} \, U_{\tilde{n}n}(\vec{K}) \right],
\end{aligned} \qquad (9.37)$$

and the **interband elements** read

$$\begin{aligned}
\langle n, \vec{K} | \hat{p}_y x | n', \vec{K}' \rangle &= \delta^{(2)}(\vec{K}' - \vec{K}) \left[ \langle n, \vec{K} | \hat{p}_y | n', \vec{K} \rangle_{\mathrm{cell}} \, \mathrm{i} \partial_{K'_x} + \hbar K_y U_{nn'}(\vec{K}) \right. \\
&\quad \left. + \sum_{\tilde{n} \in \mathbb{N}}^{\tilde{n} \neq n'} \langle n, \vec{K} | \hat{p}_y | \tilde{n}, \vec{K} \rangle_{\mathrm{cell}} \, U_{\tilde{n}n'}(\vec{K}) \right] \quad \text{for } n \neq n'.
\end{aligned} \qquad (9.38)$$

**Many-body Hamiltonian in crystal-momentum space**
After inserting the results from this subsection [Eqs. (9.28), (9.33), (9.37), and (9.38)] plus the Bloch-wave orthonormality (9.16), the full semiconductor Hamiltonian in Eq. (9.24) can be written as

$$\begin{aligned}
\hat{H}_S^{\text{full}}(t) = \iint\limits_{\text{BZ}} \sum_{n=1}^{\infty} \hat{a}_n^\dagger(t,\vec{K}) \\
\times \Bigg\{ \mathcal{E}_n(\vec{K}) - q\left[E + v_{n,y}^{\text{gr}}(\vec{K})B\right] \mathrm{i}\partial_{K_x} - \frac{qB}{m} \sum_{\tilde{n}\in\mathbb{N}}^{\tilde{n}\neq n} \langle n,\vec{K}|\hat{p}_y|\tilde{n},\vec{K}\rangle_{\text{cell}} U_{\tilde{n}n}(\vec{K}) \\
+ \frac{q^2B^2}{2m}\left[-\partial_{K_x}^2 + \sum_{\tilde{n}\in\mathbb{N}}^{\tilde{n}\neq n} |U_{n\tilde{n}}(\vec{K})|^2\right]\Bigg\} \hat{a}_n(t,\vec{K}) \\
+ \sum_{n=1}^{\infty} \sum_{n'\in\mathbb{N}}^{n'\neq n} \hat{a}_n^\dagger(t,\vec{K}) \\
\times \Bigg\{ -q\left[E + v_{n,y}^{\text{gr}}(\vec{K})B\right] U_{nn'}(\vec{K}) - \frac{qB}{m}\langle n,\vec{K}|\hat{p}_y|n',\vec{K}\rangle_{\text{cell}} \mathrm{i}\partial_{K_x} \\
+ \sum_{\tilde{n}\in\mathbb{N}}^{\tilde{n}\neq n,n'} \left[-\frac{qB}{m}\langle n,\vec{K}|\hat{p}_y|\tilde{n},\vec{K}\rangle_{\text{cell}} + \frac{q^2B^2}{2m} U_{n\tilde{n}}(\vec{K})\right] U_{\tilde{n}n'}(\vec{K}) \\
+ \frac{q^2B^2}{2m}\left[2U_{nn'}(\vec{K})\mathrm{i}\partial_{K_x} + \mathrm{i}\frac{\partial U_{nn'}(\vec{K})}{\partial K_x}\right]\Bigg\} \hat{a}_{n'}(t,\vec{K})\, \mathrm{d}^2K. \quad (9.39)
\end{aligned}$$

### 9.1.2.3. Two-band Hamiltonian

**Two-band approximation**

In the next step, we derive the two-band version of the full semiconductor Hamiltonian (9.39) by neglecting all bands besides the valence band and the conduction band ($n, n' \in$ "$\{+,-\}$" instead of $\mathbb{N}$). Additionally, we also **neglect their coupling** to other bands ($\tilde{n} \in$ "$\{+,-\}$").

In analogy to the previous chapters, we denote the **momentum/optical matrix elements** between the valence band and the conduction band by

$$\vec{\varkappa}(\vec{K}) = \begin{pmatrix} \varkappa_x(\vec{K}) \\ \varkappa_y(\vec{K}) \end{pmatrix} = \frac{\langle -,\vec{K}|\hat{\vec{p}}|+,\vec{K}\rangle_{\text{cell}}}{m} \quad (9.40)$$

and the **band-energy difference** by

$$\Delta\mathcal{E}(\vec{K}) = \mathcal{E}_+(\vec{K}) - \mathcal{E}_-(\vec{K}) > 0 \qquad \forall \vec{K}, \quad (9.41)$$

so the relevant elements of the matrix (9.27) read

$$U_{-+}(\vec{K}) = U_{+-}^*(\vec{K}) = \frac{\mathrm{i}\hbar\varkappa_x(\vec{K})}{\Delta\mathcal{E}(\vec{K})}. \quad (9.42)$$

**Two-band Hamiltonian** The resulting **two-band Hamiltonian** can be written as

$$\hat{H}_S(t) = \iint_{\mathrm{BZ}} \underline{\hat{a}}^\dagger(t,\vec{K}) \underbrace{\begin{pmatrix} \hat{\mathcal{M}}_{++}(\vec{K}) & \hat{\mathcal{M}}_{+-}(\vec{K}) \\ \hat{\mathcal{M}}_{-+}(\vec{K}) & \hat{\mathcal{M}}_{--}(\vec{K}) \end{pmatrix}}_{\hat{\mathcal{M}}(\vec{K})} \underline{\hat{a}}(t,\vec{K})\, \mathrm{d}^2K, \tag{9.43}$$

where the matrix elements [which are in fact operators acting on $\underline{\hat{a}}(t,\vec{K})$] read

$$\begin{aligned}\hat{\mathcal{M}}_{++}(\vec{K}) &= \mathcal{E}_+(\vec{K}) - q\left[E + v^{\mathrm{gr}}_{+,y}(\vec{K})B\right]\mathrm{i}\partial_{K_x} - qB\frac{\mathrm{i}\hbar\varkappa_x(\vec{K})\varkappa_y^*(\vec{K})}{\Delta\mathcal{E}(\vec{K})} \\ &\quad + \frac{q^2B^2}{2m}\left[-\partial^2_{K_x} + \frac{\hbar^2|\varkappa_x(\vec{K})|^2}{\Delta\mathcal{E}^2(\vec{K})}\right],\end{aligned} \tag{9.44}$$

$$\begin{aligned}\hat{\mathcal{M}}_{--}(\vec{K}) &= \mathcal{E}_-(\vec{K}) - q\left[E + v^{\mathrm{gr}}_{-,y}(\vec{K})B\right]\mathrm{i}\partial_{K_x} + qB\frac{\mathrm{i}\hbar\varkappa_x^*(\vec{K})\varkappa_y(\vec{K})}{\Delta\mathcal{E}(\vec{K})} \\ &\quad + \frac{q^2B^2}{2m}\left[-\partial^2_{K_x} + \frac{\hbar^2|\varkappa_x(\vec{K})|^2}{\Delta\mathcal{E}^2(\vec{K})}\right],\end{aligned} \tag{9.45}$$

$$\begin{aligned}\hat{\mathcal{M}}_{-+}(\vec{K}) &= -q\left[E + v^{\mathrm{gr}}_{-,y}(\vec{K})B\right]\frac{\mathrm{i}\hbar\varkappa_x(\vec{K})}{\Delta\mathcal{E}(\vec{K})} - qB\varkappa_y(\vec{K})\mathrm{i}\partial_{K_x} \\ &\quad + \frac{q^2B^2}{2m}\left\{2\frac{\mathrm{i}\hbar\varkappa_x(\vec{K})}{\Delta\mathcal{E}(\vec{K})}\mathrm{i}\partial_{K_x} - \hbar\frac{\partial[\varkappa_x(\vec{K})/\Delta\mathcal{E}(\vec{K})]}{\partial K_x}\right\},\end{aligned} \tag{9.46}$$

and

$$\begin{aligned}\hat{\mathcal{M}}_{+-}(\vec{K}) &= +q\left[E + v^{\mathrm{gr}}_{+,y}(\vec{K})B\right]\frac{\mathrm{i}\hbar\varkappa_x^*(\vec{K})}{\Delta\mathcal{E}(\vec{K})} - qB\varkappa_y^*(\vec{K})\mathrm{i}\partial_{K_x} \\ &\quad + \frac{q^2B^2}{2m}\left\{-2\frac{\mathrm{i}\hbar\varkappa_x^*(\vec{K})}{\Delta\mathcal{E}(\vec{K})}\mathrm{i}\partial_{K_x} + \hbar\frac{\partial[\varkappa_x^*(\vec{K})/\Delta\mathcal{E}(\vec{K})]}{\partial K_x}\right\}.\end{aligned} \tag{9.47}$$

## 9.2. Quantitative analogy in the long-wavelength limit

In order to prove the quantitative analogy between the Hamiltonians, we have to compare the $2 \times 2$ matrix in the Dirac Hamiltonian (9.12) with the matrix $\hat{\mathcal{M}}(\vec{K})$ appearing in the two-band semiconductor Hamiltonian (9.43). However, since these are both matrices of operators ($\partial_{K_x}$) for the crossed-fields profile, diagonalizing these matrices requires more sophisticated methods than in the previous two chapters (Foldy–Wouthuysen-type transformations; see, e.g., Refs. [136, 96, 163]). We only want to present a short outlook in this section, so we will simply rederive the known results on tunneling in crossed, constant fields (see Refs. [104, 97, 96, 106]) on the basis of the two-band semiconductor Hamiltonian (9.43) here.

### QED

Our explanation of the analogy for crossed fields starts with the Dirac case. It is well known that a perpendicular magnetic field reduces the Sauter–Schwinger tunneling current in QED (see Sec. 2.6 in the introduction). One way to understand this effect is via the (semiclassical) tunneling picture since the magnetic field curves the **edges of the two relativistic energy continua in space** (in addition to the tilting of the edges caused by the electric field), thereby increasing the tunneling length as shown in Fig. 2.10 on page 87. In the limit $B \nearrow E/c$, the tunneling current even vanishes completely. Let us consider the mode with the conserved wave-vector component $k_y = 0$ for simplicity. At any given point $x\vec{e}_x$ in space, the edges of the two energy continua (on which spatial turning points may lie) are given by the energy levels $\mathcal{E}_D^{\pm}(x)$ which correspond to the long-wavelength limit $k_x = 0$ (**zero momentum**). We can find these $x$-dependent energy levels by calculating the local ($\mathrm{i}\partial_{k_x} \to x$) eigenvalues of the $2 \times 2$ matrix in the Dirac Hamiltonian (9.12) for $\vec{k} = 0$:

$$\det \begin{pmatrix} mc^2 - qEx - \mathcal{E}_D^{\pm}(x) & \mathrm{i}cqBx \\ -\mathrm{i}cqBx & -mc^2 - qEx - \mathcal{E}_D^{\pm}(x) \end{pmatrix} = 0, \tag{9.48}$$

which leads to **Result**

$$\mathcal{E}_D^{\pm}(x) = -qEx \pm \sqrt{m^2c^4 + c^2q^2B^2x^2}. \tag{9.49}$$

These are precisely the relativistic "band edges" plotted in Fig. 2.10 on page 87.

### Two-band semiconductor

In the semiconductor case, we can determine the local band edges $\mathcal{E}_S^{\pm}(x)$ in exactly the same way: We arbitrarily consider the mode $K_y = 0$ and calculate the local eigenvalues of the matrix in $\hat{H}_S$ [Eq. (9.43)] for $K_x = 0$. Note that **we set $\mathrm{i}\partial_{K_x} \to x$** in analogy to the Dirac case since we have shown in the previous two chapters that $k$ is physically equivalent to $K$ in the long-wavelength regime. Let us make the following **assumptions about the band structure**: **Assumptions**

- The semiconductor should have a **direct bandgap at the center of the Brillouin zone** again, so we have

$$\Delta\mathcal{E}(\vec{K} = 0) = \mathcal{E}_g \qquad \text{and} \qquad \frac{1}{\hbar} \left. \frac{\partial \mathcal{E}_{\pm}(\vec{K})}{\partial \vec{K}} \right|_{\vec{K}=0} = \vec{v}_{\pm}^{\mathrm{gr}}(0) = 0. \tag{9.50}$$

Without loss of generality, we set

$$\mathcal{E}_{\pm}(0) = \pm\frac{\mathcal{E}_g}{2}. \tag{9.51}$$

- The band structure should be **isotropic** in the vicinity of the gap in the sense that the components of the interband momentum matrix element have the same magnitude:

$$|\varkappa_x(0)| \stackrel{!}{=} |\varkappa_y(0)| \stackrel{\text{(abbreviation)}}{=} \varkappa_0 > 0. \tag{9.52}$$

Furthermore, we already know from the 1+1-dimensional case that $\partial\varkappa_x(\vec{K})/\partial K_x = 0$ [see Eq. (7.73)] at $\vec{K} = 0$ **within the two-band approximation**. However, the matrix elements (9.44)–(9.47) still consist of many terms:

$$\begin{aligned}
\mathcal{M}_{++}(0) &= \underbrace{\frac{\mathcal{E}_g}{2} - qEx}_{\otimes} - \underbrace{qB\frac{\mathrm{i}\hbar\varkappa_x(0)\varkappa_y^*(0)}{\mathcal{E}_g}}_{\oplus} + \frac{q^2B^2}{2m}\left(\underbrace{x^2}_{\circledcirc} + \underbrace{\frac{\hbar^2\varkappa_0^2}{\mathcal{E}_g^2}}_{\oplus}\right), \\
\mathcal{M}_{--}(0) &= \underbrace{-\frac{\mathcal{E}_g}{2} - qEx}_{\otimes} + \underbrace{qB\frac{\mathrm{i}\hbar\varkappa_x^*(0)\varkappa_y(0)}{\mathcal{E}_g}}_{\oplus} + \frac{q^2B^2}{2m}\left(\underbrace{x^2}_{\circledcirc} + \underbrace{\frac{\hbar^2\varkappa_0^2}{\mathcal{E}_g^2}}_{\oplus}\right), \\
\mathcal{M}_{-+}(0) &= \underbrace{-qB\varkappa_y(0)x}_{\otimes} - \underbrace{qE\frac{\mathrm{i}\hbar\varkappa_x(0)}{\mathcal{E}_g}}_{\oplus} + \underbrace{\frac{q^2B^2}{m}\frac{\mathrm{i}\hbar\varkappa_x(0)}{\mathcal{E}_g}x}_{\ominus}, \\
\mathcal{M}_{+-}(0) &= \underbrace{-qB\varkappa_y^*(0)x}_{\otimes} + \underbrace{qE\frac{\mathrm{i}\hbar\varkappa_x^*(0)}{\mathcal{E}_g}}_{\oplus} - \underbrace{\frac{q^2B^2}{m}\frac{\mathrm{i}\hbar\varkappa_x^*(0)}{\mathcal{E}_g}x}_{\ominus}.
\end{aligned} \tag{9.53}$$

**Result** Let us just include the terms marked by "⊗" in our calculation of $\mathcal{E}_S^{\pm}(x)$ since the other terms are small (see below). This approximation yields

$$\det\begin{pmatrix} \frac{\mathcal{E}_g}{2} - qEx - \mathcal{E}_S^{\pm}(x) & -qB\varkappa_y^*(0)x \\ -qB\varkappa_y(0)x & -\frac{\mathcal{E}_g}{2} - qEx - \mathcal{E}_S^{\pm}(x) \end{pmatrix} = 0, \tag{9.54}$$

from which we get

$$\begin{aligned}
\mathcal{E}_S^{\pm}(x) &= -qEx \pm \sqrt{\left(\frac{\mathcal{E}_g}{2}\right)^2 + \varkappa_0^2 q^2B^2x^2} \\
&= -qEx \pm \sqrt{m_\star^2c_\star^4 + c_\star^2q^2B^2x^2}
\end{aligned} \tag{9.55}$$

with the **effective constants**

$$c_\star = \varkappa_0 = |\varkappa_x(0)| = |\varkappa_y(0)| \qquad \text{and} \qquad m_\star = \frac{\mathcal{E}_g}{2\varkappa_0^2}. \tag{9.56}$$

This coincides with the QED result (9.49), and the same effective constants were also found in Refs. [104, 96, 97].

**Smallness of the neglected terms**

However, in order to confirm this analogy, we have to show that the other terms in the matrix elements (9.53) are negligible. As usual, we assume a **subcritical electric field** and a perpendicular magnetic field $B \lesssim E/c_\star$ (i.e., approximately up to the point at which tunneling is completely suppressed), so we have

$$E \ll E_{\text{crit}}^{\text{SC}} \qquad \text{and} \qquad B \lesssim \frac{E}{c_\star} = \frac{E}{\varkappa_0} \ll \frac{E_{\text{crit}}^{\text{SC}}}{\varkappa_0} = B_{\text{crit}}^{\text{SC}}, \tag{9.57}$$

where we have defined the **critical magnetic field strength** $B_{\text{crit}}^{\text{SC}} = E_{\text{crit}}^{\text{SC}}/c_\star$. Note that

$$E_{\text{crit}}^{\text{SC}} = \frac{m_\star^2 c_\star^3}{\hbar q} = \frac{\mathcal{E}_g^2}{4\hbar q \varkappa_0} \qquad \Rightarrow \qquad B_{\text{crit}}^{\text{SC}} = \frac{\mathcal{E}_g^2}{4\hbar q \varkappa_0^2}. \tag{9.58}$$

Let us now consider the absolute values of the constant ($x$-independent) terms (Ⓘ) we ignored in the diagonal elements $\mathcal{M}_{\pm\pm}(0)$ in Eq. (9.53) above and compare them to $\mathcal{E}_g/2$, the constant we did not neglect:

$$\frac{\left|qB\mathrm{i}\hbar\varkappa_x(0)\varkappa_y^*(0)/\mathcal{E}_g\right|}{\mathcal{E}_g/2} = \frac{2\hbar qB\varkappa_0^2}{\mathcal{E}_g^2} = \frac{1}{2}\underbrace{\frac{B}{B_{\text{crit}}^{\text{SC}}}}_{\ll 1} \ll 1 \tag{9.59}$$

[due to Eq. (9.57)] and

$$\begin{aligned}\frac{q^2B^2\hbar^2\varkappa_0^2/(2m\mathcal{E}_g^2)}{\mathcal{E}_g/2} = \frac{\hbar^2q^2B^2\varkappa_0^2}{m\mathcal{E}_g^3} = \left(\frac{B}{B_{\text{crit}}^{\text{SC}}}\right)^2\frac{\mathcal{E}_g}{16m\varkappa_0^2} = \left(\frac{B}{B_{\text{crit}}^{\text{SC}}}\right)^2\frac{2m_\star c_\star^2}{16mc_\star^2}\\ = \left(\frac{B}{B_{\text{crit}}^{\text{SC}}}\right)^2\underbrace{\frac{m_\star}{8m}}_{\ll 1\,\text{(typically)}} \ll 1,\end{aligned} \tag{9.60}$$

so neglecting these terms was justified. In the off-diagonal elements, the same is true for the constant terms (Ⓘ) since

$$\frac{\left|qE\mathrm{i}\hbar\varkappa_x(0)/\mathcal{E}_g\right|}{\mathcal{E}_g/2} = \frac{2\hbar qE\varkappa_0}{\mathcal{E}_g^2} = \frac{E}{2E_{\text{crit}}^{\text{SC}}} \ll 1, \tag{9.61}$$

and the additional linear terms (⊖) are always much smaller than $|qB\varkappa_y(0)x|$ because

$$\left|\frac{q^2B^2\mathrm{i}\hbar\varkappa_x(0)/(m\mathcal{E}_g)}{qB\varkappa_y(0)}\right| = \frac{\hbar qB}{m\mathcal{E}_g} = \frac{B}{B_{\mathrm{crit}}^{\mathrm{SC}}}\frac{\mathcal{E}_g}{4m\varkappa_0^2} = \frac{B}{B_{\mathrm{crit}}^{\mathrm{SC}}}\frac{m_\star}{2m} \ll 1. \tag{9.62}$$

**Quadratic terms in $x$**

The only remaining terms we neglected are the quadratic terms (◎) in the diagonal elements $\mathcal{M}_{\pm\pm}(0)$. In $\mathcal{E}_S^\pm(x)$ in Eq. (9.55), they would appear as additional contributions $-q^2B^2x^2/(2m)$ next to $-qEx$, so these **quadratic terms will always dominate for large $x$**! However, even if we consider large enough $x$ such that we are in the "relativistic regime" of $\mathcal{E}_S^\pm(x)$ (i.e., where $c_\star^2q^2B^2x^2$ dominates the square root),

$$c_\star^2q^2B^2x^2 \gg \left(\frac{\mathcal{E}_g}{2}\right)^2 \qquad \Leftrightarrow \qquad x^2 \gg \frac{\mathcal{E}_g^2}{4c_\star^2q^2B^2}, \tag{9.63}$$

there is still an "intermediate" range of not *too* large $x$ such that $q^2B^2x^2/(2m)$ is still smaller than $\mathcal{E}_g/2$ (and thus negligible) because

$$\frac{\mathcal{E}_g}{2} > \frac{q^2B^2x^2}{2m} \overset{\text{Eq. (9.63)}}{\gg} \frac{\mathcal{E}_g^2}{8mc_\star^2} = \frac{\mathcal{E}_g}{2}\underbrace{\frac{m_\star}{2m}}_{\ll 1}. \tag{9.64}$$

That is, even though the quadratic terms (◎) will eventually dominate for large enough $x$, the QED-like form (9.55) of $\mathcal{E}_S^\pm(x)$ is correct for smaller $x$, including the "relativistic range" (9.63).

In summary, we have confirmed the well-known analogy for tunneling in two-band semiconductors exposed to crossed fields. Note, however, that we just compared the local band edges $\mathcal{E}_D^\pm(x)$ and $\mathcal{E}_S^\pm(x)$ with each other, *not* the underlying Hamiltonians $\hat{H}_D$ and $\hat{H}_S$.

### 9.2.1. Example: GaAs

We conclude this chapter by providing some experimentally interesting values for tunneling in GaAs in crossed fields. The **critical magnetic field** in GaAs reads

$$B_{\mathrm{crit}}^{\mathrm{GaAs}} = \frac{E_{\mathrm{crit}}^{\mathrm{GaAs}}}{c_\star^{\mathrm{GaAs}}} = 374\,\mathrm{T} \tag{9.65}$$

(see Table 7.2 on page 187 for the values of $E_{\mathrm{crit}}^{\mathrm{GaAs}}$ and $c_\star^{\mathrm{GaAs}}$). Hence, for an electric field $E = E_{\mathrm{crit}}^{\mathrm{GaAs}}/100 = 5.7\,\mathrm{MV/m}$ (approximately one order of magnitude below the breakdown field), tunneling should be completely suppressed if the perpendicular magnetic field attains the value

$$B = \frac{E}{c_\star^{\mathrm{GaAs}}} = \frac{B_{\mathrm{crit}}^{\mathrm{GaAs}}}{100} = 3.7\,\mathrm{T}. \tag{9.66}$$

Even a field of $B = 1\,\mathrm{T}$ should significantly reduce the tunneling current in this case.

# Part IV.

# Conclusion

We began this thesis by introducing the physical framework which is required to understand nonperturbative electron–positron pair creation from the Dirac vacuum via classical electromagnetic fields in Ch. 1. We then explained the Sauter–Schwinger effect (pair creation in a constant electric field) and reviewed some known results on how this mechanism is affected in nonconstant fields and under which conditions its tunneling-like (nonperturbative) nature is preserved in such fields (Ch. 2).

**Part I (Chs. 1–2): introduction**

In Part II of this thesis, we studied QED pair creation induced by homogeneous, **time-dependent external electric fields** ($\ll E_{\text{crit}}^{\text{QED}}$) in 1+1 spacetime dimensions via the semiclassical solution [contour-integral representation (2.110) of $R_k^{\text{out}}$] of the Riccati equation.

**Part II: pulse-shape dependence in the dynamically assisted Sauter–Schwinger effect**

**Ch. 3**

We first considered the pure Sauter–Schwinger effect in a constant $E$ field and calculated the full contour integral for an arbitrary conserved $k$. We found that the contributions from the singularity (residuum) and from the branch cut are of the same order of magnitude in this case [Eq. (3.43)]; that is, both contributions are exponentially suppressed by $\exp[-\pi E_{\text{crit}}^{\text{QED}}/(2E)]$ and just differ by a (nearly) constant prefactor. On the basis of this finding, we argued that **neglecting the branch cuts** in the case of more complicated field profiles (considered in the two subsequent chapters) merely affects the nonexponential prefactor in $R_k^{\text{out}}$, not the exponents, which are crucial in the context of dynamically assisted tunneling as has recently been shown in Ref. [41].

**Ch. 4**

We then studied pair creation in a strong constant ("background") field $E_{\text{strong}}$ plus an assisting temporal Gauss pulse $E_{\text{weak}} \exp[-(\omega t)^2]$ of weak amplitude ($\varepsilon = E_{\text{weak}}/E_{\text{strong}} \ll 1$). In analogy to the "original" dynamically assisted Sauter–Schwinger effect [67] (assisting Sauter pulse), the Gauss pulse effectively enhances the tunneling exponent for $\gamma_c \propto \omega/E_{\text{strong}}$ above a certain $\gamma_c^{\text{crit}}$ (assuming fixed field strengths) [Eq. (4.23) and Fig. 4.4 on page 118]—but while $\gamma_c^{\text{crit}}$ is constant for $\varepsilon \ll 1$ in the case of a temporal Sauter pulse due to the first pole of the corresponding vector potential $\propto \tanh(\omega t)$ in the complex plane, we found a pulse-amplitude-dependent $\gamma_c^{\text{crit}} \sim \sqrt{|\ln \varepsilon|}$ for a temporal Gauss pulse, whose vector potential is free of singularities. Hence, although both pulse shapes appear to be qualitatively very similar on the real axis, their **pole structures** deviate from one another, which leads to **crucial physical differences** regarding dynamical assistance. As another example for these differences, dynamical assistance by a Sauter pulse can be described well by first-order perturbation theory while the Gauss pulse requires higher orders in general. The reason for this difference lies in the asymptotic behaviors of the Fourier spectra (which are related to the pole structures), as has recently been shown in Ref. [71]. However, both profiles have in common that the main singularity (corresponding to the single Sauter–Schwinger singular-ity in the case of a constant field) generates the leading-order contribution to $R^{\text{out}}_{k=0}$ (dominating mode), respectively, at least if we are not too far above the critical threshold (Fig. 4.7 on page 125).

**Ch. 5**

As another example, we analyzed dynamical assistance via a weak oscillation $E_{\text{weak}}\cos(\omega t)$, which should be a good approximation for **counterpropagating laser beams**. Even though this assisting field is not pulse shaped, its analytic structure is similar to that of the Gauss pulse (no singularities), and we found $\gamma_c^{\text{crit}} \sim |\ln\varepsilon|$ for small $\varepsilon$ [Eqs. (5.9) and (5.14) and Fig. 5.3 on page 132]. Again, the main singularity dominates $R^{\text{out}}_{k=0}$ not too far above the threshold (Fig. 5.7 on page 142). We furthermore showed that the singularity equation for $k = 0$ can be solved graphically (Fig. 5.4 on page 136), which makes the systematic calculation of all additional singularities easier than for the Gauss-pulse profile, where we could not establish such a method.

**Part III: quantitative analogy between QED and direct-bandgap semiconductors**

Part III has been dedicated to the analogy between the Dirac equation and nonrelativistic Bloch electrons in a two-band semiconductor (both bands treated as nondegenerate) with a direct bandgap at the center of the Brillouin zone. Our goal was to derive the **quantitative analogy** between both systems under the influence of external electric or electromagnetic fields, with special regard to **long-wavelength processes**, which should describe (dynamically assisted) tunneling transitions (nonperturbative pair creation) between the energy continua/bands well. Such condensed-matter analogs could be useful to study high-energy QED effects like those considered in Part II experimentally. It should be emphasized that the analogy for long wavelengths (small momenta) does not imply that the trajectories of created particles, which are then accelerated by the external field, are the same in both systems.

**Ch. 6**

We started this part by explaining the analogy (see, e.g., Fig. 6.1 on page 152) and summarized the **known results** (scale equivalents $c \leftrightarrow c_\star$ and $m \leftrightarrow m_\star$), which were all derived on the basis of **constant external fields**. In the following chapters, we were able to confirm these scale equivalents for nonconstant fields. We also discussed the most important aspects of semiconductors which are not in accordance with the perfect analogy: anisotropies of material properties and the coupling with other energy bands. Especially the coupling between the conduction band and the split-off band could lead to deviations of a few percents from the ideal analogy according to Ref. [130] (see Sec. 6.1.2).

**Ch. 7**

The first detailed derivation of the analogy applies to **time-dependent electric fields** in 1+1 spacetime dimensions (no magnetic field). We chose the temporal gauge $E(t) = \dot{A}(t)$, which conserves a canonical $k$ or $K$ for each electron, and we showed that each semiconductor $K$ mode is formally equivalent to a Dirac mode $k(K)$, but with $K$-dependent effective scales $m_\star(K)$ and $c_\star(K)$.

However, these values are approximately constant around $K = 0$ for sufficiently small $K$ [Eq. (7.72)] according to first-order $\boldsymbol{K} \cdot \boldsymbol{p}$ perturbation theory and when neglecting the coupling with other bands (**two-band approximation**). These modes should be most suitable to describe tunneling processes from the valence band to the conduction band, which confirms the analogy. The effective constants $m_\star$ and $c_\star$ for these small canonical momenta **coincide with the previously known results** [Eq. (7.63)]. Within the two-band approximation, we found that $m_\star$ can be expressed as the harmonic mean of the effective band masses (curvatures) $m_{\star,e}$ and $m_{\star,h}$ at $K = 0$ for the homogeneous, but arbitrarily time-dependent electric fields considered here [Eq. (7.87)]—a result which has been known before for constant fields only [128].

We then provided an analogous derivation for **spacetime-dependent electric fields** $E(t,x) = \partial_x \Phi(t,x)$ in 1+1 dimensions and found that the analogy does still work for small momenta (with the same $m_\star$ and $c_\star$ as before). However, we had to make two additional assumptions to show this: First, the **effective band masses** $m_{\star,e}$ and $m_{\star,h}$ at the gap $K = 0$ **must coincide** [Eq. (8.13)]. This condition corresponds to the fact that the electron mass equals the positron mass in QED. In the inhomogeneous fields considered here, the total energy of created particles depends on the positions of *all* constituents in space (which depend on the way they are accelerated), and thus the analogy can only hold for $m_{\star,e} = m_{\star,h}$. This assumption is only met approximately in many conventional semiconductors (GaAs: $m_{\star,h}$ deviates from $m_{\star,e}$ by about 20% [126]). The second assumption we had to make is that $\Phi$ (and thus $E$) must **vary slowly on the length scale set by the lattice constant**. We showed that this condition should always be satisfied in typical semiconductors [Eq. (8.20)] if the external field incorporates only photon energies far below the bandgap, which is true for all mechanisms of dynamical assistance considered in this thesis. Based on these findings, we concluded that the dynamically assisted Sauter–Schwinger effect with a Sauter-pulse background field [90] applies to semiconductors as well, and we argued that the scaling $\omega_{\text{crit}} \sim \sqrt{q\Delta\Phi - \mathcal{E}_g}$ should be universal for exponentially decaying pulse-shaped background fields in the no-tunneling limit ($q\Delta\Phi \searrow \mathcal{E}_g$); see Sec. 8.3.1. We also estimated how the **critical threshold** for dynamical assistance by an **oscillation** $E_{\text{weak}} \cos(\omega t)$ **decreases in a Sauter-pulse background field** [Eq. (8.40) and Fig. 8.2 on page 206]. **Ch. 8**

In the final chapter, we derived the **exact two-band Hamiltonian** for Bloch electrons in 2+1 dimensions for **crossed constant electric and magnetic fields** [Eq. (9.43)]. The fields were described by the purely $x$-dependent potentials $\Phi(x) = Ex$ and $\vec{A}(x) = -Bx\vec{e}_y$, according to which nonperturbative pair creation can be understood in the tunneling picture. Since the **Ch. 9** single-body Hamiltonians of both systems (Dirac theory and semiconductor) contain derivative operators even in reciprocal space in this case, a detailed comparison between the Hamiltonians is more complicated here and requires techniques such as the Foldy–Wouthuysen transformation (cf., e.g., Refs. [136, 96, 163]). This could be a worthwhile route for future research on the analogy in electromagnetic fields. We just derived the local energy-band edges (long-wavelength limit) from both Hamiltonians and found them to be approximately equivalent for subcritical fields and if the band structure is isotropic around the gap (again, the **effective scales** $m \leftrightarrow m_\star$ and $c \leftrightarrow c_\star$ **are the same as before**) [Eq. (9.55)], thus confirming the known results [104, 97, 96, 106]. However, we also found the quadratic $x$ terms [see Eq. (9.53)] in the local band edges of the semiconductor to cause deviations from the QED-like form for large $x$—but in the semiclassical range $E \ll E_{\text{crit}}^{\text{SC}}$ and $B \ll B_{\text{crit}}^{\text{SC}} = E_{\text{crit}}^{\text{SC}}/c_\star$, these deviations should not spoil the analogy [Eq. (9.63)].

**Analogy in gallium arsenide**

Throughout Part III, we proposed various ways to **simulate dynamically assisted nonperturbative QED pair creation in GaAs** and estimated the corresponding expected equivalents of the QED scales (Secs. 7.3, 8.3, and 9.2.1, Table 7.2 on page 187). These results are summarized in Table 9.1. Gallium arsenide appears to be a good candidate for such a laboratory analog since it is a standard semiconductor with a direct bandgap at the zone center, and the effective band masses differ by about 20% only, which is a quite good agreement when compared to other typical direct-bandgap semiconductors [126].

| QED | | GaAs |
|---|---|---|
| $m$ | $\leftrightarrow$ | $m_\star = 0.058m$ |
| $c$ | $\leftrightarrow$ | $c_\star = 0.0050c$ |
| $2mc^2 = 1\,\text{MeV}$ | $\leftrightarrow$ | $\mathcal{E}_g = 1.5\,\text{eV}$ |
| $E_{\text{crit}} = 1.3 \times 10^{18}\,\text{V/m}$ | $\leftrightarrow$ | $E_{\text{crit}} = 565\,\text{MV/m}$ |
| $B_{\text{crit}} = E_{\text{crit}}/c = 4.4 \times 10^9\,\text{T}$ | $\leftrightarrow$ | $B_{\text{crit}} = E_{\text{crit}}/c_\star = 374\,\text{T}$ |
| **Dynamically assisted Sauter–Schwinger effect (constant $E_{\text{strong}}$) ...** | | |
| $E_{\text{strong}} = E_{\text{crit}}/10 \approx 10^{17}\,\text{V/m}$ | $\leftrightarrow$ | $E_{\text{crit}}/10 = 57\,\text{MV/m}$ |
| **...with assisting $E_{\text{weak}}/\cosh^2(\omega t)$** | | |
| $\hbar\omega_{\text{crit}} = 80\,\text{keV}$ | $\leftrightarrow$ | $\hbar\omega_{\text{crit}} = 0.12\,\text{eV}$ |
| **...with assisting $E_{\text{weak}}\cos(\omega t)$** | | |
| $\hbar\omega = 80\,\text{keV}$ | $\leftrightarrow$ | $\hbar\omega = 0.117\,\text{eV}$ ($CO_2$ laser) |
| $= 0.078 \times 2mc^2$ | $\leftrightarrow$ | $= 0.078\,\mathcal{E}_g$ |
| $I_{\text{crit}} = c\varepsilon_0(E^{\text{crit}}_{\text{weak}})^2/2$ | $\leftrightarrow$ | $I_{\text{crit}} = c\varepsilon_0(E^{\text{crit}}_{\text{weak}})^2/2$ |
| $= 1.3 \times 10^{23}\,\text{W/cm}^2$ | $\leftrightarrow$ | $= 24\,\text{kW/cm}^2$ |
| | | *Reduction of $I_{\text{crit}}$ in pulse-shaped background field: see Fig. 8.2 on page 206.* |
| **Perpendicular $B$ field stops Sauter–Schwinger tunneling ($E$)** | | |
| $E = E_{\text{crit}}/100 \approx 10^{16}\,\text{V/m}$ | $\leftrightarrow$ | $E = E_{\text{crit}}/100 = 5.7\,\text{MV/m}$ |
| $B = E/c = 44\,\text{MT}$ | $\leftrightarrow$ | $B = E/c_\star = 3.7\,\text{T}$ |

**Table 9.1.**: Comparison between QED scales and their counterparts in the gallium-arsenide analog.

# Bibliography

[1] M. F. Linder, C. Schneider, J. Sicking, N. Szpak, and R. Schützhold, "Pulse shape dependence in the dynamically assisted Sauter-Schwinger effect," Phys. Rev. D **92**, 085009 (2015), arXiv:1505.05685.

[2] M. F. Linder, A. Lorke, and R. Schützhold, "Analog Sauter-Schwinger effect in semiconductors for spacetime-dependent fields," Phys. Rev. B **97**, 035203 (2018), arXiv:1503.07108.

[3] M. F. Linder, R. Schützhold, and W. G. Unruh, "Derivation of Hawking radiation in dispersive dielectric media," Phys. Rev. D **93**, 104010 (2016), arXiv:1511.03900.

[4] E. Schrödinger, "An Undulatory Theory of the Mechanics of Atoms and Molecules," Phys. Rev. **28**, 1049–1070 (1926).

[5] G. E. Uhlenbeck and S. Goudsmit, "Ersetzung der Hypothese vom unmechanischen Zwang durch eine Forderung bezüglich des inneren Verhaltens jedes einzelnen Elektrons," Naturwissenschaften **13**, 953–954 (1925).

[6] G. E. Uhlenbeck and S. Goudsmit, "Spinning Electrons and the Structure of Spectra," Nature **117**, 264–265 (1926).

[7] W. Pauli, "Zur Quantenmechanik des magnetischen Elektrons," Z. Phys. **43**, 601–623 (1927).

[8] P. A. M. Dirac, "The Quantum Theory of the Electron," Proc. Royal Soc. A **117**, 610–624 (1928).

[9] P. A. M. Dirac, "The Quantum Theory of the Electron. Part II." Proc. Royal Soc. A **118**, 351–361 (1928).

[10] P. A. M. Dirac, "A Theory of Electrons and Protons," Proc. Royal Soc. A **126**, 360–365 (1930).

[11] H. Bethe and W. Heitler, "On the Stopping of Fast Particles and on the Creation of Positive Electrons," Proc. Royal Soc. A **146**, 83–112 (1934).

[12] G. Breit and J. A. Wheeler, "Collision of Two Light Quanta," Phys. Rev. **46**, 1087–1091 (1934).

[13] C. D. Anderson, "The Apparent Existence of Easily Deflectable Positives," Science **76**, 238–239 (1932).

[14] C. D. Anderson, "The Positive Electron," Phys. Rev. **43**, 491–494 (1933).

[15] O. Klein, "Quantentheorie und fünfdimensionale Relativitätstheorie," Z. Phys. **37**, 895–906 (1926).

[16] W. Gordon, "Der Comptoneffekt nach der Schrödingerschen Theorie," Z. Phys. **40**, 117–133 (1926).

[17] O. Klein, "Die Reflexion von Elektronen an einem Potentialsprung nach der relativistischen Dynamik von Dirac," Z. Phys. **53**, 157–165 (1929).

[18] F. Sauter, "Über das Verhalten eines Elektrons im homogenen elektrischen Feld nach der relativistischen Theorie Diracs," Z. Phys. **69**, 742–764 (1931).

[19] F. Sauter, "Zum 'Kleinschen Paradoxon'," Z. Phys. **73**, 547–552 (1932).

[20] A. D. Alhaidari, "Resolution of the Klein paradox," Phys. Scr. **83**, 025001 (2011), arXiv:0907.5588.

[21] S. Weinberg, *The Quantum Theory of Fields. Volume I: Foundations* (Cambridge University Press, Cambridge, England, 1995).

[22] W. Heisenberg and H. Euler, "Folgerungen aus der Diracschen Theorie des Positrons," Z. Phys. **98**, 714–732 (1936).

[23] V. Weisskopf, "Über die Elektrodynamik des Vakuums auf Grund der Quantentheorie des Elektrons," Math.-fys. Medd. **XIV**, 1–39 (1936).

[24] ATLAS Collaboration, "Evidence for light-by-light scattering in heavy-ion collisions with the ATLAS detector at the LHC," Nat. Phys. **13**, 852–858 (2017), arXiv:1702.01625.

[25] G. V. Dunne, "Heisenberg-Euler Effective Lagrangians: Basics and Extensions," in *From Fields to Strings: Circumnavigating Theoretical Physics*, edited by M. Shifman, A. Vainshtein, and J. Wheater (World Scientific, Singapore, 2005) pp. 445–522.

[26] G. V. Dunne, "The Heisenberg-Euler Effective Action: 75 years on," Int. J. Mod. Phys. A **27**, 1260004 (2012), arXiv:1202.1557.

[27] J. Schwinger, "On Gauge Invariance and Vacuum Polarization," Phys. Rev. **82**, 664–679 (1951).

[28] T. D. Cohen and D. A. McGady, "Schwinger mechanism revisited," Phys. Rev. D **78**, 036008 (2008), arXiv:0807.1117.

[29] A. I. Nikishov, "Barrier scattering in field theory removal of Klein paradox," Nucl. Phys. B **21**, 346–358 (1970).

[30] F. Gelis and N. Tanji, "Schwinger mechanism revisited," Prog. Part. Nucl. Phys. **87**, 1–49 (2016), arXiv:1510.05451.

[31] R. P. Feynman, "Mathematical Formulation of the Quantum Theory of Electromagnetic Interaction," Phys. Rev. **80**, 440–457 (1950).

[32] I. K. Affleck, O. Alvarez, and N. S. Manton, "Pair production at strong coupling in weak external fields," Nucl. Phys. B **197**, 509–519 (1982).

[33] S. P. Kim and D. N. Page, "Schwinger pair production via instantons in strong electric fields," Phys. Rev. D **65**, 105002 (2002), arXiv:hep-th/0005078.

[34] G. V. Dunne and C. Schubert, "Worldline instantons and pair production in inhomogenous fields," Phys. Rev. D **72**, 105004 (2005), arXiv:hep-th/0507174.

[35] S. P. Kim and D. N. Page, "Schwinger pair production in electric and magnetic fields," Phys. Rev. D **73**, 065020 (2006), arXiv:hep-th/0301132.

[36] G. V. Dunne, Q.-h. Wang, H. Gies, and C. Schubert, "Worldline instantons and the fluctuation prefactor," Phys. Rev. D **73**, 065028 (2006), arXiv:hep-th/0602176.

[37] G. V. Dunne, "Worldline instantons, vacuum pair production and Gutzwiller's trace formula," J. Phys. A: Math. Theor. **41**, 164041 (2008).

[38] C. K. Dumlu and G. V. Dunne, "Complex worldline instantons and quantum interference in vacuum pair production," Phys. Rev. D **84**, 125023 (2011), arXiv:1110.1657.

[39] C. Schubert, "Lectures on the Worldline Formalism," in *School on Spinning Particles in Quantum Field Theory: Worldline Formalism, Higher Spins and Conformal Geometry* (Morelia, Michoacán, Mexico, 2012).

[40] C. Schneider, *Sauter-Schwinger effect in spacetime-dependent fields*, Master's thesis, Universität Duisburg-Essen (2014).

[41] C. Schneider and R. Schützhold, "Prefactor in the dynamically assisted Sauter-Schwinger effect," Phys. Rev. D **94**, 085015 (2016), arXiv:1603.00864.

[42] A. I. Nikishov, "Pair Production by a Constant External Field," Sov. Phys. JETP **30**, 660–662 (1970).

[43] E. Brezin and C. Itzykson, "Pair Production in Vacuum by an Alternating Field," Phys. Rev. D **2**, 1191–1199 (1970).

[44] V. S. Popov, "Pair Production in a Variable and Homogeneous Electric Field as an Oscillator Problem," Sov. Phys. JETP **35**, 659–666 (1972).

[45] H. Kleinert, R. Ruffini, and S.-S. Xue, "Electron-positron pair production in space- or time-dependent electric fields," Phys. Rev. D **78**, 025011 (2008), arXiv:0807.0909.

[46] H. Gies and K. Klingmüller, "Pair production in inhomogeneous fields," Phys. Rev. D **72**, 065001 (2005), arXiv:hep-ph/0505099.

[47] S. P. Kim and D. N. Page, "Improved approximations for fermion pair production in inhomogeneous electric fields," Phys. Rev. D **75**, 045013 (2007), arXiv:hep-th/0701047.

[48] A. Ilderton, G. Torgrimsson, and J. Wårdh, "Nonperturbative pair production in interpolating fields," Phys. Rev. D **92**, 065001 (2015), arXiv:1506.09186.

[49] V. S. Popov, "Production of $e^+e^-$ Pairs in an Alternating External Field," JETP Lett. **13**, 185–187 (1971).

[50] V. S. Popov, "The 'Imaginary-time' Method in Problems Concerning the Ionization of Atoms and Pair Production," Sov. Phys. JETP **36**, 840–846 (1973).

[51] V. S. Popov, "Resonant pair production in a strong electric field," JETP Lett. **18**, 255–258 (1973).

[52] L. V. Keldysh, "Ionization in the Field of a Strong Electromagnetic Wave," Sov. Phys. JETP **20**, 1307–1314 (1965).

[53] G. R. Mocken, M. Ruf, C. Müller, and C. H. Keitel, "Nonperturbative multiphoton electron-positron–pair creation in laser fields," Phys. Rev. A **81**, 022122 (2010).

[54] D. Blaschke, N. T. Gevorgyan, A. D. Panferov, and S. A. Smolyansky, "Schwinger effect at modern laser facilities," J. Phys. Conf. Ser. **672**, 012020 (2016).

[55] D. L. Burke, R. C. Field, G. Horton-Smith, J. E. Spencer, D. Walz, S. C. Berridge, W. M. Bugg, K. Shmakov, A. W. Weidemann, C. Bula, K. T. McDonald, E. J. Prebys, C. Bamber, S. J. Boege, T. Koffas, T. Kotseroglou, A. C. Melissinos, D. D. Meyerhofer, D. A. Reis, and W. Ragg, "Positron Production in Multiphoton Light-by-Light Scattering," Phys. Rev. Lett. **79**, 1626–1629 (1997).

[56] F. Hebenstreit, R. Alkofer, and H. Gies, "Pair production beyond the Schwinger formula in time-dependent electric fields," Phys. Rev. D **78**, 061701 (2008), arXiv:0807.2785.

[57] H. Taya, H. Fujii, and K. Itakura, "Finite pulse effects on $e^+e^-$ pair creation from strong electric fields," Phys. Rev. D **90**, 014039 (2014), arXiv:1405.6182.

[58] M. S. Marinov and V. S. Popov, "Electron-Positron Pair Creation from Vacuum Induced by Variable Electric Field," Fortschr. Phys. **25**, 373–400 (1977).

[59] C. K. Dumlu, "Quantum kinetic approach and the scattering approach to vacuum pair production," Phys. Rev. D **79**, 065027 (2009), arXiv:0901.2972.

[60] C. K. Dumlu, "Schwinger vacuum pair production in chirped laser pulses," Phys. Rev. D **82**, 045007 (2010), arXiv:1006.3882.

[61] C. K. Dumlu and G. V. Dunne, "Interference effects in Schwinger vacuum pair production for time-dependent laser pulses," Phys. Rev. D **83**, 065028 (2011), arXiv:1102.2899.

[62] C. Fey and R. Schützhold, "Momentum dependence in the dynamically assisted Sauter-Schwinger effect," Phys. Rev. D **85**, 025004 (2012), arXiv:1110.5499.

[63] E. Akkermans and G. V. Dunne, "Ramsey Fringes and Time-Domain Multiple-Slit Interference from Vacuum," Phys. Rev. Lett. **108**, 030401 (2012), arXiv:1109.3489.

[64] L. C. Baird, "New Integral Formulation of the Schrödinger Equation," J. Math. Phys. **11**, 2235–2242 (1970).

[65] J. P. Davis and P. Pechukas, "Nonadiabatic transitions induced by a time-dependent Hamiltonian in the semiclassical/adiabatic limit: The two-state case," J. Chem. Phys. **64**, 3129–3137 (1976).

[66] S. Massar and R. Parentani, "Particle creation and non-adiabatic transitions in quantum cosmology," Nucl. Phys. B **513**, 375–401 (1998), arXiv:gr-qc/9706008.

[67] R. Schützhold, H. Gies, and G. Dunne, "Dynamically Assisted Schwinger Mechanism," Phys. Rev. Lett. **101**, 130404 (2008), arXiv:0807.0754.

[68] F. Hebenstreit, R. Alkofer, G. V. Dunne, and H. Gies, "Momentum Signatures for Schwinger Pair Production in Short Laser Pulses with a Subcycle Structure," Phys. Rev. Lett. **102**, 150404 (2009), arXiv:0901.2631.

[69] C. K. Dumlu and G. V. Dunne, "Stokes Phenomenon and Schwinger Vacuum Pair Production in Time-Dependent Laser Pulses," Phys. Rev. Lett. **104**, 250402 (2010), arXiv:1004.2509.

[70] M. Orthaber, F. Hebenstreit, and R. Alkofer, "Momentum spectra for dynamically assisted Schwinger pair production," Phys. Lett. B **698**, 80–85 (2011), arXiv:1102.2182.

[71] G. Torgrimsson, C. Schneider, J. Oertel, and R. Schützhold, "Dynamically assisted Sauter-Schwinger effect — non-perturbative versus perturbative aspects," J. High Energy Phys. **2017**, 1–26 (2017), arXiv:1703.09203.

[72] A. Nuriman, B.-S. Xie, Z.-L. Li, and D. Sayipjamal, "Enhanced electron–positron pair creation by dynamically assisted combinational fields," Phys. Lett. B **717**, 465–469 (2012).

[73] A. Nuriman, B.-S. Xie, Z.-L. Li, and D. Sayipjamal, "Electron-Positron Pair Production in a Strong Laser Field Enhanced by an Assisted High Frequency Weak Field," Commun. Theor. Phys. **59**, 331–334 (2013).

[74] N. Abdukerim, Z.-L. Li, and B.-S. Xie, "Effects of laser pulse shape and carrier envelope phase on pair production," Phys. Lett. B **726**, 820–826 (2013).

[75] A. Otto, D. Seipt, D. Blaschke, B. Kämpfer, and S. A. Smolyansky, "Lifting shell structures in the dynamically assisted Schwinger effect in periodic fields," Phys. Lett. B **740**, 335–340 (2015), arXiv:1412.0890.

[76] A. Otto, D. Seipt, D. Blaschke, S. A. Smolyansky, and B. Kämpfer, "Dynamical Schwinger process in a bifrequent electric field of finite duration: Survey on amplification," Phys. Rev. D **91**, 105018 (2015), arXiv:1503.08675.

[77] A. D. Panferov, S. A. Smolyansky, A. Otto, B. Kämpfer, D. B. Blaschke, and Ł. Juchnowski, "Assisted dynamical Schwinger effect: pair production in a pulsed bifrequent field," Eur. Phys. J. D **70**, 56 (2016), arXiv:1509.02901.

[78] A. Otto, T. Nousch, D. Seipt, B. Kämpfer, D. Blaschke, A. D. Panferov, S. A. Smolyansky, and A. I. Titov, "Pair production by Schwinger and Breit–Wheeler processes in bi-frequent fields," J. Plasma Phys. **82**, 655820301 (2016), arXiv:1604.00196.

[79] A. I. Nikishov and V. I. Ritus, "Quantum Processes in the Field of a Plane Electromagnetic Wave and in a Constant Field. I," Sov. Phys. JETP **19**, 529–541 (1964).

[80] A. I. Nikishov and V. I. Ritus, "Pair Production by a Photon and Photon Emission by an Electron in the Field of an Intense Electromagnetic Wave and in a Constant Field," Sov. Phys. JETP **25**, 1135–1142 (1967).

[81] T. N. Tomaras, N. C. Tsamis, and R. P. Woodard, "Back reaction in light cone QED," Phys. Rev. D **62**, 125005 (2000), arXiv:hep-ph/0007166.

[82] H. M. Fried and R. P. Woodard, "The one loop effective action of QED for a general class of electric fields," Phys. Lett. B **524**, 233–239 (2002), arXiv:hep-th/0110180.

[83] F. Hebenstreit, A. Ilderton, and M. Marklund, "Pair production: The view from the lightfront," Phys. Rev. D **84**, 125022 (2011), arXiv:1109.3712.

[84] V. Dinu, T. Heinzl, A. Ilderton, M. Marklund, and G. Torgrimsson, "Vacuum refractive indices and helicity flip in strong-field QED," Phys. Rev. D **89**, 125003 (2014), arXiv:1312.6419.

[85] A. Ilderton, "Localisation in worldline pair production and lightfront zero-modes," J. High Energy Phys. **2014**, 166 (2014), arXiv:1406.1513.

[86] M. Ruf, G. R. Mocken, C. Müller, K. Z. Hatsagortsyan, and C. H. Keitel, "Pair Production in Laser Fields Oscillating in Space and Time," Phys. Rev. Lett. **102**, 080402 (2009), arXiv:0810.4047.

[87] F. Hebenstreit, R. Alkofer, and H. Gies, "Schwinger pair production in space- and time-dependent electric fields: Relating the Wigner formalism to quantum kinetic theory," Phys. Rev. D **82**, 105026 (2010), arXiv:1007.1099.

[88] F. Hebenstreit, R. Alkofer, and H. Gies, "Particle Self-Bunching in the Schwinger Effect in Spacetime-Dependent Electric Fields," Phys. Rev. Lett. **107**, 180403 (2011), arXiv:1106.6175.

[89] M. Jiang, W. Su, Z. Q. Lv, X. Lu, Y. J. Li, R. Grobe, and Q. Su, "Pair creation enhancement due to combined external fields," Phys. Rev. A **85**, 033408 (2012).

[90] C. Schneider and R. Schützhold, "Dynamically assisted Sauter-Schwinger effect in inhomogeneous electric fields," J. High Energy Phys. **2016**, 164 (2016), arXiv:1407.3584.

[91] C. K. Dumlu, "Multidimensional quantum tunneling in the Schwinger effect," Phys. Rev. D **93**, 065045 (2016), arXiv:1507.07005.

[92] C. Kohlfürst and R. Alkofer, "On the effect of time-dependent inhomogeneous magnetic fields in electron–positron pair production," Phys. Lett. B **756**, 371–375 (2016), arXiv:1512.06668.

[93] I. A. Aleksandrov, G. Plunien, and V. M. Shabaev, "Electron-positron pair production in external electric fields varying both in space and time," Phys. Rev. D **94**, 065024 (2016), arXiv:1606.06313.

[94] H. Gies and G. Torgrimsson, "Critical Schwinger Pair Production," Phys. Rev. Lett. **116**, 090406 (2016), arXiv:1507.07802.

[95] H. Gies and G. Torgrimsson, "Critical Schwinger pair production. II. Universality in the deeply critical regime," Phys. Rev. D **95**, 016001 (2017), arXiv:1612.00635.

[96] A. G. Aronov and G. E. Pikus, "Tunnel Current in a Transverse Magnetic Field," Sov. Phys. JETP **24**, 188–197 (1967).

[97] M. H. Weiler, W. Zawadzki, and B. Lax, "Theory of Tunneling, Including Photon-Assisted Tunneling, in Semiconductors in Crossed and Parallel Electric and Magnetic Fields," Phys. Rev. **163**, 733–742 (1967).

[98] D. Cangemi, E. D'Hoker, and G. Dunne, "Effective energy for (2+1)-dimensional QED with semilocalized static magnetic fields: A solvable model," Phys. Rev. D **52**, R3163–R3167 (1995), arXiv:hep-th/9506085.

[99] G. Dunne and T. M. Hall, "An exact $\mathrm{QED}_{3+1}$ effective action," Phys. Lett. B **419**, 322–325 (1998), arXiv:hep-th/9710062.

[100] M. Jiang, Q. Z. Lv, Y. Liu, R. Grobe, and Q. Su, "Pair creation in localized electromagnetic fields of different spatial extensions," Phys. Rev. A **90**, 032101 (2014).

[101] A. Di Piazza, C. Müller, K. Z. Hatsagortsyan, and C. H. Keitel, "Extremely high-intensity laser interactions with fundamental quantum systems," Rev. Mod. Phys. **84**, 1177–1228 (2012), arXiv:1111.3886.

[102] A. Ringwald, "Pair production from vacuum at the focus of an X-ray free electron laser," Phys. Lett. B **510**, 107–116 (2001), arXiv:hep-ph/0103185.

[103] R. Alkofer, M. B. Hecht, C. D. Roberts, S. M. Schmidt, and D. V. Vinnik, "Pair Creation and an X-Ray Free Electron Laser," Phys. Rev. Lett. **87**, 193902 (2001), arXiv:nucl-th/0108046.

[104] W. Zawadzki and B. Lax, "Two-Band Model for Bloch Electrons in Crossed Electric and Magnetic Fields," Phys. Rev. Lett. **16**, 1001–1003 (1966).

[105] A. G. Aronov and G. E. Pikus, "Light Absorption in Semiconductors in Crossed Electric and Magnetic Fields," Sov. Phys. JETP **24**, 339–346 (1967).

[106] W. Zawadzki, "Magnetotunneling Effects in Semiconductors," in *Tunneling Phenomena in Solids*, edited by E. Burstein and S. Lundqvist (Springer, Boston, MA, 1969) Ch. 16, pp. 219–231.

[107] J. Sicking, *Pulsformabhängigkeit im dynamisch verstärkten Sauter-Schwinger-Effekt*, Bachelor's thesis, Universität Duisburg-Essen (2012).

[108] "NIST Digital Library of Mathematical Functions," `http://dlmf.nist.gov/`, Release 1.0.16 of 2017-09-18, edited by F. W. J. Olver, A. B. Olde Daalhuis, D. W. Lozier, B. I. Schneider, R. F. Boisvert, C. W. Clark, B. R. Miller, and B. V. Saunders.

[109] R. P. Feynman, R. B. Leighton, and M. Sands, *The Feynman Lectures on Physics, Volume II* (California Institute of Technology, M. A. Gottlieb, and R. Pfeiffer, 2013) freely accessible at `http://www.feynmanlectures.caltech.edu/`.

[110] N. Szpak and R. Schützhold, "Quantum simulator for the Schwinger effect with atoms in bichromatic optical lattices," Phys. Rev. A **84**, 050101 (2011), arXiv:1103.0541.

[111] N. Szpak and R. Schützhold, "Optical lattice quantum simulator for quantum electrodynamics in strong external fields: spontaneous pair creation and the Sauter–Schwinger effect," New J. Phys. **14**, 35001 (2012), arXiv:1109.2426.

[112] V. Kasper, F. Hebenstreit, M. K. Oberthaler, and J. Berges, "Schwinger pair production with ultracold atoms," Phys. Lett. B **760**, 742–746 (2016), arXiv:1506.01238.

[113] E. A. Martinez, C. A. Muschik, P. Schindler, D. Nigg, A. Erhard, M. Heyl, P. Hauke, M. Dalmonte, T. Monz, P. Zoller, and R. Blatt, "Real-time dynamics of lattice gauge theories with a few-qubit quantum computer," Nature **534**, 516–519 (2016), arXiv:1605.04570.

[114] D. Allor, T. D. Cohen, and D. A. McGady, "Schwinger mechanism and graphene," Phys. Rev. D **78**, 096009 (2008), arXiv:0708.1471.

[115] B. Dóra and Roderich Moessner, "Nonlinear electric transport in graphene: Quantum quench dynamics and the Schwinger mechanism," Phys. Rev. B **81**, 165431 (2010), arXiv:0909.2528.

[116] G. L. Klimchitskaya and V. M. Mostepanenko, "Creation of quasiparticles in graphene by a time-dependent electric field," Phys. Rev. D **87**, 125011 (2013), arXiv:1305.5700.

[117] F. Fillion-Gourdeau and S. MacLean, "Time-dependent pair creation and the Schwinger mechanism in graphene," Phys. Rev. B **92**, 035401 (2015).

[118] I. Akal, R. Egger, C. Müller, and S. Villalba-Chávez, "Low-dimensional approach to pair production in an oscillating electric field: Application to bandgap graphene layers," Phys. Rev. D **93**, 116006 (2016), arXiv:1602.08310.

[119] K. S. Novoselov, A. K. Geim, S. V. Morozov, D. Jiang, M. I. Katsnelson, I. V. Grigorieva, S. V. Dubonos, and A. A. Firsov, "Two-dimensional gas of massless Dirac fermions in graphene," Nature **438**, 197–200 (2005), arXiv:cond-mat/0509330.

[120] A. H. Castro Neto, F. Guinea, N. M. R. Peres, K. S. Novoselov, and A. K. Geim, "The electronic properties of graphene," Rev. Mod. Phys. **81**, 109–162 (2009), arXiv:0709.1163.

[121] M. S. Nevius, M. Conrad, F. Wang, A. Celis, M. N. Nair, A. Taleb-Ibrahimi, A. Tejeda, and E. H. Conrad, "Semiconducting Graphene from Highly Ordered Substrate Interactions," Phys. Rev. Lett. **115**, 136802 (2015), arXiv:1505.00435.

[122] P. Y. Yu and M. Cardona, *Fundamentals of Semiconductors: Physics and Materials Properties*, 4th ed. (Springer, Berlin, Heidelberg, 2010).

[123] W. Zawadzki, S. Klahn, and U. Merkt, "Semirelativistic Behavior of Electrons in InSb in Crossed Magnetic and Electric Fields," Phys. Rev. Lett. **55**, 983–986 (1985).

[124] E. O. Kane, "Energy band structure in p-type germanium and silicon," J. Phys. Chem. Solids **1**, 82–99 (1956).

[125] E. O. Kane, "Band structure of indium antimonide," J. Phys. Chem. Solids **1**, 249–261 (1957).

[126] S. M. Sze and K. K. Ng, *Physics of Semiconductor Devices*, 3rd ed. (Wiley, New York, 2006).

[127] E. O. Kane, "Theory of Tunneling," J. Appl. Phys. **32**, 83–91 (1961).

[128] E. O. Kane, "Zener tunneling in semiconductors," J. Phys. Chem. Solids **12**, 181–188 (1959).

[129] J. B. Krieger and G. J. Iafrate, "Time evolution of Bloch electrons in a homogeneous electric field," Phys. Rev. B **33**, 5494–5500 (1986).

[130] A. Pan and C. O. Chui, "Modeling direct interband tunneling. I. Bulk semiconductors," J. Appl. Phys. **116**, 054508 (2014).

[131] D. Kim, T. Krishnamohan, L. Smith, H.-S. P. Wong, and K. C. Saraswat, "Band to Band Tunneling Study in High Mobility Materials : III-V, Si, Ge and strained SiGe," in *2007 65th Annual Device Research Conference* (IEEE, 2007) pp. 57–58.

[132] L. D. Landau, "Zur Theorie der Energieübertragung. II," Phys. Z. Sowjet. **2**, 46–51 (1932).

[133] C. Zener, "Non-Adiabatic Crossing of Energy Levels," Proc. Royal Soc. A **137**, 696–702 (1932).

[134] C. Zener, "A Theory of the Electrical Breakdown of Solid Dielectrics," Proc. Royal Soc. A **145**, 523–529 (1934).

[135] W. V. Houston, "Acceleration of Electrons in a Crystal Lattice," Phys. Rev. **57**, 184–186 (1940).

[136] E. N. Adams, "Motion of an Electron in a Perturbed Periodic Potential," Phys. Rev. **85**, 41–50 (1952).

[137] E. N. Adams, "The Crystal Momentum as a Quantum Mechanical Operator," J. Chem. Phys. **21**, 2013–2017 (1953).

[138] F. Sauter and J. Weisse, "Zur Theorie der inneren Feldemission," Z. Phys. **140**, 150–155 (1955).

[139] G. Eilenberger, "Ein Beitrag zur Theorie der inneren Feldemission," Z. Phys. **164**, 59–77 (1961).

[140] J. B. Krieger, "Theory of electron tunneling in semiconductors with degenerate band structure," Ann. Phys. **36**, 1–60 (1966).

[141] J. Rau and B. Müller, "From reversible quantum microdynamics to irreversible quantum transport," Phys. Rep. **272**, 1–59 (1996).

[142] W. Zawadzki, "Zitterbewegung and its effects on electrons in semiconductors," Phys. Rev. B **72**, 085217 (2005), arXiv:cond-mat/0411488.

[143] T. Oka and H. Aoki, "Ground-State Decay Rate for the Zener Breakdown in Band and Mott Insulators," Phys. Rev. Lett. **95**, 137601 (2005), arXiv:cond-mat/0503503.

[144] S. A. Smolyansky, A. V. Tarakanov, and M. Bonitz, "Vacuum Particle Creation: Analogy with the Bloch Theory in Solid State Physics," Contrib. Plasma Phys. **49**, 575–584 (2009).

[145] P. A. Wolff, "Matrix elements and selection rules for the two-band model of bismuth," J. Phys. Chem. Solids **25**, 1057–1068 (1964).

[146] Ľ. Hrivnák, "Relativistic analogies in direct-gap semiconductors," Prog. Quant. Electr. **17**, 235–271 (1993).

[147] W. Zawadzki and T. M. Rusin, "Zitterbewegung (trembling motion) of electrons in semiconductors: a review," J. Phys. Condens. Matter **23**, 143201 (2011), arXiv:1101.0623.

[148] W. Zawadzki, "Electron Dynamics in Crystalline Semiconductors," Acta Phys. Pol. A **123**, 132–138 (2013), arXiv:1209.3235.

[149] W. Franz, "Zur Theorie der inneren Feldemission aus dem Valenzband," Z. Naturforsch. A **14**, 415–418 (1959).

[150] E. I. Blount, "Bloch Electrons in a Magnetic Field," Phys. Rev. **126**, 1636–1653 (1962).

[151] B. A. Foreman, "Theory of the effective Hamiltonian for degenerate bands in an electric field," J. Phys. Condens. Matter **12**, R435–R461 (2000).

[152] J. M. Luttinger and W. Kohn, "Motion of Electrons and Holes in Perturbed Periodic Fields," Phys. Rev. **97**, 869–883 (1955).

[153] B. Gu, N. H. Kwong, and R. Binder, "Relation between the interband dipole and momentum matrix elements in semiconductors," Phys. Rev. B **87**, 125301 (2013).

[154] N. W. Ashcroft and N. D. Mermin, *Solid state physics* (Brooks/Cole, Cengage Learning, Belmont, CA, 2008).

[155] J. L. Smith, "Surface damage of GaAs from 0.694- and 1.06-$\mu$ laser radiation," J. Appl. Phys. **43**, 3399–3402 (1972).

[156] J. L. Smith, "Effects of Laser Flux on GaAs," in *Laser Induced Damage In Optical Materials: 1973*, edited by A. J. Glass and A. H. Guenther (National Bureau of Standards, Boulder, CO, 1973) pp. 103–106.

[157] J. L. Smith and G. A. Tanton, "Intense laser flux effects on GaAs," Appl. Phys. **4**, 313–315 (1974).

[158] M. N. Polyanskiy, "Refractive index database," `https://refractiveindex.info`, accessed: 2018-01-13.

[159] G. D. Mahan, "Many-Body Theory of Tunneling: Polarons in Schottky Junctions," in *Tunneling Phenomena in Solids*, edited by E. Burstein and S. Lundqvist (Springer, Boston, MA, 1969) Ch. 22, pp. 305–313.

[160] J. Zak, "Berry's phase for energy bands in solids," Phys. Rev. Lett. **62**, 2747–2750 (1989).

[161] D. Xiao, M.-C. Chang, and Q. Niu, "Berry phase effects on electronic properties," Rev. Mod. Phys. **82**, 1959–2007 (2010), arXiv:0907.2021.

[162] M. Gradhand, D. V. Fedorov, F. Pientka, P. Zahn, I. Mertig, and B. L. Györffy, "First-principle calculations of the Berry curvature of Bloch states for charge and spin transport of electrons," J. Phys. Condens. Matter **24**, 213202 (2012).

[163] W. Zawadzki, "On the $v^2/c^2$ expansion of the Dirac equation with external potentials," Am. J. Phys. **73**, 756–758 (2005), arXiv:quant-ph/0408065.